\documentclass[11pt,twoside,a4paper,pdf]{article}  
\usepackage{graphicx,color}
\usepackage{times}
\usepackage{xspace}
\usepackage{booktabs}
\usepackage[english]{babel}
\usepackage{subfig}
\usepackage{amssymb}
\usepackage{cite}
\usepackage{xspace}
\usepackage{a4wide}
\usepackage{amsmath,multicol}
\usepackage[toc]{multitoc}

\usepackage{afterpage}
\usepackage{multirow}
\usepackage{url}

\usepackage[colorlinks=true, pdftex, citecolor=red, pdftitle={},
  pdfauthor={P. Skands}, pdfsubject={Monte Carlo generators}, pdfkeywords={Monte Carlo, event generators, QCD, HEP, Rivet, HEPDATA, validation, tuning}]{hyperref}

\newcommand{\be}{\begin{equation}}
\newcommand{\ee}{\end{equation}}
\newcommand{\bea}{\begin{eqnarray}}
\newcommand{\eea}{\end{eqnarray}}
\newcommand{\bi}{\begin{itemize}}
\newcommand{\ei}{\end{itemize}}
\newcommand{\ben}{\begin{enumerate}}
\newcommand{\een}{\end{enumerate}}

\def\frac#1#2{{{#1}\over {#2}}}
\def\gsim{\mathrel{\rlap{\lower4pt\hbox{\hskip1pt$\sim$}}
    \raise1pt\hbox{$>$}}}         
\def\lsim{\mathrel{\rlap{\lower4pt\hbox{\hskip1pt$\sim$}}
    \raise1pt\hbox{$<$}}}         

\newcommand{\draft}[1]{}

\def\beq{\begin{equation}}  
\def\eeq{\end{equation}}  

\def \n0{N_j^{(0)}}

\def\lapprox{\lower .7ex\hbox{$\;\stackrel{\textstyle <}{\sim}\;$}}
\def\gapprox{\lower .7ex\hbox{$\;\stackrel{\textstyle >}{\sim}\;$}}

\newcommand{\mrm}[1]{\ensuremath{\mathrm{#1}}\xspace}
\newcommand{\ttt}[1]{\texttt{#1}}

\newcommand{\MeV}{\ensuremath{\,\mbox{MeV}}\xspace}
\newcommand{\GeV}{\ensuremath{\,\mbox{GeV}}\xspace}
\newcommand{\Ecm}{\ensuremath{E_\mrm{cm}}}

\newcommand{\appRef}[1]{appendix~\ref{#1}\xspace}
\newcommand{\AppRef}[1]{Appendix~\ref{#1}\xspace}
\newcommand{\eqRef}[1]{eq.~(\ref{#1})\xspace}
\newcommand{\EqRef}[1]{Eq.~(\ref{#1})\xspace}

\newcommand{\secRef}[1]{section~\ref{#1}\xspace}
\newcommand{\SecRef}[1]{Section~\ref{#1}\xspace}
\newcommand{\secsRef}[1]{sections~\ref{#1}\xspace}

\newcommand{\tabRef}[1]{tab.~\ref{#1}\xspace}

\newcommand{\figRef}[1]{fig.~\ref{#1}\xspace}

\newcommand{\tabsRef}[1]{tabs.~\ref{#1}\xspace}

\newcommand{\figsRef}[1]{figs.~\ref{#1}\xspace}

\renewcommand{\and}{, }

\newcommand{\cname}[1]{\MakeUppercase{#1}}

\newcommand{\Hw}{\cname{herwig}\xspace}

\newcommand{\Pw}{\cname{powheg}\xspace}
\newcommand{\Py}{\cname{pythia}\xspace}
\newcommand{\Pp}{\cname{pythia~8}\xspace}
\newcommand{\Sh}{\cname{sherpa}\xspace}
\newcommand{\Vc}{\cname{vincia}\xspace}

\newcommand{\mcplots}{\href{http://mcplots.cern.ch}{MCPLOTS}\xspace}

\newcommand{\dscale}{0.375}

\begin{document}

\vspace*{-1.75cm}\begin{minipage}{\textwidth}
\flushright\small
CERN-PH-TH-2014-069\\
MCNET-14-08\\
OUTP-14-05P \\
\end{minipage}
\vskip1.25cm
{\Large\bf
\begin{center}
Tuning PYTHIA 8.1: the Monash 2013 Tune
\end{center}}
\vskip5mm
{\begin{center}
{\large 
P.~Skands$^1$, S.~Carrazza$^2$, J.~Rojo$^{1,3}$
}\end{center}
$^1$:~\parbox[t]{0.985\textwidth}{Theoretical Physics, CERN, CH-1211,
Geneva 23, Switzerland}\\
$^2$:~\parbox[t]{0.985\textwidth}{Dipartimento di Fisica, Universit\`a
  di Milano and INFN, Sezione 
di Milano, Via Celoria 16, I-20133 Milano, Italy}\\
$^3$:~\parbox[t]{0.985\textwidth}{Rudolf Peierls Centre for
    Theoretical Physics, 1 Keble Road, University of Oxford, UK}\\[3mm]
\vskip5mm
\begin{center}
\parbox{0.88\textwidth}{
\begin{center}
\textbf{Abstract}
\end{center}\small
We present an updated set of parameters for the PYTHIA 8 event
generator. We reevaluate the constraints imposed by LEP and SLD on hadronization,
in particular with regard to heavy-quark fragmentation and strangeness
production. For hadron collisions, we combine the updated
fragmentation parameters with the new NNPDF2.3 LO PDF set. We
use minimum-bias, Drell-Yan, and underlying-event data from the LHC to
constrain the initial-state-radiation and multi-parton-interaction
parameters, combined with data 
from SPS and the Tevatron to constrain the energy scaling. Several
distributions show significant improvements with respect to the current
defaults, for both $ee$ and $pp$ collisions, though we emphasize that
interesting discrepancies remain in particular for
strange particles and baryons. The updated
parameters are available as an option starting from \Py 8.185, by
setting \texttt{Tune:ee = 7} and \texttt{Tune:pp = 14}.
}
\end{center}\vspace*{2mm}}


\section{Introduction \label{sec:intro}}

A truly impressive amount of results on QCD has been produced by the
first run of the LHC. Most of these are already 
available publicly, e.g.\ via the data preservation site 
HEPDATA~\cite{Buckley:2010jn}. A
large fraction has also been encoded in the analysis preservation tool
RIVET\footnote{In particular, RIVET ensures that
any (current or future) Monte Carlo event-generator codes can be
compared consistently to the data, with exactly the same cuts,
definitions, etc., as the original
analysis.}~\cite{Buckley:2010ar}. Especially in the area of  
soft QCD, many of the experimental results have spurred further
modelling efforts in the theory community (nice
summaries of some of the current challenges can be found in 
\cite{Sjostrand:2013cya,Sjostrand:2013sma}), while there is also 
significant activity dedicated to improving (``tuning'') the parameters of
the existing models to better describe some or all of the available
new data (see, e.g., the recent review in \cite{Katzy:2013lea}). 

The \Py\ event generator~\cite{Sjostrand:2006za,Sjostrand:2007gs} has
been extensively compared to LHC data, and several tuning efforts have
already incorporated data from Run
1~\cite{Skands:2010ak,Corke:2010yf,ATLAS:2010caa,ATLAS:2011gmi,ATLAS:2011zja,ATLAS:2012uec,Field:2012kd,AlcarazMaestre:2012vp,Firdous:2013noa,Katzy:2013lea}.  
However, in particular for the newest version of the model,
\Py~8~\cite{Sjostrand:2007gs}, it has been some time since the 
constraints imposed by $ee$ colliders were revised (in 2009), and then only via
an undocumented tuning effort (using the PROFESSOR
tool~\cite{Buckley:2009bj}). One of the main aims of this paper is
therefore first to take a critical look at the constraints arising from LEP,
SLD, and other $e^+e^-$ experiments, reoptimize the
final-state radiation and hadronization parameters, and 
document our findings. We do this manually, rather than in an
automated setup, in order to better explain the reasoning
behind each parameter adjustment.  This writeup
is thus also intended to function as an aid to others wishing to
explore the \Py~8 parameter space.

We then consider the corresponding case for hadron colliders, and use
the opportunity to try out a new PDF set, an LO fit produced by the
NNPDF collaboration~\cite{Ball:2011uy,Ball:2013hta,Carrazza:2013axa} which has
recently been introduced in \Py~8 (NLO and NNLO sets are also
available, for people that want to check the impact of using LO vs (N)NLO
PDFs in hard-scattering events). In a spirit 
similar to that of the so-called ``Perugia tunes'' of
\Py~6~\cite{Skands:2010ak,Cooper:2011gk}, we choose the same value of 
$\alpha_s(M_Z)=0.1365$ for both initial- and final-state
radiation. (Though we do regard this choice as somewhat arbitrary,
it may facilitate matching applications~\cite{Cooper:2011gk}.)
Again, we adjust  
parameters manually and attempt to give brief explanations for each
modification. We also choose the $\alpha_s(M_Z)$ value for
hard-scattering matrix elements to be the same as that in the PDFs,
here $\alpha_s(M_Z)=0.13$. (The difference between the value used for
radiation and that used for hard-scattering MEs may be interpreted as
an artifact of translations between the CMW and $\overline{\mrm{MS}}$
schemes, see \secRef{sec:isr}.) 

Below, in \SecRef{sec:plotLegend}, we begin by giving a brief general
explanation 
of the plots and $\chi^2$ values that are used throughout the paper.
Next, in \secRef{sec:hadronization}, we describe the physics, parameters, and
constraints governing fragmentation in hadronic $Z$ decays
(final-state radiation and string fragmentation). We turn to
hadron colliders in \secRef{sec:hadronColliders} (PDFs, initial-state
radiation, and multi-parton interactions). We then focus on the  
energy scaling between different $ee$ and $pp$ ($p\bar{p}$) collider
energies in \secRef{sec:energyScaling}, including in particular 
the recently published high-statistics data from the Tevatron 
energy scan from 300 to 1960 GeV~\cite{CDFnote10874,rickscan}. We
round off with  
conclusions and a summary of recommendations for future efforts in
\secRef{sec:proposals}.  

A complete listing of the Monash 2013 tune parameters is given in
\appRef{app:tunes}. \AppRef{app:plots} contains a few sets of additional
plots, complementing those presented in the main body of the paper.

\subsection{Plot Legends and $\mathbf{\chi^2}$ Values \label{sec:plotLegend}}
In several places, we have chosen to use data sets / constraints 
that differ from the standard ones available e.g.\ through
RIVET (as documented below). Since our tuning setup is furthermore
manual, rather than 
automated, we have in fact not relied on RIVET in this work (though we
have made extensive use of HEPDATA~\cite{Buckley:2010jn}). Instead,
we use the VINCIAROOT plotting tool~\cite{Giele:2011cb}, which we
have here upgraded to include a simple $\chi^2$ calculation, the
result of which is shown on each plot. 

Note that we include a blanket
5\% ``theory uncertainty'' in the definition of the $\chi^2$ value, 
representing a baseline sanity limit for the achievable accuracy of
the modeling\footnote{We note that a similar convention is used on the MCPLOTS
validation web site~\cite{Karneyeu:2013aha}.} that also gives a basic
protection against overfitting. Note 
also that, rather than letting the MC uncertainty enter in the definition of
the $\chi^2$ value (and thereby risking that low statistics generate
artificially low $\chi^2$ values), we use the generated MC statistics to
compute a $\pm$ uncertainty on the calculated $\chi^2$ value, which is
also shown on the plots. Our definition of $\chi^2$ is thus:
\begin{equation}
\left<\chi^2_{5\%}\right> \ = \ \frac{1}{N_\mrm{bins}}
\sum_{i=1}^{N_\mrm{bins}} \frac{(\mrm{MC}_i -
  \mrm{Data}_i)^2}{\sigma_{\mrm{Data},i}^2 + (0.05 \mrm{MC}_i)^2}~,
\label{eq:chi2}
\end{equation}
with the corresponding MC uncertainty, $\sigma_\mrm{MC,i}$, used to compute the
statistical uncertainty on the $\chi^2$ computation, as mentioned
above. As is shown 
here, the normalization  
is always 1/$N_\mrm{bins}$, regardless of whether the distributions
are normalized to a fixed number or not, and we do not attempt to take
into account correlations between the different observables. Since our
tuning is not directly driven by a $\chi^2$ minimization, we regard this as
acceptable; the $\chi^2_{5\%}$ values are intended merely to
give an overall 
indication of the level of agreement or disagreement for each
observable. 

\begin{figure}[t!p]
\centering
\includegraphics*[scale=\dscale]{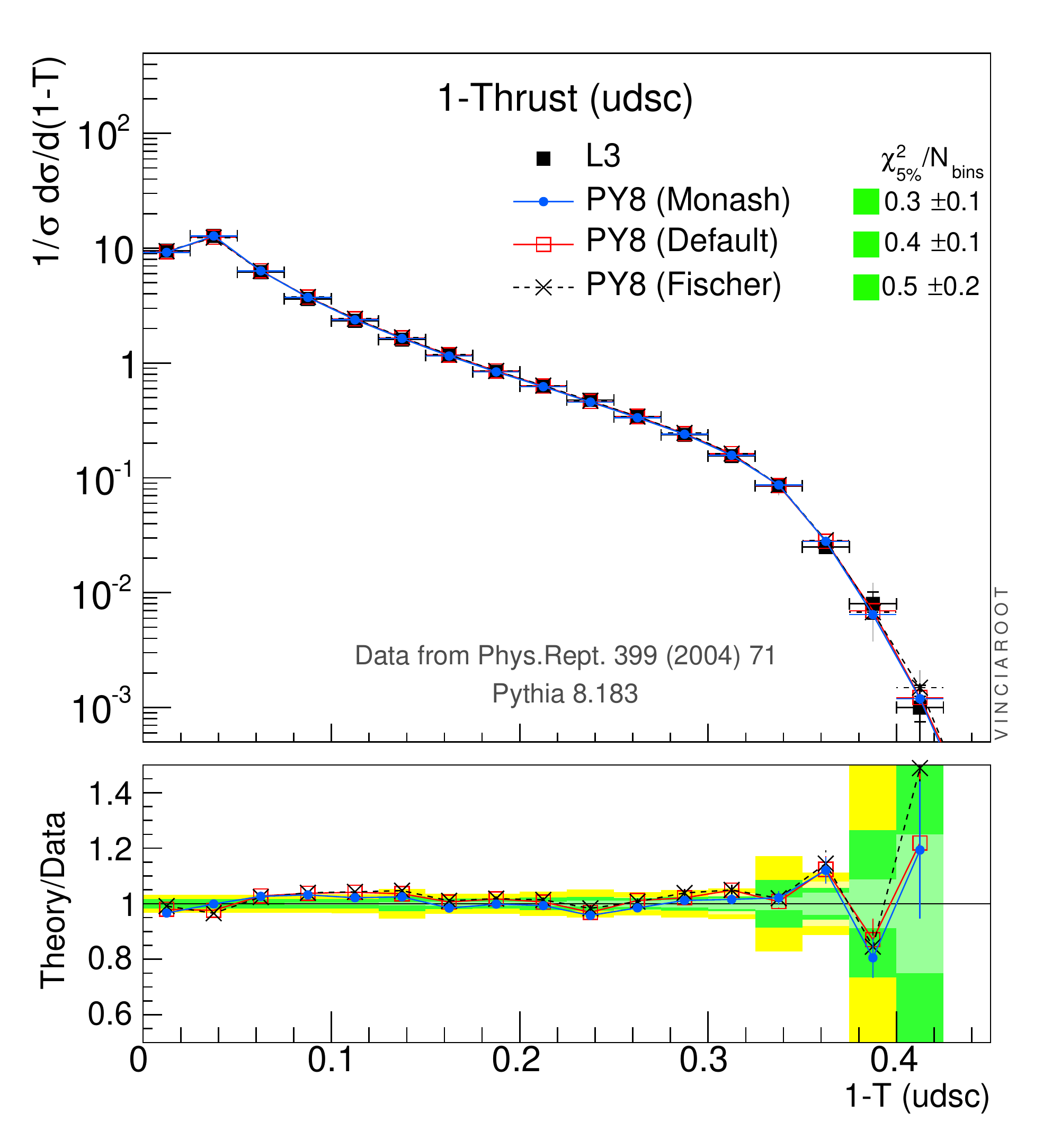}\vskip-3mm
\caption{Hadronic $Z$ decays at $\sqrt{s}=91.2\GeV$. The Thrust
  distribution in light-flavour tagged events, compared with L3 data~\cite{Achard:2004sv}.
\label{fig:thrust}}
\end{figure}

The resulting plots look as illustrated in \figRef{fig:thrust}, with a
main pane (top) showing the distribution itself and a bottom pane showing ratios. 
In the top pane, experimental data is always shown with filled black square
symbols, with vertical black lines indicating the one-sigma
uncertainties (with two separate black crossbars if separate statistical and
systematic uncertainties are given). Lighter (grey) extensions of the
vertical lines are used to indicate two-sigma uncertainties. In the
ratio pane, the green shaded region indicates the one-sigma
uncertainty region, while yellow is used to denote the two-sigma
one. An internal lighter/darker shading variation in each band is used
to denote the breakdown into statistical-only (inner) and
statistical+systematic uncertainties (outer), whenever separate values for
each of these are given. Finally, next to each MC legend the
$\chi^2_{5\%}$ value defined above is printed, along with its MC
uncertainty. A colour-coded box next to the $\chi^2$ value is shaded
green ($\chi^2<1$), yellow ($1<\chi^2<4$), orange ($4<\chi^2<9$), or
red ($9<\chi^2$), depending on the level of agreement or
disagreement. This functionality will be included in a forthcoming
update of the \Vc\ plug-in to \Py~8.

\section[FSR and Hadronization]{Final-State Radiation and Hadronization \label{sec:hadronization}}

The main parameter governing final-state radiation is the effective
value of the strong coupling, which in \Py~8 is specified by giving
the value of $\alpha_s(M_Z)$. 
We follow the strategy of~\cite{Giele:2011cb} and use a  
set of light-flavour ($udsc$) tagged $e^+e^-$ event shapes provided by the L3
experiment~\cite{Achard:2004sv} to extract a best-fit value for
$\alpha_s(M_Z)$. (This
prevents $B$ decays from contaminating this step of the analysis. Heavy-quark
fragmentation will be treated 
separately, below.) The renormalization scale for final-state shower
emissions in \Py\ is fixed to be~\cite{Sjostrand:2004ef}:
\begin{equation}
\mbox{FSR:~~~}\mu_R^2 = p_{\perp\mrm{evol}}^2 = z(1-z)Q^2~,
\end{equation}
with $Q^2=p^2 - m_0^2$ the offshellness of the emitting parton (with
on-shell mass $m_0$), and $z$ the energy fraction appearing in the
DGLAP splitting kernels, $P(z)$. (To 
estimate the shower uncertainties associated with this choice of
renormalization scale, we recommend using $\ln(\mu_R^2) \pm \ln(2)$,
corresponding to a factor $\sqrt{2}$ variation of $\mu_R$.) 

Theoretically, a set of formally subleading terms can be resummed by 
using 2-loop running of $\alpha_s$ in the so-called MC (a.k.a.\ CMW)
scheme~\cite{Catani:1990rr}. However, in a leading-order code like
\Py, this produces too little hard radiation in practice, due to
missing NLO ``K'' factors for hard emissions (see, e.g., the study of
NLO corrections 
in~\cite{Hartgring:2013jma}).   
Empirically, we find that a better overall description is achieved with
one-loop running, which, for a fixed value of $\Lambda_\mrm{QCD}$, 
can effectively mimic the effect of missing $K$ factors via its
relatively slower pace of running, leading to values of $\alpha_s(M_Z)$ in
the range $0.135 - 0.140$, consistent with other LO
extractions of the same quantity. (See~\cite{Hartgring:2013jma} for
an equivalent extraction at NLO.)

For this study, we did not find any
significant advantage in reinterpreting this value in the CMW
scheme\footnote{One slight \emph{disadvantage} is that the CMW scheme
  produces somewhat larger $\Lambda_\mrm{QCD}$ values. Since the
  current formulation of the shower algorithm does not include 
  a non-perturbative regularization of $\alpha_s$, a higher
  $\Lambda_\mrm{QCD}$ value necessitates a larger IR cutoff in the
  shower, which can leave an undesirable gap between the
  transverse kicks generated by shower emissions and those generated
  by non-perturbative string splittings.} and hence merely settled on an
effective $\alpha_s(M_Z) = 0.1365$ (to be compared with the current
default value of $0.1383$).

For the infrared shower cutoff, we choose a value close
to\footnote{The IR shower cutoff must still remain somewhat above the
  Landau pole of $\alpha_s$; a lower cutoff scale would activate a
  hardcoded protection mechanism implemented in the \Py\ shower,
  forcing it to be higher than $\Lambda_\mrm{QCD}$.}
$\Lambda_\mrm{QCD}$, in order
to have a  smooth transition between 
low-$p_\perp$ perturbative emissions and non-perturbative string
breaks, the latter of which involve $p_\perp$ kicks of order
$\Lambda_\mrm{QCD}$. (In principle, the perturbative evolution could
be continued to even lower scales, if combined with a non-perturbative
regularization of $\alpha_s$, but such low cutoff values could risk
generating problems at the fragmentation stage 
since the technical implementation of the 
string model becomes complicated if there are too many small
gluon ``kinks'' spaced closely along the strings.)
The set of relevant parameters in the code is:

\noindent{\small\begin{verbatim}
# FSR: Strong Coupling
  TimeShower:alphaSvalue  = 0.1365
  TimeShower:alphaSorder  = 1
  TimeShower:alphaSuseCMW = off
# FSR: IR cutoff
  TimeShower:pTmin        = 0.50   ! for QCD radiation
  TimeShower:pTminChgQ    = 0.50   ! for QED radiation off quarks
# FSR: Spin Correlations
  TimeShower:phiPolAsym   = on     ! approximate FSR polarization effects
\end{verbatim}
}

The resulting distribution of the Thrust event-shape variable was shown
in \figRef{fig:thrust}, comparing the Monash 2013 tune
to the current default tune and to an alternative contemporary tune by
N.~Fi\-scher~\cite{Fischer:2014bja}. To avoid clutter, the other event-shape
variables ($C$, $D$, $B_W$, and $B_T$) are collected in
\appRef{app:LEP}. 
There are no significant changes to any of the light-flavour tagged
event shapes in our tune as compared to the current default one. 

\subsection{Light-Flavour Fragmentation \label{sec:light}}

Given a set of post-shower partons, resolved at a scale of
$Q_\mathrm{had}\sim$ 1 GeV, the non-perturbative stage of the
fragmentation modeling now takes over, to convert the partonic
state into a set of on-shell hadrons. In the 
leading-colour approximation, each perturbative
dipole is dual to a non-perturbative string piece~\cite{gustafson86}. 
Quarks thus become string endpoints, while gluons become transverse
kinks, connecting two string pieces~\cite{andersson80}. The Lund
string fragmentation model~\cite{Andersson:1998tv} describes the
fragmentation of such string systems into on-shell hadrons. 

Since the shower has already resolved all the
(perturbative) physics down to a transverse-momentum 
scale of $p_{T\mrm{min}} = 0.5$ GeV
(for the Monash 2013 tune), we find it reasonable that the
$p_\perp$ kicks involved in string breaking should effectively average
over dynamics in roughly the range $250~\mrm{MeV}=\sqrt{\kappa/\pi} < \sigma_\perp <
p_{T\mrm{min}}$, with the lower bound given by Fermi motion (with
$\kappa$ the string tension,
see~\cite{oai:arXiv.org:1101.2599}). Further, since we here choose 
$p_{T\mrm{min}}$ to be only slightly greater than $\Lambda_\mrm{QCD}$,
  the size of the non-perturbative corrections is naturally 
  limited to kicks/corrections appropriate for non-perturbative
  dynamics (in contrast, e.g., to the 
  cluster model~\cite{webber84}, which can generate substantially
  larger kicks, of 
  order the largest allowed cluster mass, which can be several
  GeV~\cite{Fischer:2014bja}).  
For the Monash 2013 tune, we have settled on a value
of $\sigma_\perp = 0.335$ GeV, with a small (1\%) tail of breaks
involving higher $p_\perp$ values carried over from the default
settings.  

\noindent{\small\begin{verbatim}
  StringPT:sigma            = 0.335
  StringPT:enhancedFraction = 0.01
  StringPT:enhancedWidth    = 2.0
\end{verbatim}
}

\noindent This value is obtained essentially from the first two bins
of the Thrust distribution, \figRef{fig:thrust}, and from the bins
near zero of the other event shapes, see
\appRef{app:LEP}. 
Note that the $\sigma_\perp$ value is interpreted as 
the width of a Gaussian distribution in the total $p_\perp$ (measured
transversely to the local string direction, which may differ from the
global event axis), such that each of the $p_x$ and $p_y$ components
have a slightly smaller average value, $\sigma_{x,y}^2 =
\frac12\sigma_\perp^2 = (0.237\,\mrm{GeV})^2$. 
Also
note that each non-leading hadron will receive two $p_\perp$ kicks,
one from each of the breaks surrounding it, hence
$\left<p_{\perp\mrm{had}}^2\right> = 2 \sigma_\perp^2 =
(0.474\,\mrm{GeV})^2$.  

For massless quarks, the longitudinal component of the 
energy carried by a hadron formed in the string-breaking process
$\mrm{string} \to \mrm{hadron} + \mrm{string'}$ is governed by the
Lund symmetric fragmentation function:
\begin{equation}
f(z) \propto \frac{z^{(a_i-a_j)}(1-z)^{a_j}}{z}
\exp\left(\frac{-b m_\perp^2}{z}\right)~,
\label{eq:fMassless}
\end{equation}
where $z$ is the energy carried by the newly formed $(ij)$ hadron,
expressed as a fraction of the (lightcone) energy of the quark (or
antiquark) endpoint, $i$, of the fragmenting string. (The
remaining energy fraction, $(1-z)$, goes to the new 
$\mrm{string'}$ system, from which another hadron can be split off in the
same manner, etc., until all the energy is used up.) The transverse
mass of the produced $(ij)$ hadron is defined by 
$m_\perp^2 = m^2_\mrm{had} + p_{\perp,\mrm{had}}^2$, hence heavier
hadrons have harder spectra.  
The proportionality sign in \eqRef{eq:fMassless} 
indicates that the function is to be
normalized to unity. 

The $a$ and $b$ parameters govern the shape of
the fragmentation function, and must be constrained by fits to
data. \EqRef{eq:fMassless} expresses the most general form of the fragmentation
function, for which the $a$ parameters of the original 
string-endpoint quark, $a_i$,
and that of the (anti-)quark produced in the string break, $a_j$, can
in principle be different, while the $b$ parameter is universal. 
Within the Lund model, the $a$ value is
normally also taken to be universal, the same for all quarks, with the only
freedom being that a larger $a$ parameter can be assigned to
diquarks~\cite{andersson81}, 
from which baryons are formed, and hence meson and baryon spectra can
be decoupled somewhat.  (See \texttt{StringZ:aExtraDiquark} below.)

Roughly speaking, large $a$ parameters 
suppress the hard region $z\to 1$, while a large $b$ parameter
suppresses the soft region $z\to 0$. By adjusting them independently, 
both the average hardness and the width  of the resulting
fragmentation spectra can be modified. For example,   
increasing both $a$ and $b$ yields a narrower distribution,
while changing them in opposite directions moves the average.
\begin{figure}[t!p]
\centering
\begin{tabular}{ccc}
\underline{The $a$ parameter} & & \underline{The $b$ parameter}\\[2mm]
 \hspace*{1.8cm}$\color{red}a=0.9$ \hspace*{4mm}
$\color{blue}a=0.1$
 & & \hspace*{1cm}$\color{red}b=0.5$ \hspace*{7mm} $\color{blue}b=2.0$ 
\\
\includegraphics*[scale=1]{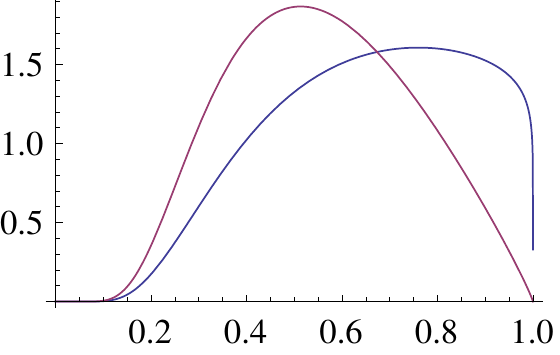} & \hspace*{1cm} &
\includegraphics*[scale=1]{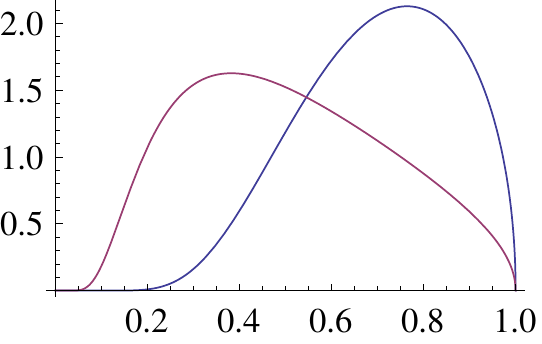}\\
$b=1\GeV^{-2}$, $m_\perp=1\GeV$ & &
$a=0.5$, $m_\perp=1\GeV$ 
\end{tabular}
\caption{Illustration of the Lund symmetric fragmentation
function (normalized to unity), for $a_i=a_j\equiv a$. 
{\sl Left:} variation of the $a$ parameter, from 0.1 (blue)
to 0.9 (red), with fixed $b$. {\sl Right:}
variation of the $b$ parameter, from 0.5 (red) to 2 (blue) GeV$^{-2}$, 
with fixed $a$. \label{fig:LundFF}}
\end{figure}
An illustration of the effect of
varying the $a$ 
and $b$ parameters, for $a_i=a_j\equiv a$, is given in \figRef{fig:LundFF}; see
also the lecture notes in~\cite{Skands:2012ts}.  Note that the
$\sigma_\perp$ parameter 
also affects the hardness, with larger $\sigma_\perp$ values
generating harder hadrons, the difference being that the
$\sigma_\perp$ parameter acts mainly in the direction transverse to the
string\footnote{Explicitly, $\sigma_\perp$ expresses the $p_\perp$
  broadening transverse to the string direction, but implicitly its
  size also enters in the logitudinal fragmentation function, via the
  $m_\perp^2$ term in \eqRef{eq:fMassless}, causing higher-$p_\perp$
  hadrons to have relatively harder longitudinal spectra as well.}
  (and is an absolute scale expressed in GeV), while the
$a$ and $b$ parameters act longitudinally (with $z$ a relative scale
expressed as a fraction of the endpoint's energy). 

In the context of this work, we 
included the possibility of letting the $a$
parameter for strange quarks be slightly different from that of $u$
and $d$ quarks, but did not find any significant
advantages. The relevant parameters in the code we settled on for the
Monash tune are: 

\noindent{\small\begin{verbatim}
  StringZ:aLund         =  0.68
  StringZ:bLund         =  0.98
  StringZ:aExtraDiquark =  0.97
  StringZ:aExtraSquark  =  0.00
\end{verbatim}
}

\begin{figure}[t!p]
\centering
\begin{tabular}{ccc}
\includegraphics*[scale=\dscale]{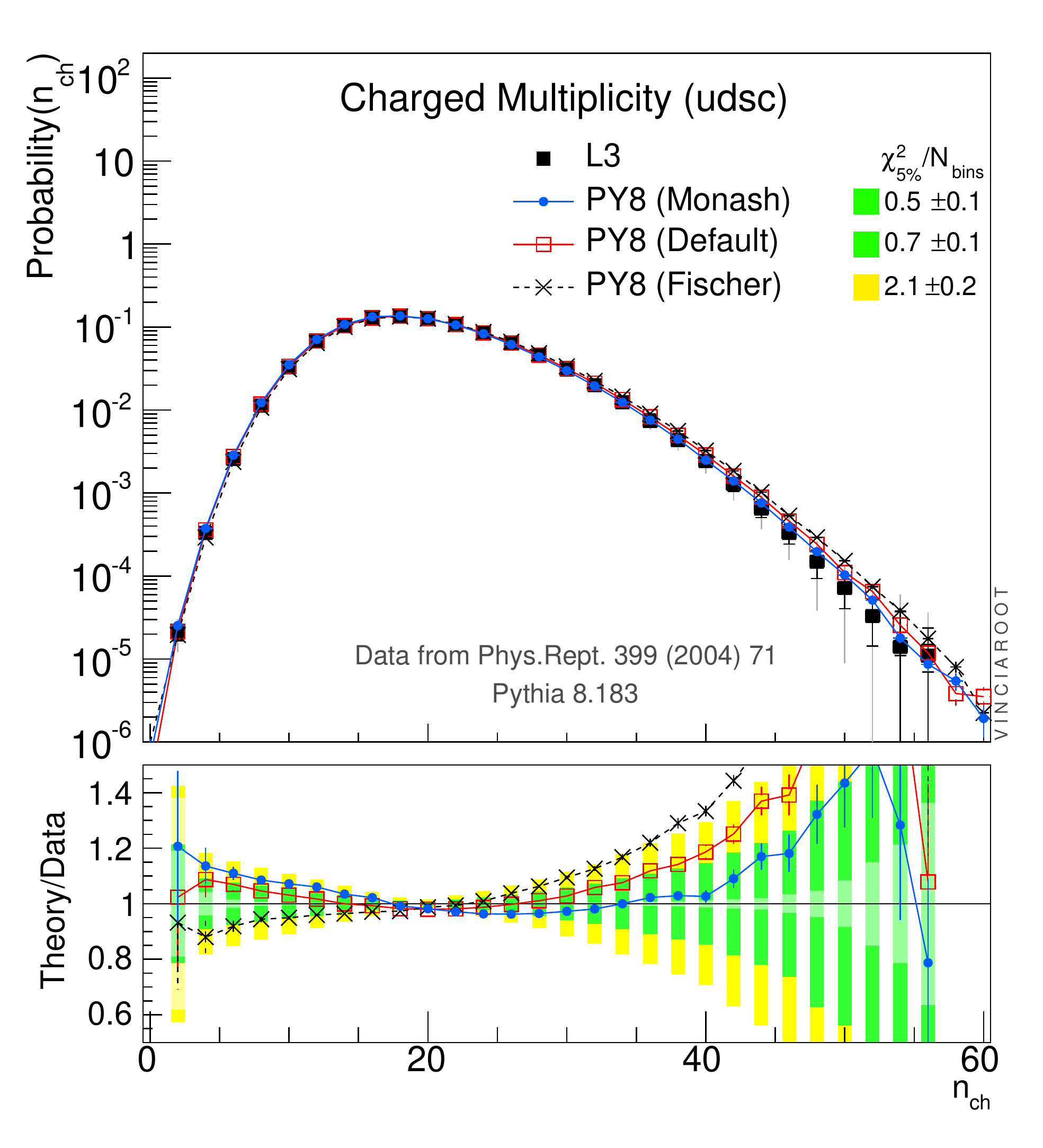}
\includegraphics*[scale=\dscale]{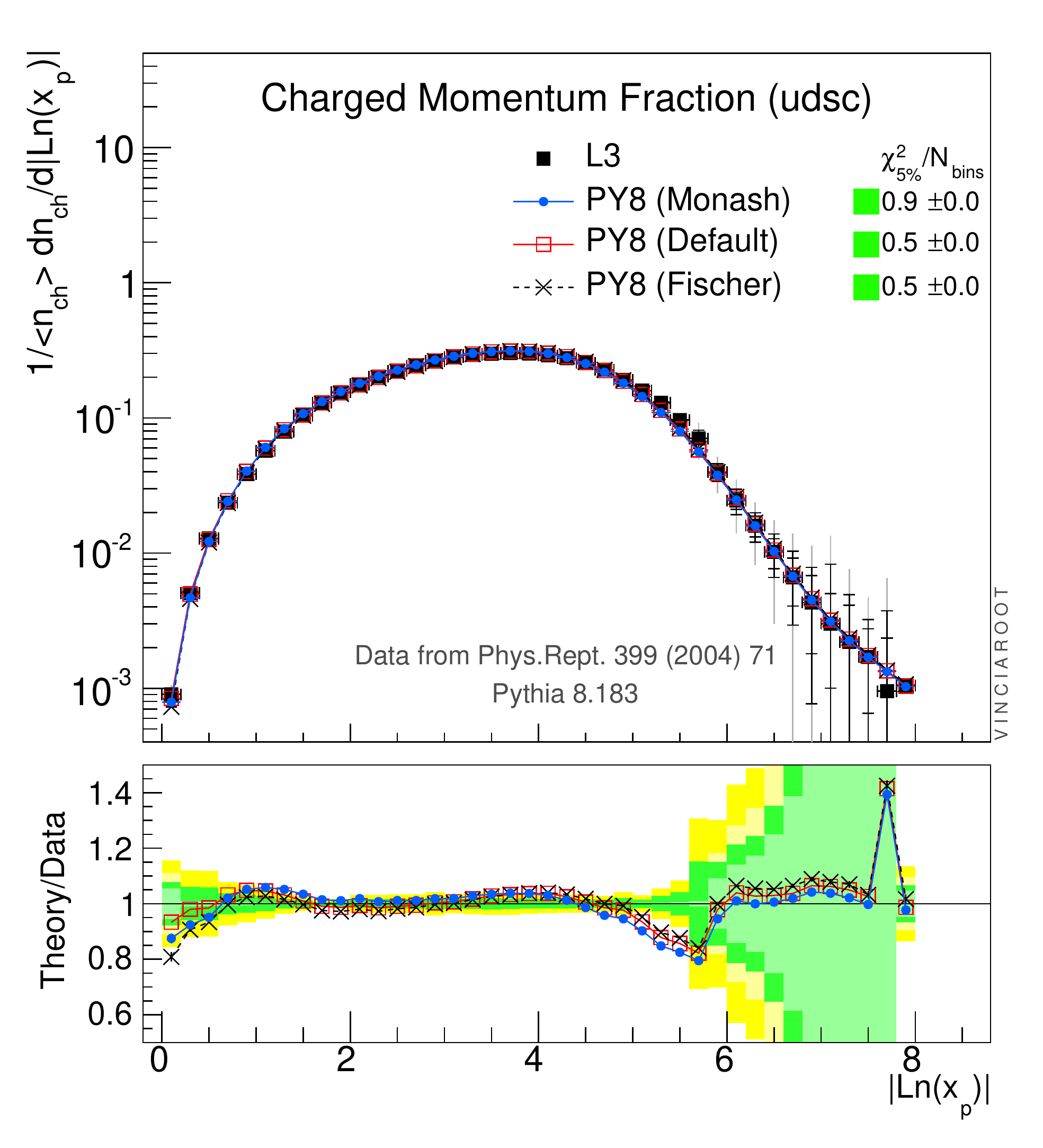}
\end{tabular}\vskip-3mm
\caption{Hadronic $Z$ decays at $\sqrt{s}=91.2\GeV$.
Charged-particle multiplicity (left) and momentum-fraction
  (right) spectra.
\label{fig:multLight}
}
\end{figure}

The average hardness of the produced hadrons 
is tightly (anti-)correlated with the average
multiplicity, via momentum conservation: if each hadron takes a lot of
energy, then fewer hadrons 
must be made, and vice versa. Thus, the $\sigma_\perp$ value and the 
$a$ and $b$ parameters of the
fragmentation function can be well constrained by simultaneously considering
both momentum and multiplicity spectra. In order to be as universal as
possible, one normally uses the inclusive charged-particle
spectra for this purpose. 
These are shown in \figRef{fig:multLight}. 
(Note: the Fischer tune only included the average
  particle multiplicity as a constraint, so the full $n_{\mrm{ch}}$
  distribution is not expected to be reproduced
  perfectly~\cite{Fischer:2014bja}.) The momentum fraction in the right-hand
plot is defined by:
\begin{equation}
x_p = \frac{2|p|}{\Ecm}~.
\end{equation}
As above, the experimental data come from a 
measurement by L3~\cite{Achard:2004sv} which only includes the four
lightest flavours, thus excluding $b$ quarks (which will be treated
separately below). 

Both of the earlier tunes exhibit a somewhat too broad multiplicity
distribution in comparison with the L3 data. The relatively large Lund
$a$ and $b$ values used for the Monash 
tune, combined with its large $\sigma_\perp$ value, 
produce a narrower $n_\mrm{Ch}$ spectrum, with in particular a smaller
tail towards large multiplicities.
All the tunes produce a sensible momentum spectrum. The dip around
$\left\vert\ln (x)\right\vert \sim 5.5$ corresponds to the extreme soft-pion tail, 
with momenta at or below $\Lambda_\mrm{QCD}$. We did not find it
possible to remove it by retuning, since a smaller $b$ parameter would
generate significantly too high particle multiplicities and a smaller
$\sigma_\perp$ would lead to conflict with the event-shape distributions. 

\begin{figure}[t!p]
\centering
\begin{tabular}{ccc}
\includegraphics*[scale=\dscale]{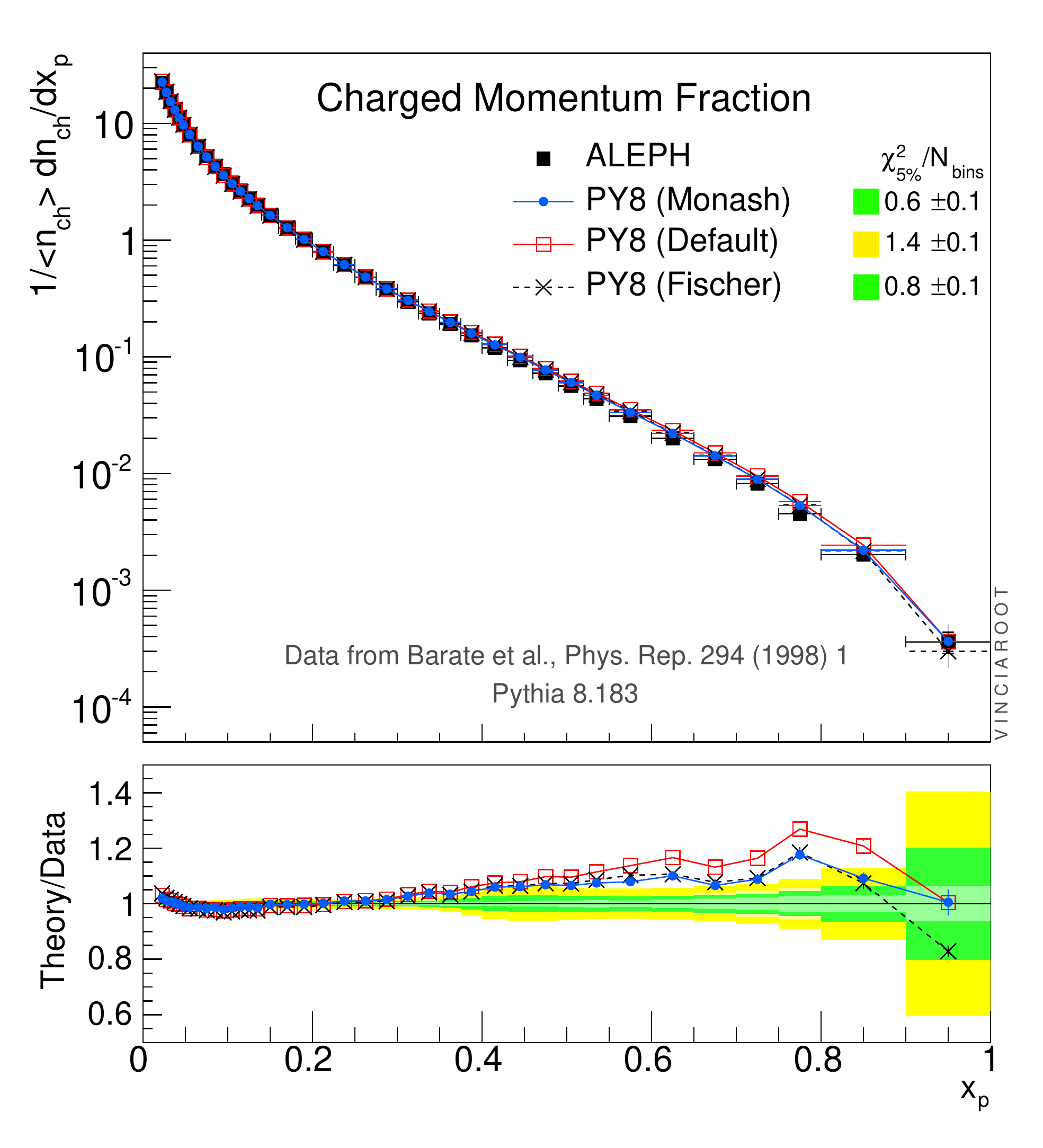}
\includegraphics*[scale=\dscale]{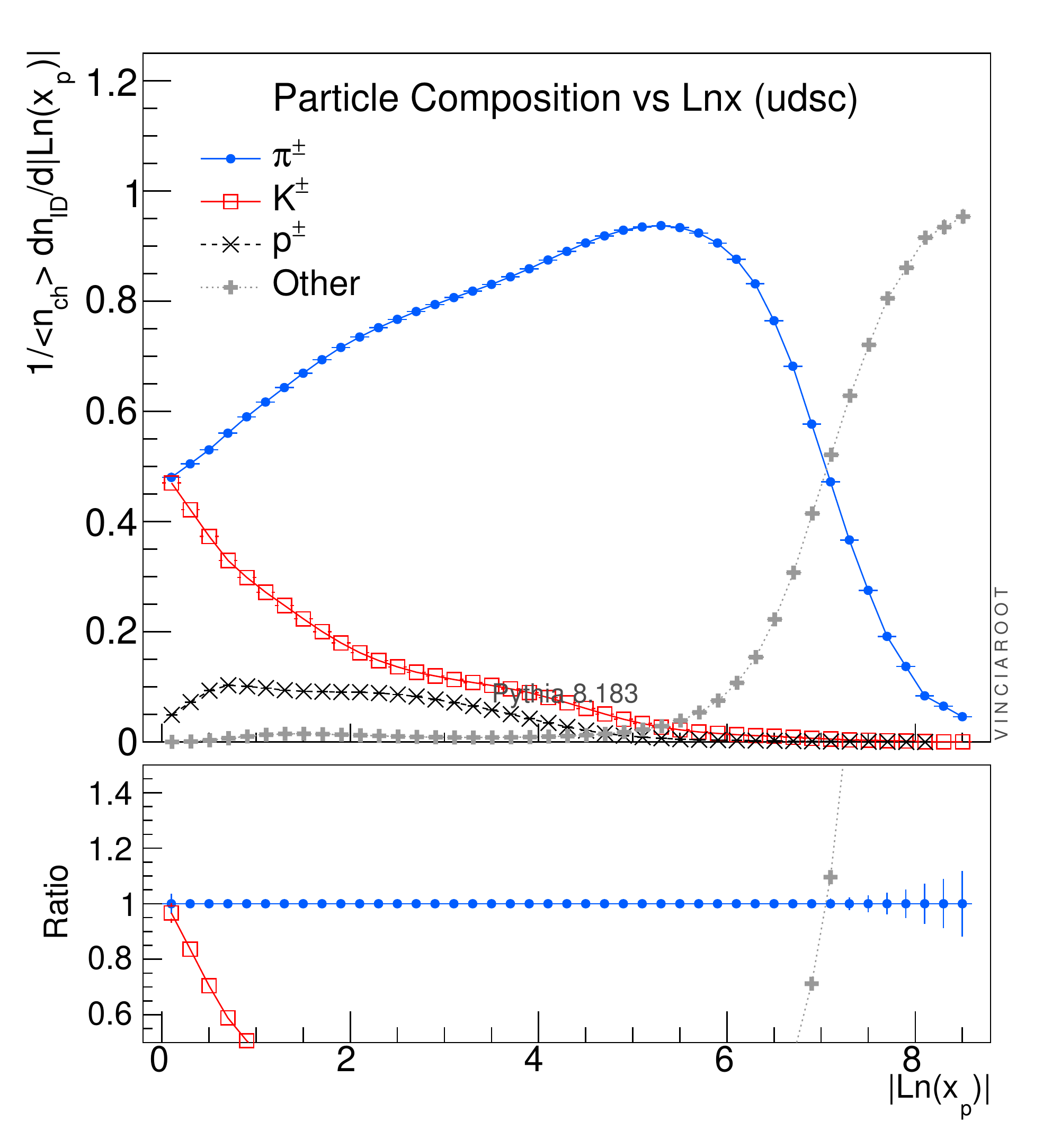}
\end{tabular}\vskip-3mm
\caption{Hadronic $Z$ decays at $\sqrt{s}=91.2\GeV$.
Charged-particle momentum fraction $x_p$, on a linear scale 
(left) and relative particle composition (right) 
for the log-scale distribution shown in \figRef{fig:multLight}. 
\label{fig:xLight}
}
\end{figure}

A zoom on the high-momentum tail is provided by the left-hand plot 
in \figRef{fig:xLight}, which shows a comparison on a linear momentum
scale, to a measurement 
by ALEPH~\cite{Barate:1996fi} (now including $Z\to b\bar{b}$ events as well as
light-flavour ones). 
All the tunes exhibit a mild 
overshooting of the data in the region $0.5<x_p<0.8$, corresponding to
$0.15<|\ln(x)|<0.7$, in which no similar excess was present in the L3
comparison. We therefore do not regard this as a significant
issue\footnote{One might worry whether the effect could be due solely to the
  $Z\to b\bar{b}$ events which are only present in the ALEPH
  measurement, and if so, whether this could indicate a
  significant mismodeling of the momentum distribution
  in $b$ events. 
However, as we show below in the section 
  on $b$ fragmentation, the charged-particle momentum distribution in
  $b$-tagged events shows no excess in that region 
  (in fact, it shows an undershooting).}
but note that the excess is 
somewhat milder in the Fischer and Monash tunes. 

Further information
to elucidate the structure of the momentum distribution is provided by 
the plot in the right-hand pane of \figRef{fig:xLight}, which 
uses the same $\left\vert\ln (x)\right\vert$ axis as the right-hand plot in
\figRef{fig:multLight} and shows the relative particle composition in
the Monash tune for each histogram bin. (The category ``Other''
contains electrons and muons from weak decays.)
An interesting observation is that the relatively harder spectrum of
Kaons implies that, for the highest-momentum bins, the
charged tracks 
are made up of an almost exactly equal mixture of Kaons and pions,
despite Kaons on average only making up about 10\% of the charged
multiplicity.

\subsection{Identified Particles \label{sec:idLEP}} 

Continuing on the topic of identified particles, we note that 
the extraction of the $a$ and $b$ parameters from the inclusive
charged-particle distributions is made slightly more complicated by the
fact that not all observed particles are ``primary'' (originating
directly from string breaks); many lower-mass particles are
``secondaries'', 
produced by prompt decays of more massive states (e.g., $\rho \to
\pi\pi$), whose relative rates and decay kinematics therefore
influence the spectra. In the $e^+e^-$ measurements we include here, particles
with $c\tau < 100\,\mbox{mm}$ were treated as unstable, hence leading
to secondaries. (For completeness, we note that the equivalent
standard cut at the LHC is normally $10\,\mbox{mm}$.)

The particle composition in \Pp\ was already tuned to a set of reference
values provided by the PDG~\cite{Beringer:1900zz}, and the default
parameters do reasonably well, certainly for the most copiously
produced sources of 
secondaries. Nonetheless, we have here reoptimized the flavour-selection
parameters of the string-fragmentation model using a slightly
different set of reference data, combining the PDG tables 
with information provided directly by the LEP experiments via
HEPDATA~\cite{Buckley:2010jn}. 
Based on the level of agreement or disagreement between
different measurements of the same particles, we have made our 
own judgement as to the level of uncertainty for a few of the
particles, 
as follows. (Unless otherwise stated, we use the value from the PDG. 
Particles and antiparticles are implicitly summed over, and secondaries from 
particles with $c\tau < 100\,\mrm{mm}$ are included.)  

\begin{itemize}
\item The various LEP and SLD measurements of the $\phi$ meson rate on
  HEPDATA are barely compatible. E.g., OPAL~\cite{Ackerstaff:1998ue}
  reports $\left<n_{\phi}\right> = 0.091 \pm 0.002 \pm 0.003$ while
  ALEPH~\cite{Barate:1996fi} quotes $\left<n_{\phi}\right> = 0.122 \pm
  0.004 \pm 0.008$, a difference of 30\% with uncertainties supposedly
  less than 10\%. DELPHI~\cite{Abreu:1996sn} and SLD~\cite{Abe:1998zs}
  fall in between. The PDG value is  $\left<n_{\phi}\right> = 0.0963
  \pm 0.003$, i.e., with a combined uncertainty of just
  3\%. We choose to inflate the systematic uncertainties and arrive at
  $\left<n_{\phi}\right> = 0.101 \pm 0.007$. 
\item  For $\Lambda$ production, we use the most 
precise of the LEP measurements, by OPAL\footnote{We note
  that HEPDATA incorrectly gives the systematic error as $0.002$ while
  the value in the OPAL paper is
  $0.010$~\cite{Alexander:1996qj}. This has been communicated to the
  HEPDATA maintainers.}~\cite{Alexander:1996qj},  
$\left<n_\Lambda\right> =
0.374\pm0.002\pm0.010$, about 5\% lower than 
the corresponding PDG value. 
\item For $\Sigma^\pm$ baryons,
we use a combination of the two most recent LEP measurements, by
L3~\cite{Acciarri:2000zf} for $\Sigma^+ +
\overline{\Sigma}^-$ and by DELPHI~\cite{Abreu:2000nu} 
for $\Sigma^- + \overline{\Sigma}^+$, for an estimated
$\left<n_{\Sigma^\pm}\right> = 0.195 \pm 0.018$, which is roughly 10\%
  higher than the PDG value. 
\item For $\Sigma^0$ baryons, we use the
 most recent measurement, by L3~\cite{Acciarri:2000zf}, 
$\left<n_{\Sigma^0}\right> = 0.095
\pm 0.015 \pm 0.013$; this is about 20\% larger than the PDG
value. The L3 paper comments on their relatively high value by 
noting that L3 had the best coverage for
low-momentum baryons, hence smaller model-dependent 
correction factors. 
\item For $\Delta^{++}$ baryons, there are only two
measurements in HEPDATA~\cite{Alexander:1995gq,Abreu:1995we}, 
which are mutually discrepant by about $2\sigma$. The DELPHI
measurement is nominally the most precise, but OPAL gives a
much more serious discussion of systematic uncertainties. 
We choose to increase the estimated extrapolation errors of the 
DELPHI measurement
by 50\% and obtain a weighted average\footnote{Even with the inflated
  error, the uncertainty on the DELPHI measurement is still less than
  a third that of the OPAL one. DELPHI therefore still dominates the average.}
  of $\left<n_{\Delta^{++}}\right>
= 0.09 \pm 0.017$, 5\% larger than the PDG value, with a
20\% larger uncertainty. 
\item For $\Sigma^*$, the three measurements on
HEPDATA~\cite{Barate:1996fi,Abreu:1995qx,Alexander:1996qj}  
are likewise discrepant by
$2\sigma-3\sigma$. We inflate the systematic uncertainties and arrive 
at $\left<n_{\Sigma^{*\pm}}\right> = 0.050 \pm 0.006$, which is 
again 5\% higher than the PDG value, with twice as much
uncertainty. 
\item The measurements for $\Xi^\pm$ are in good
agreement~\cite{Abreu:1995qx,Alexander:1996qj,Barate:1996fi}, 
with a weighted average of $\left<n_{\Xi^\pm}\right> = 0.0266 \pm
0.0012$, slightly larger than the PDG value.
\item For $\Xi^{*0}$, however,
the DELPHI measurement~\cite{Abreu:1995qx} gives a far lower number than the
OPAL~\cite{Alexander:1996qj} and ALEPH~\cite{Barate:1996fi} ones, and
the weighted average 
differs by more than 
10\% from the PDG value, despite the latter claiming an uncertainty
smaller than 10\%. Our weighted average is $\left<n_{\Xi^{*0}}\right>
= 0.0059\pm 0.0012$. 
\item Finally, for the $\Omega$ baryon, the
DELPHI~\cite{Adam:1996hw} and
OPAL~\cite{Alexander:1996qj} measurements are in agreement, and we use
the PDG value, 
$\left<n_{\Omega}\right> = 0.0016 \pm 0.0003$. 
\end{itemize} 
We summarize the constraints on the light-meson and baryon rates used
here in \tabRef{tab:idparticles}. 
Note that we express them as percentages of the average charged
multiplicity,
\begin{equation}
\left<n_{\mrm{Ch}}\right> = 20.7~,
\end{equation}
obtained as a weighted average over MARK-II~\cite{Abrams:1989rz}, 
ALEPH~\cite{Barate:1996fi}, DELPHI~\cite{Abreu:1998vq},
OPAL~\cite{Ackerstaff:1998hz}, 
and L3~\cite{Adeva:1992gv} measurements.

\begin{table}[t!p]
\centering\small
\begin{tabular}{llll}
\toprule
\mbox{\bf Mesons}
 & \multicolumn{2}{l}{Our Reference} & Our \\
$\left<n\right>/\left<n_{\mrm{Ch}}\right>$ & \multicolumn{2}{l}{Value (in \%)}
& Source
\\\midrule
   $\pi^{+}+\pi{-}$ &  $82.2~$ &$ \pm 0.9$ & P
\\ $\pi^0$    & $45.5~$  &$\pm 1.5$   & P 
\\ $K^++K^-$    & $10.8~$   &$\pm 0.3$ & P
\\ $\eta$    & $~5.06$   &$\pm 0.38$ & P
\\ $\eta'$   &  $~0.73$  &$\pm0.09$  & P
\\ $\rho^++\rho^-$ & $11.6~$  &$\pm 2.1$ & P
\\ $\rho^0$  &  $~5.95$   &$\pm 0.47$ & P 
\\ $K^{*+}+K^{*-}$ & $~3.45$  &$\pm 0.28$ &P 
\\ $\omega$  & $~4.90$   &$\pm 0.31$ & P 
\\ $\phi$    & $~0.49$   &$\pm 0.035$ & ADOS
\\ \bottomrule
\end{tabular}
\begin{tabular}{llll}
\toprule
\mbox{\bf Baryons} 
  & \multicolumn{2}{l}{Our Reference} & Our \\ 
$\left<n\right>/\left<n_{\mrm{Ch}}\right>$ &
\multicolumn{2}{l}{Value (in \%)} & Source
\\\midrule
   $p+\bar{p}$ &   $5.07$ & $\pm 0.16$ & P
\\ $\Lambda + \bar{\Lambda}$    &$1.81$ & $\pm 0.32$ & O
\\ $\Sigma^+ + \Sigma^- + \bar{\Sigma}^+ + \bar{\Sigma}^-$ & $0.942$ &
$\pm0.087$ & DL
\\ $\Sigma^0 + \bar{\Sigma}^0$  & $0.459$ & $\pm 0.096$ & L
\\ $\Delta^{++} + \bar{\Delta}^{--}$   & $0.434$ & $\pm 0.082$ & DO
\\ $\Sigma^{*+} + \Sigma^{*-} + \bar{\Sigma}^{*+} + \bar{\Sigma}^{*-}$ 
     & $0.242$ & $\pm 0.029$& ADO
\\ $\Xi^{+} + \bar{\Xi}^{-}$  & $0.125$ & $\pm 0.0050$ & ADO
\\ $\Xi^{*0} + \bar{\Xi}^{*0}$ & $0.0285$ & $\pm 0.0058$ & ADO
\\ $\Omega^- + \bar{\Omega}^+$  & $0.0077$ & $\pm 0.0015$ & P
\\
\\ \bottomrule
\end{tabular}
\caption{Hadronic $Z$ decays at $\sqrt{s}=91.2\GeV$.
Measured rates of light-flavour mesons and baryons, 
expressed as percentages of the
  average charged-particle multiplicity, as
  used in this work. Multiply the numbers
 by 20.7/100 to translate
  the percentages to corresponding production rates.
Source labels indicate: A (ALEPH), D (DELPHI), L (L3), O (OPAL), S
(SLD), P (PDG). 
\label{tab:idparticles}
}
\end{table} 

The light-flavour-selection parameters for the Monash tune are (see
\appRef{app:tunes} for a comparison of these values to the current
default ones):

\noindent{\small\begin{verbatim}
# Light-Meson Sector
 StringFlav:ProbStoUD     = 0.217
 StringFlav:mesonUDvector = 0.5
 StringFlav:mesonSvector  = 0.55
 StringFlav:etaSup        = 0.60  
 StringFlav:etaPrimeSup   = 0.12
# Baryon Sector
 StringFlav:probQQtoQ     = 0.081
 StringFlav:probSQtoQQ    = 0.915
 StringFlav:probQQ1toQQ0  = 0.0275
 StringFlav:suppressLeadingB = off
 StringFlav:popcornSpair  = 0.9
 StringFlav:popcornSmeson = 0.5
\end{verbatim}
}

\begin{figure}[t!p]
\centering
\includegraphics*[scale=\dscale]{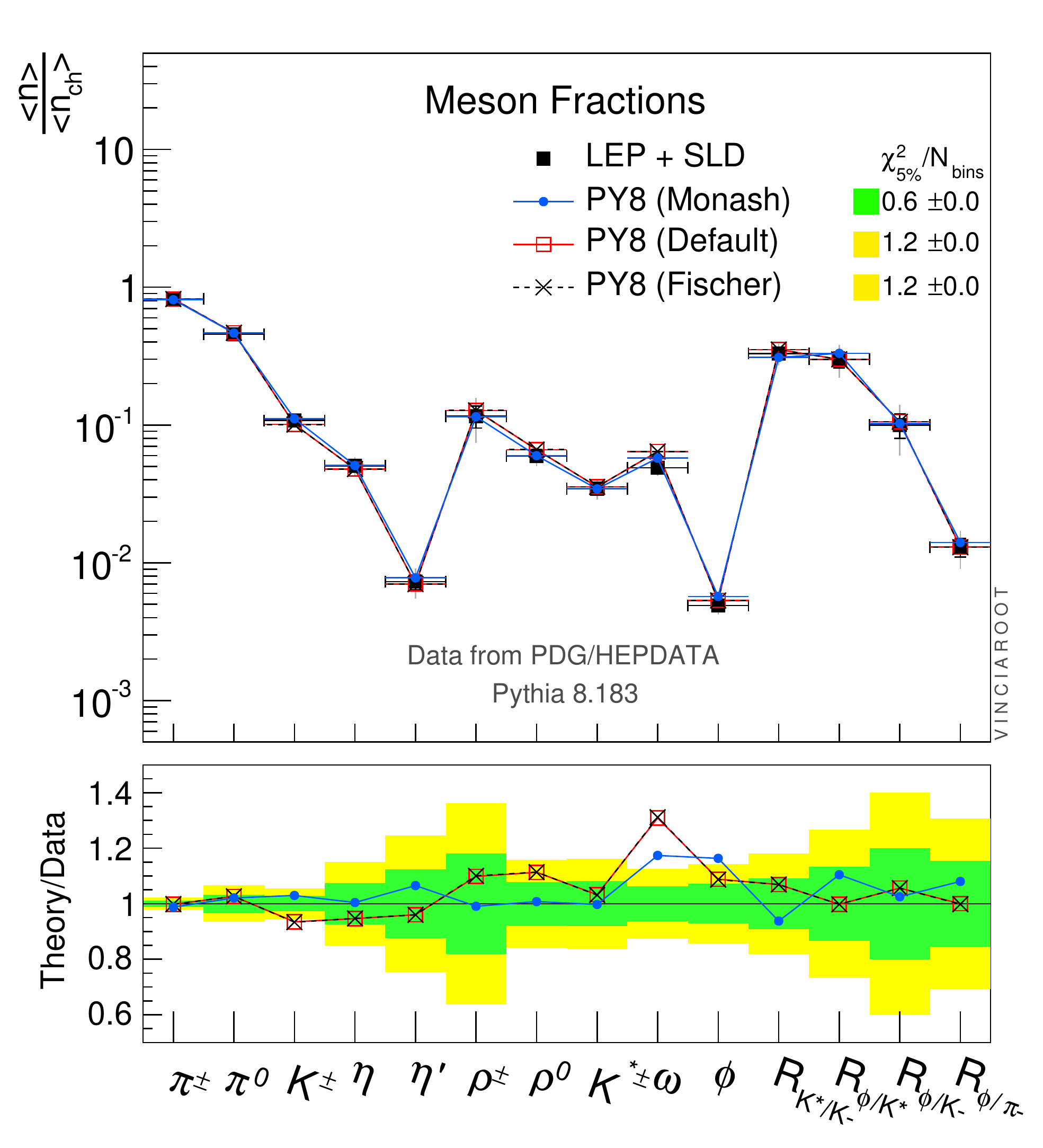}
\includegraphics*[scale=\dscale]{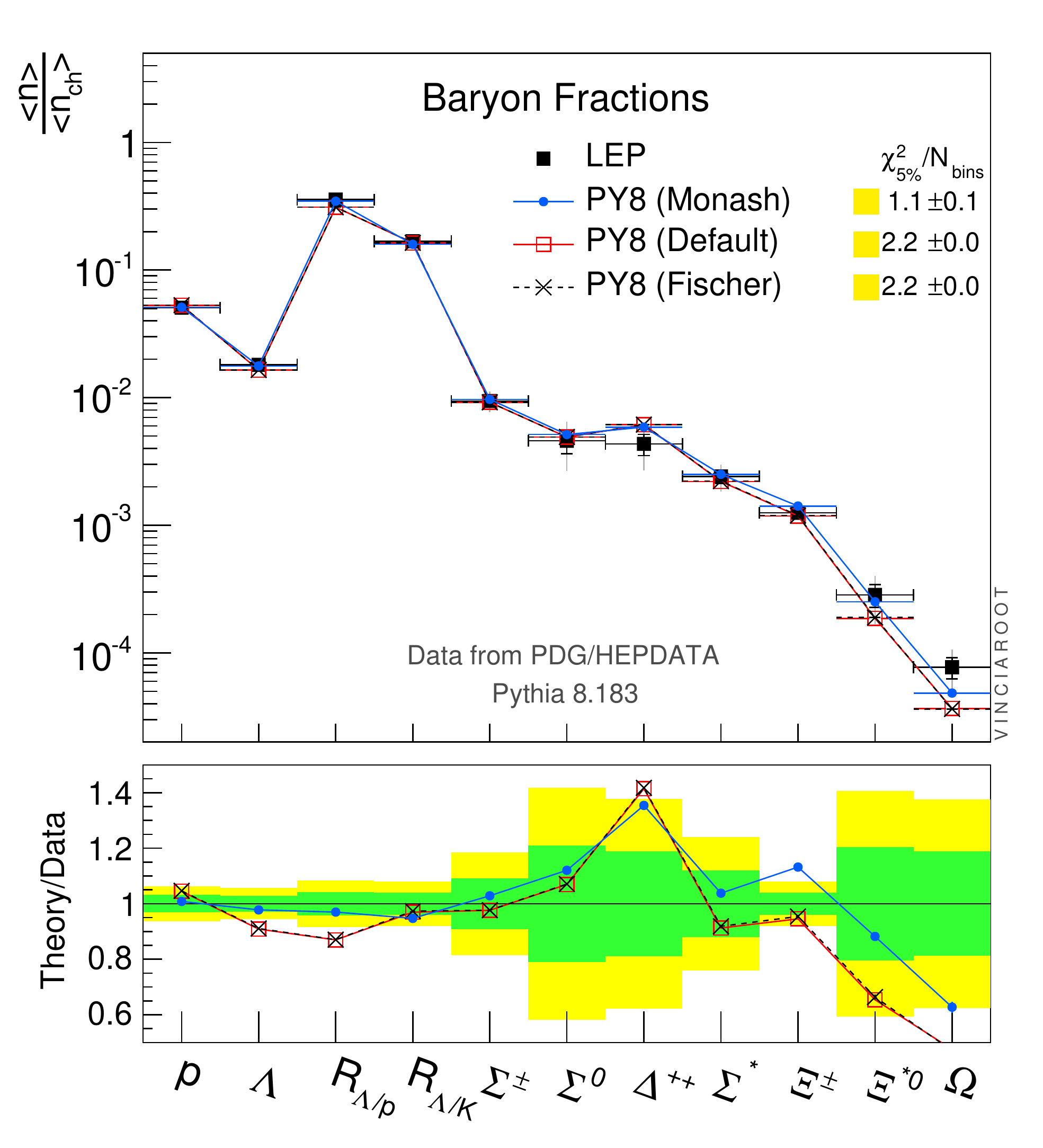}
\caption{Hadronic $Z$ decays at $\sqrt{s}=91.2\GeV$. 
Identified-meson and -baryon rates, expressed as fractions of the average
charged-particle multiplicity.
\label{fig:idParticles}}
\end{figure}

Since strange-particle and baryon spectra at the LHC exhibit interesting
differences with respect to existing models (see below), we paid
particular attention to first obtaining a good 
description of these sectors in $e^+e^-$ collisions. 
Specifically, we have increased the overall amount of strangeness by
about 10\%, while decreasing the rate of vector mesons by a similar
amount\footnote{For reference, the current default value of \ttt{ProbStoUD} is
  0.19 while ours is 0.217. The increased value also improves the
  agreement with the $D_s$ and $B_s$ rates, see \secRef{sec:b}. 
The default values of \ttt{mesonUDvector}
  and \ttt{mesonSvector} are 0.62 and 0.725 respectively, while ours
  are 0.5 and 0.55.}
 (these two effects largely cancel for $K^*$).  
This improves the total $K^\pm$, $\rho^0$, $\omega$,   
$\Lambda$, $\Xi^*$, and $\Omega$ yields on our combined LEP
estimates discussed above. The price is that we now
overshoot the measured rate of $\Xi^\pm$ baryons by 10\%. 
The resulting identified-meson and
-baryon rates, expressed as fractions of the average charged-particle
multiplicity are plotted in \figRef{fig:idParticles}.
Note that the last four bins of the meson plot and the third and fourth
bins of the baryon plot are not
$\left<n\right>/\left<n_\mrm{Ch}\right>$ fractions, 
but rather the $K^*/K$, $\phi/K^*$, $\phi/K$, $\phi/\pi$, $\Lambda/p$
and $\Lambda/K$ ratios, 
respectively. Note also that \secRef{sec:energyScaling} on energy
scaling below includes a comparison to the average Kaon and
Lambda rates as a function of $ee$ CM energy (\figRef{fig:LEPscaling}). 

\begin{figure}[t!p]
\centering
\includegraphics*[scale=\dscale]{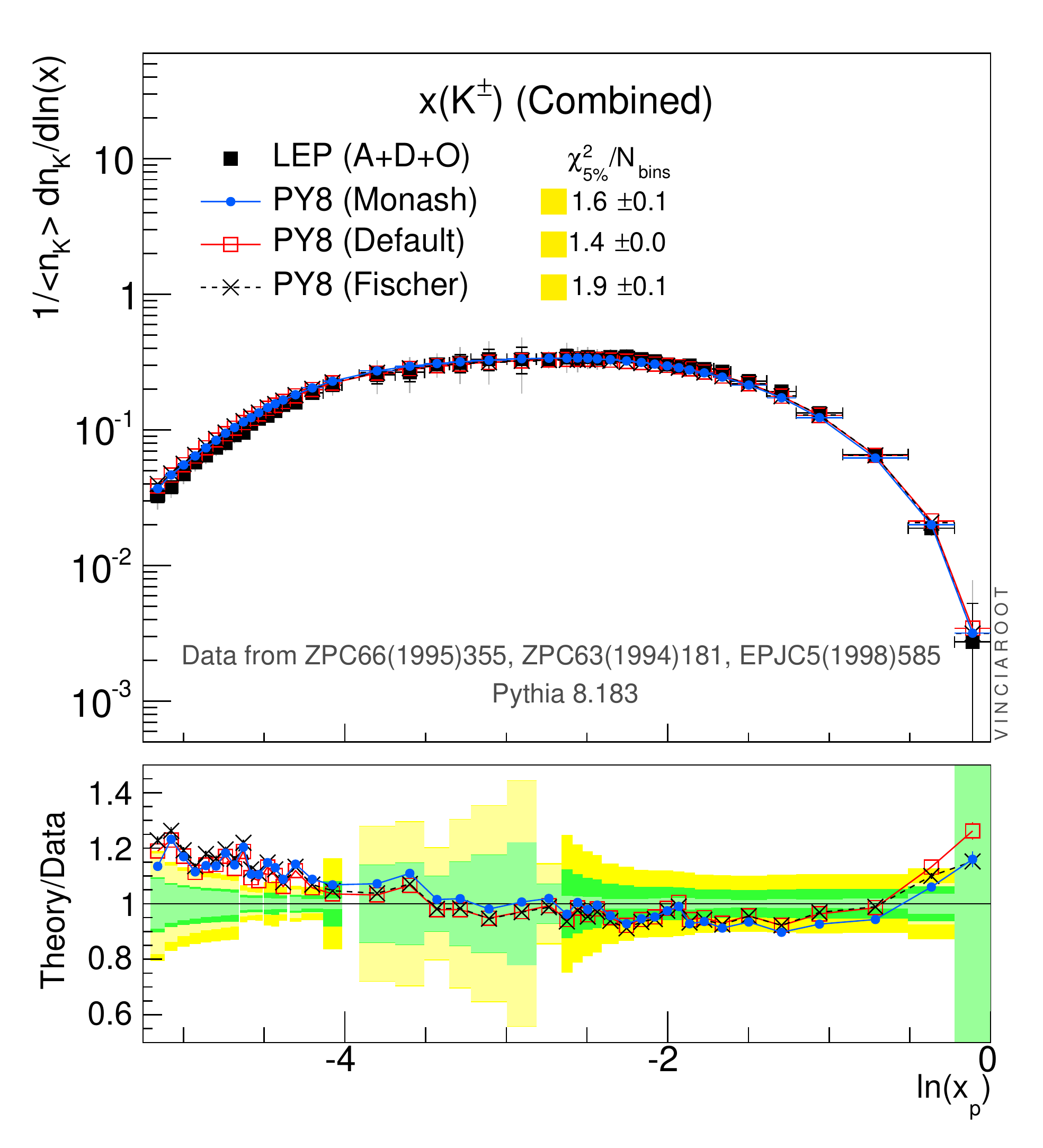}
\includegraphics*[scale=\dscale]{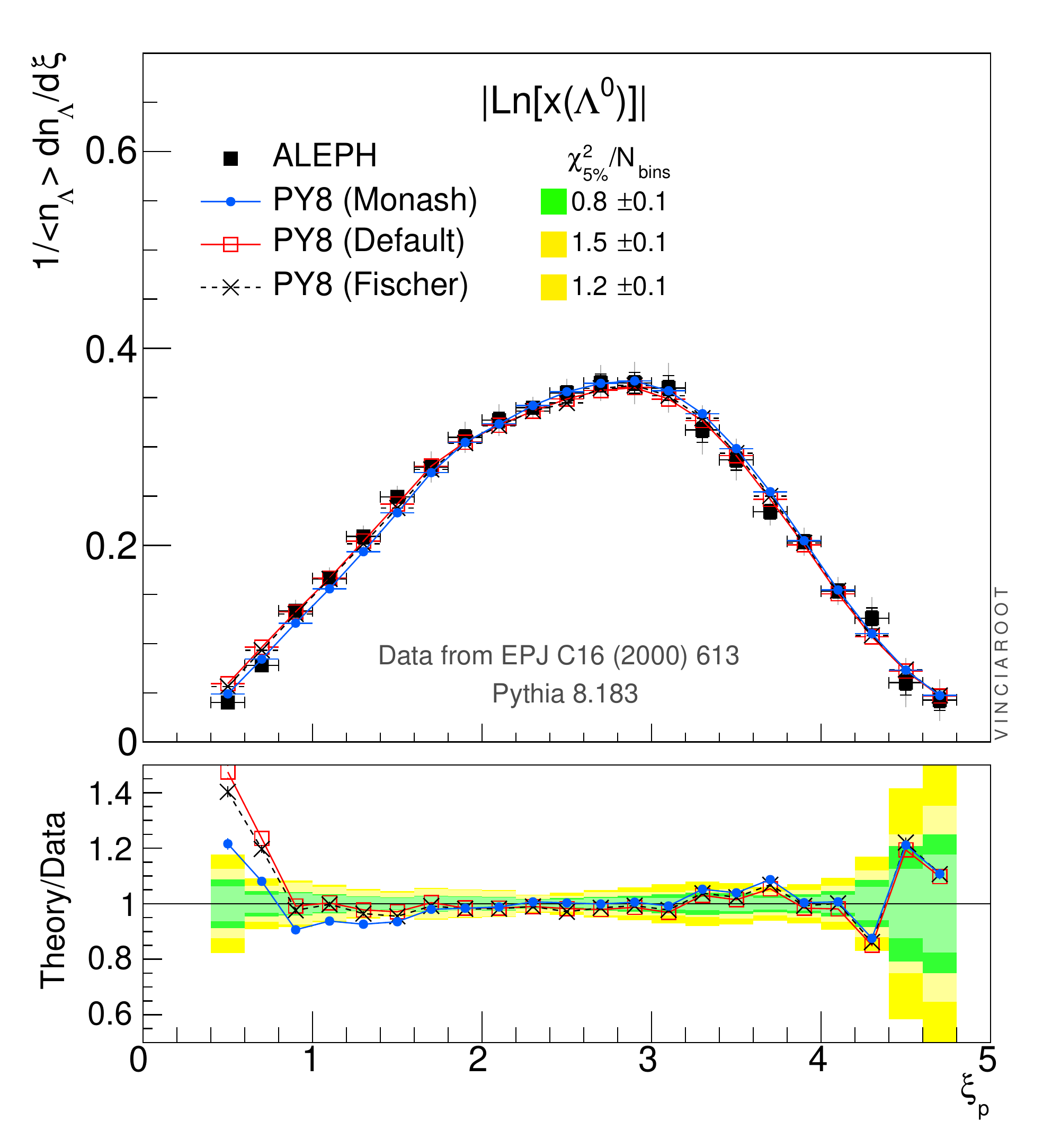}
\caption{Hadronic $Z$ decays at $\sqrt{s}=91.2\GeV$. 
$K^\pm$ and $\Lambda$ momentum-fraction spectra.
\label{fig:idSpectra}}
\end{figure}

To provide further information on identified particles, 
we include a limited comparison to momentum spectra of 
$K^\pm$, $p$, $\Lambda$, and $\Xi^\pm$, which are the
states of most immediate interest in the context of similar
comparisons now being made at LHC. The spectra of $K^\pm$ mesons and
$\Lambda$ baryons are shown in 
\figRef{fig:idSpectra}, while the $p^\pm$ and $\Xi^\pm$ spectra are
relegated to \appRef{app:id}. 
The modified parameters of the Monash tune have virtually no effect on
the Kaon distribution, which still exhibits too many very
soft Kaons (with $\ln(x)<-4$, corresponding to $x<0.018$, so momentum
scales below $\sim$ 1\GeV), while the significant increase in the 
value of \ttt{aExtraDiquark} from 0.5 (Default) to 0.97 (Monash,
cf.~\secRef{sec:light})  
produces a desirable suppression of very hard $\Lambda$ baryons. The
corresponding change in the measured parts of the $p$ and 
$\Xi^\pm$ spectra (cf.~\appRef{app:id}) are small compared with the
experimental uncertainties.  

It is interesting, however, to note that all of these spectra
indicate, or are at least consistent with,  a modelling 
excess of soft identified-particle 
production below $\ln(x)\sim -4.5$, corresponding to absolute momentum
scales around $500\MeV$, while we recall that the
inclusive $\ln(x)$ spectrum above showed an underproduction around
$\ln(x) \sim -5.5$. Within the constraints of the current
theory model, we have not managed to find a way to
mitigate these features while remaining consistent with the rest of
the data. Nonetheless, it should be mentioned that these 
observations could have relevance also in the context of understanding
identified-particle spectra at LHC, a possibility which to our
knowledge has so far been ignored.

\subsection{Heavy-Quark Fragmentation \label{sec:b}}

Similarly to above, we first discuss the inclusive rates of hadrons
containing heavy quarks, before we discuss their
spectra. Unfortunately, there are also here substantial disagreements
between different pieces of information. We have made the following
choices. 

\begin{figure}[t!p]
\centering
\includegraphics*[scale=\dscale]{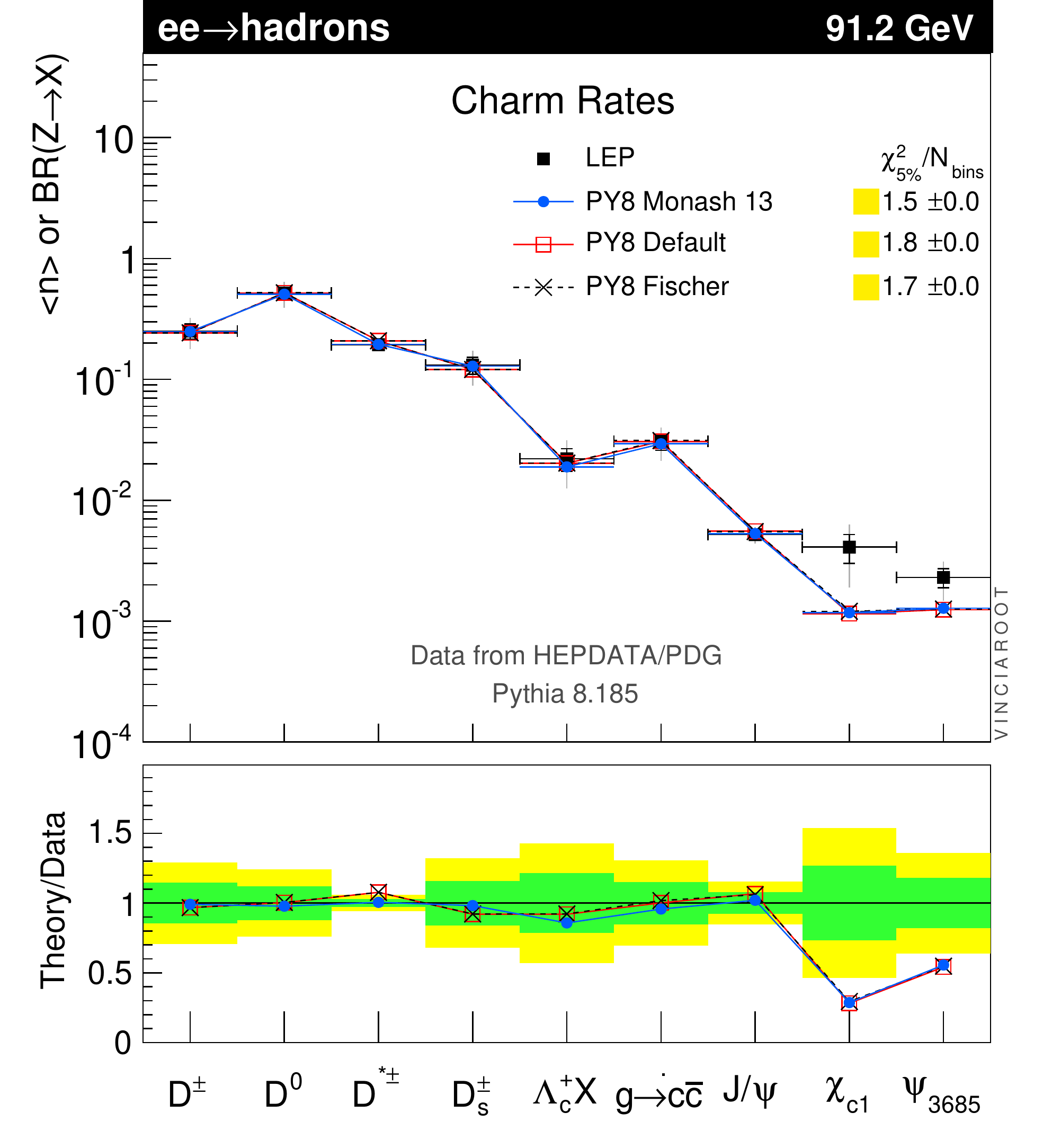}
\includegraphics*[scale=\dscale]{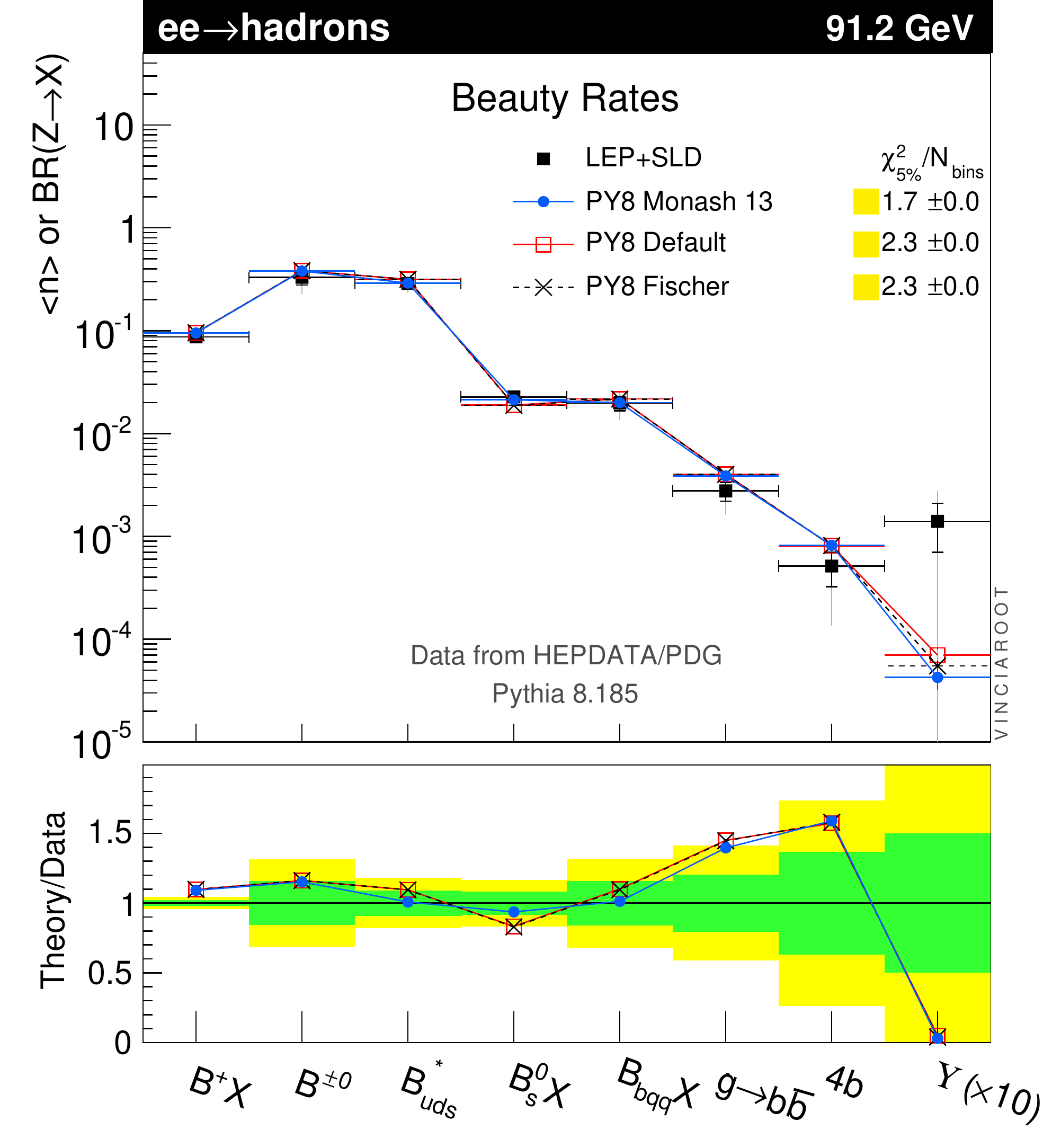}
\caption{Hadronic $Z$ decays at $\sqrt{s}=91.2\GeV$. 
Rates and inclusive $Z\to X$ branching fractions (normalized
  to  $Z\to\mrm{hadrons}$) of particles containing $c$ and
  $b$ quarks
\label{fig:heavymesons}}
\end{figure}

\begin{itemize}
\item For $D$ mesons, the average 
 $D^{\pm}$ rate given in sec.~46 of the PDG (0.175) is equal
to the inclusive branching fraction for $Z\to D^\pm X$ 
 given in the $Z$ boson summary table 
in the same \emph{Review} (after normalizing the latter to the hadronic $Z$
fraction of $69.91\%$~\cite{Beringer:1900zz}). However, the former ought to be
substantially larger given that some $Z\to c\bar{c}$ 
events will contain two $D^{\pm}$
mesons (counting once in the $Z\to D^\pm X$ branching fraction 
but twice in the average
$D^\pm$ multiplicity). We therefore here use a measurement by
ALEPH~\cite{Buskulic:1993iu} to fix the $D^\pm$ and $D^0$ rates,
resulting in a reference value for the 
average $D^\pm$ multiplicity almost twice as large as that given by
sec.~46 in the PDG. 
\item For $\Lambda_c^+$, the average multiplicity given in sec.~46 of
the PDG is twice as large as that indicated by the branching fraction
$\mrm{BR}(Z\to \Lambda_c^+ X)$ in the $Z$ boson summary table in the
same \emph{Review}. We here use the branching from the $Z$ boson
summary table as our constraint on the $\Lambda_c^+$ rate, normalized
to the total branching fraction $\mrm{BR}(Z\to \mrm{hadrons})$.
\item We also include the average rate of $g\to
  c\bar{c}$ splittings, obtained by combining an ALEPH~\cite{Barate:1999bg}
  and an OPAL measurement~\cite{Abbiendi:1999sxa}, but with an additional 10\%
  systematic uncertainty added to both measurements to account for
  possibly larger mismodeling effects in the correction
  factors~\cite{Miller:1998ig,Biebel:2001ka}.  
\item For $B$ particles, we use the quite precise inclusive $Z\to
  B^+X$ branching 
fraction from the $Z$ boson summary in the PDG. 
\item We also use 
the sum of $B^\pm$ and $B^0(\bar{B}^0)$ in sec.~46
of the PDG\footnote{Note that we have a factor 2 relative to the PDG,
  since it appears the PDG quotes the average, rather than the
  sum. Note also that all the average $B$ meson
multiplicities in sec.~46 of the PDG are accompanied by a note,
``(d)'', stating that the SM $B(Z\to b\bar{b}) = 0.217$ was used for
the normalization. For
completeness, the reader should be aware that this is
the fraction normalized to hadronic $Z$ decays; the branching
fraction relative to all $Z$ decays, is 0.151~\cite{Beringer:1900zz}.}. 
\item The  
$B_s^0$ multiplicity given in sec.~46 of the PDG ($0.057\pm 0.013$) is
more than twice the inclusive $\mrm{BR}(Z\to
B_s^0X)/\mrm{BR}(Z\to\mrm{hadrons})$ branching fraction 
($0.0227\pm 0.0019$) quoted in the $Z$ boson summary table. We find
these two numbers difficult to reconcile and choose to use the inclusive 
$\mrm{BR}(Z\to B_s^0X)/\mrm{BR}(Z\to\mrm{hadrons})$ branching fraction
as our main constraint. 
\item We also include the inclusive branching
fractions for $B$-baryons (summed over baryons and antibaryons), 
the rate of $g\to b\bar{b}$ splittings obtained by combining
ALEPH~\cite{Barate:1998vs}, 
DELPHI~\cite{Abreu:1997nf}, and SLD~\cite{Abe:1999qg} measurements (including an
additional 10\% systematic to account for larger possible mismodeling
effects in the correction factors~\cite{Miller:1998ig,Biebel:2001ka}) 
and the rate of $Z\to bb\bar{b}\bar{b}$ from the
PDG $Z$ boson summary table~\cite{Beringer:1900zz}.  
\end{itemize}
Our constraints on the heavy-quark particle rates are summarized 
in \tabRef{tab:heavyrates}. 
\begin{table}[t!p]
\centering\small
\begin{tabular}{llll}
\toprule
\mbox{\bf Charm} & \multicolumn{2}{l}{Our Reference} & \hspace*{-2mm}Our\\
$\left<n\right>\mbox{~or~}\mrm{BR}$ & \multicolumn{2}{l}{Value} & \hspace*{-2mm}Source
\\\midrule
   $D^{+} + D^-$ & $0.251$ & $\pm 0.047$ & A
\\ $D^0 + \bar{D}^0 $    & $0.518$ & $\pm 0.063$ & A
\\ $D^{*+} + D^{*-}$ & $0.194$ & $\pm0.0057$ & P
\\ $D_s^+ + D_s^-$  & $0.131$ & $\pm0.021$ & P
\\ $\mrm{BR}(Z\to \Lambda_c^+X)$  & $0.0220$ & $\pm 0.0047$ & Z
\\ $\mrm{BR}(Z\to X+c\bar{c})$\!\!& $0.0306$ & $\pm 0.0047$ & AO
\\ $J/\psi$ & $0.0052$ & $\pm 0.0004$ & P
\\ $\chi_{c1}$  & $0.0041$ & $\pm 0.0011$ & P
\\ $\psi'$ & $0.0023$ & $\pm 0.0004$ & P
\\ \bottomrule
\end{tabular}
\begin{tabular}{llll}
\toprule
\mbox{\bf Beauty} & 
  \multicolumn{2}{l}{Our Reference} &  \hspace*{-2mm}Our
\\ $\left<n\right>\mbox{~or~}\mrm{BR}$ & \multicolumn{2}{l}{Value}
& \hspace*{-2mm}Source
\\\midrule
 $\mrm{BR}(Z\to B^{+}X)$   & $0.087$ & $\pm 0.002$ & Z
\\ $B^{+} + B^0 + \bar{B}^0 + B^-$\!& $0.330$ & $\pm 0.052$ & P
\\ $B^{*}_{u}+B^{*}_d+B^*_s$ & $0.288$ & $\pm 0.026$ & P
\\ $\mrm{BR}(Z\to B_s^0X)$  & $0.0227$ & $\pm 0.0019$ & Z
\\ $\mrm{BR}(Z\to B_\mrm{baryon}X)$\!& $0.0197$ & $\pm 0.0032$ & Z
\\ $\mrm{BR}(Z\to X+b\bar{b})$ & $0.00288$ & $\pm 0.00061$ & ADS
\\ $\mrm{BR}(Z\to bb\bar{b}\bar{b}X)$ & $0.00051$ & $\pm 0.00019$ & Z
\\ $\Upsilon~(\times 10)$ & $0.0014$ & $\pm 0.0007$ & P
\\
\\ \bottomrule
\end{tabular}
\caption{Hadronic $Z$ decays at $\sqrt{s}=M_Z$. 
Measured rates and inclusive branching fractions 
of particles containing $c$ and $b$ quarks, as
  used in this work. Note: the 
  branching fractions are normalized to $Z\to \mrm{hadrons}$,
  and hence should be interpreted as, e.g., $\mrm{BR}(Z\to B^+
  X)/\mrm{BR}(Z\to\mrm{hadrons})$. Note 2: the sum over $B^*$ states includes
  both particles and anti-particles. Note 3: the $\Upsilon$ rate is
  multiplied by a factor 10.
Source labels indicate: A (ALEPH), D (DELPHI), O (OPAL), P (PDG,
section 46), S (SLD), Z (PDG Z Boson Summary Table).  \label{tab:heavyrates} 
}
\end{table}
Comparisons to these rates are shown in 
\figRef{fig:heavymesons}, now without normalizing to the
average charged-particle multiplicity.
Given that most of the $c$ and $b$ quarks come directly from $Z\to
c\bar{c}$ and $Z\to b\bar{b}$ decays, there is not a lot of room for
tuning to these numbers, apart from the relative rates of vector
mesons vs.\ pseudoscalars,
 which is  controlled by the parameters:

\noindent{\small\begin{verbatim}
# Heavy Mesons
  StringFlav:mesonCvector = 0.88
  StringFlav:mesonBvector = 2.2   
\end{verbatim}
}

\noindent Our parameters are slightly smaller than the current default
values, leading to slightly smaller $D^*$ and $B^*$ rates, as can be
seen from the plots in \figRef{fig:heavymesons}. Note also that the
increased overall amount of strangeness in the fragmentation leads to slightly
higher $D_s$ and $B_s$ fractions, in better agreement with the data. 
Uncertainties are, however, large, and 
some exotic onium states, like
$\chi_{c1}$, $\psi'$, and $\Upsilon$ are not well described by the
default modeling. (It is encouraging that at least the multiplicity of
$J/\psi$ mesons is well described, though a substantial 
fraction of this likely owes 
to the feed-down from $B$ decays, and hence does not depend directly
on the string-fragmentation model itself.) 
 
We also note that it would be desirable to reduce the rate of $g\to
b\bar{b}$ and $Z\to bb\bar{b}\bar{b}$ events, while the $g\to
c\bar{c}$ one appears consistent with the LEP constraints. We suspect
that this issue may be tied to the fixed choice of using $p_\perp$ as the
renormalization scale for both gluon emissions and for $g\to q\bar{q}$
splittings in the current version of \Py. A more natural choice for
$g\to q\bar{q}$ could be $\mu_R\propto m_{q\bar{q}}$, as used e.g.\ in
the \Vc\ shower model~\cite{Hartgring:2013jma}.

We now turn to the dynamics of heavy-quark
fragmentation, focusing mainly on the $b$ quark. 

For heavy quarks, the Lund fragmentation function is modified due to
the (massive) endpoints not moving along straight lightcones: as the
string pulls on them, they slow down, resulting in the string tracing
out a smaller space-time area than it would for massless quarks. This
modifies the implications of the string area law, in a manner captured
by the so-called Bowler modification of the 
fragmentation function~\cite{Bowler:1981sb} 
\begin{equation}
f_{\mrm{massive}}(z,m_Q) \propto \frac{f(z)}{z^{b r_Q m_Q^2}}~,
\label{eq:fMassive}
\end{equation}
with $m_Q$ the heavy-quark mass, $b$ the same universal parameter that
appears in the massless fragmentation function, \eqRef{eq:fMassless}, 
and $r_Q$ a tuning parameter
which is unity in the original derivation of Bowler but can be
assigned values different from unity to reduce ($r_Q\to 0$) 
or emphasize ($r_Q > 1$) the effect. Since $r_Q$ multiplies the
heavy-quark mass (squared), it can also be viewed as an effective
rescaling of the mass value. The net result is a suppression of 
the region $z\to 1$, hence a relative softening of the 
fragmentation spectrum for heavy flavours (\emph{relative} since 
the presence of $m_\perp^2$ in the exponent of \eqRef{eq:fMassless}
still implies an overall harder fragmentation for higher hadron masses.)

We emphasize that this is the only fragmentation function that
is self-consistent  within the
string-fragmentation
model~\cite{Bowler:1981sb,Andersson:1998tv}. Although a few
alternative 
forms of the fragmentation functions for massive quarks are available 
in the code, we therefore here work only with the Bowler type. As for
the massless function, the proportionality sign in \eqRef{eq:fMassive}
indicates that the function is normalized to unity.

In \Py, separate $r_Q$ parameters are provided for $c$ and $b$
quarks. We consider the one for $b$ quarks first. 
Its default value is $r_b=0.67$, but this appears to give too
hard $b$ fragmentation spectra when compared to LEP and SLD data, see below. 
For the Monash tune, we instead use 

\noindent{\small\begin{verbatim}
  StringZ:rFactB = 0.855
\end{verbatim}
}

\noindent which produces
softer $B$ spectra and simultaneously agrees better with the
theoretically preferred value ($r_b=1$). 

\begin{figure}[t!p]
\centering
\includegraphics*[scale=\dscale]{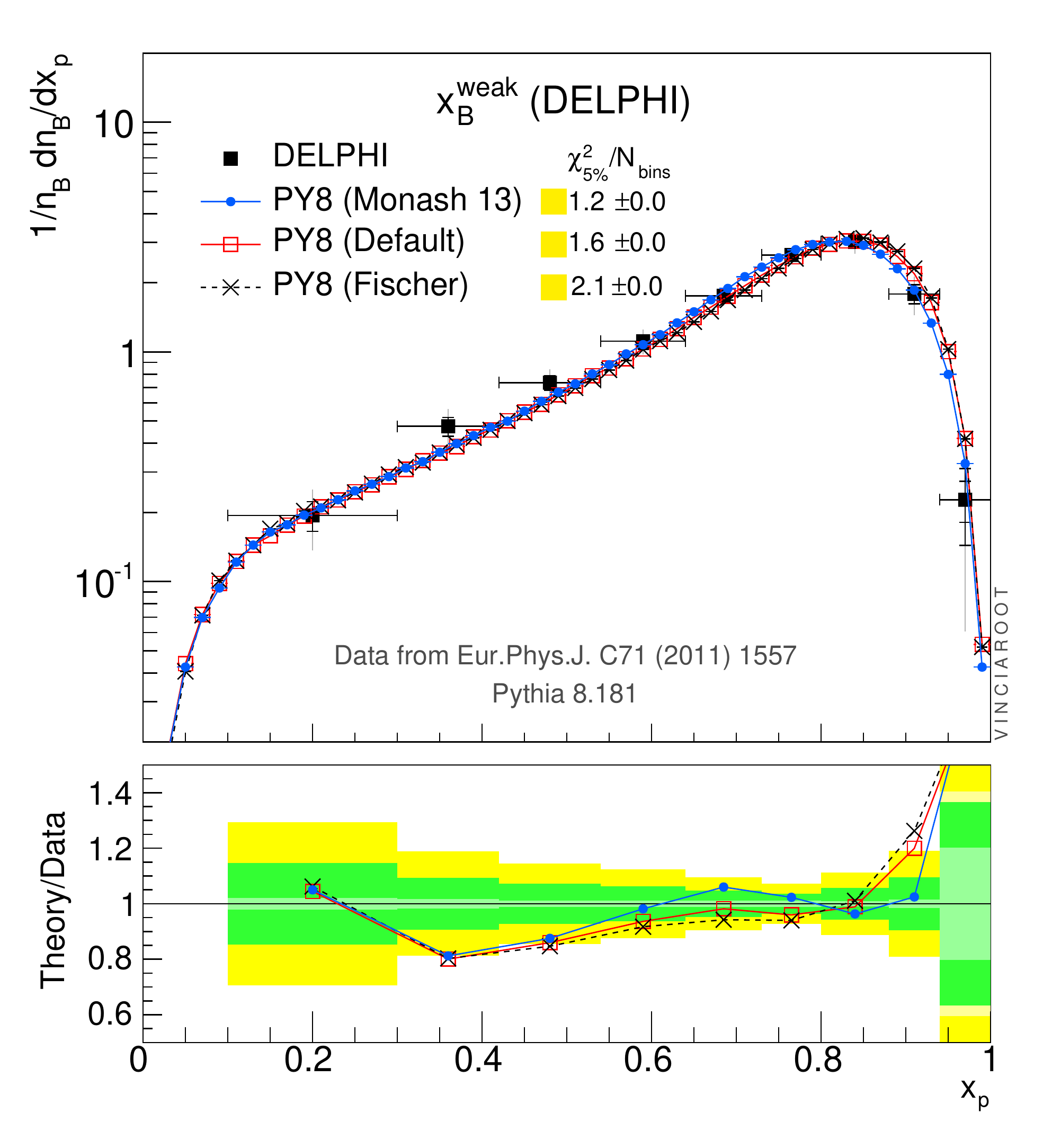}
\includegraphics*[scale=\dscale]{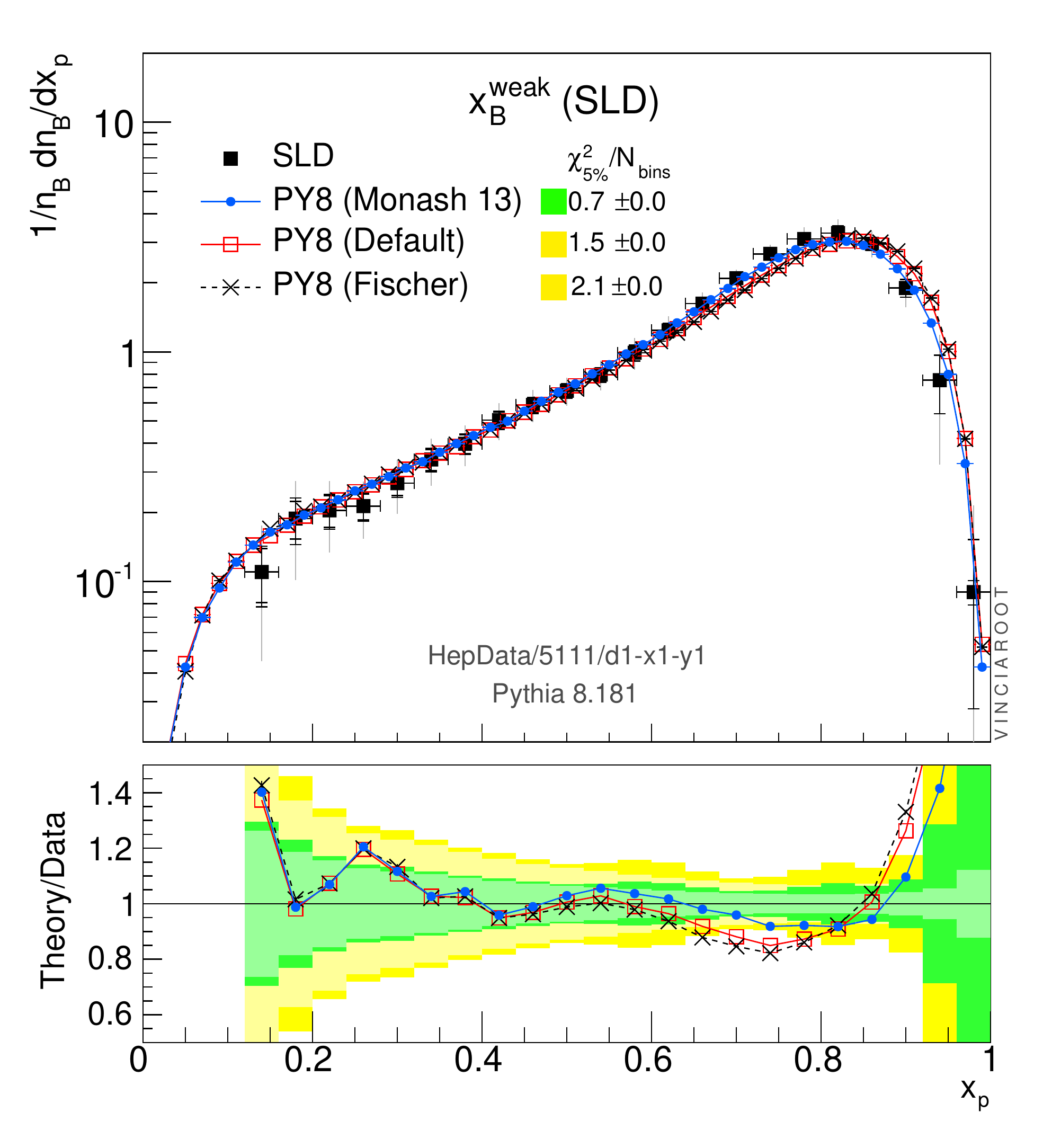}
\caption{Hadronic $Z$ decays at $\sqrt{s}=91.2\GeV$. 
Momentum ($x_B$) spectra of weakly decaying $B$ hadrons,
  compared to data from DELPHI~\cite{DELPHI:2011aa} (left) and
  SLD~\cite{Abe:2002iq} (right) 
\label{fig:bSpectra}}
\end{figure}

\begin{figure}[t!p]
\centering
\includegraphics*[scale=\dscale]{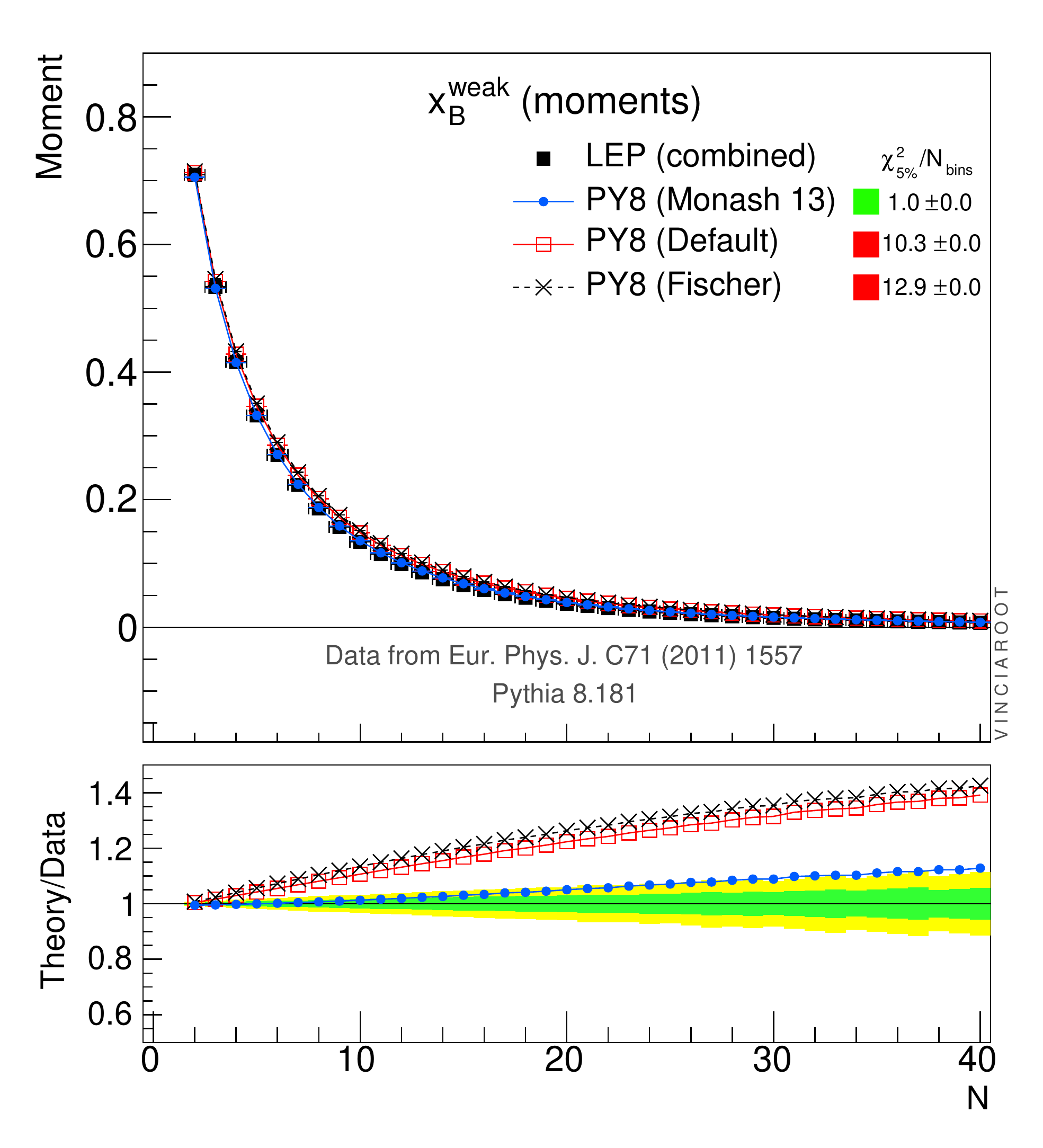}\vskip-3mm
\caption{Hadronic $Z$ decays at $\sqrt{s}=91.2\GeV$.
Moments of the $B$ fragmentation function, compared to a
  combined analysis of LEP+SLD data by DELPHI~\cite{DELPHI:2011aa}
\label{fig:bMoments}}
\end{figure}

A comparison to the scaled-momentum
spectra ($x_B = 2|p_B|/\Ecm$) of weakly decaying
$B$ hadrons from both 
DELPHI~\cite{DELPHI:2011aa} and SLD~\cite{Abe:2002iq} is given in
\figRef{fig:bSpectra} (due to small 
differences between the two measured results, we choose to show both). The
dampening of the hardest part of the spectrum caused by the increase
in the $r_b$ 
parameter is visible in the right-most two bins of the distributions
and in the smaller $\chi^2_{5\%}$ values for the Monash tune.
The effects of the modification can be further emphasized by an
analysis of the moments of the distribution, in which the higher
moments are increasingly dominated by the
region $x_B\to 1$. A comparison to a combined LEP analysis of the
moments of the $x_B$ distribution~\cite{DELPHI:2011aa} is given in
\figRef{fig:bMoments}, 
further emphasizing that the high-$x_B$ part of the distribution is
now under better control. 

\begin{figure}[tp]
\centering
\includegraphics*[scale=\dscale]{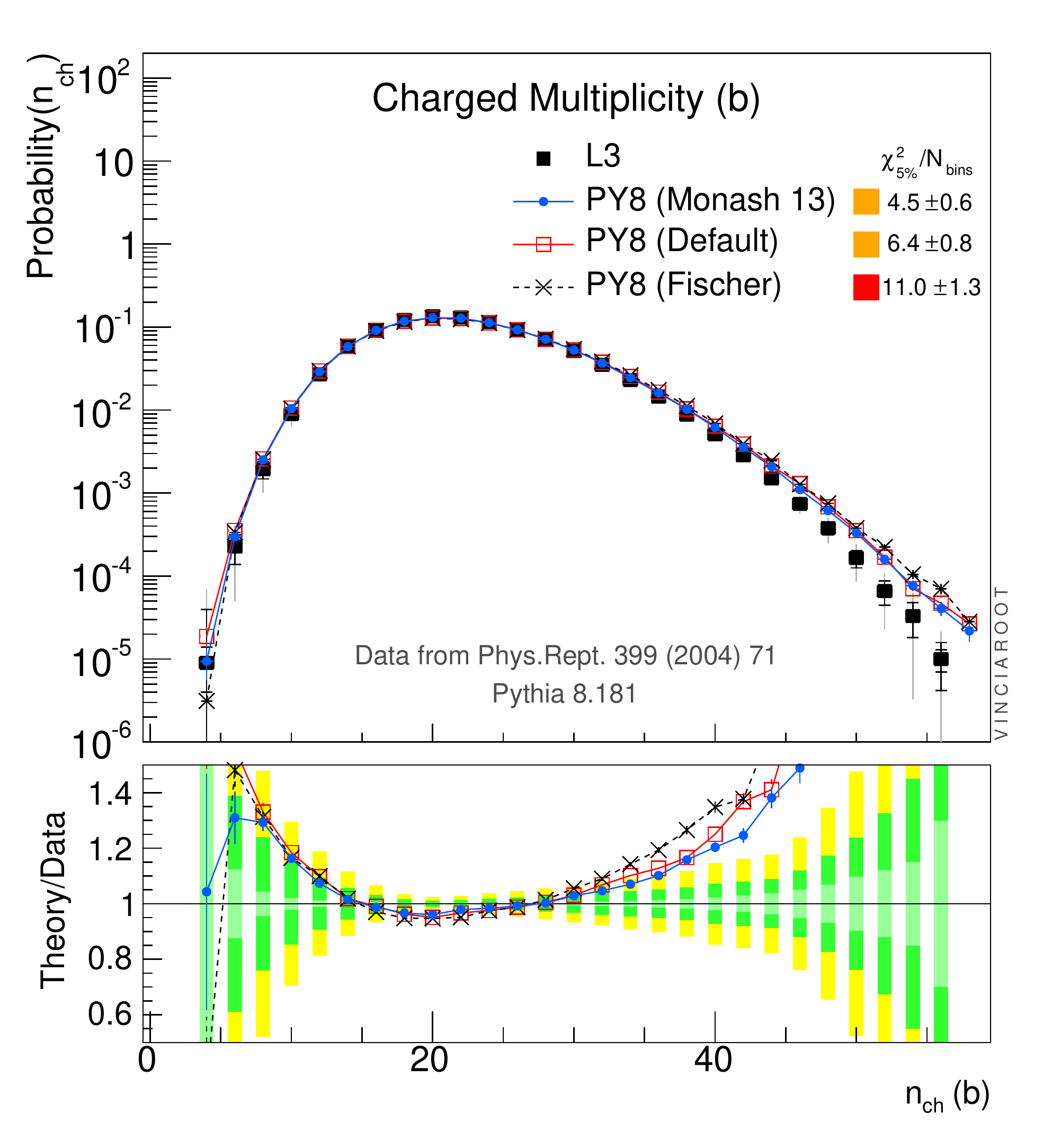}
\includegraphics*[scale=\dscale]{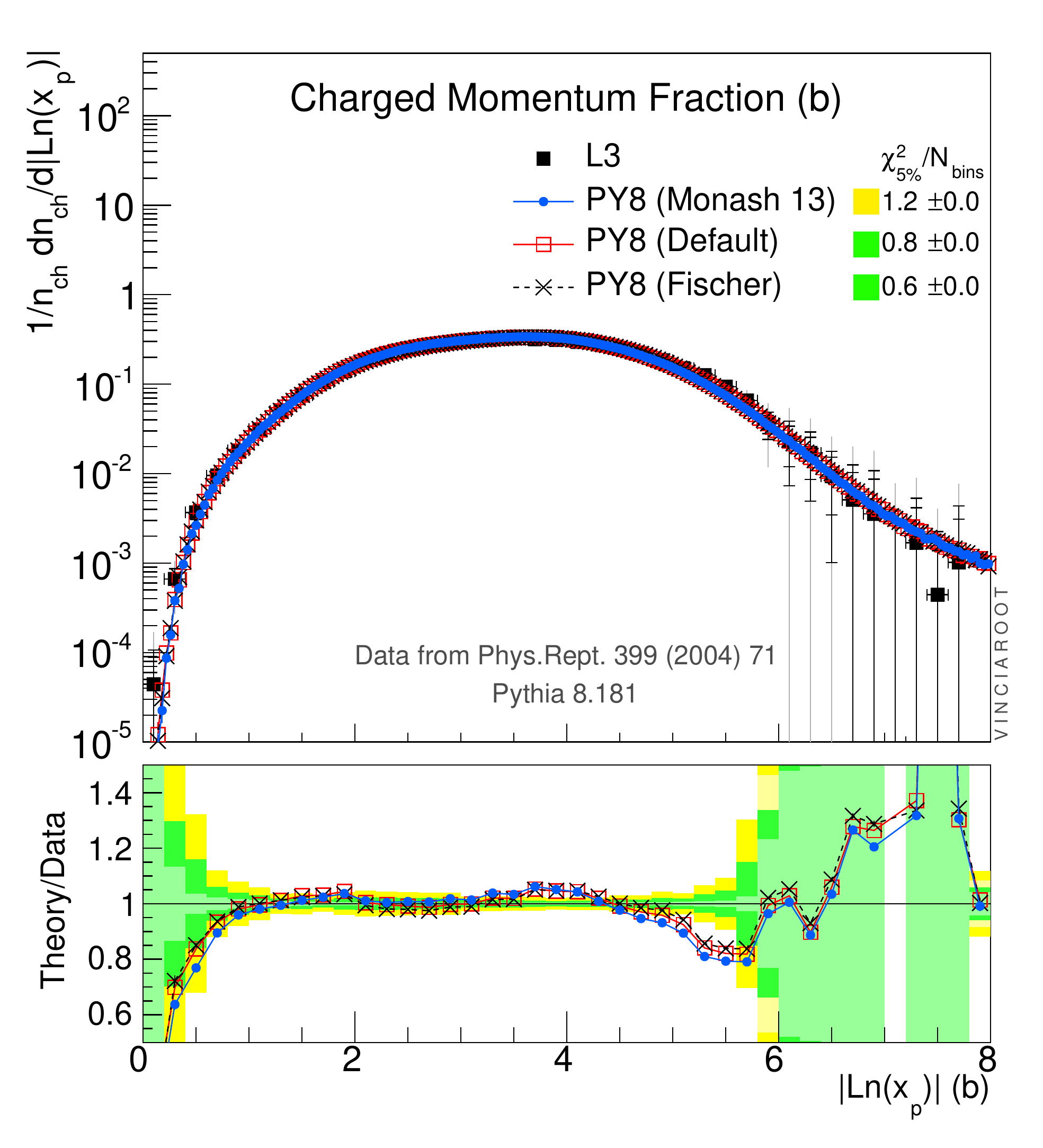}
\caption{Hadronic $Z$ decays at $\sqrt{s}=91.2\GeV$.
Charged-hadron multiplicity (left) and momentum-fraction (right) spectra in
  $b$-tagged events.
\label{fig:bMult}}
\end{figure}

The reason we have not increased the $r_b$ parameter further is that
it comes at a price. If the $B$ hadrons are taking less energy, then
there is more energy left over to produce other particles, and the
generated multiplicity distribution in $b$ events already exhibits a slightly
high tail towards large multiplicities. Nonetheless, since the revised 
light-flavour fragmentation parameters produce an overall narrower
fragmentation function, the end result is still a slight improvement
in the multiplicity distribution also for $b$ events. This
is illustrated, 
together with the inclusive momentum distribution for $b$-tagged events, in
\figRef{fig:bMult}, compared to measurements by L3~\cite{Achard:2004sv}.
Interestingly, the multiplicity distribution still appears to be
too wide, but within the constraints of the present study, 
we were unable to obtain further improvements. As a point of
speculation, we note that the distribution of the 
number of partons before hadronization is also quite wide in \Py, and
this may be playing a role in effectively setting a lower limit on
the width that can be achieved for the hadron-level distribution.

Comparisons to L3 event shapes in $b$-tagged events are
collected in \appRef{app:LEP} (the left column of plots contains light-flavour
tagged event shapes, the right column $b$-tagged ones). In
particular, the Monash tune gives a significant improvement in the soft
region of the jet-broadening parameters in
$b$-tagged events, while no significant changes are observed for 
the other event shapes. These small improvements are presumably a direct
consequence of the softening of the $b$ fragmentation function; it 
is now less likely to find an isolated ultra-hard $B$ hadron.

\begin{figure}[tp]
\centering
\includegraphics*[scale=\dscale]{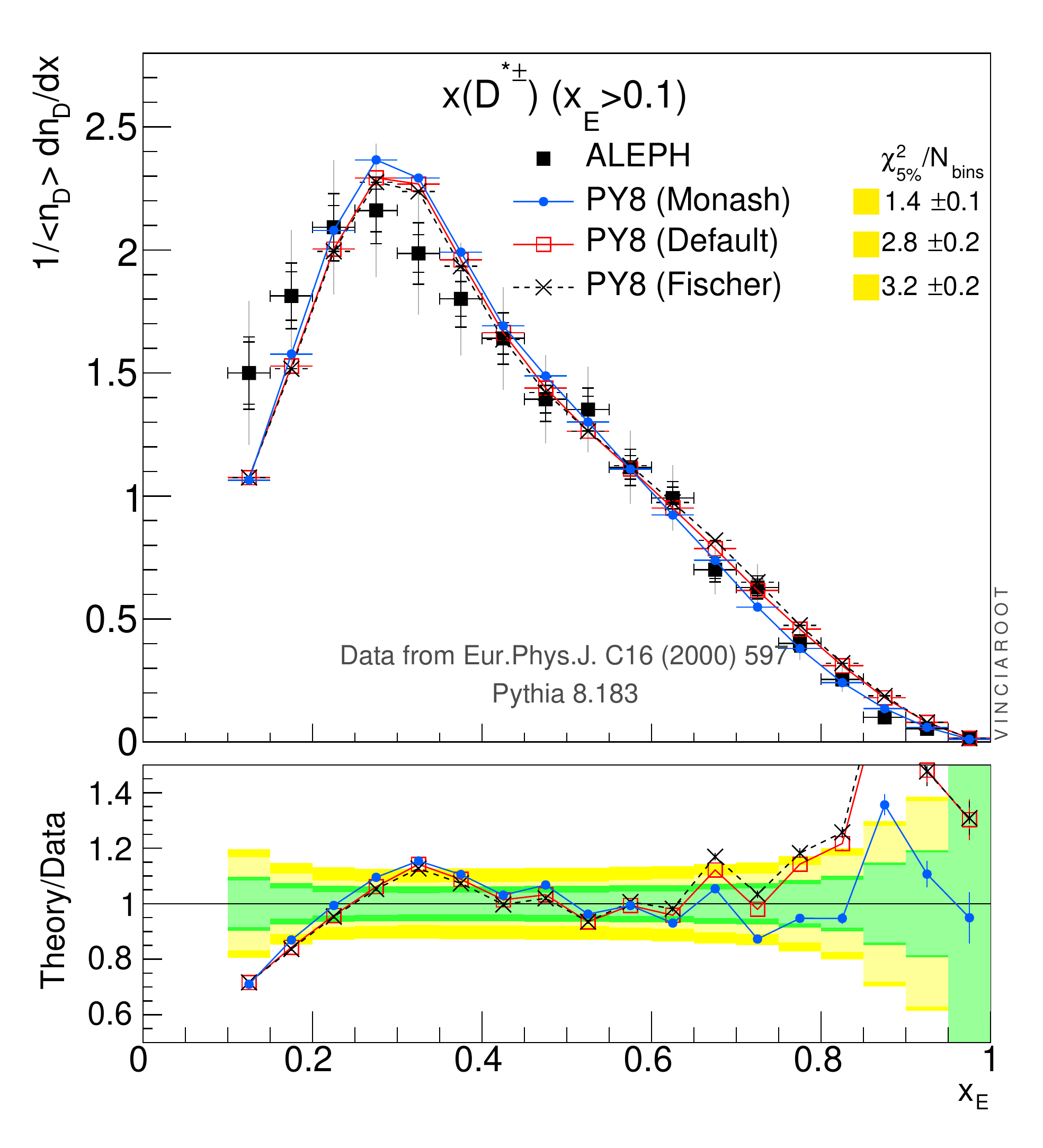} \ 
\includegraphics*[scale=0.33]{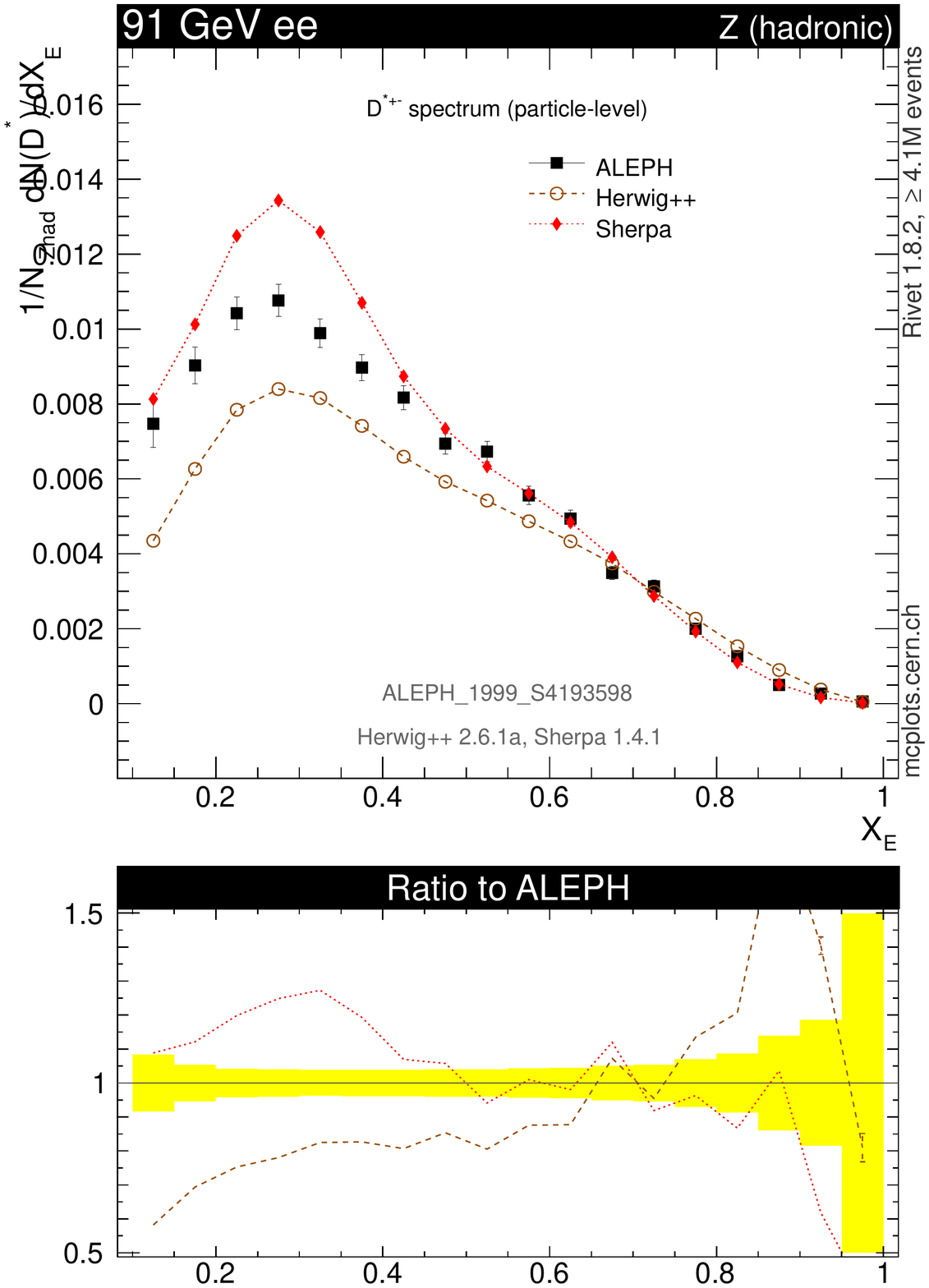}
\caption{The inclusive $D^*$ spectrum in hadronic $Z$
  decays~\cite{Barate:1999bg}. {\sl Left:} Monash 2013 tune compared
  with default \Py~8 and the Fischer tune. {\sl Right:} comparison
  with \Hw (dashed) and \Sh (dotted), from
  \mcplots~\cite{Karneyeu:2013aha}. 
  Note that the plot in the
  left-hand pane is normalized to unity, while the one in the
  right-hand pane is normalized to the number of hadronic $Z$ decays.
\label{fig:Dstar}}
\end{figure}
We round off the discussion of heavy-quark fragmentation by noting
that a similarly comprehensive study of charm-quark fragmentation
would be desirable. However, charm-quark tagged multiplicity
and event-shape data is not available to our knowledge, and most of
the $D$ meson spectra on HEPDATA concern only specific
decay chains (hence depend on the decay modeling),
and/or are limited to restricted fiducial regions (limiting their
generality). Experimentally, the cleanest measurement is obtained 
from $D^*$ decays, and an inclusive momentum spectrum for
$D^*$ mesons has been measured by ALEPH~\cite{Barate:1999bg}. 
From this distribution, shown in \figRef{fig:Dstar}, 
we determine a
value for $r_c$ of:

{\small\begin{verbatim}
  StringZ:rFactC = 1.32
\end{verbatim}
}

\noindent We note
that the low-$x$ part of the $D^*$ spectrum originates from 
$g\to c\bar{c}$ shower splittings, while the high-$x$ tail
represents prompt $D^*$ production from leading charm in $Z\to
c\bar{c}$ (see~\cite{Barate:1999bg} for a nice figure illustrating this).
The intermediate range contains a large component of 
feed-down from $b\to c$ decays, hence this distribution is also
indirectly sensitive to the $b$-quark sector. The previous default
tune had
a harder spectrum for both $b$- and $c$-fragmentation, leading to an
overestimate of the high-$x$ part of the $D^*$ distribution. The
undershooting at low $x_{D^*}$ values, which remains unchanged in the
Monash tune, most likely indicates an underproduction of $g\to c\bar{c}$
branchings in the shower. We note that such an underproduction 
may also be reflected in the LHC data on $D^*$ production, see
e.g.~\cite{Aad:2011td}. We return to this issue in the discussion of
identified particles at LHC, \secRef{sec:id}. 

For completeness, the
right-hand pane of \figRef{fig:Dstar} shows the $D^*$ spectra from 
the two other general-purpose MC models, \Hw~\cite{Bahr:2008pv} and
\Sh~\cite{Gleisberg:2008ta}. The \Hw spectrum (dashed lines) is similar to the
default \Py one,  
with a deficit in the $g\to c\bar{c}$ dominated
region at low $x_E$ and a significant overshooting in the
hard leading-charm region, $x_E\to 1$. Interestingly, the $D^*$
spectrum in \Sh~(dotted lines) exhibits an \emph{excess} at small
$x_E$ values, suggesting relatively larger contributions from $b$
decays and from $g\to c\bar{c}$ splittings. 

\section{Hadron Collisions \label{sec:hadronColliders}}

We discuss PDFs in \secRef{sec:pdfs}, the choice of
strong coupling (and total cross sections) in \secRef{sec:alphaS},
initial-state radiation (and primordial $k_T$) in
\secRef{sec:isr}, minimum-bias and underlying event in
\secRef{sec:uemb}, and finally identified-particle spectra in
\secRef{sec:id}. Energy scaling is discussed separately, in
\secRef{sec:energyScaling}. 

\subsection{Parton Distributions \label{sec:pdfs}}

In the context of MC models, a highly important role is played by the
small-$x$ gluon PDF, which has a strikingly different behavior between
LO and NLO/NNLO fits.
\begin{figure}[t]
\centering
\includegraphics*[width=0.70\textwidth]{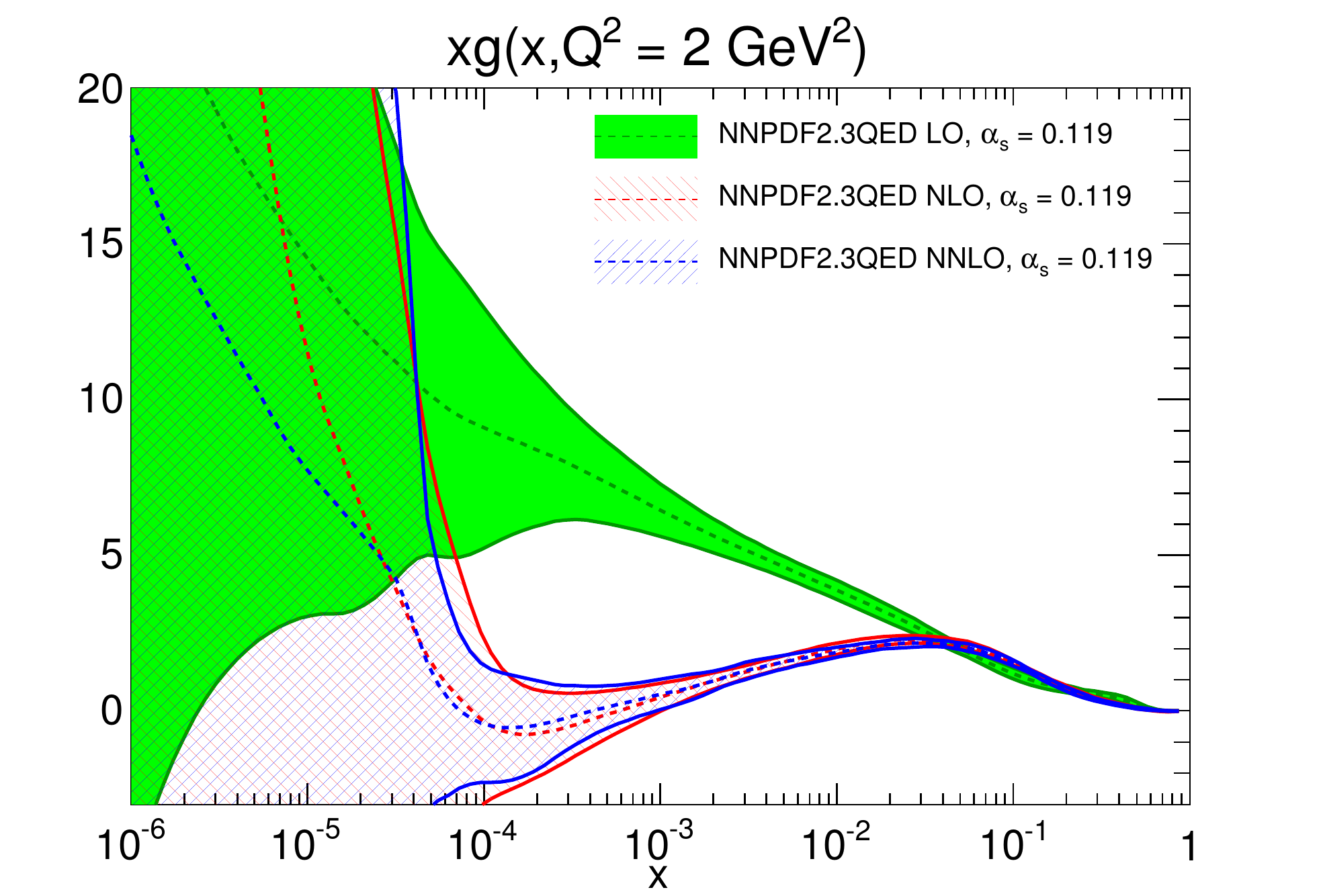}
\caption{Comparison of the gluon PDF at $Q^2=2$ GeV$^2$
between the LO, NLO and NNLO fits of the NNPDF2.3QED family.
 \label{fig:pdf_xg_log_2gev2}}
\end{figure}
This effect is illustrated in Fig.~\ref{fig:pdf_xg_log_2gev2}, obtained
from the NNPDF2.3QED PDF sets~\cite{Ball:2013hta} (see also the useful
plot of colour-weighted parton fluxes, fig.~2 in \cite{ATLAS:2012uec}).
The origin of this different small-$x$ behavior is the missing large
higher-order corrections to the DIS splitting functions and matrix
elements (represented by cofficient functions) in the LO fit.
Another source of the differences between LO and N(N)LO is related to the 
positivity of PDFs.
Indeed, while at LO PDFs have a probabilistic interpretation and are thus
positive-definite, starting from NLO they are scheme-dependent quantities and
thus can become negative~\cite{Altarelli:1998gn}.
(Of course, physical observables like structure functions are 
positive-definite to all orders in the perturbative expansion.)

In recent years there has been some discussion about 
possible modifications of the vanilla LO PDFs that could
lead to improved predictions from LO event generators.
Some possibilities for these improvements that have been explored
include the use of the LO value of $\alpha_s$ but with two-loop running,
or relaxing the momentum sum rules constraint from the LO fits.
These and other related ideas underlie recent attempts to produce
modified LO PDFs such as  MRST2007lomod PDFs~\cite{Sherstnev:2007nd} and the
CT09MC1/MC2~\cite{Lai:2009ne} PDFs.
The claim was that such improved LO (also called LO*) PDFs lead to a better
 agreement between data and theory in the LO fit and
that their predictions for some important collider observables
are closer to the results using the full NLO calculation.  We note,
however, that in the context of earlier
multi-parton-interaction-model tuning studies 
undertaken by us~\cite{Skands:2010ak} and by
ATLAS~\cite{ATLAS:2012uec}, the large gluon component in
LO* PDFs has been problematic (driving very high inclusive-jet and
MPI rates).

In the context of the NNPDF fits, which we shall use for the Monash 2013
tune, the above modifications were also studied. In particular, in the
study of the NNPDF2.1LO fits in Ref.~\cite{Ball:2011uy}, 
it was found that, from the point of view of the agreement
between data and theory,
the standard LO PDFs provided as good a description as the other
possible variations, including a different value of $\alpha_s(M_Z)$, using the
one- or two-loop running or relaxing the momentum sum rule.
The different results found by previous studies could be related to the
limited flexibility in the input gluon PDFs in the
CTEQ/MSRT LO fits: indeed,
with a flexible enough parametrization such as that used in the NNPDF fits,
the differences between these theory choices can always be absorbed into the
initial condition.

Therefore, we have settled on an unmodified LO PDF set 
for the Monash 2013 tune, the NNPDF2.3 LO
set~\cite{Ball:2013hta,Carrazza:2013axa}, 
which combines the NNPDF2.1 LO PDFs with a determination
of the photon PDF and a combined QCD+QED 
evolution~\cite{Ball:2013hta,Bertone:2013vaa}.
The relevant parameter in the code is: 

{\small\noindent\begin{verbatim}
# Choice of PDF set (NNPDF2.3 LO alphaS(mZ)=0.13)
  PDF:pSet = 13
\end{verbatim}
}

Note that the NNPDF2.3 LO sets are provided for two values of the strong
coupling, $\alpha_s(M_Z)=0.119$ and $0.130$; we use the latter here. 
The sets have also been
extended in order to have a wider validity range,
in particular they are valid down to $x=10^{-9}$ and $Q=$ 1 GeV$^2$,
precisely with the motivation of using them in LO event generators.

\begin{figure}[t]
\centering
\includegraphics*[width=0.70\textwidth]{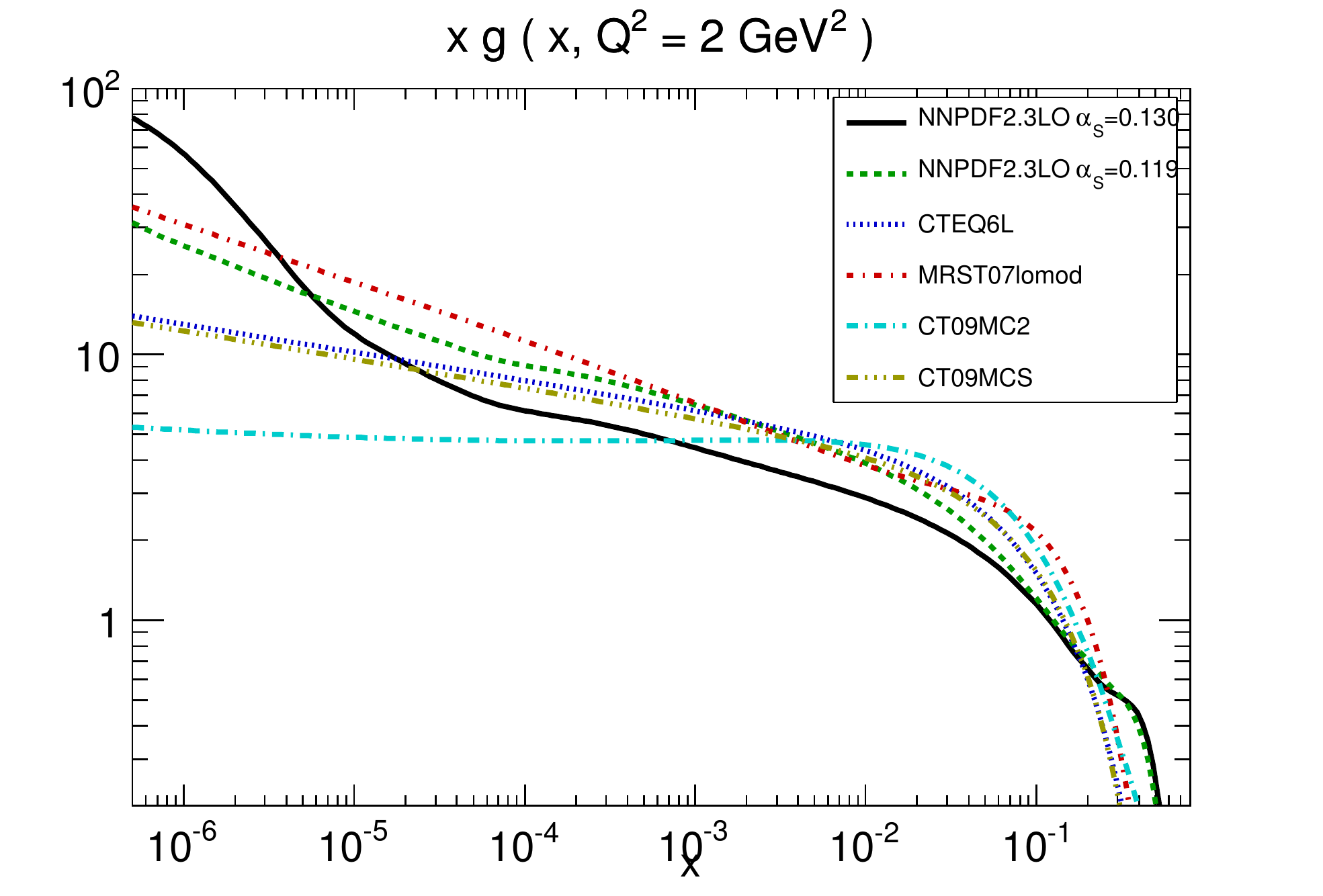}
\caption{\small Comparison of the gluon PDF at $Q^2=2$ GeV$^2$
between recent LO and LO* PDF determinations.
For NNPDF2.3LO, results for both $\alpha_s(M_Z)=0.130$ and
 $\alpha_s(M_Z)=0.119$ are shown.
 \label{fig:xg-comp2-q2-2gev2}}
\end{figure}

In Fig.~\ref{fig:xg-comp2-q2-2gev2}, we compare the 
gluon PDF $xg(x,Q^2)$ for the two NNPDF2.3 LO fits (central
values only) with other recent LO and LO* PDFs. There is a significant 
spread between the various LO/LO* PDF determinations, reflecting the substantial
theoretical uncertainties in LO fits.
These differences are further enhanced at small $x$ due to the
lack of experimental constraints in this region.
For instance, the CTEQ LO sets have a smaller gluon at small $x$ than
the other sets.
The NNPDF2.3 LO PDF set for $\alpha_s(M_Z)=0.130$ is the largest
at small $x$, beginning in $x\sim 5\times 10^{-6}$, and is smaller
than the other sets in the middle-$x$ region. 
These differences will translate into different
phase-space populations for the multi-parton-interaction processes
relevant for the tuning of event generators.

\subsection{The Strong Coupling and Total Cross Sections\label{sec:alphaS}}

\noindent For hard QCD matrix elements in \Py (including those for MPI), 
we use the same strong-coupling value as in the PDF set\footnote{The
  difference between this $\alpha_s$ value and that used for ISR/FSR
  will be discussed in \secRef{sec:isr}.}, $\alpha_s(M_Z)=0.130$: 

{\small\noindent\begin{verbatim}
  SigmaProcess:alphaSvalue = 0.130
  MultipartonInteractions:alphaSvalue = 0.130
\end{verbatim}
}

\noindent This is slightly lower than
the current default value of $\alpha_s(M_Z)=0.135$, 
which however tends to produce too high inclusive jet rates, cf.\ the
\mcplots web
site~\cite{Karneyeu:2013aha}. Reducing the $\alpha_s$ value also
for MPI seems a reasonable first assumption; it should 
result in a slightly less  ``jetty'' underlying event, with
activity shifted to lower $p_\perp$ scales. 

\begin{figure}[t]
\centering
\includegraphics*[scale=0.36]{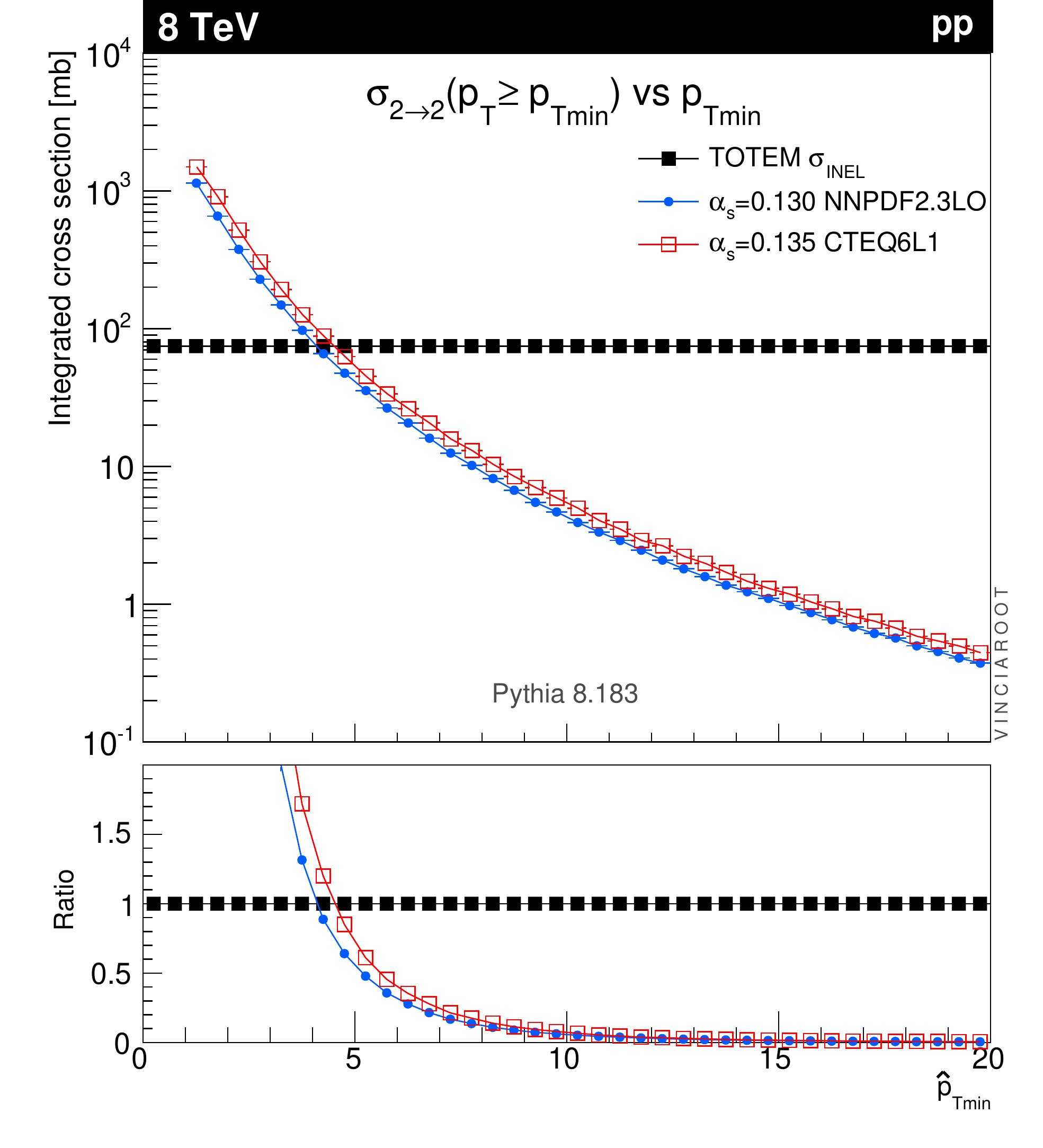}\vspace*{-2mm}
\caption{Integrated LO QCD $2\to 2$ cross section vs $p_{T\mrm{min}}$ for 8 TeV
  $pp$ collisions, with two different $\alpha_s$ and PDF choices,
  compared with the measured $\sigma_\mrm{inel}$. \label{fig:sigma8} 
}
\end{figure}

Already at this level, before considering any details of the MPI modelling,
we can show one of the main theoretical reference distributions for multi-parton
interactions: the integrated partonic QCD $2\to 2$ cross section
(integrated above some $p_{T\mrm{min}}$ scale), as a function of
$p_{T\mrm{min}}$. All that is required to compute this are the PDFs,
the value of $\alpha_s(M_Z)$, and the simple QCD LO $d\sigma_{2\to 2}$
differential cross sections. There is no
dependence on other model parameters at this stage. Due to the
$1/p_T^4$ singularity of the differential Rutherford cross section\footnote{
$t$-channel gluon exchange gives an amplitude squared proportional to
$1/t^2$, which for small $p_T$ goes to $1/p_T^4$.}, this
distribution diverges at low $p_{T\mrm{min}}$, an effect which is
further amplified by the running of $\alpha_s$ (which blows up at low scales)
and the PDFs (which become large at low $x$). 
MPI models reconcile the calculated divergent parton-parton cross
section with the measured (or parametrized) total inelastic
hadron-hadron cross section, by interpreting the 
divergence as a consequence of each hadron-hadron collision containing several
parton-parton ones, with 
\begin{equation}
\left<n\right>_\mrm{MPI}(p_T \ge p_{T\mrm{min}}) \approx \frac{\sigma_{2\to
    2}(p_T\ge p_{T\mrm{min}})}{\sigma_\mrm{inel}}~. 
\end{equation}
Note that there is some ambiguity whether to normalize to the total
inelastic cross section, or to a diffraction-subtracted smaller
number. To be conservative, we show a comparison to the full
$\sigma_\mrm{inel}$ in \figRef{fig:sigma8}.
We compare two different $\alpha_s$ and PDF settings, corresponding to the
choices made in the Monash 2013 tune (filled blue dots) and the
current default 4C tune (open red squares), to the highly precise
measurement of the total inelastic cross section at 8 TeV 
by the TOTEM collaboration~\cite{Antchev:2013paa}, 
\begin{equation}
\sigma_{\mrm{inel}}(8~\mrm{TeV}) = (74.7\pm
1.7)~\mrm{mb}.
\end{equation}
For reference, the value obtained from the default Donnachie-Landshoff
and Schuler-Sj\"ostrand 
parametrizations currently used in PYTHIA ($\propto s^{0.0808}$ at high
energies~\cite{Donnachie:1992ny,Schuler:1993td}) is 73~mb, consistent
with the TOTEM 
measurement\footnote{We note, however, that the value obtained for the
  8-TeV elastic cross section in PYTHIA is 20~mb, whereas the value
  measured by TOTEM is $27.1\pm1.4$~mb~\cite{Antchev:2013paa}. While 
this discrepancy does not influence the normalization or modelling of
inelastic events and hence is a non-issue in that context, an update
of the total cross-section expressions in PYTHIA may be timely in the
near future, e.g.\ using the updated Donnachie-Landsgoff analysis
in~\cite{Donnachie:2013xia}. We also note that the decomposition of
the inelastic cross section into individual non-diffractive and
diffractive components, which follows
Schuler-Sj\"ostrand~\cite{Schuler:1993td}, may also be due for an update.}.   
The fact that the
curves cross each other at a value of $p_{T\mrm{min}}\sim 5~\mrm{GeV}$
means that we can make a relatively model-independent statement that 
every inelastic event will, on average, contain at least one 5-GeV
partonic subprocess. (This value agrees with that found by earlier
analyses~\cite{Sjostrand:1987su,oai:arXiv.org:hep-ph/0402078,oai:arXiv.org:0806.2949}). The
corresponding $p_{T\mrm{min}}$ scales at $\sqrt{s}=200$ or $900$ GeV 
are just 1 -- 2 GeV (see plots included \appRef{app:sigma}), hence
the expected presence of ``semi-hard'' partonic substructure, at a
scale of 5 GeV, in min-bias events is a qualitatively new feature at
LHC energies; for completeness the corresponding scale at the Tevatron
was about 2.5 GeV~\cite{oai:arXiv.org:hep-ph/0402078}. The  
plots in  \appRef{app:sigma} also show extrapolations to higher
energies. At 100 TeV, we expect the partonic cross section to saturate the
total inelastic one at a $p_{T}$ scale of 10 GeV.

\subsection{Initial-State Radiation and Primordial kT\label{sec:isr}}
We follow the approach of the Perugia tunes 
of \Py~6~\cite{Skands:2010ak,Sjostrand:2006za} and use the same 
$\alpha_s(M_z)$ value for initial-state radiation 
as that obtained for final-state radiation. That is, we use one-loop
running with $\alpha_s(M_Z) = 0.1365$ for both FSR and ISR.   
This choice is made essentially to facilitate matching applications,
see e.g.~\cite{Cooper:2011gk}. Nonetheless, we emphasize that we 
do not regard this choice as mandatory, for the following
reasons. 

Firstly, since each collinear direction is associated with
its own singular (set of) diagram(s), one can consistently
associate at least the collinear radiation components with separate
well-defined $\alpha_s$ values without violating gauge invariance. 
Secondly, while the LO splitting
functions for ISR and FSR 
  are identical, they differ at higher
  orders (beyond the shower accuracy), and there are important
  differences between the collinear (DGLAP) evolution performed in PDF
  fits and the (coherent, momentum-conserving) evolution performed by
  parton showers; these differences could well 
  be desired to be reflected in slightly different effective scale
  choices for ISR with respect to FSR, one possibility then being to 
  absorb this  in a
  redefinition of the effective value of $\alpha_s(M_Z)$. Thirdly,
  and perhaps most importantly, while   
we agree that maintaining
  separate $\alpha_s$ values (equivalent to making slightly
  different effective scale choices) for ISR and FSR is
  ambiguous for wide-angle radiation, we emphasize that merely using the same
  $\alpha_s(M_Z)$ value for the two algorithms does \emph{not} remove
  this fundamental ambiguity. This is because, in the context of a
  shower algorithm, the value of the renormalization scale depends upon 
  \emph{which} parton is branching, and
  that assignment is fundamentally ambiguous outside the collinear limit. 
  For instance, an emitted gluon with a certain
  momentum will have a different $p_\perp$ with respect to
  the beam (ISR), than it will with respect to a final-state parton (FSR), and
  hence the argument of $\alpha_s$, typically taken to be
  proportional to some measure of $p_\perp$, will be different, depending on who
  the emitter was. This effect is present in all parton-based shower algorithms
   and is \emph{not} cured by arbitrarily setting $\alpha_s(M_Z)$ to
   be the same for ISR and FSR. Using the same $\alpha_s(M_Z)$ for
   both ISR and FSR (as we do here) should therefore not be perceived
   of as being more rigorous than not doing so; it is a choice we make
   purely for convenience. (The situation is slightly better in 
    antenna-based showers~\cite{Gustafson:1987rq,Giele:2007di,Ritzmann:2012ca},
    where there is 
    no distinction 
  between radiator and recoiler in the soft limit, hence the
  renormalization-scale choice is unique, at leading colour.)

The difference between the value $\alpha_s(M_Z) = 0.130$ used for QCD
matrix elements (and in the PDF evolution) and that used for ISR/FSR
may be interpreted as follows. The former is specified 
in the $\overline{\mrm{MS}}$ scheme, while the effective ISR/FSR one should
presumably be interpreted in something closer to the so-called MC
(CMW) scheme~\cite{Catani:1990rr}.  
Taking the translation into account (corresponding 
roughly to a factor 1.6 on the value of
$\Lambda_\mrm{QCD}$), the PDF value comes out slightly lower than the 
shower one.  
Given the ambiguities caused by   
the non-identical nature of PDF and shower evolutions, however, we
nonetheless regard this small difference as acceptable, in particular
since the shower evolution is intrinsically somewhat slower than the PDF one,
due to coherence effects and a more restrictive phase space that are
not taken into account in the PDF evolution. For
completeness, we note that the renormalization scale for ISR in
\Py\ is~\cite{Sjostrand:2004ef}:
\begin{equation}
\mbox{ISR:~~~}\mu_R^2 = p_{\perp\mrm{evol}}^2 = (1-z)Q^2~,
\end{equation}
with $Q^2=-p^2$ the virtuality of the (spacelike) emitting
parton (defined so that $Q^2$ is positive; note that $Q^2=-p^2 +
m_0^2$ is used for $g\to Q\bar{Q}$ splittings) and $z$ the energy 
fraction appearing in the DGLAP splitting kernels, $P(z)$, which in
\Py\ is defined as the ratio of $\hat{s}$ values before and after the
branching in question. (To
estimate the shower uncertainties associated with this choice of
renormalization scale, we recommend using $\ln(\mu_R^2) \pm \ln(2)$,
corresponding to a factor $\sqrt{2}$ variation of $\mu_R$, similarly
to what was recommended for final-state radiation in
\secRef{sec:hadronization}.)  

The remaining settings for the
ISR evolution are taken over from the previous default tune. 
The relevant parameters in the code are:

{\small\noindent\begin{verbatim}
# ISR: Strong Coupling (same as FSR)
  SpaceShower:alphaSvalue   = 0.1365
  SpaceShower:alphaSuseCMW  = off
  SpaceShower:alphaSorder   = 1
# ISR: Infrared Cutoff (fixed value at 2.0 GeV)
  SpaceShower:samePTasMPI   = off 
  SpaceShower:pT0Ref        = 2.0
  SpaceShower:ecmRef        = 7000.0  
  SpaceShower:ecmPow        = 0.0     
# ISR: Coherence and Spin Correlations
  SpaceShower:rapidityOrder = on
  SpaceShower:phiPolAsym    = on 
  SpaceShower:phiIntAsym    = on  
\end{verbatim}
}

\noindent We choose a fixed ISR cutoff, rather than one that scales
with CM energy, in order to maintain a correspondence between the ISR
cutoff and the ``primordial $k_T$'' component which 
parametrizes additional non-perturbative and/or unresolved motion in
the beam remnant. This latter component does not scale with the CM
energy (though it may depend on the $Q^2$ scale of the hard process),
hence we believe it is most consistent to keep the ISR cutoff fixed as
well. Since we choose an ISR cutoff of $2\GeV$ (see the ISR parameter list
above), there are no
perturbative (ISR) corrections generated below that scale, and soft
processes involving momentum transfers less than $2\GeV$ do not
receive any perturbative corrections at all. To represent the combined
effects of unresolved radiation and non-perturbative Fermi motion, we
add a Gaussian-distributed primordial-$k_T$ component to the partons
extracted from the proton at the low-$Q$ end of the ISR cascade. In
the Monash tune, the
width of the Gaussian starts at $0.9\GeV$, for an infinitely soft
process, and gradually rises to an asymptotic value of $1.8\GeV$, with
a characteristic ``half-scale'' of $Q = 1.5\GeV$:

{\small\noindent\begin{verbatim}
  BeamRemnants:primordialKTsoft = 0.9
  BeamRemnants:primordialKThard = 1.8
  BeamRemnants:halfScaleForKT   = 1.5
\end{verbatim}
}

\begin{figure}[tp]
\centering
\includegraphics*[scale=\dscale]{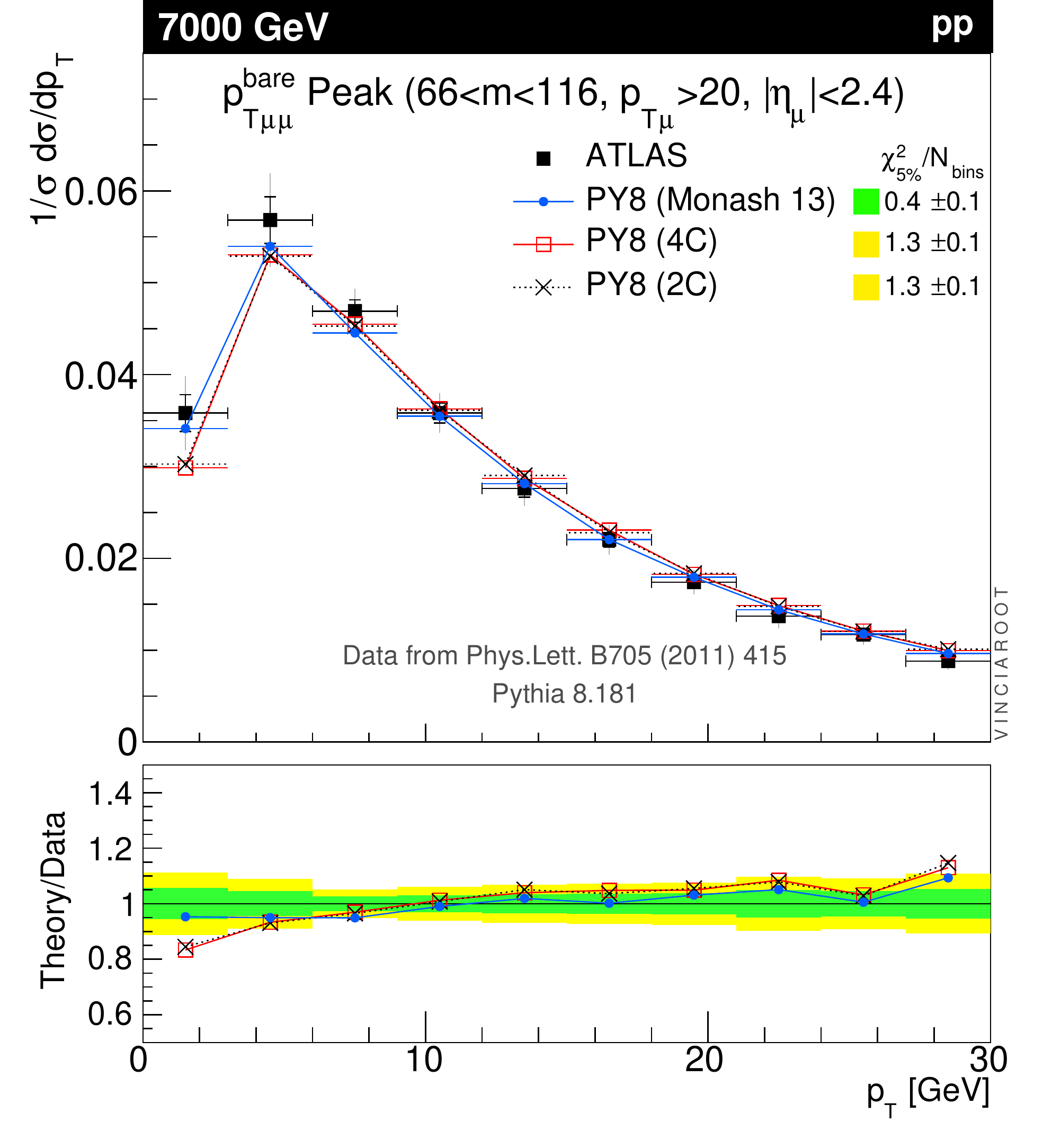}
\includegraphics*[scale=\dscale]{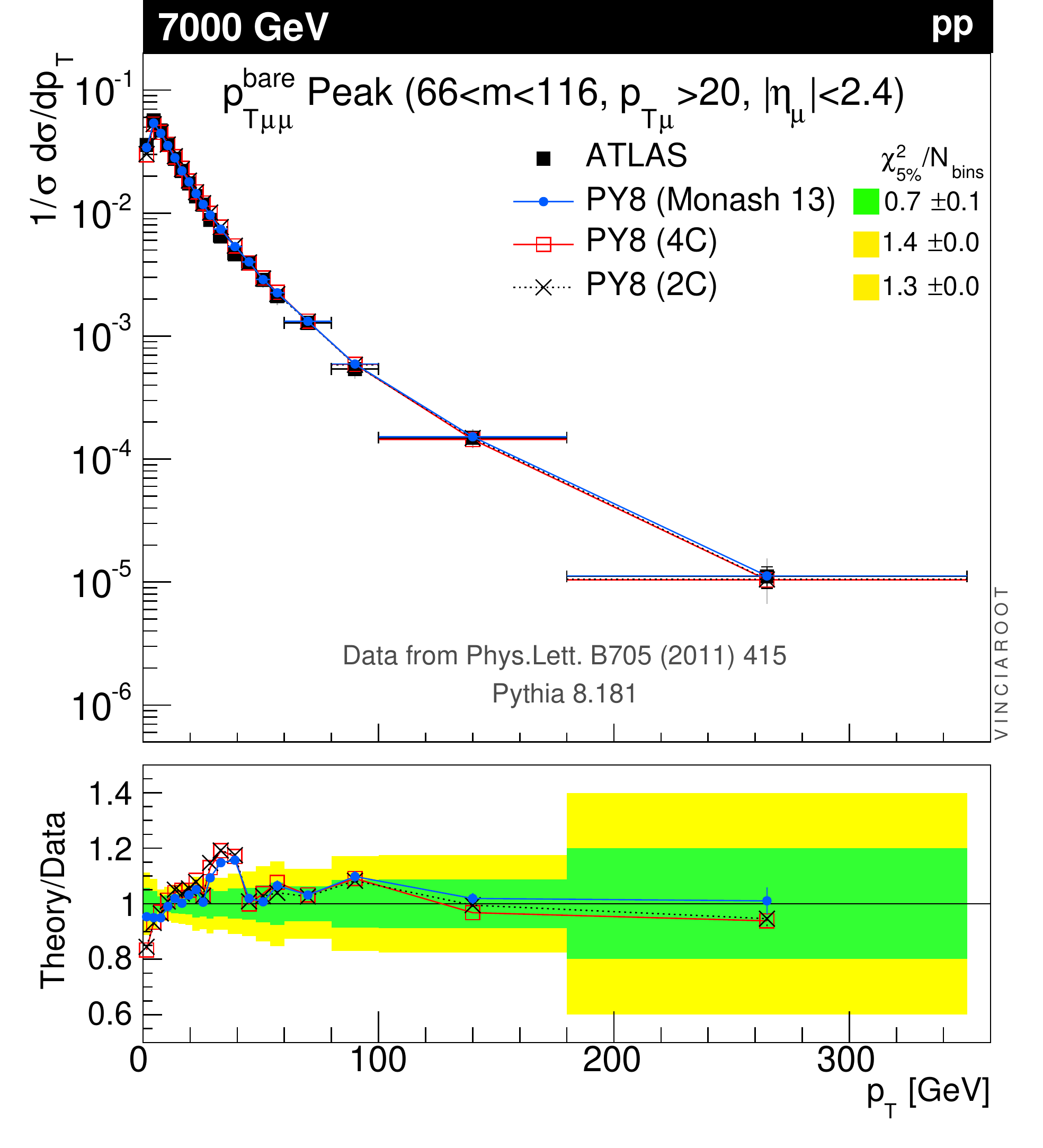}\\
\includegraphics*[scale=\dscale]{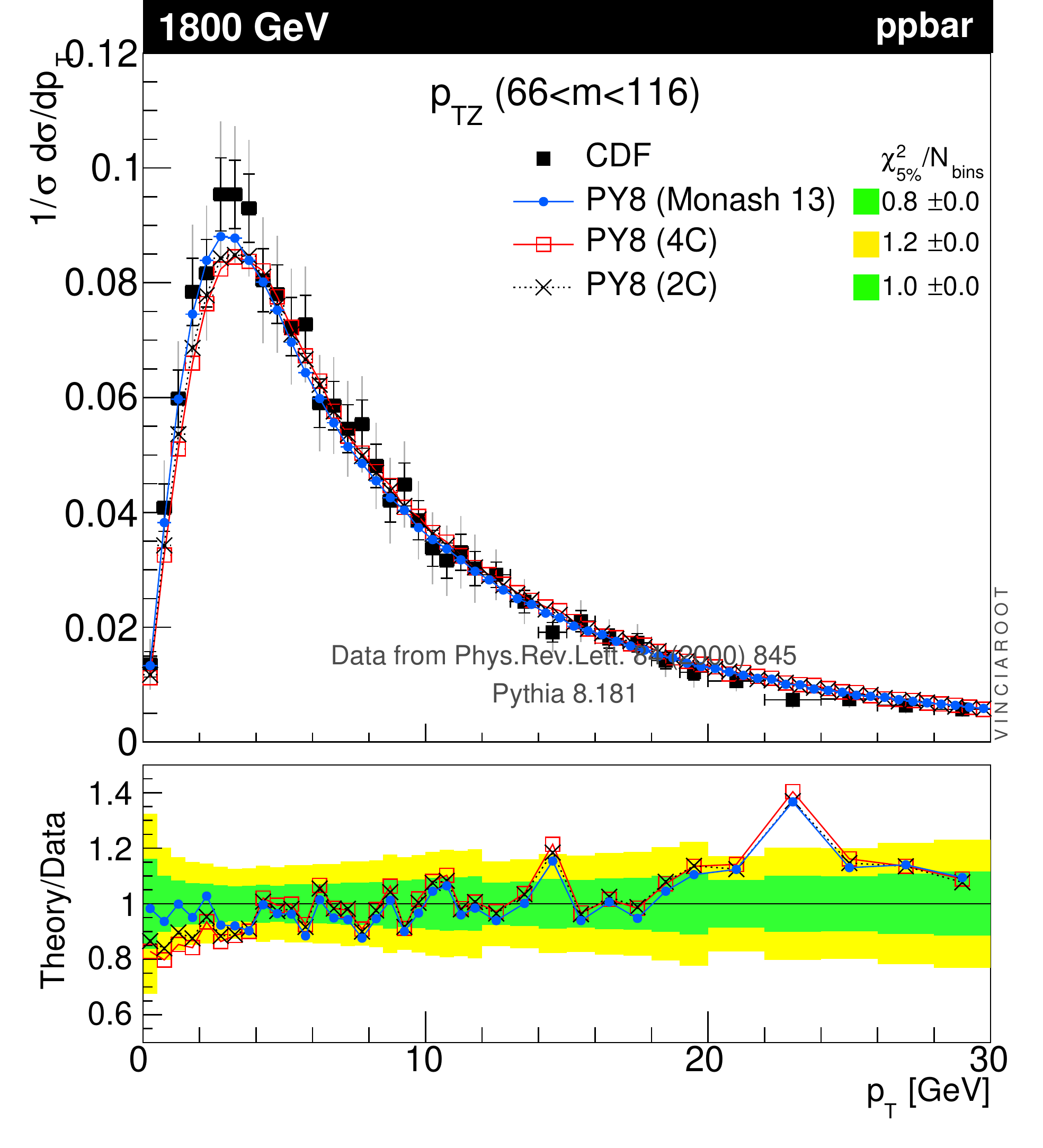}
\includegraphics*[scale=\dscale]{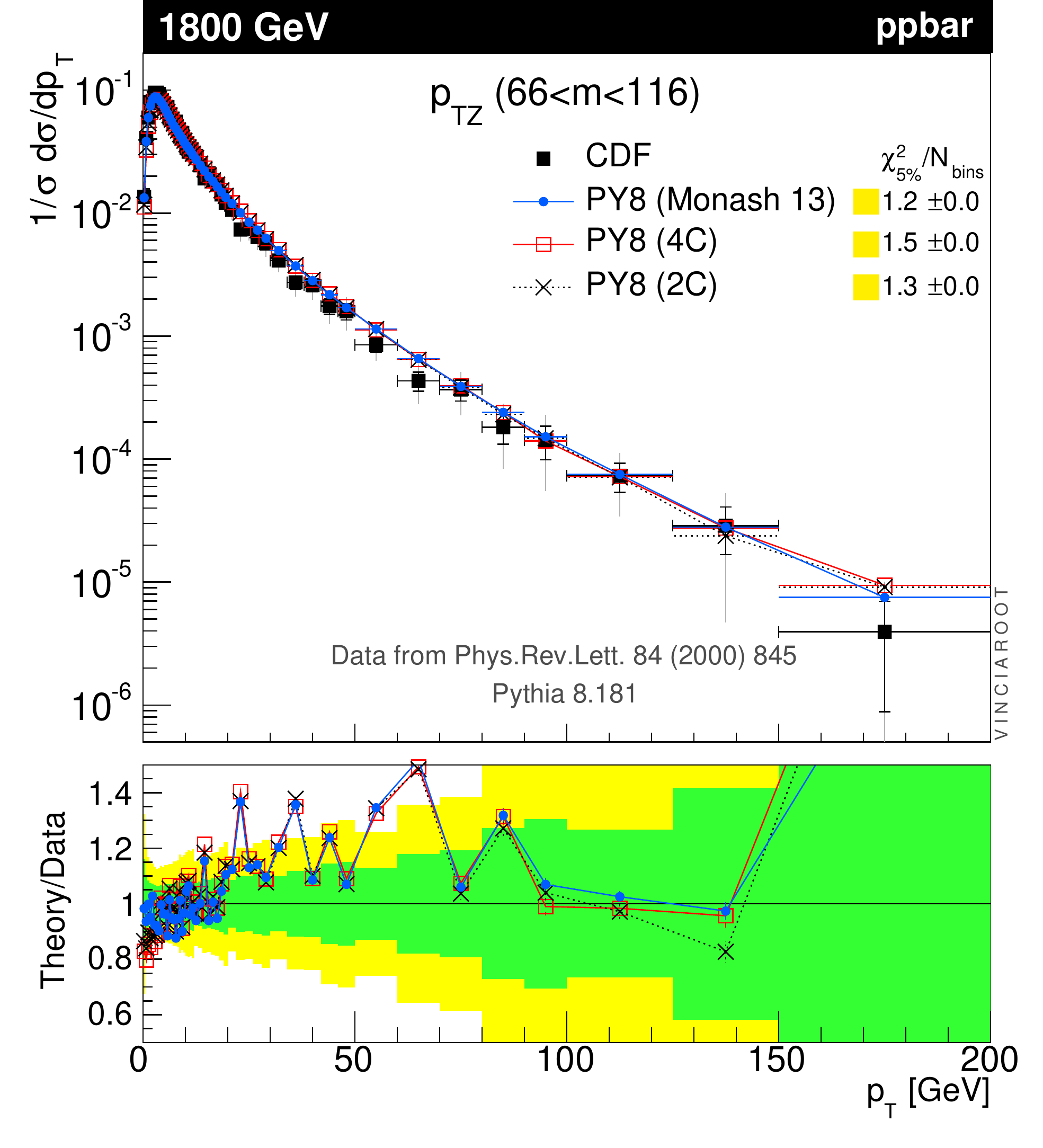}\vspace*{-2mm}
\caption{The peak (left) and tail (right) of the $Z$ $p_\perp$
  distribution, as measured at 7 TeV (using ``bare'' muon
  pairs)~\cite{Aad:2011gj} and 1.8 TeV (corrected to unphysical 
 generator-level, see~\cite{hesketh})~\cite{Affolder:1999jh}.
\label{fig:ZpT}}
\end{figure}
\noindent The half-scale of $Q=1.5\GeV$ was chosen in order to prevent
the primordial-$k_T$ component from generating momentum kicks larger 
than that of the ``hard'' process, for low-scale processes. 
The asymptotic value of $1.8\GeV$ was chosen by comparing to
the $p_\perp$ spectrum of the lepton pair in $pp \to Z \to \ell^+\ell^-$ events
measured by the ATLAS and CDF
experiments~\cite{Aad:2011gj,Affolder:1999jh}. Note that \Py's parton
shower is automatically corrected to reproduce the full LO
$Z+\mrm{jet}$ matrix element~\cite{Miu:1998ju,Sjostrand:2004ef}, in a
manner highly similar to (but predating) that of \Pw~\cite{Frixione:2007vw}.  
Our value for primordial $k_T$ ($1.8\GeV$) is slightly lower than the current
default ($2\GeV$) and gives a better 
agreement with the low-$p_\perp$ part of the lepton-pair $p_\perp$ spectrum,
as is illustrated in \figRef{fig:ZpT}, for 7 TeV (top row) and 1800
GeV (bottom row) $pp$ ($p\bar{p}$) collisions. Note that the left-hand
panes show a ``closeup'' of the peak 
region at low $p_\perp$ while the right-hand panes show the full
spectrum. (Note also that these $p_\perp$ spectra are normalized to
unity, so the normalization of the inclusive $Z$ cross section drops out.) 

In the ATLAS spectra, the feature around $p_\perp^{\mu\mu} \sim
35\GeV$ is repeated by  
all MCs in the comparisons shown on the
\href{http:mcplots.cern.ch}{MCPLOTS} web site~\cite{Karneyeu:2013aha},
hence we regard it as an artifact of the data. We note however 
that there is a tendency for \Py\ to overshoot the data between
$p_\perp$ values of 
roughly $20\GeV$ to  $100\GeV$, at both CM energies. This is an
interesting region intermediate between low-$p_\perp$ bremsstrahlung 
and high-$p_\perp$ $Z$+jet processes, which will be particularly
relevant to reconsider in the context of matrix-element corrections at
the ${\cal O}(\alpha_s^2)$ level and beyond~\cite{Hamilton:2012rf}. 

\subsection[Min-Bias and Underlying Event]{Minimum Bias and Underlying Event \label{sec:uemb}}

The Monash 2013 tune has been constructed to give a reasonable
description of both soft-inclusive (``minimum-bias'') physics as well
as underlying-event (UE) type observables. The difference between the two
is sensitive to the shape of the hadron-hadron overlap profile in
impact-parameter space (the UE probes the most ``central'' collisions 
while min-bias (MB) is more inclusive) and to the modeling of colour
reconnections (CR). 
Most previous tunes, including the current default
Tune 4C~\cite{Corke:2010yf}, have used a Gaussian
assumption~\cite{Sjostrand:1987su} for the transverse matter
distribution, but this appears to 
give a slightly too low UE level (for a given average MB level).

For the Monash tune, we have chosen a slightly more peaked transverse
matter profile~\cite{Sjostrand:2004ef}, thus generating a relatively
larger UE for the same 
average MB quantities. We note, however, that there are still several
indications that the dynamics are not well understood, in particular
when it comes to very low multiplicities (overlapping with
diffraction), very high multiplicities (e.g., the so-called CMS
``ridge'' effect~\cite{Khachatryan:2010gv}), and to 
identified-particle spectra (e.g., possible modifications 
by re-scattering~\cite{Corke:2009tk}, string boosts from colour
reconnections~\cite{Ortiz:2013yxa}, or other collective effects).    

For the 7-TeV reference energy we focus on here (energy scaling will
be studied in the following subsection), the relevant
parameters in the code are:

{\small\noindent\begin{verbatim}
# Hadron transverse mass overlap density profile
  MultipartonInteractions:bProfile = 3 
  MultipartonInteractions:expPow = 1.85
# IR regularization scale for MPI and energy scaling
  MultipartonInteractions:pT0Ref = 2.28
  MultipartonInteractions:ecmRef = 7000.
  MultipartonInteractions:ecmPow = 0.215
\end{verbatim}
}

\begin{figure}[tp]
\centering
\includegraphics*[scale=\dscale]{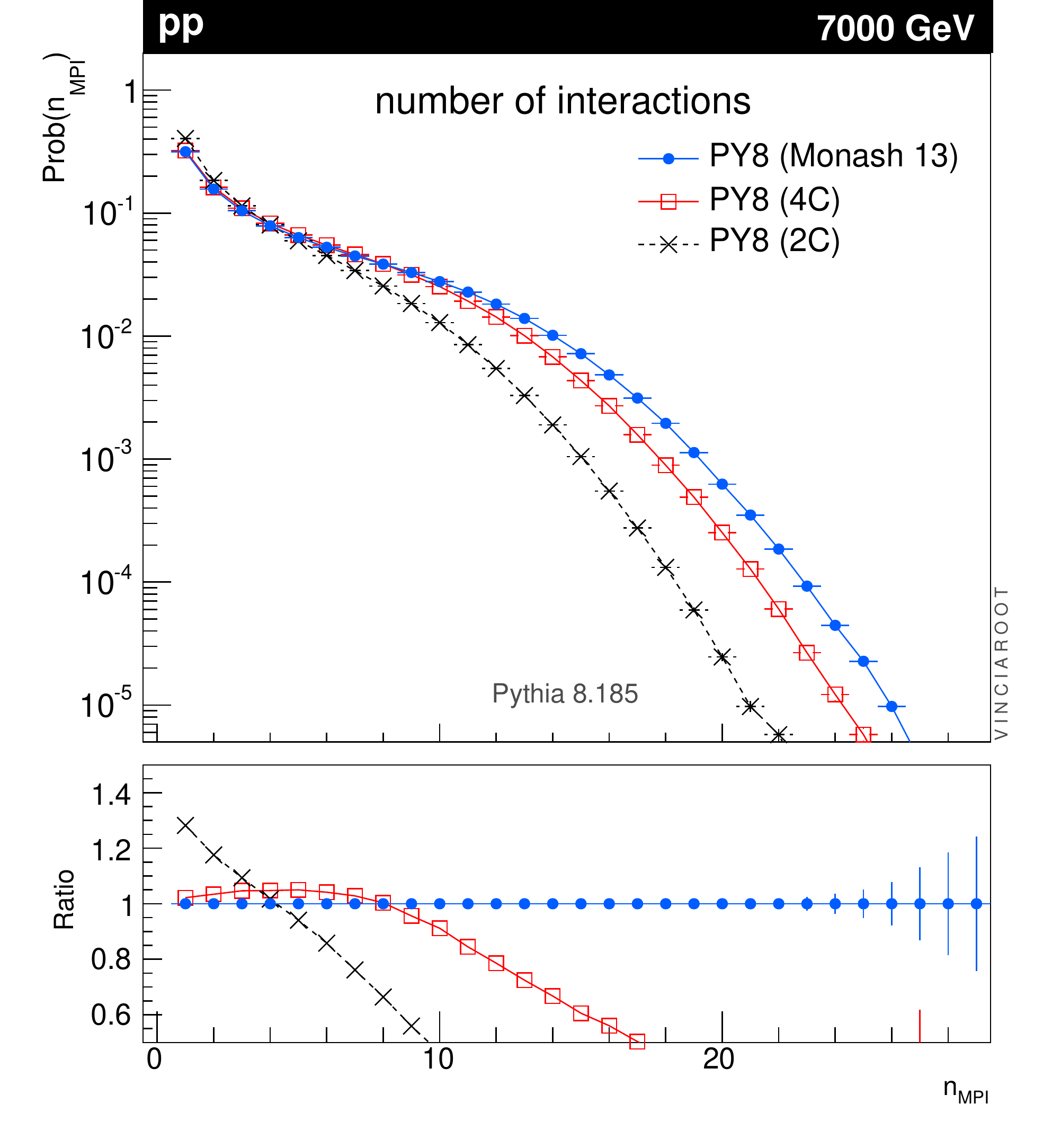}\vspace*{-2mm}
\caption{$pp$ collisions at 7 TeV. Number of MPI in inelastic events.
\label{fig:nmpi}}
\end{figure}
The slightly more peaked matter distribution, combined with a
relatively low $p_{\perp 0}$ value, produces an intrinsically
broader distribution in the number of parton-parton interactions
(MPI), illustrated by the theory-level plot in \figRef{fig:nmpi}. 

\begin{figure}[tp]
\centering
\includegraphics[scale=\dscale]{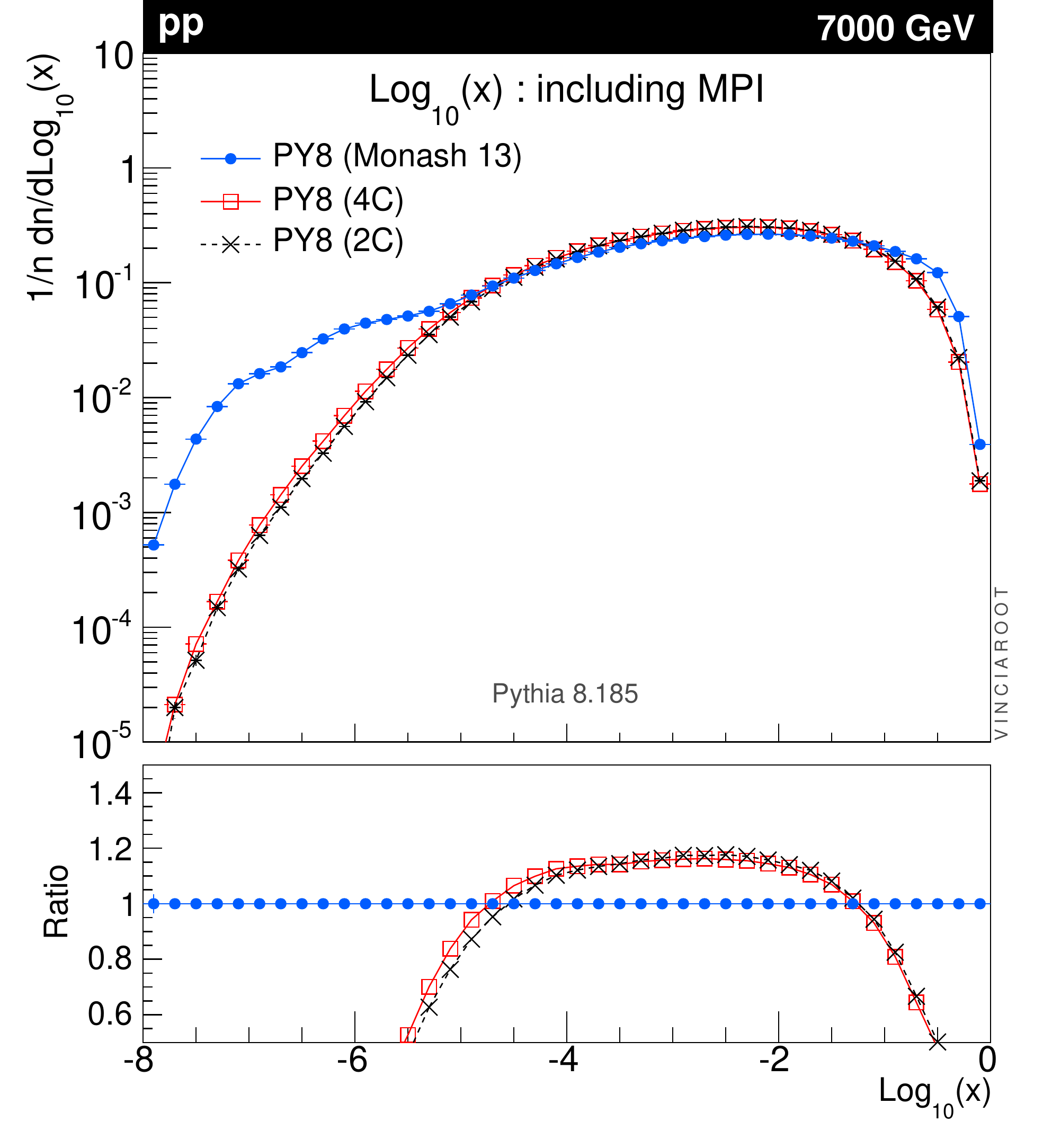}
\includegraphics[scale=\dscale]{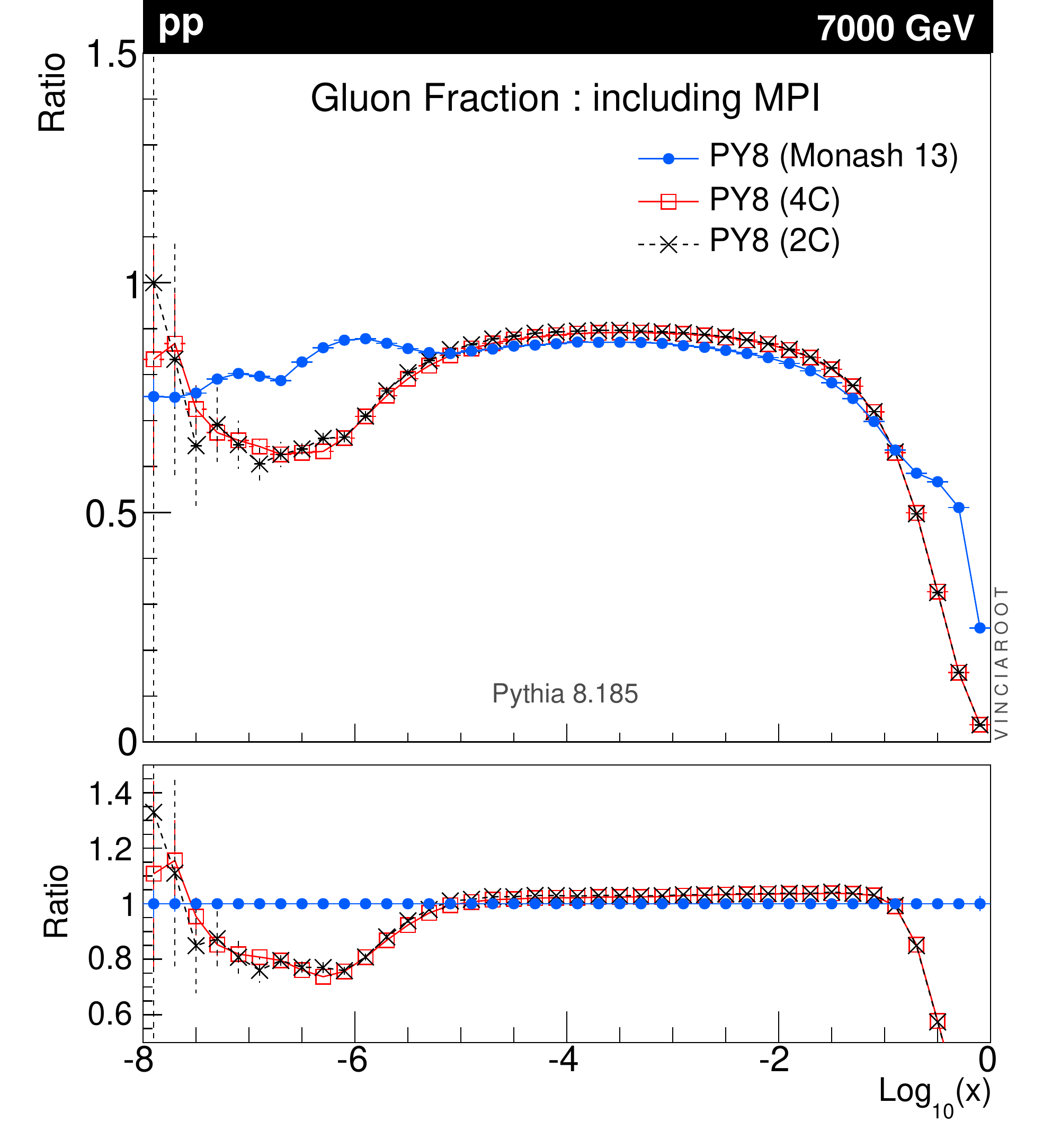}\\
\includegraphics[scale=\dscale]{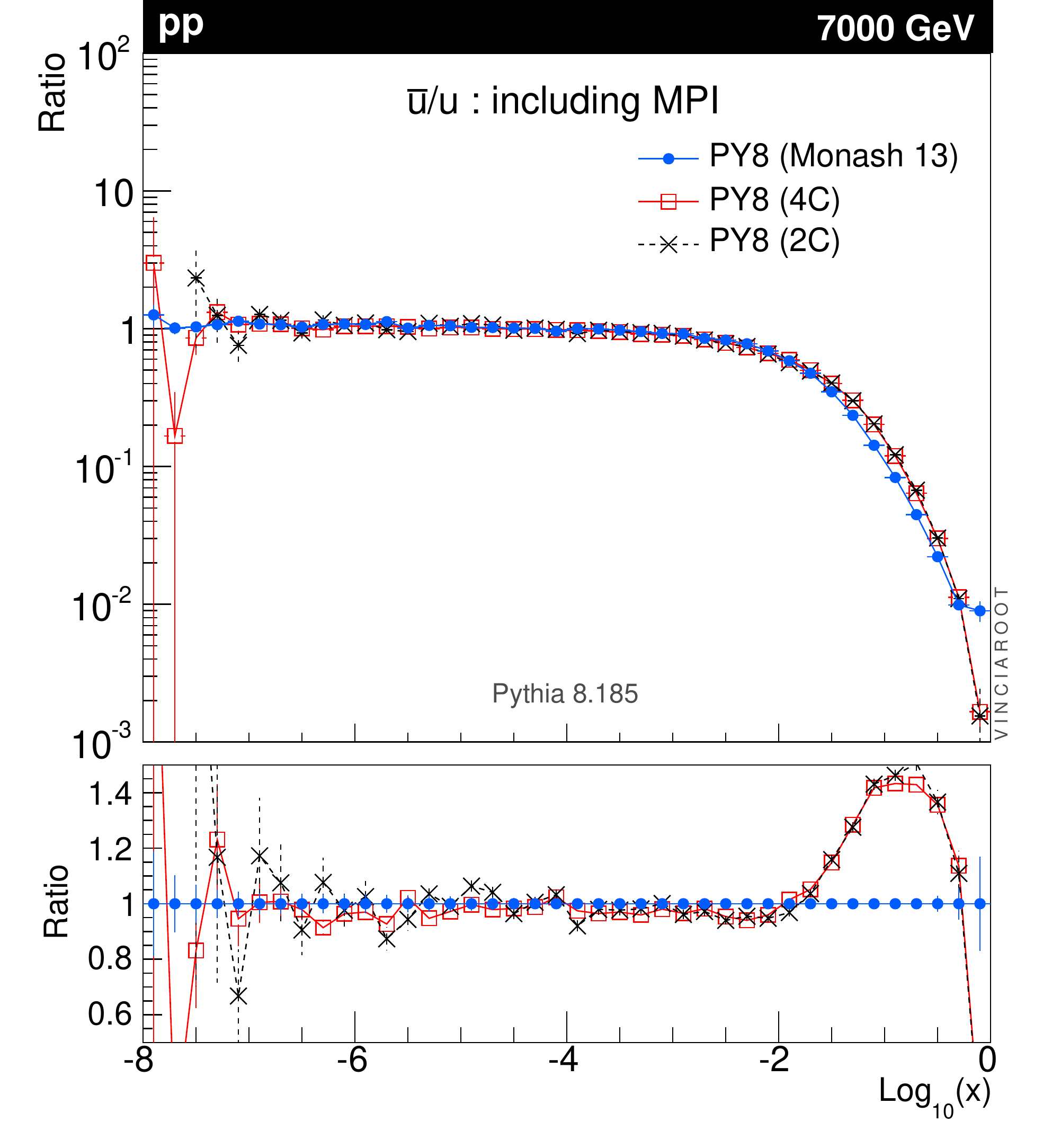}
\includegraphics[scale=\dscale]{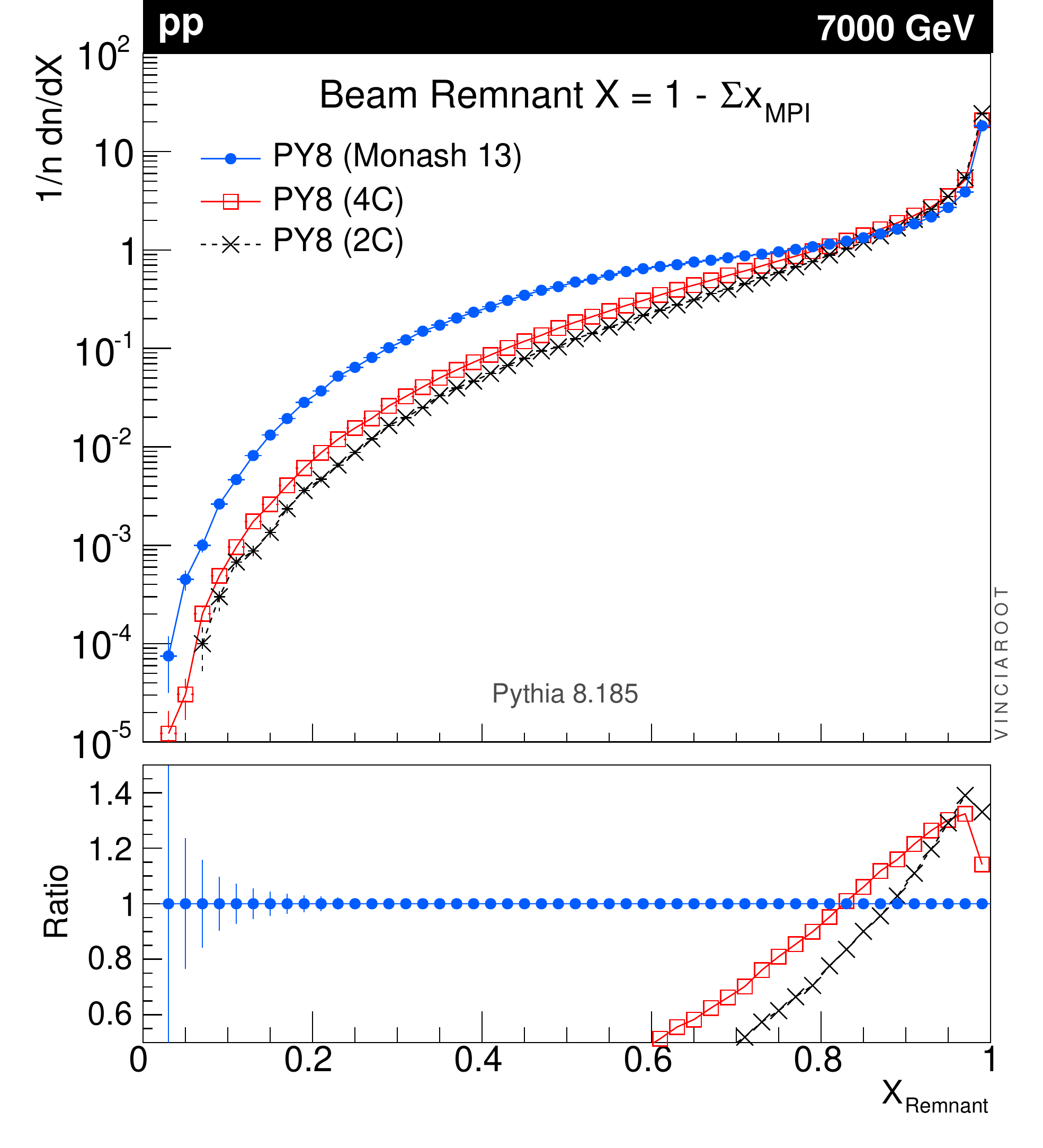}
\caption{PDF sampling by MPIs in inelastic non-diffractive $pp$
  collisions at 7 TeV. {\sl Top Left:} the $x$
  distribution of all MPI initiators (including the hardest
  scattering). {\sl Top Right:}  the fraction 
  of MPI initiators which are gluons, as a function of $x$. {\sl
    Bottom Left:} the $\bar{u}/u$ ratio. {\sl Bottom Right:} 
the distribution of the amount of $x$
  left in the beam remnant, after MPI (note: linear scale in
  $x$). \label{fig:pdfplots}} 
\end{figure}

The sampling of the PDFs by MPI initiators (including also the hardest
scattering in our definition of ``MPI''), as a function of parton $x$
values, is illustrated in \figRef{fig:pdfplots}, for the three tunes
considered in this paper.  
The top left-hand pane shows the most inclusive quantity, simply the
probability distribution of the $x$ value of all MPI initiators
(again, we emphasize that we include the hardest-interaction
initiators in our definition of ``MPI'' here), on a logarithmic $x$ axis. 
Here we see that the NNPDF tune has a
harder distribution both at large and small $x$ as compared to the
CTEQ6L1 tunes. The effect is particularly marked at small
$x$. Since MPI is dominated by the low-$Q$ gluon PDF,
cf.~\figRef{fig:pdf_xg_log_2gev2}, this is 
precisely what we expect; the shape of the distribution of sampled $x$
values follows that of the PDFs themselves. 
Indeed, the NNPDF2.3 gluon is harder than
the CTEQ6L one for $x > 0.2$ and for $x < 10^{-5}$.

The relative dominance of the gluon PDF is illustrated by the bottom
right-hand pane of \figRef{fig:pdfplots}, showing the gluon fraction
(relative to all MPI initiators) as a function of
$\log_{10}(x)$. Below $x\sim 0.1$, the NNPPDF sampling is 80\%
gluon-dominated, and the gluon fraction is higher than in CTEQ6L1 for
both very small $x<10^{-5}$ as well as for very large $x>0.2$.

A further consistency check is provided by the $\bar{u}/u$ ratio,
shown in the bottom left-hand pane of \figRef{fig:pdfplots}. This is
consistent with unity (as expected for sea quarks) in the entire
small-$x$ region $x<10^{-2}$. The valence bump appears to be slightly
more pronounced in the NNPDF tune (relative to the sea), since the
$\bar{u}/u$ ratio drops off more quickly above $10^{-2}$. This trend
persists until the very highest bin, at $x\sim 1$, where the
experimental uncertainties are extremely large. The CTEQ6L1
parametrization there forces the $\bar{u}$ PDF to zero, while
the NNPDF parametrization allows for a small amount of 
$\bar{u}$ to remain even at the largest $x$ values, though we note
that they are still outnumbered by $u$ quarks at a level of hundred-to-one. 

The last pane of \figRef{fig:pdfplots} shows the amount of $x$
remaining in the beam remnant, after all MPI (including both the
hardest interaction and additional MPI) have been
considered, i.e., 
\begin{equation}
X_\mrm{rem} = 1 - \sum_{i\in\mrm{MPI}} x_i ~.
\end{equation}
Note the linear scale in $x$ on this plot, and the highly
logaritmic axis. In the vast majority of cases, the beam remnant thus still
retains over 90\% of the initial hadron energy. But there is a class
of events, at the level of $10^{-4}$ or $10^{-5}$ of the total cross
section (depending on the
tune), in which the beam remnant retains less than 10\% of the incoming
hadron energy. Experiments studying the amount and distribution of
forward scattered energy in particular may be able to tell us about
whether this class of events, which we term ``Catastrophic Energy Loss''
events, really exists, and at what level. Note that these events are
typically \emph{not} caused by a single hard partonic scattering process,
due to the high penalty associated with accessing PDFs in the region
$x>0.5$. Rather, they are an intrinsic consequence of MPI. A
straightforward extrapolation, requiring a catastrophic energy loss on
both sides of the event --- more than 90\% of the energy scattered out of
\emph{both} beams, which we term ``Total Inelastic Scattering'' ---
may occur at a level of $10^{-10} - 10^{-8}$ of the cross
section, or between 10 - 1000~pb (though we of course only have \Py's
word for it). This would be an extremely
interesting part of hadron-hadron collision physics to
study, very far from the single-interaction dominated limit, and hence
potentially very sensitive to the existence of possible collective effects. 
Designing efficient triggers for this class of events would be a great 
accomplishment.

\begin{figure}[tp]
\centering
\includegraphics*[scale=\dscale]{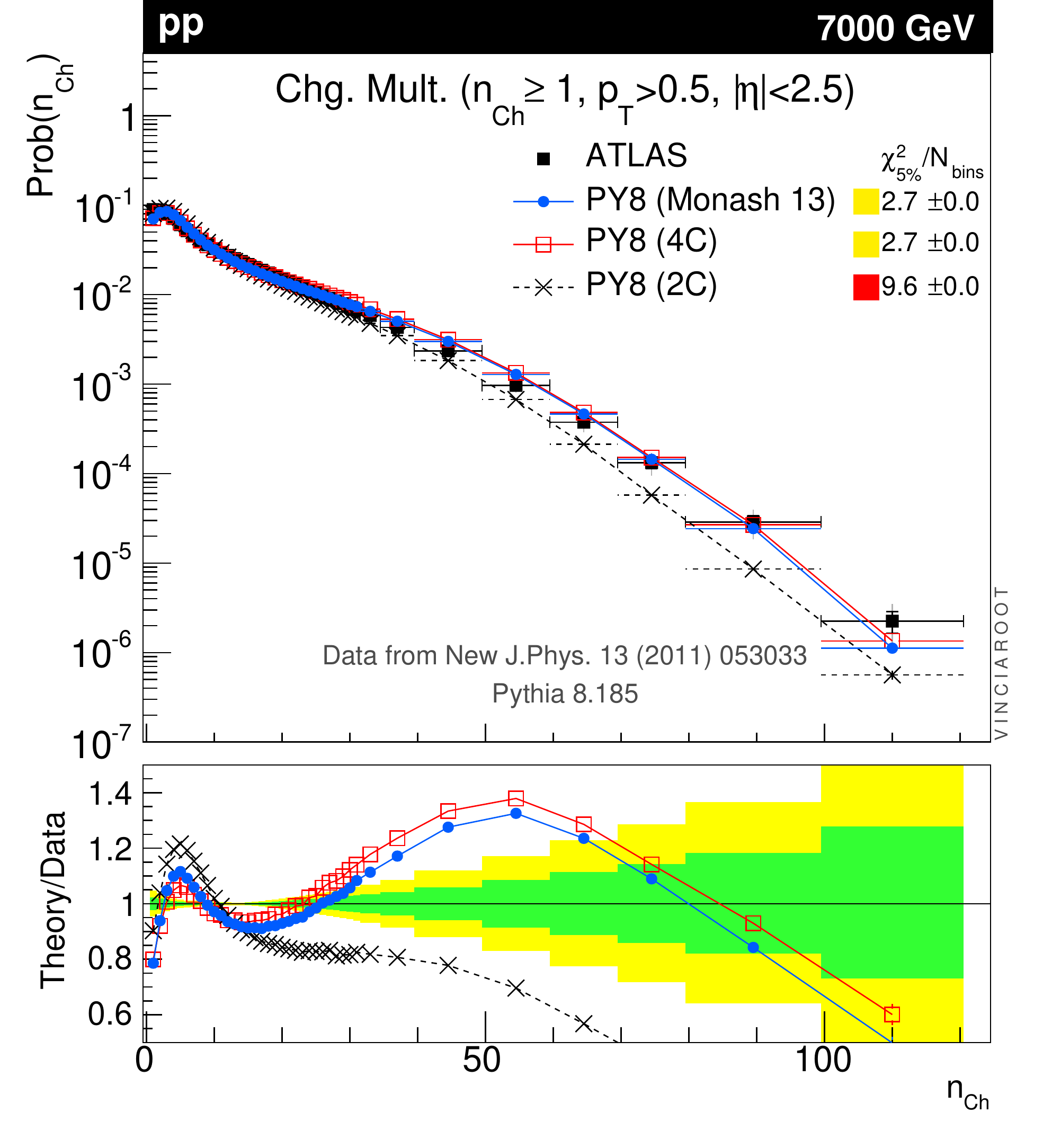}
\includegraphics*[scale=\dscale]{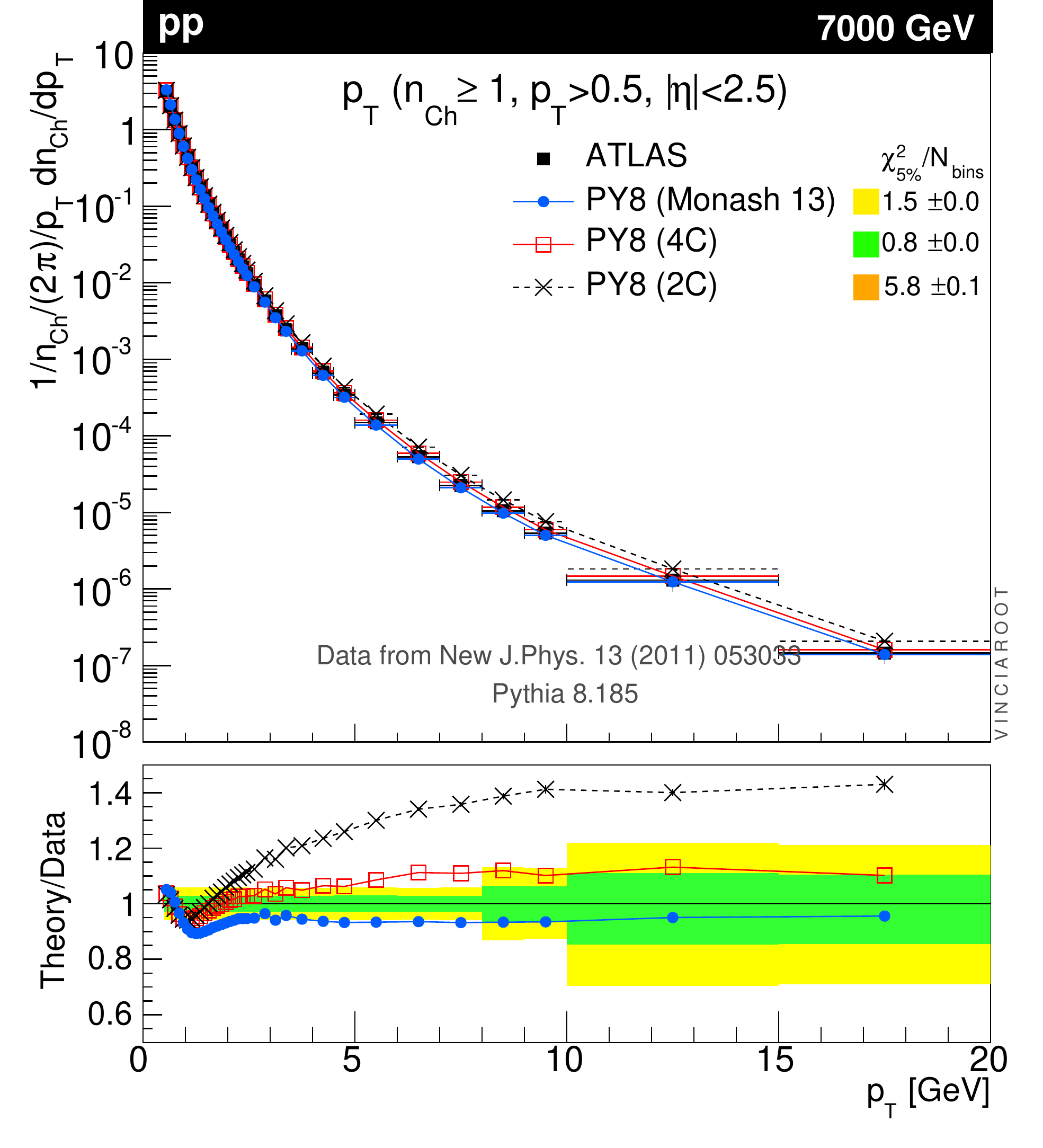}\\
\includegraphics*[scale=\dscale]{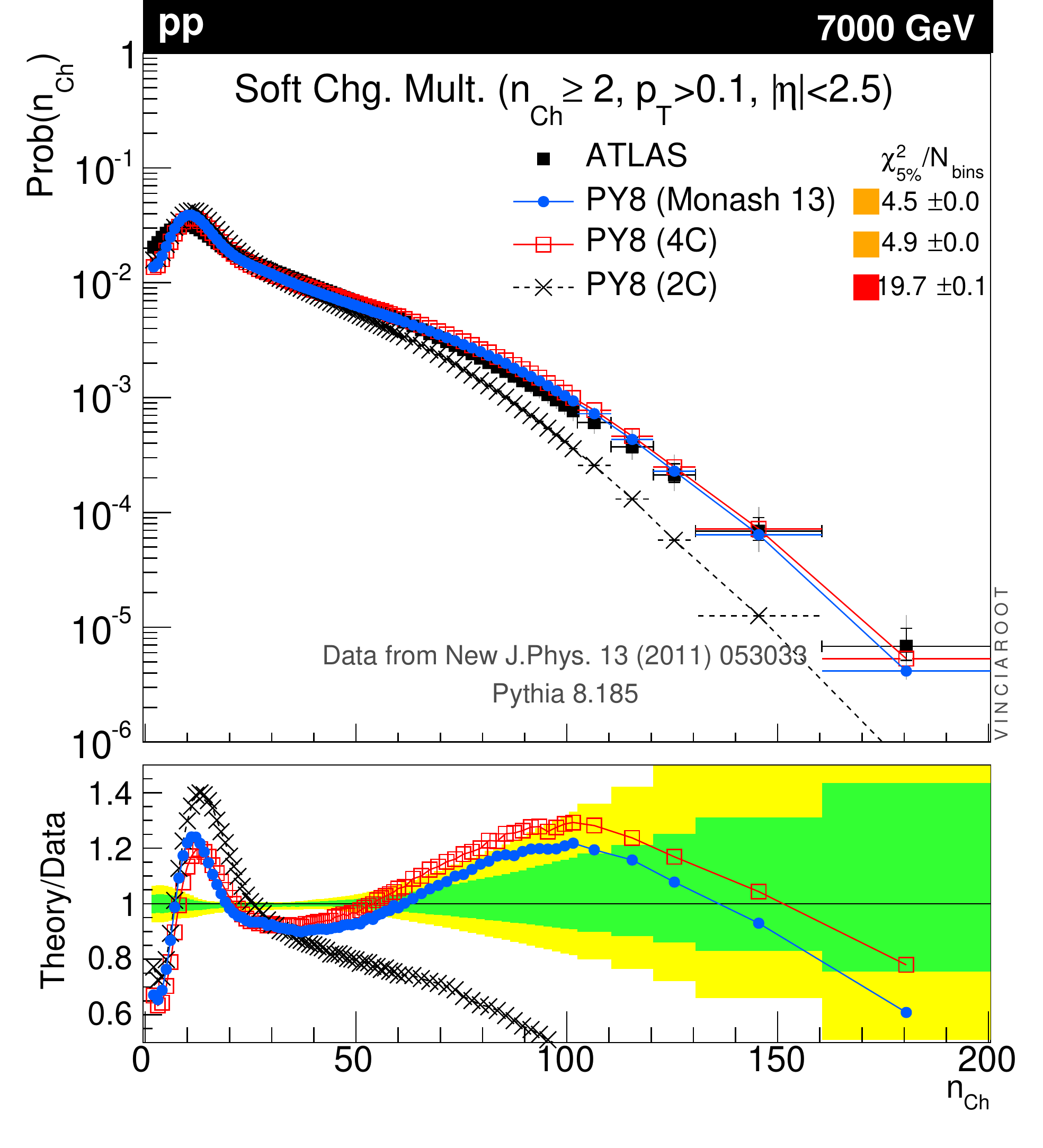}
\includegraphics*[scale=\dscale]{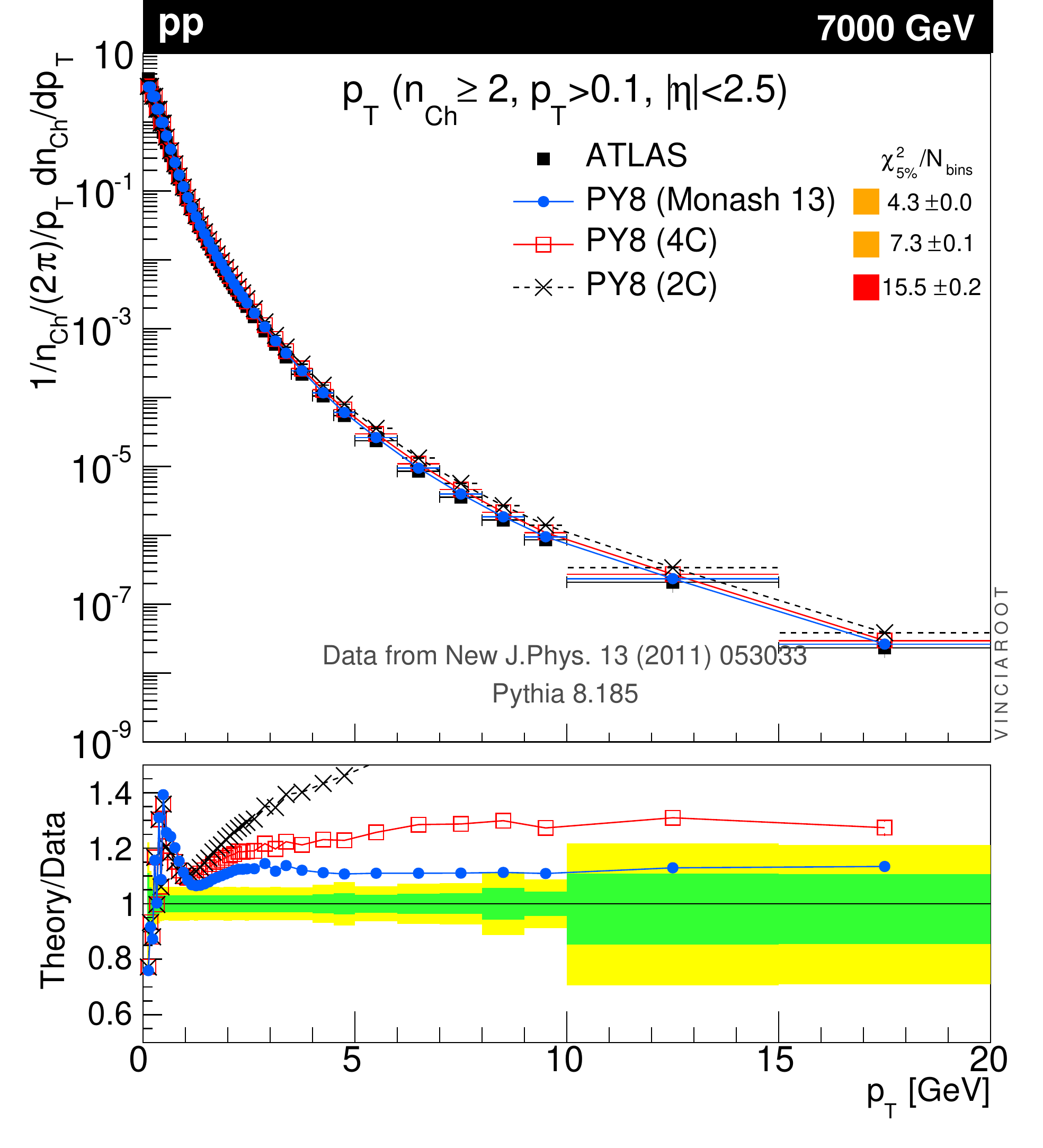}\vspace*{-2mm}
\caption{Min-bias
  $pp$ collisions at 7 TeV. Charged-multiplicity and $p_\perp$ distributions, with standard (top row) and soft (bottom
  row) fiducial cuts, compared to ATLAS
  data~\cite{Aad:2010ac}.
\label{fig:nch}}
\end{figure}
Turning now to physics distributions in min-bias events, the broader
MPI distribution in the Monash tune translates to a broader
charged-multiplicity spectrum, though the 
effect is modulated by the colour-reconnection model. The resulting
multiplicity and $p_\perp$ spectra are shown in \figRef{fig:nch}, for
``standard'' fiducial cuts (top row: $p_{\perp}\ge 500\MeV$,
$|\eta|<2.5$, $n_\mrm{Ch}\ge 1$) and ``soft'' fiducial cuts (bottom
row: $p_\perp\ge 100\MeV$, $|\eta|<2.5$, $n_\mrm{Ch}\ge 2$), with the
latter representing the most inclusive phase-space region accessible
with the ATLAS detector. For both of the $n_\mrm{Ch}$ distributions, 
we note that a significant ``double-crested wave''
pattern is still present in the ratio panes, though it has been dampened
slightly. The $p_\perp$ spectra in the right-hand panes are a bit
below the data for the standard fiducial cuts and above it for the
soft cuts, hence we regard the Monash tune as a reasonable
compromise. 

\begin{figure}[tp]
\centering
\includegraphics*[scale=\dscale]{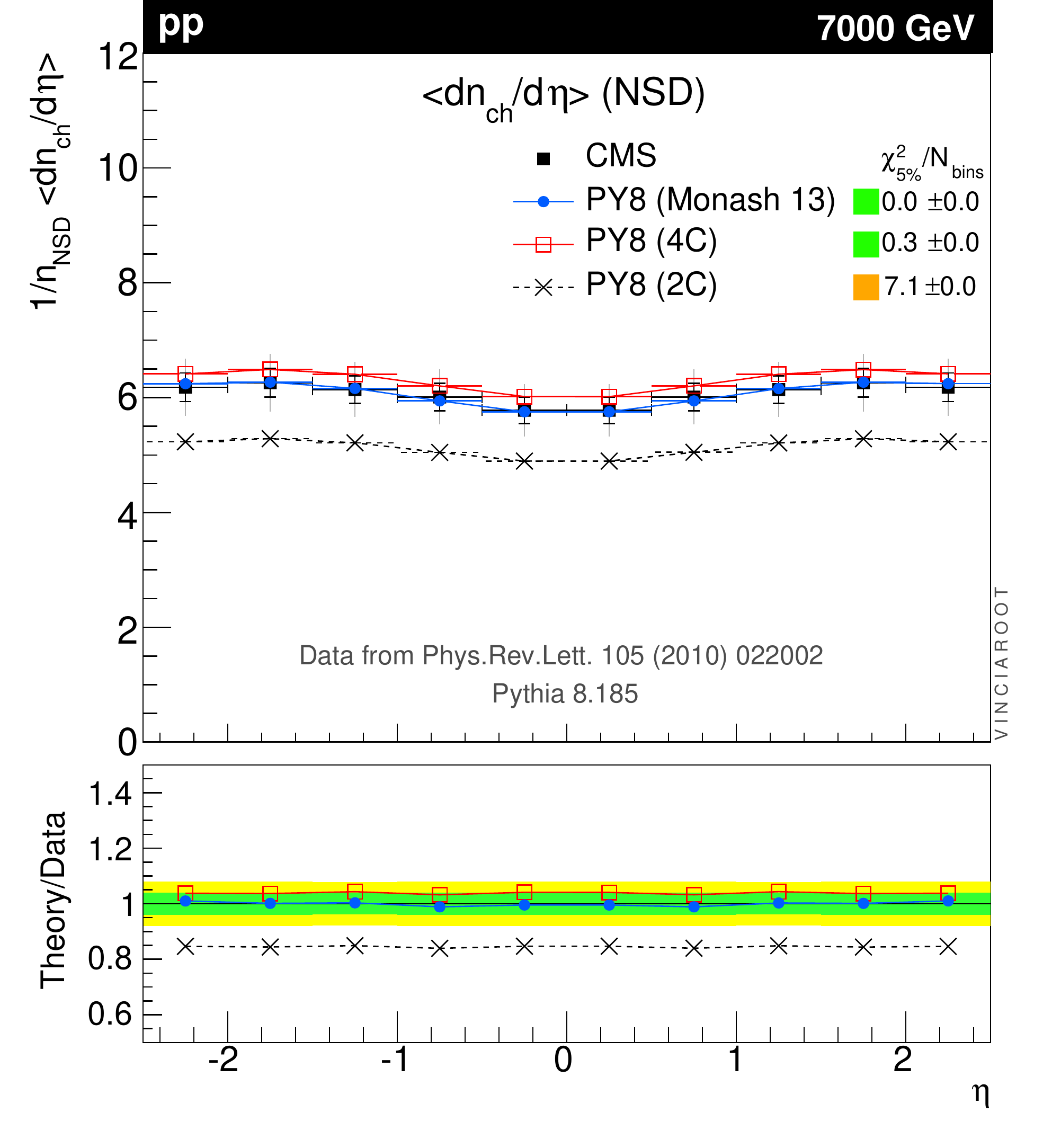}\\
\includegraphics*[scale=\dscale]{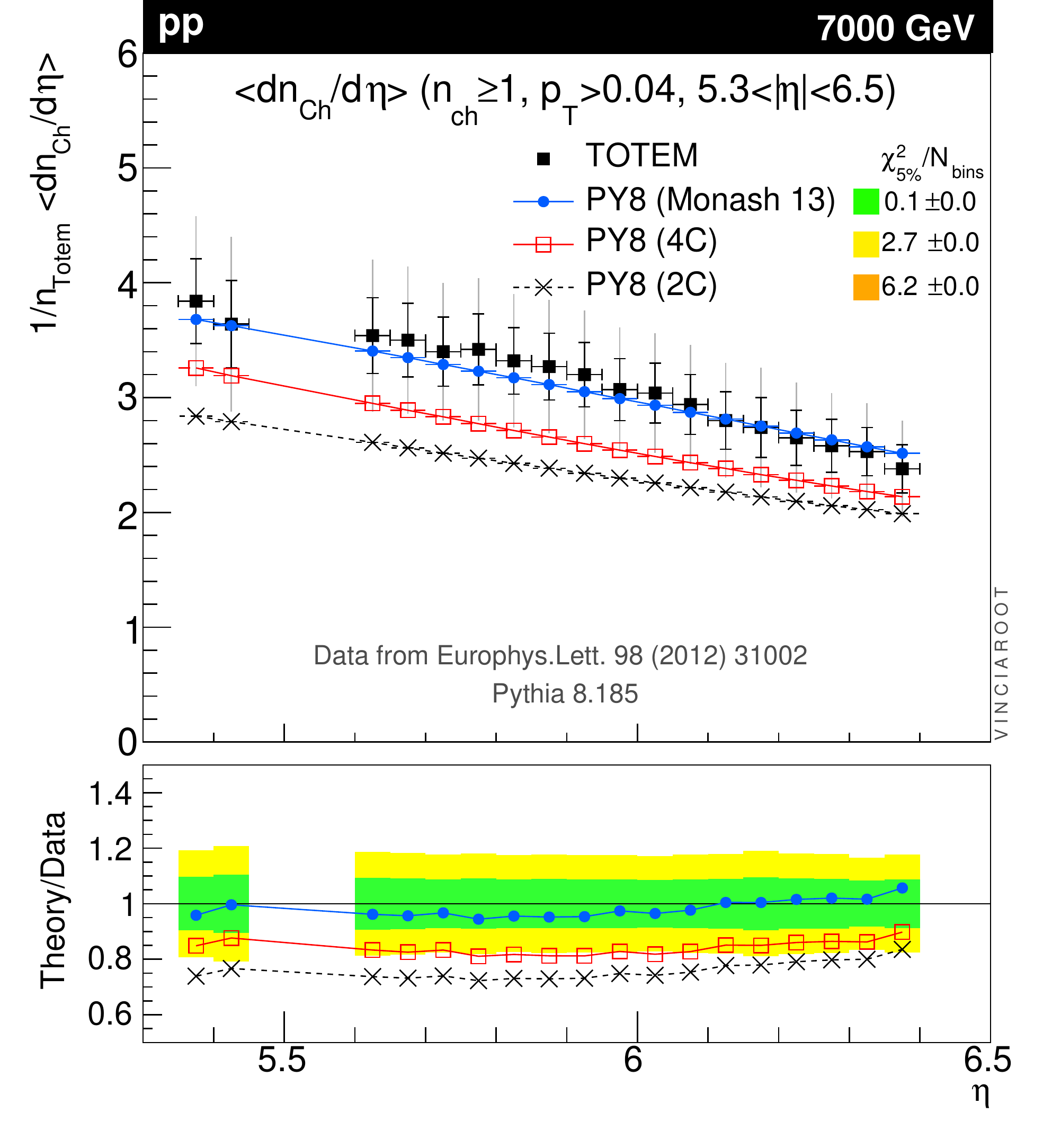}
\includegraphics*[scale=\dscale]{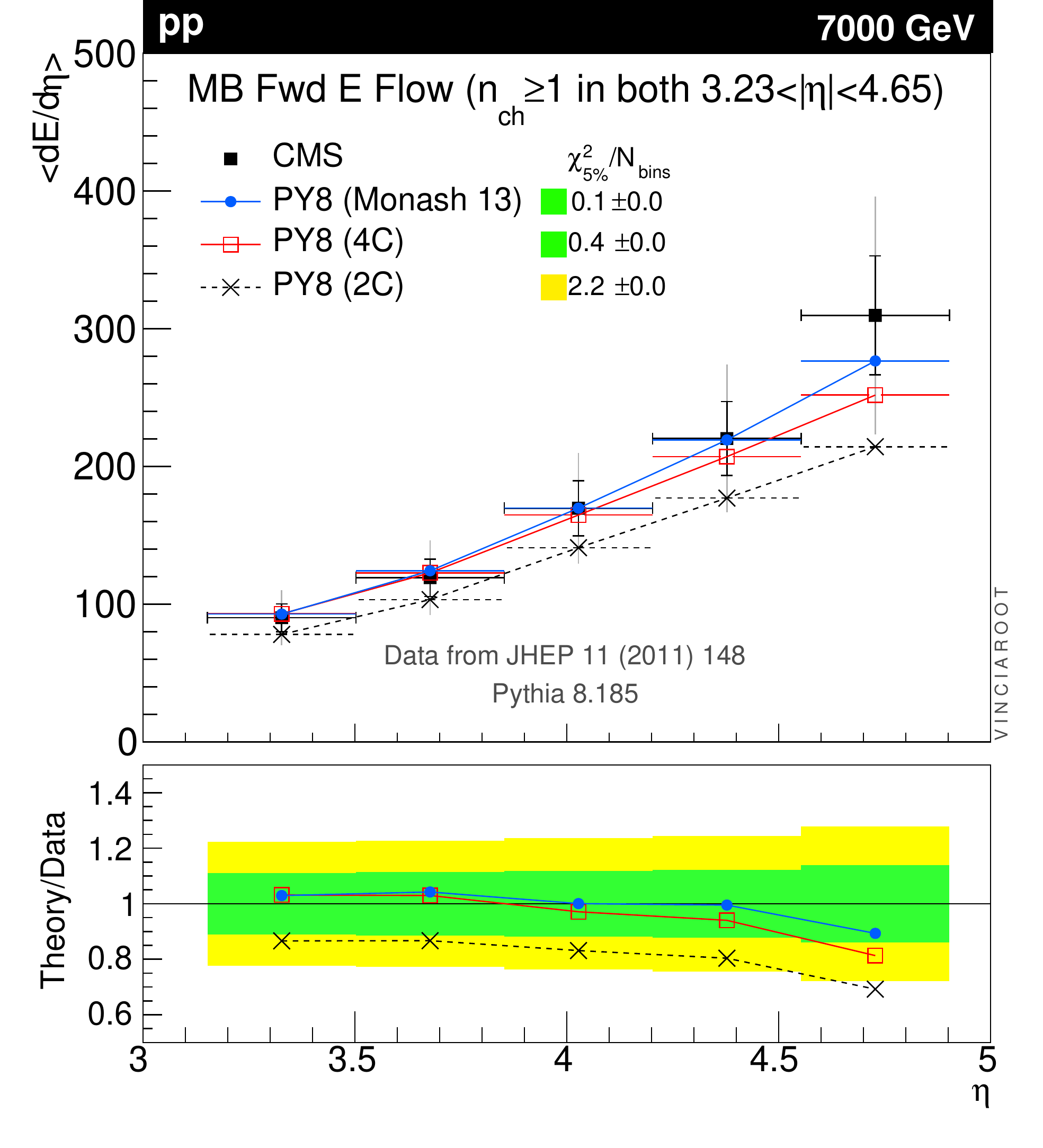}\vspace*{-2mm}
\caption{Charged-particle pseudorapidity distributions and forward
  energy flow in min-bias $pp$ collisions at 7 TeV, compared to
  CMS~\cite{Khachatryan:2010us,Chatrchyan:2011wm} 
  and TOTEM~\cite{Aspell:2012ux} data. 
\label{fig:eta}}
\end{figure}
Pseudorapidity distributions are shown in \figRef{fig:eta}. However, 
due to the
complicated interplay between diffractive contributions at 
low multiplicity and high-multiplicity multi-parton interactions (with
associated questions of transverse matter density profile and colour
reconnections), the average multiplicity by itself is a very difficult
quantity to extract reliable conclusions from. 
Note also that the CMS 
measurement~\cite{Khachatryan:2010us}  shown in the top pane of
\figRef{fig:eta} was corrected to
an unphysical ``non-single diffractive'' event definition which essentially
amounts to switching off single-diffractive contributions in the MC
generator. (We note that later CMS measurements instead use a physical
observable related to the diffractive mass to define NSD.) For the
comparisons to CMS NSD data shown here, the single-diffractive
contributions were switched off in the generator. With these caveats
in mind, we note that both the 4C and Monash 2013 tunes are in good
agreement with the CMS measurement, with the Monash one giving a
slightly lower central charged-track density (by about 5\%). This is
closer to the values observed in data, though as already noted in
\secRef{sec:plotLegend} we do not regard differences at the 5\% level as
significant. 

In the bottom two panes of \figRef{fig:eta}, we focus on the forward
region (with physical event selections). In particular, we see that 
the NNPDF set~\cite{Carrazza:2013axa} generates a broader rapidity
spectrum, so that while the 
activity in the central region (top pane) is reduced slightly, the
activity in the very forward region actually increases, and comes into 
agreement with the TOTEM measurement~\cite{Aspell:2012ux}, covering the range
$5.3<|\eta|<6.4$. The bottom right-hand pane shows the forward energy
flow measured by CMS~\cite{Chatrchyan:2011wm}, in the intermediate region
$3.23<|\eta|<4.65$. The dependence 
on $\eta$ is a bit steeper in the Monash tune than in the previous
one, and more similar to that seen in the data. 

A complementary observable, which is highly sensitive to 
interconnection effects between the MPI 
(and hence, e.g., to the effects of ``colour
reconnections''~\cite{Skands:2007zg}),  
is the average charged-particle $p_\perp$ as a function of the number
of charged particles. In a strict leading-colour picture, each MPI
would cause one or two new strings to be stretched between the
remnants, but each such string would be independent (modulo endpoint
effects); therefore 
(modulo jets) the $p_\perp$ spectrum of the hadrons produced by each
of these strings would be independent of the number of strings. The
result would be a flat $\left<p_\perp\right>(n_\mrm{Ch})$ spectrum. 
\begin{figure}[t!p]
\centering
\includegraphics*[scale=\dscale]{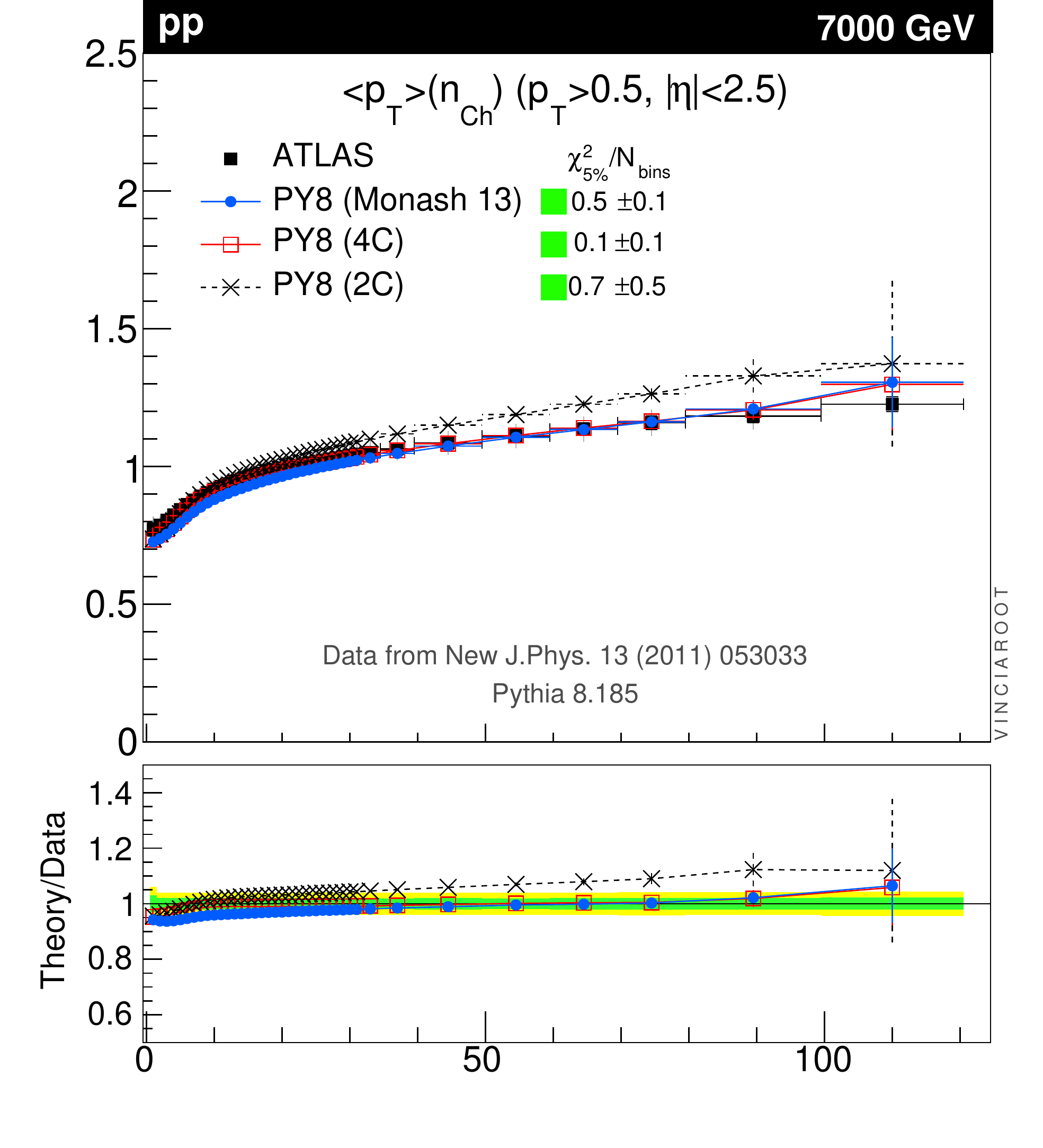}
\includegraphics*[scale=\dscale]{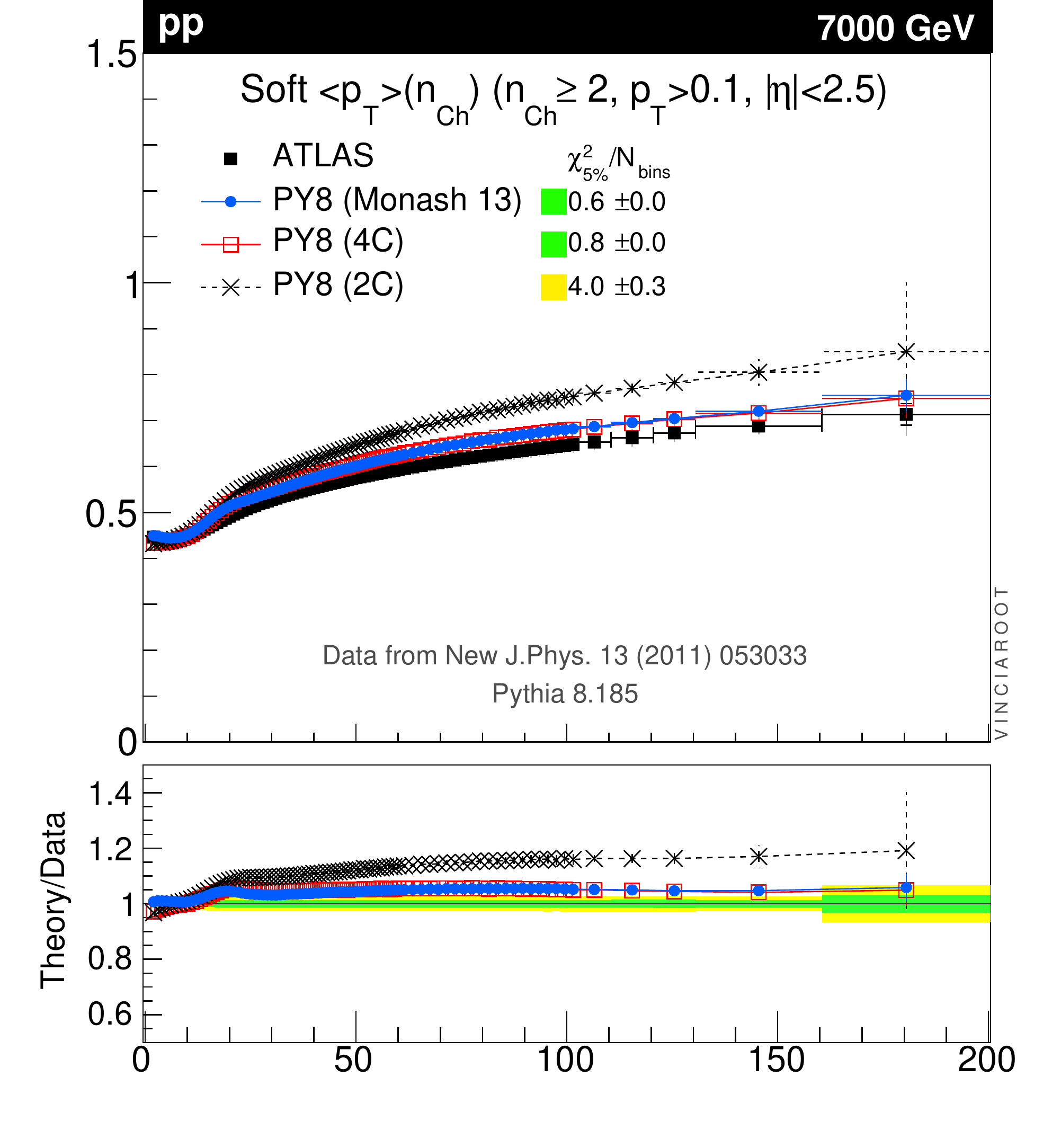}
\caption{Average-$p_\perp$ vs.~ charged-multiplicity distributions in min-bias
  $pp$ collisions at 7 TeV, with standard (left) and soft (right)
  fiducial cuts, compared to ATLAS data~\cite{Aad:2010ac}.
\label{fig:ptnch}}
\end{figure}
Jets and colour
reconnections both produce a rising spectrum. The spectra observed by 
ATLAS~\cite{Aad:2010ac} are compared to the Monash, 2C, and 4C tunes in
\figRef{fig:ptnch}, for standard (left) and soft (right) fiducial
cuts. 
Both of the Monash and 4C tunes reproduce the data
quite well, with $\chi_{5\%}^2<1$, while the older tune 2C had a
higher CR strength optimized to describe Tevatron
data~\cite{Aaltonen:2009ne}. We certainly consider the energy scaling
of the effective CR strength among the most uncertain parameters of
the current min-bias/underlying-event modelling (a similar conclusion
was reached for the CR modelling in \Py~6 in~\cite{Schulz:2011qy}),
and intend to study the physics aspects of this issue more closely in
a forthcoming paper. 

\begin{figure}[t!p]
\centering
\includegraphics*[scale=\dscale]{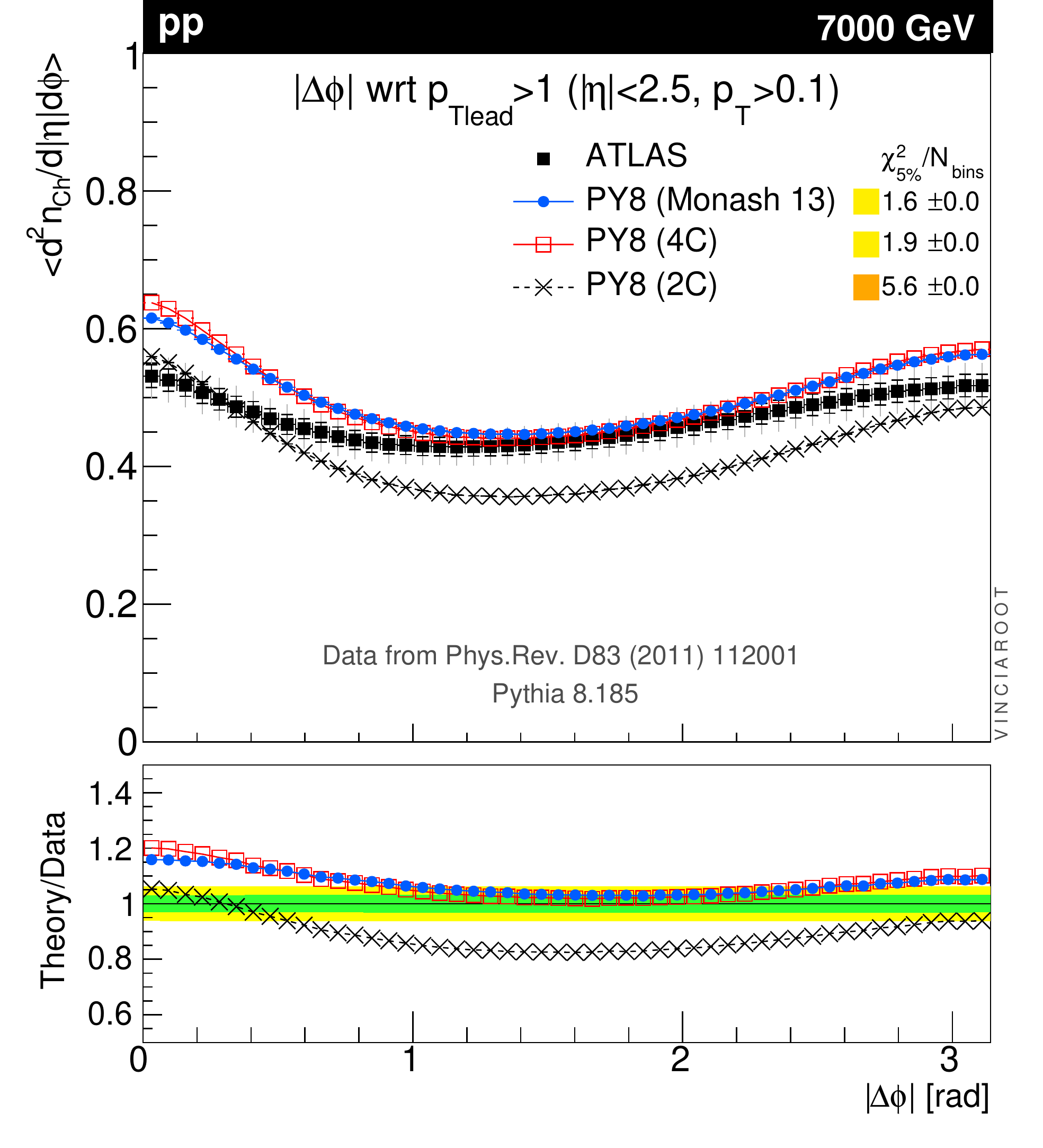}
\includegraphics*[scale=\dscale]{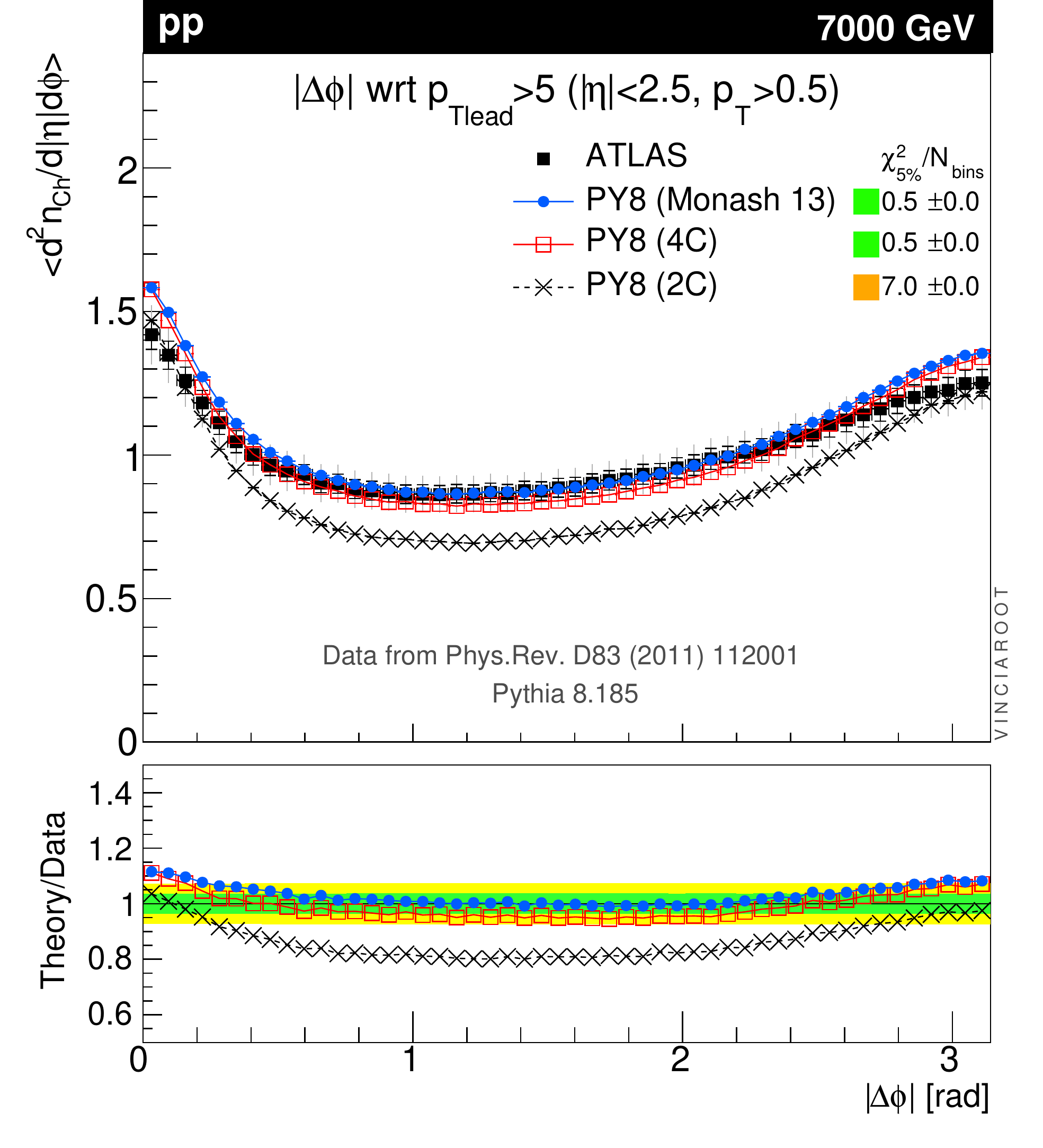}\vspace*{-2mm}
\caption{$pp$ collisions at 7 TeV. $\Delta\phi$ of charged particles
  with respect to the hardest track, for two different hardest-track
  triggers, compared with ATLAS data~\cite{Aad:2010fh}.
\label{fig:uePhi}}
\end{figure}
For a more differential look at the event structure, we consider the
charged-track $\Delta\phi$ distributions with respect to the azimuthal
angle of the leading track, in \figRef{fig:uePhi}, compared with ATLAS
data~\cite{Aad:2010fh}. The plot in the
left-hand pane corresponds to a requirement of $p_{\perp\mrm{lead}}\ge
1\GeV$, while the one in the right-hand pane is for a harder trigger,
$p_{\perp\mrm{lead}}\ge 5\GeV$. The former can roughly be taken as
characteristic of min-bias events, while the latter is related to the
differential distribution of the underlying event. In both cases, the
activity in the wide-angle region near $\pi/2$ is significantly better
described by the 4C and Monash 2013 tunes (which agrees with their
improved description of the overall activity), while there is a too
strong peaking at low $\Delta\phi$, especially for the lowest
$p_{\perp\mrm{lead}}$ cut (left), possibly indicating that the
structure of the min-bias events is still slightly too ``lumpy''
(i.e., jetty). For the higher $p_{\perp\mrm{lead}}$ cut (right), the
overcounting at very low $\Delta\phi$ is already significantly milder,
and we observe a good agreement with the data.

\begin{figure}[t!p]
\centering
\includegraphics*[scale=\dscale]{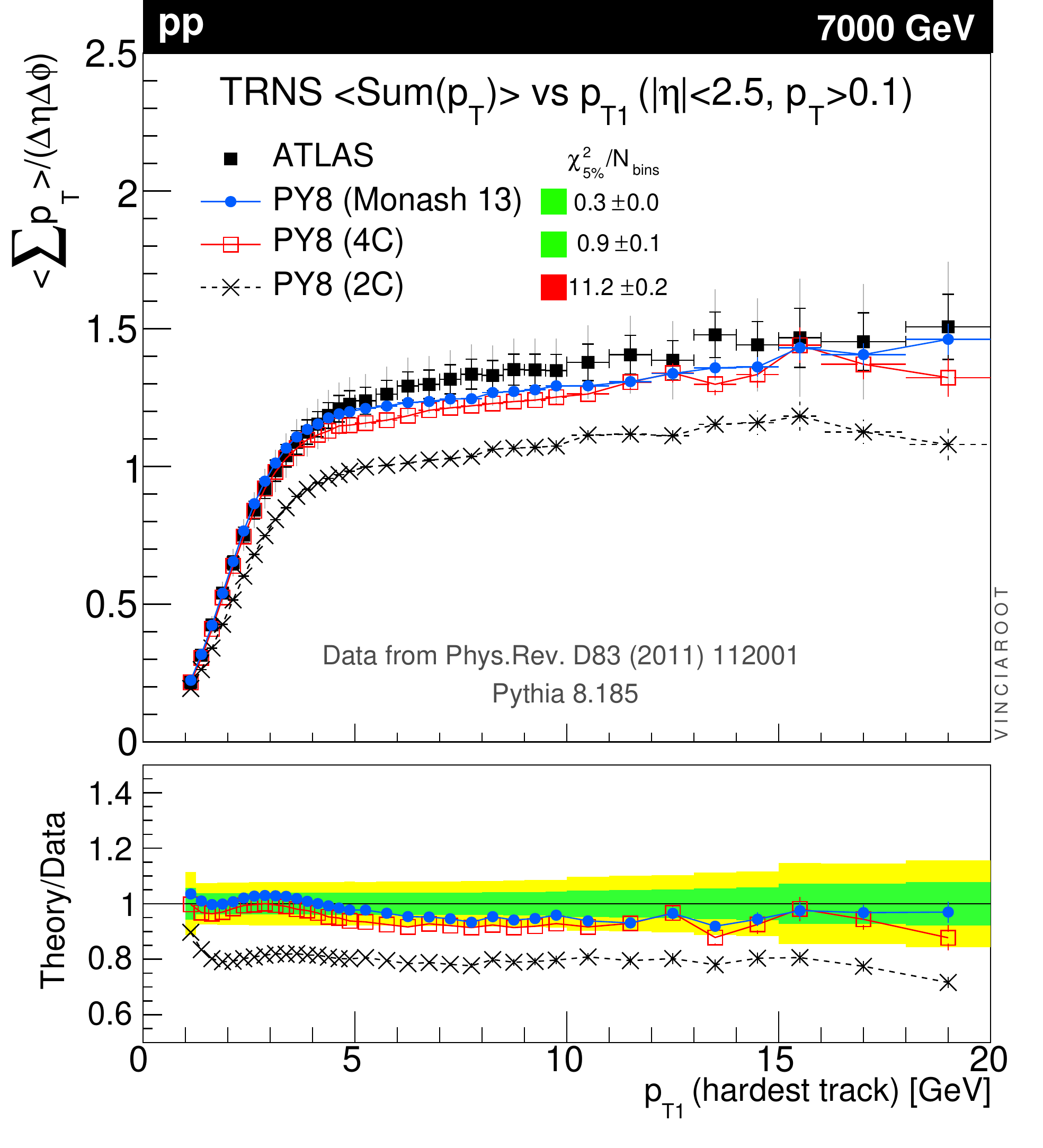}
\includegraphics*[scale=\dscale]{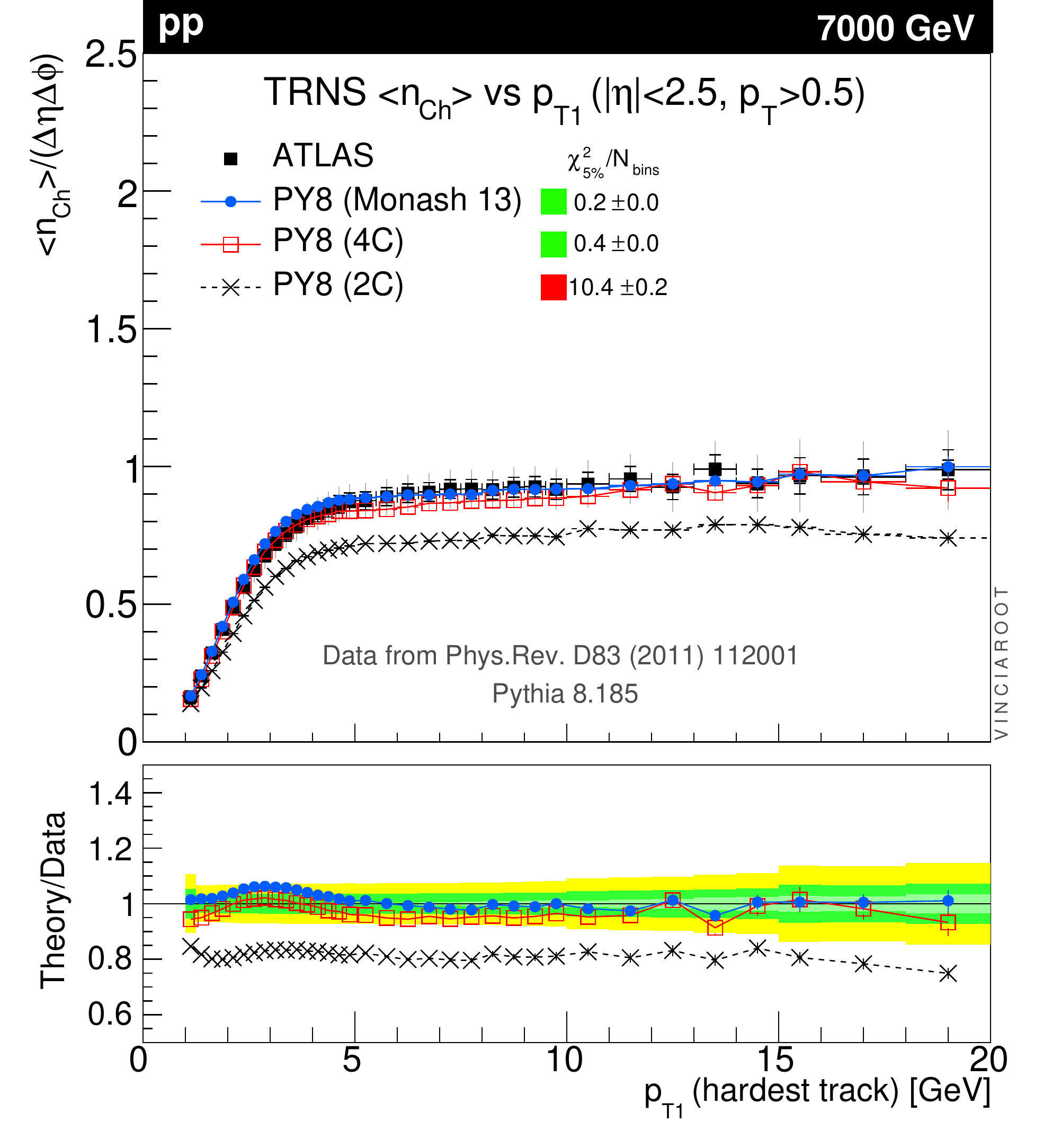}\vspace*{-2mm}
\caption{$pp$ collisions at 7 TeV. UE (``Transverse region'')
  transverse-momentum sum density (left) and charged-track density
  (right), compared with ATLAS data~\cite{Aad:2010fh}.
\label{fig:ueTrns}}
\end{figure}

Turning now to the underlying event (UE), what matters most for
high-$p_\perp$ jet studies is that the MC models describe the UE
contamination per $\Delta R$ jet area. The most important UE
observable from this perspective is thus the $p_\perp$ sum 
density in the UE, and its fluctuations. For charged particles at LHC,
typically a $p_\perp$ cut of 500 MeV is relevant, since softer tracks
will form helices and hence not contribute to calorimetric jet
energies. Neutral particles are of course relevant across all
$p_\perp$ scales. In \figRef{fig:ueTrns}, we show the charged $p_\perp$ sum
density (left, with the lowest possible $p_\perp$ cut of 100 MeV) and
the charged-track density (right, with a $p_\perp$ cut of 500 MeV), in
the so-called ``Transverse Region'' (defined by
$60^\circ<\Delta\phi<120^\circ$ with respect to the leading track),
inside the ATLAS acceptance of $|\eta|<2.5$~\cite{Aad:2010fh}. 
As is now well known the Tevatron extrapolations (represented here by Tune
2C) predicted a UE level which was 10\% -- 20\% below the LHC
data. Both the current default tune 4C (which included LHC data) and
the Monash 2013 tune exhibit significantly better agreement with the LHC
measurements, with the Monash one giving a slight additional
improvement in the $\chi^2_{5\%}$ values. We conclude that the Monash
2013 tune parameters are appropriate for both min-bias and UE studies. 

\subsection{Identified Particles at LHC \label{sec:id}}

\begin{figure}[t]
\centering
\includegraphics*[scale=\dscale]{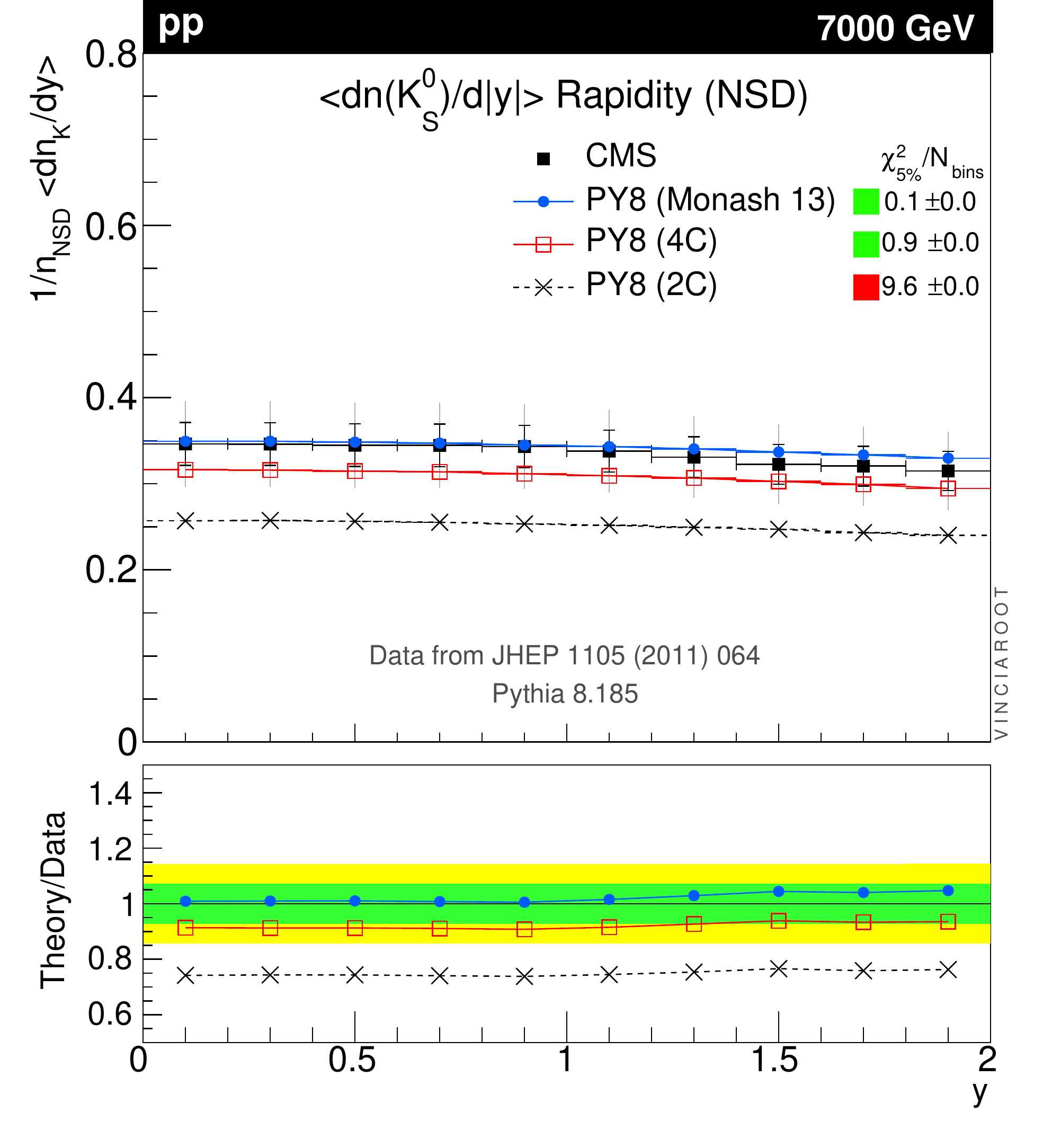}
\includegraphics*[scale=\dscale]{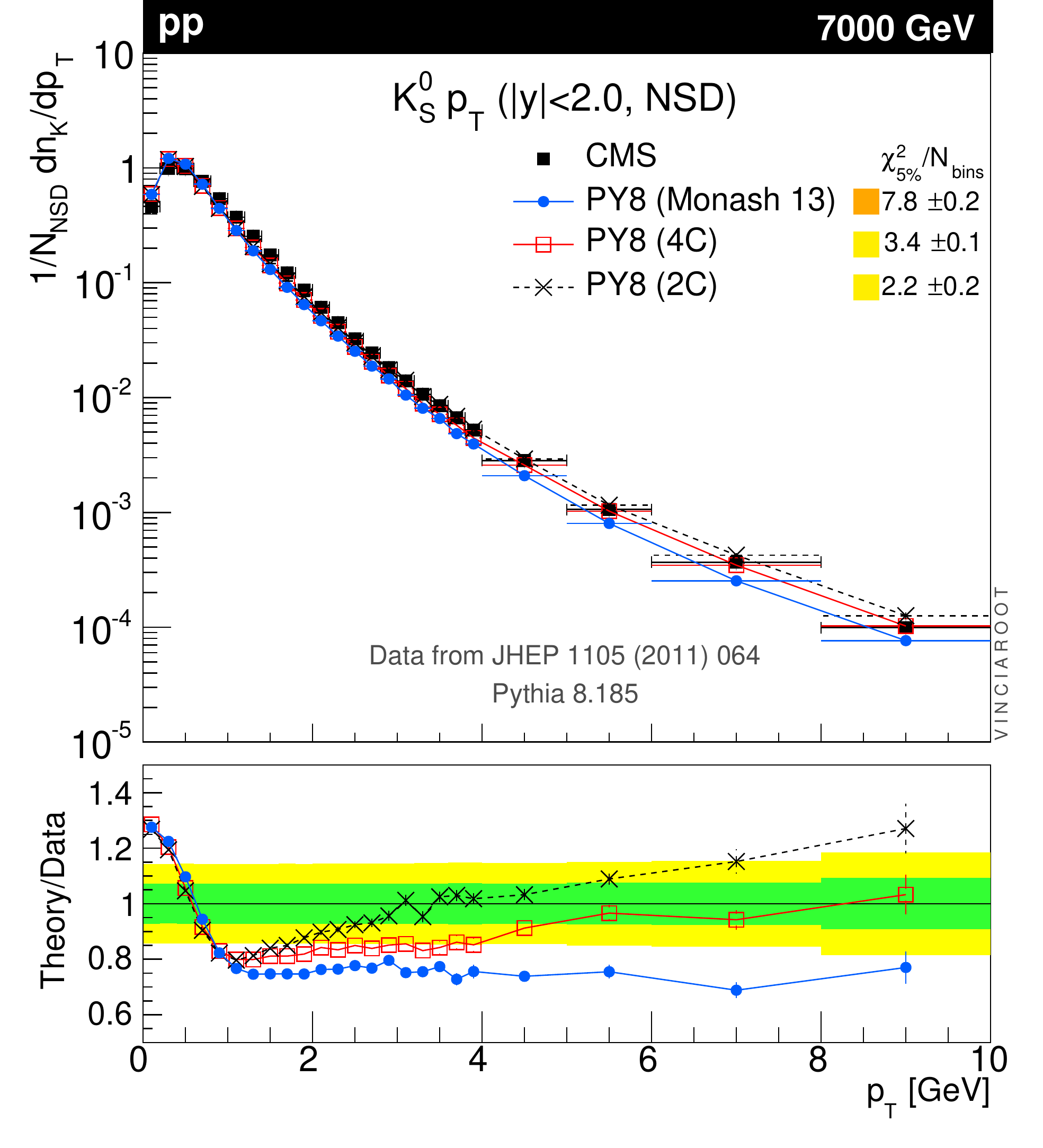}\vspace*{-2mm}
\caption{$pp$ collisions at 7 TeV. $K^0_S$ rapidity and $p_\perp$
  spectrum, compared with CMS data~\cite{Khachatryan:2011tm}.
\label{fig:LHCK0S}} 
\end{figure}

\begin{figure}[t!p]
\centering
\includegraphics*[scale=\dscale]{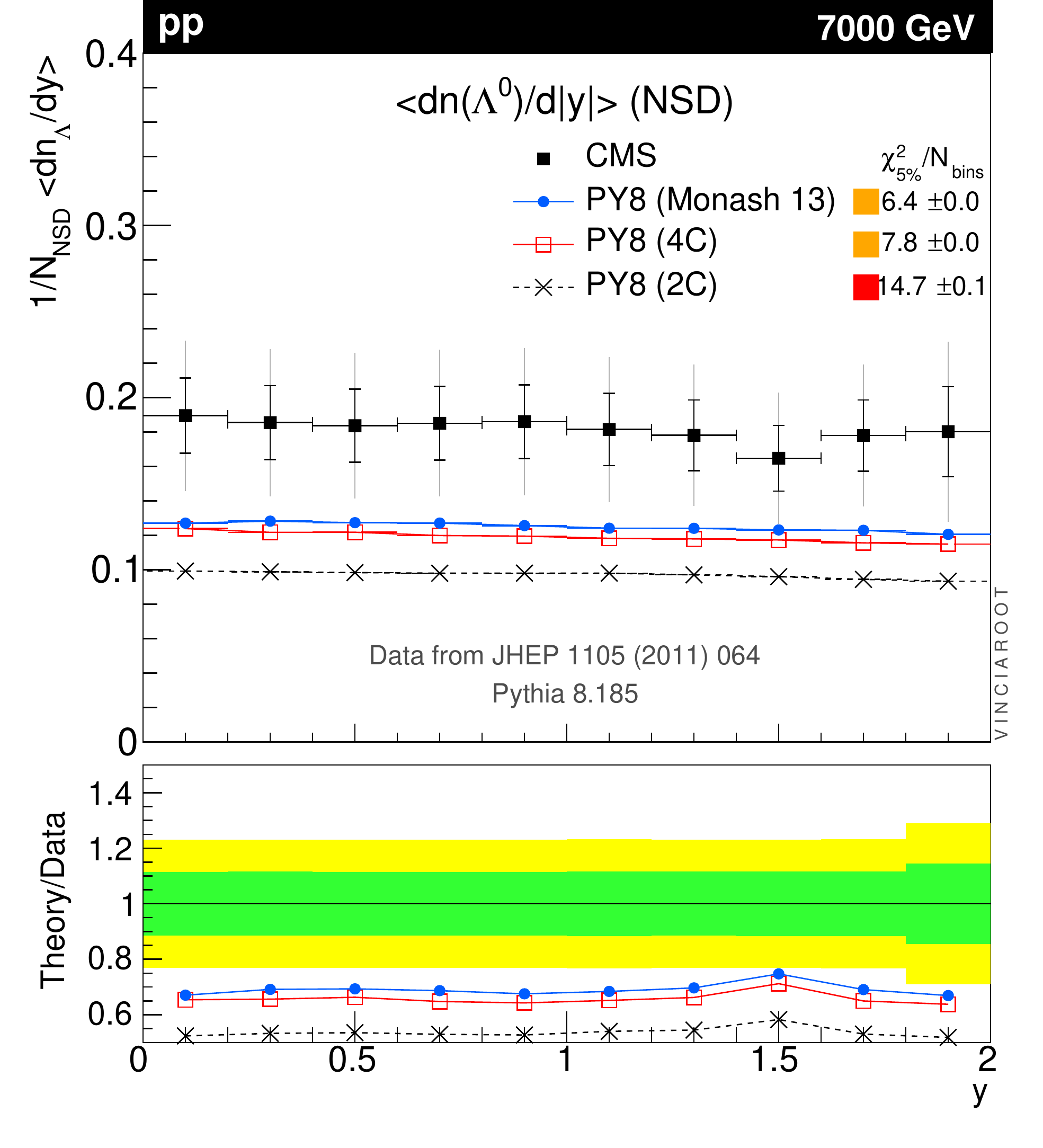}
\includegraphics*[scale=\dscale]{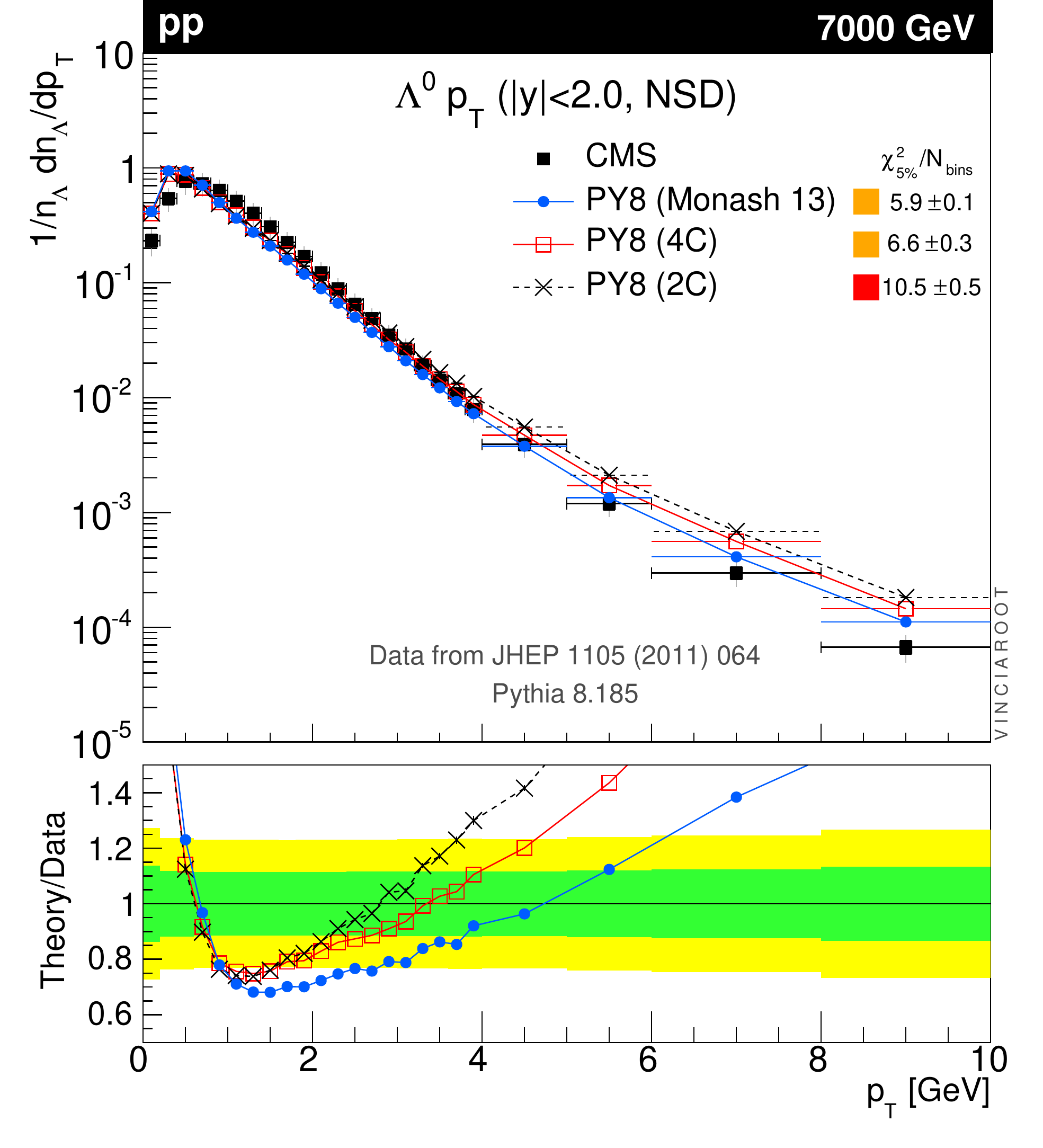}\vspace*{-2mm}
\caption{$pp$ collisions at 7 TeV. $\Lambda^0$ rapidity and $p_\perp$
  spectrum, compared with CMS data~\cite{Khachatryan:2011tm}. 
\label{fig:LHCLam}} 
\end{figure}

While the description of inclusive charged particles, discussed in the
previous section, is acceptable, larger discrepancies emerge when we
consider the spectra of identified particles. We here focus on strange
particles, in particular $K^0_S$ mesons and $\Lambda^0$ hyperons in
\figsRef{fig:LHCK0S} and \ref{fig:LHCLam}, respectively. The
experimental measurements come from CMS~\cite{Khachatryan:2011tm}. 
Additional comparisons to strange-particle spectra ($K^*$, $\phi$, and
$\Xi$) are collected in \appRef{app:id}. 

In the $K^0_S$ rapidity distribution, shown in the left-hand pane of
\figRef{fig:LHCK0S}, we observe that tune 4C exhibits a mild 
underproduction, of about 10\%. Though it might be tempting to speculate
whether this could indicate some small reduction of strangeness
suppression in $pp$ collisions, however, we already noted in
\secRef{sec:light} that the strangeness production in $ee$ collisions also
needed to be increased by about 10\%. After this adjustment, we see
that the overall $K^0_S$ yield in the Monash 2013 tune is fully 
consistent with the CMS measurement. Nonetheless, we note that the momentum
distribution is still not satisfactorily described, as shown in
the right-hand pane of \figRef{fig:LHCK0S}. Our current best guess is
therefore that the overall rate of strange quarks is consistent, at
least in the \emph{average} min-bias collision (dedicated comparisons
in high-multiplicity samples would still be interesting), but that the
phase-space distribution of strange hadrons needs more work. Similarly
to the case in $ee$ collisions, cf.~\figRef{fig:idSpectra}, the model
predicts too many very soft kaons, though we do not currently 
know whether there is a dynamic link between the $ee$ and $pp$ observations.

For strange baryons, we note that the increase in the $\Lambda^0$
fraction in $ee$ collisions (cf.~\figRef{fig:idParticles}) does
\emph{not} result in an equivalent improvement of the $\Lambda^0$ rate
in $pp$ collisions, shown in \figRef{fig:LHCLam}. The Monash 2013 tune
still produces only about 2/3 of the observed $\Lambda^0$ rate (and
just over half of the observed $\Xi^-$ rate, cf.~\appRef{app:id}). We
therefore believe it to be likely that an additional source of net
baryon production is needed (at least within the limited context of the
current \Py\ modelling), in 
order to describe the LHC data. The 
momentum spectrum  
is likewise quite discrepant, exhibiting an excess at very low
momenta (stronger than that for kaons), a dip between 1--4 GeV, and then an  
excess of very hard $\Lambda^0$ production. The latter hard tail is
somewhat milder in
the Monash 2013 tune than previously, and it may
be consistent with the trend also seen in the $\Lambda^0$ spectrum at LEP,
cf.~\figRef{fig:idSpectra}. We conclude that baryon production 
still requires further modelling and tuning efforts.  

\section{Energy Scaling \label{sec:energyScaling}}

\begin{figure}[t!p]
\centering
\includegraphics*[height=0.38\textheight]{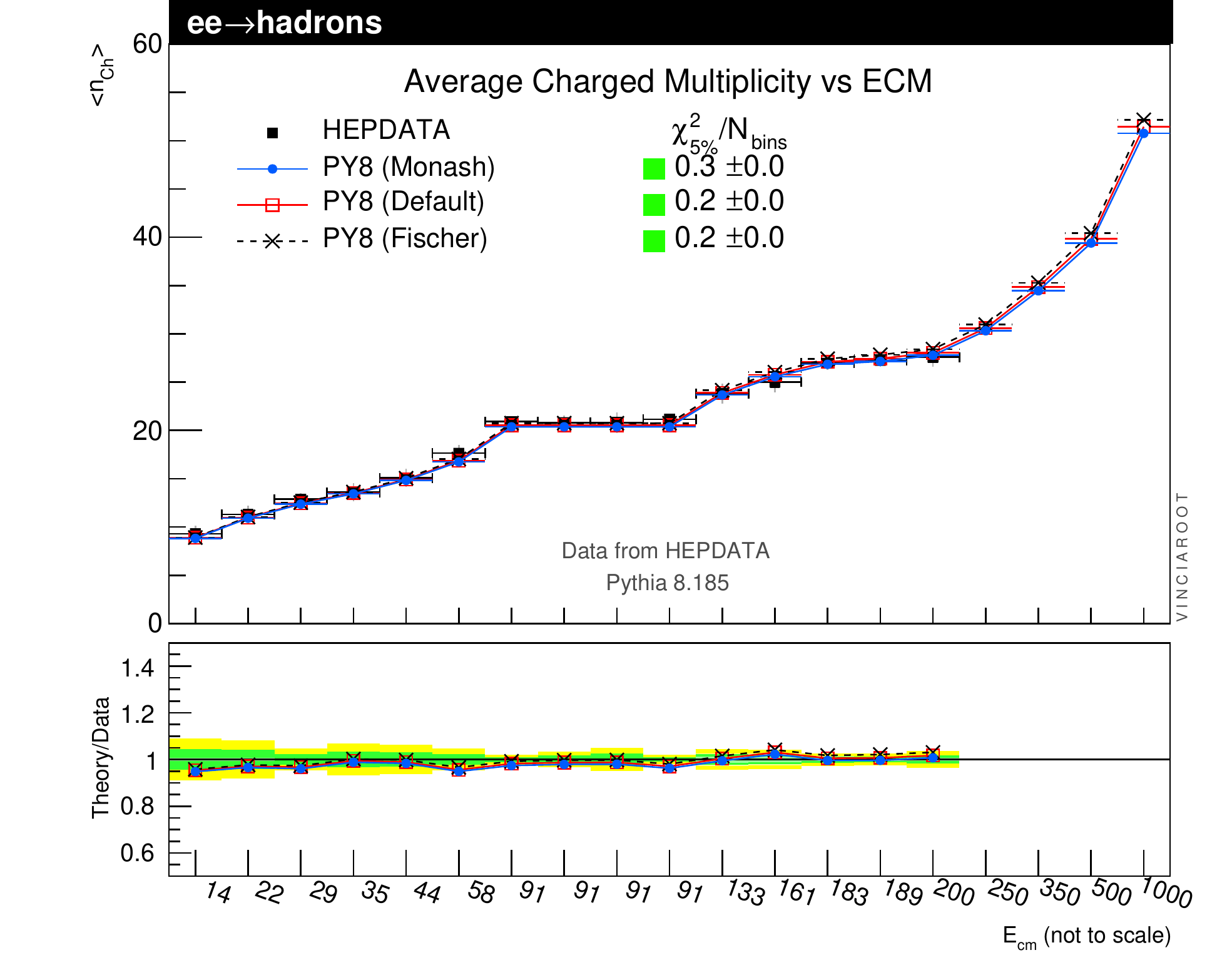}\\
\includegraphics*[height=0.34\textheight]{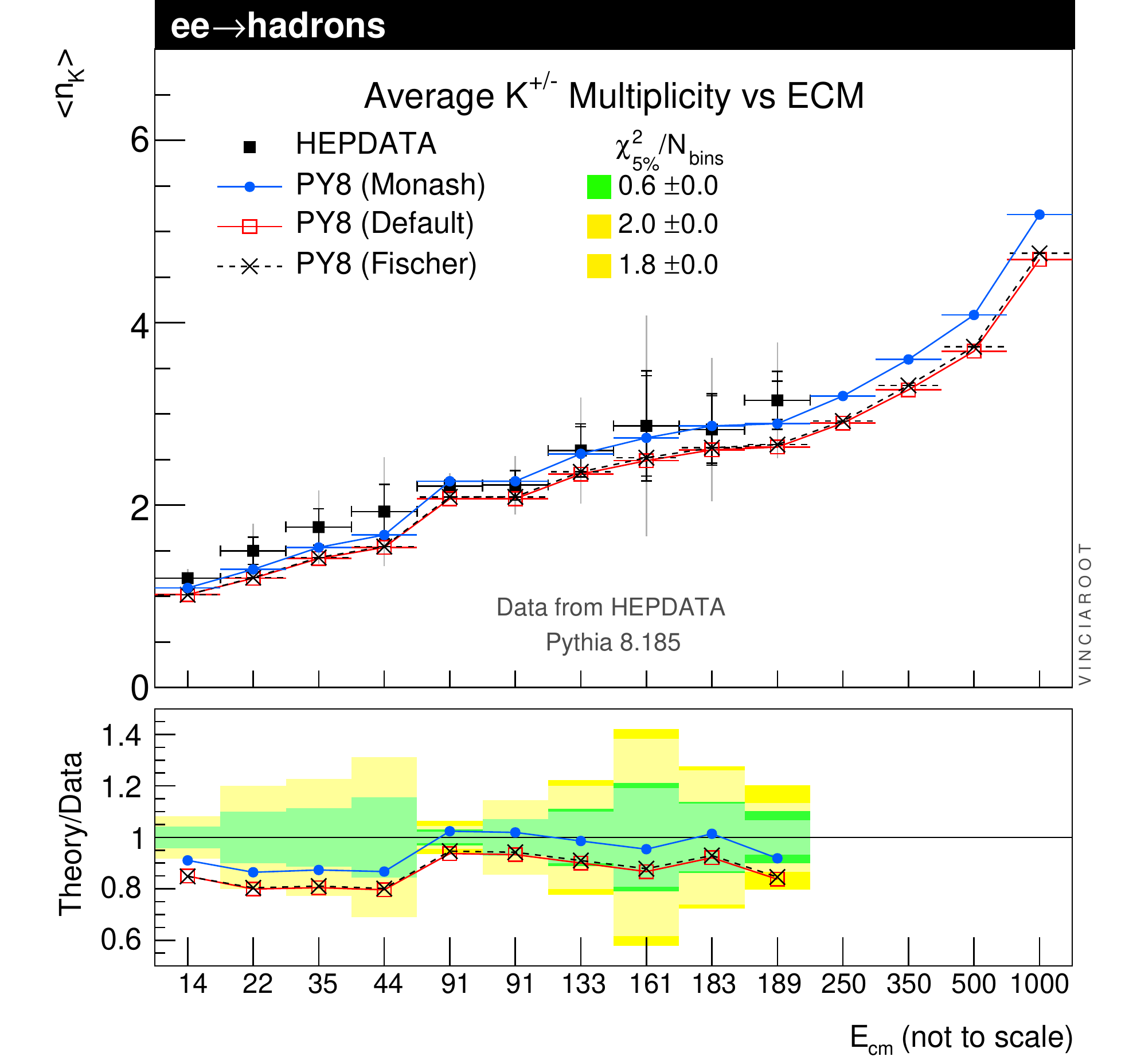}
\includegraphics*[height=0.34\textheight]{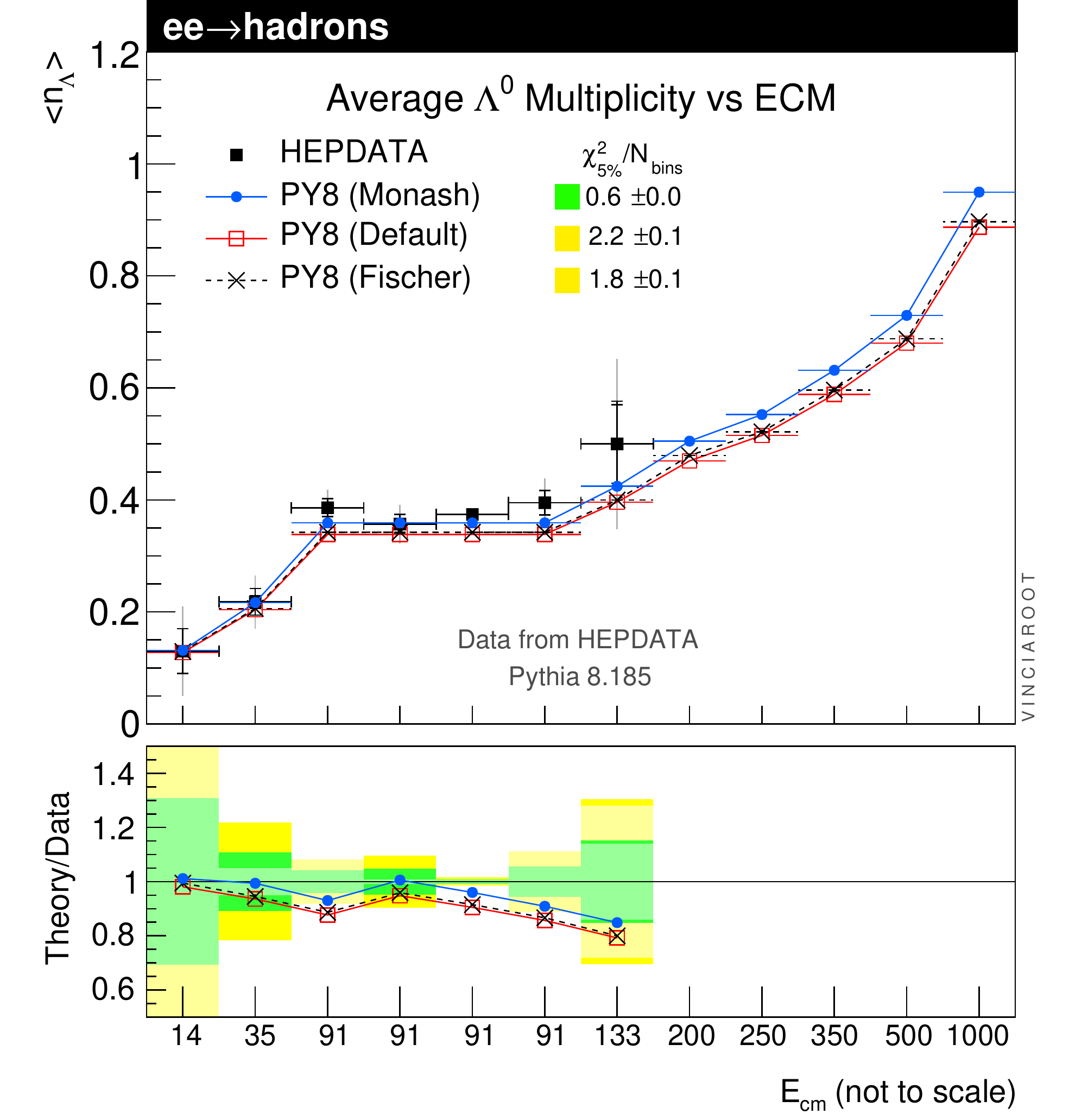}\vspace*{-2mm}
\caption{$e^+e^-\to\mrm{hadrons}$. Energy scaling of $\left<n_\mrm{Ch}\right>$,
  $\left<n_{K^\pm}\right>$, and
  $\left<n_\mrm{\Lambda}\right>$, in $e^+e^-\to q\bar{q}$ events, including
comparisons to measurements from HEPDATA for CM energies from 14 GeV
to 200 GeV. Also shown are model extrapolations up to 1000 GeV. 
\label{fig:LEPscaling}} 
\end{figure}

Though energy scaling these days mostly refers to the scaling of $pp$
collisions (see e.g., \cite{Schulz:2011qy}), an important first step
is to consider the scaling of 
observables in $ee\to\gamma^*/Z\to\mrm{hadrons}$. This scaling
contains information on the 
relative contributions of perturbative and non-perturbative
fragmentation. Thus, at low $ee$ energies, the non-perturbative
components of the fragmentation model dominate, 
while perturbative bremsstrahlung increases in importance towards higher $ee$
energies. In
\figRef{fig:LEPscaling}, we consider the scaling of the average
charged-particle multiplicity and that of charged Kaons and Lambda
baryons from CM energies of 14 GeV to 200 GeV, obtained from
measurements available at HEPDATA. 
Below the $Z$ pole, the measurements we include mostly come from
TASSO~\cite{Braunschweig:1989bp}, though a few points on
$\left<n_\mrm{Ch}\right>$ come from HRS (at 29 GeV~\cite{Derrick:1986jx})
and TOPAZ (at 57.8 GeV~\cite{Nakabayashi:1997hr}). 
At the $Z$ pole, the data come from the four LEP
experiments~\cite{Barate:1996fi,Abreu:1998vq,Adeva:1992gv,Ackerstaff:1998hz}, 
with the latter extending also to energies above
$M_Z$~\cite{Buskulic:1996tt,Alexander:1996kh,Abreu:1997dm,Ackerstaff:1997kk,Abbiendi:1999sx,Abreu:2000gw}. For
completeness and as reference for future $ee$ collider studies, model
extrapolations for CM energies up to 1000 GeV are also shown (though
still only including the 
$ee\to\gamma^*/Z\to\mrm{hadrons}$ component, as usual with photon ISR
switched off). 

From the plots in \figRef{fig:LEPscaling}, it is clear that there are
no significant differences between the energy scaling of the three
$ee$ tunes considered here (mainly reflecting that they have been
tuned to same reference point, at 91.2 GeV, and that their scaling is
dictated by the same underlying physics model), and that their
energy dependence closely matches that observed in 
data. However, the 
increased amount of non-perturbative strangeness production in the
Monash tune leads to a better agreement with the overall normalization
of the $K^\pm$ and $\Lambda$ rates at all energies. 

\begin{figure}[tp]
\centering
\includegraphics*[scale=0.31]{main04-7000-Monash-etaChCms.pdf}
\includegraphics*[scale=0.31]{main04-7000-Monash-nCh1.pdf}\\
\includegraphics*[scale=0.31]{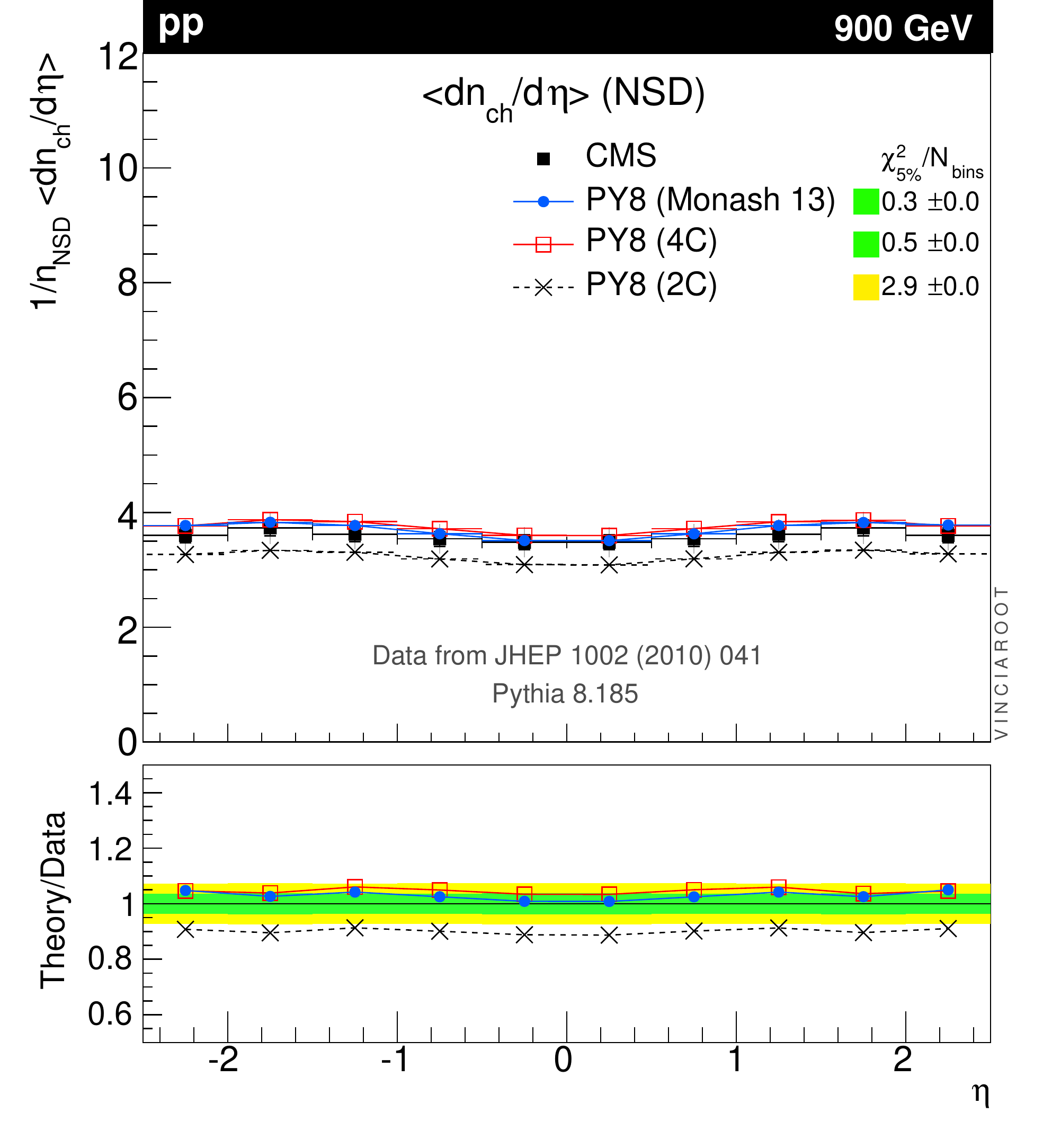}
\includegraphics*[scale=0.31]{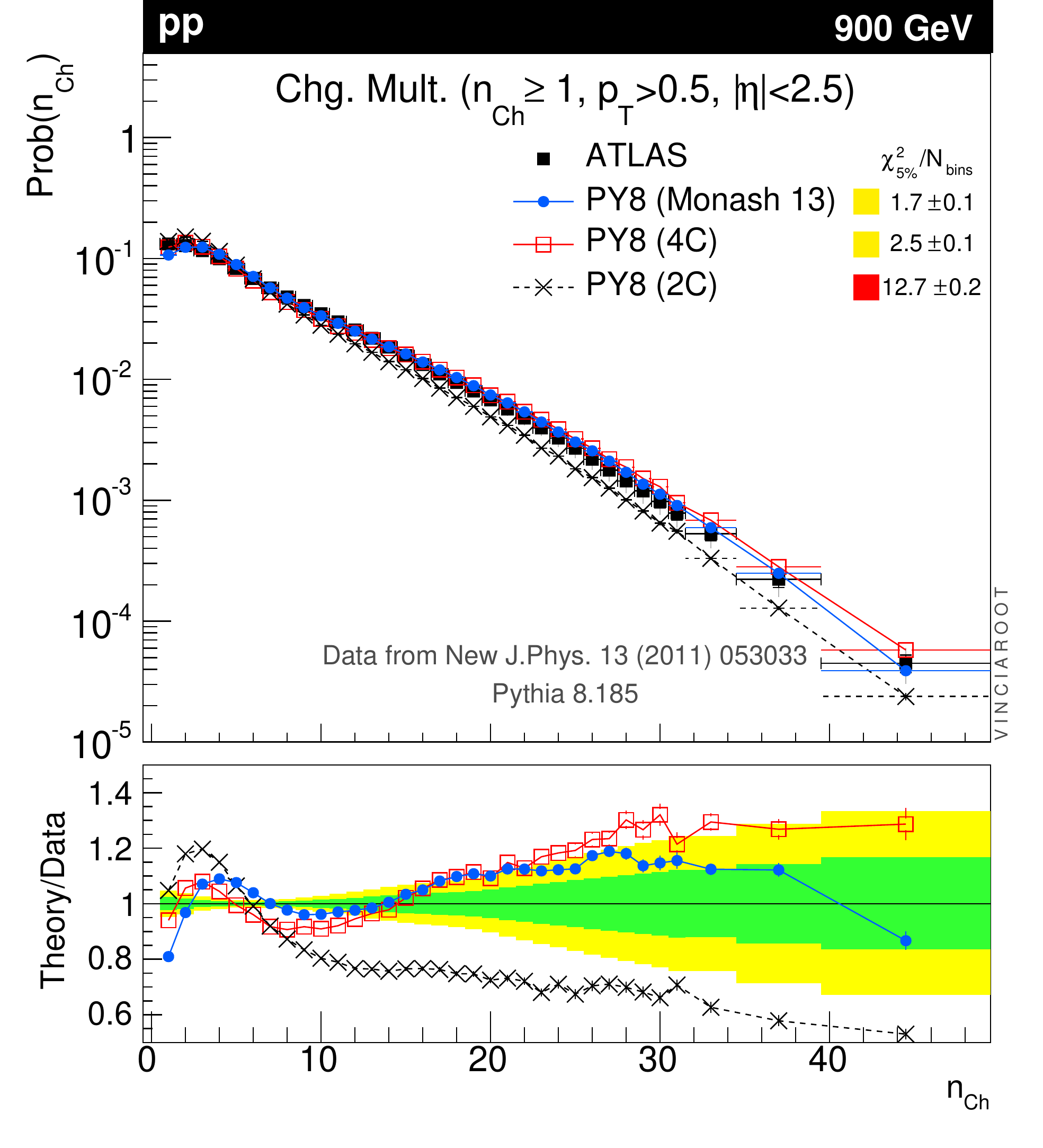}\\
\includegraphics*[scale=0.31]{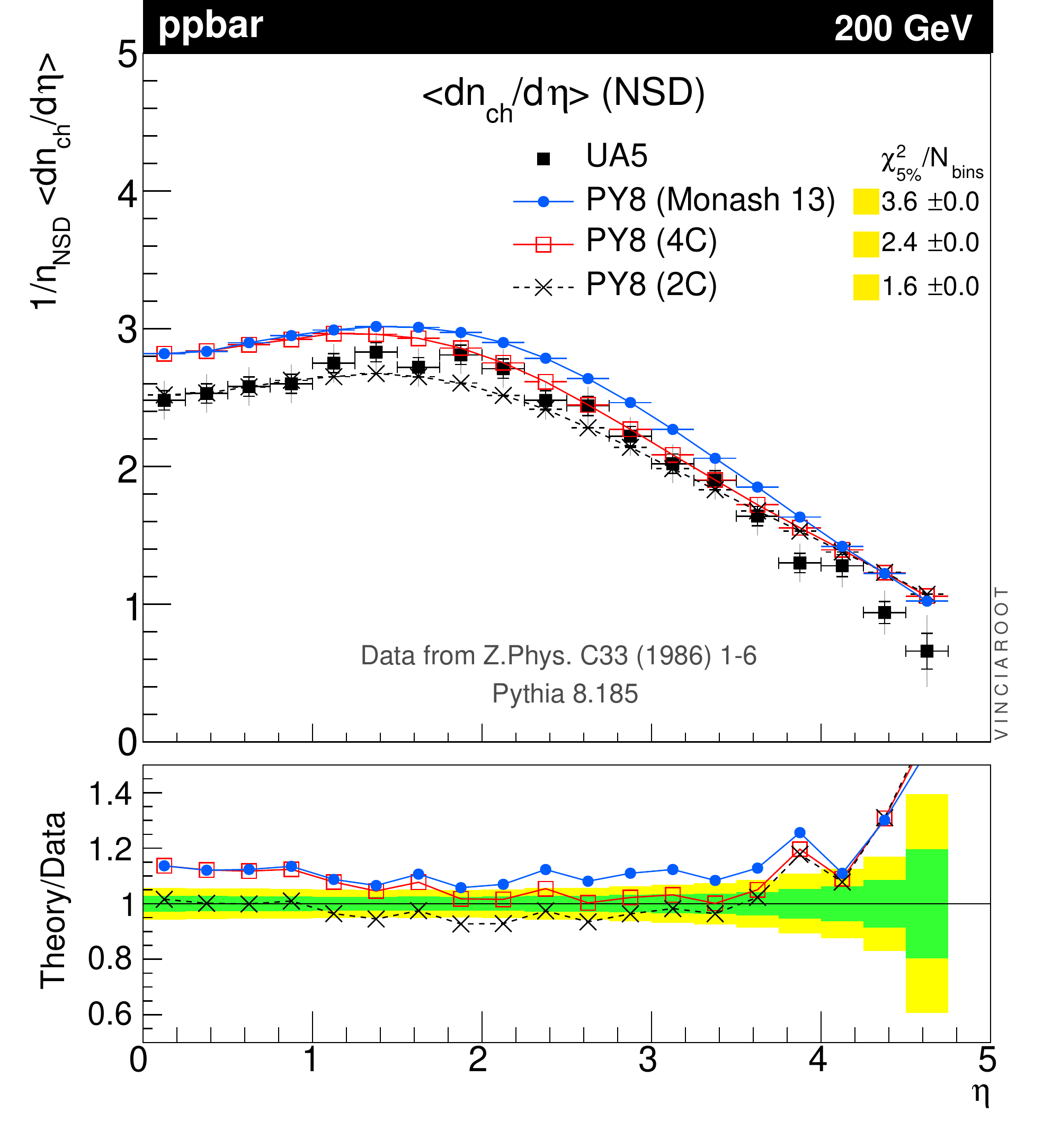}
\includegraphics*[scale=0.31]{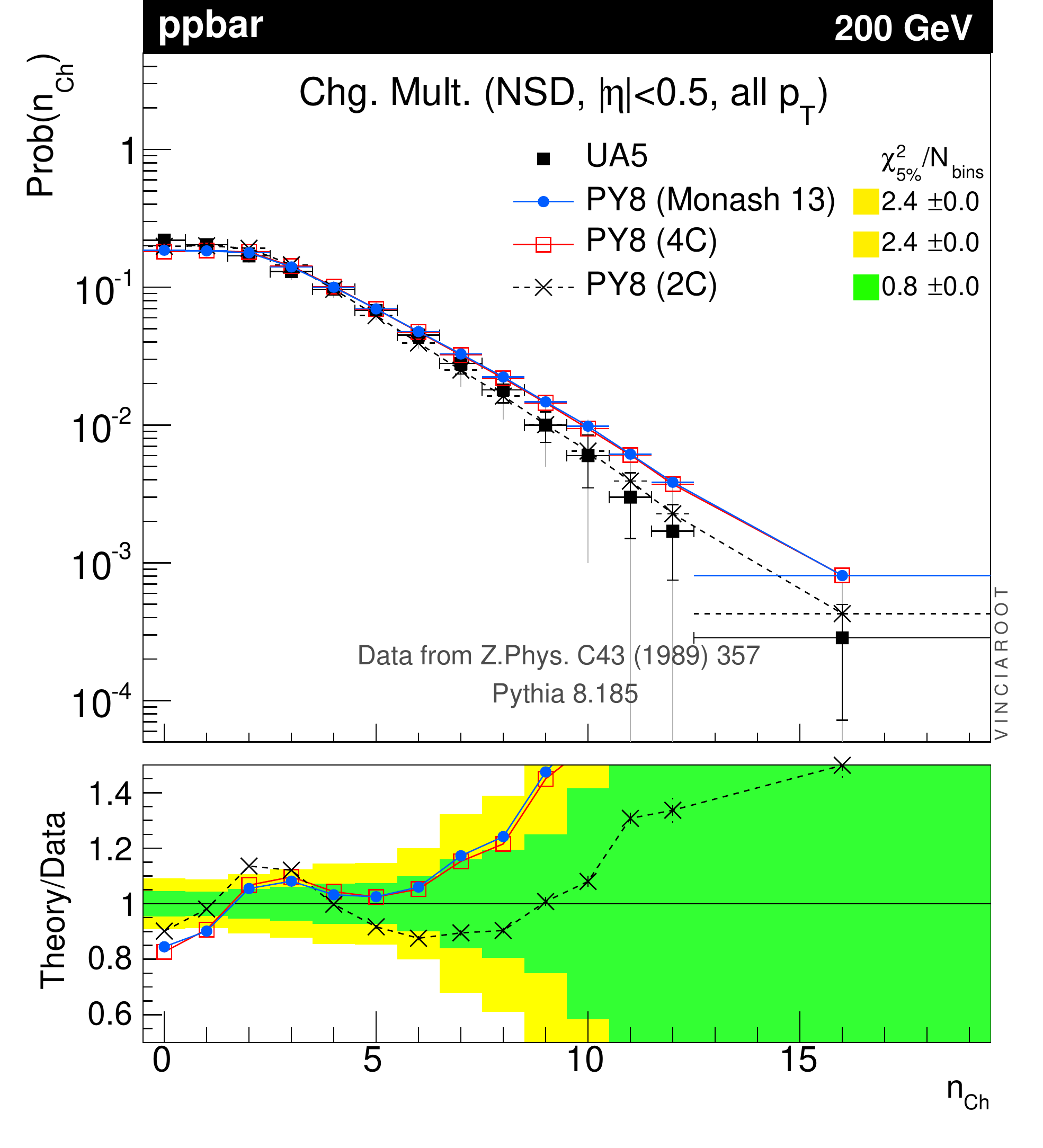}\vspace*{-2mm}
\caption{Min-bias $pp$ events, from 200 to 7000 GeV. 
Energy scaling of $\left<dn_{Ch}/d\eta\right>$ (left) and $P(n_{Ch})$
  (right). 
\label{fig:scalingNch}}
\end{figure}
Moving to $pp$ collisions, the plots in \figRef{fig:scalingNch} show
the scaling of the average charged multiplicity (left column) and
multiplicity distributions (right column) in min-bias collisions 
from 7000 GeV (top row) to 900 GeV (middle row) and 200 GeV (bottom
row), compared with data from
CMS~\cite{Khachatryan:2010us,Khachatryan:2010xs},
ATLAS~\cite{Aad:2010ac}, and
UA5~\cite{Alner:1986xu,Ansorge:1988kn}. We regret 
the omission of additional relevant min-bias measurements from the
Tevatron and RHIC experiments here, but have chosen to focus in this
paper mainly on the LHC. The comparisons at 7 TeV were
already discussed in the main section on $pp$ collisions,
\secRef{sec:hadronColliders}. At 900 GeV, the Monash 2013 tune again
gives a roughly 5\% lower average central charged multiplicity than the 4C one,
with a better description of the tail towards high multiplicities. At 200 GeV, the UA5 measurement we include here 
extends over the full rapidity and $p_\perp$ range, hence the interplay between 
diffraction and low-multiplicity non-diffractive processes is presumably
(much) more important. We believe imperfections in this modelling to
be the likely cause of the significant discrepancies observed at high $\eta$
and for $n_{\mrm{Ch}}\le 20$ at these energies. Since a dedicated study of
this interplay is beyond the scope of this study, we limit ourselves
merely to stating this observation, as a point for future
studies to help clarify.  

\begin{figure}[tp]
\centering
\includegraphics*[scale=0.31]{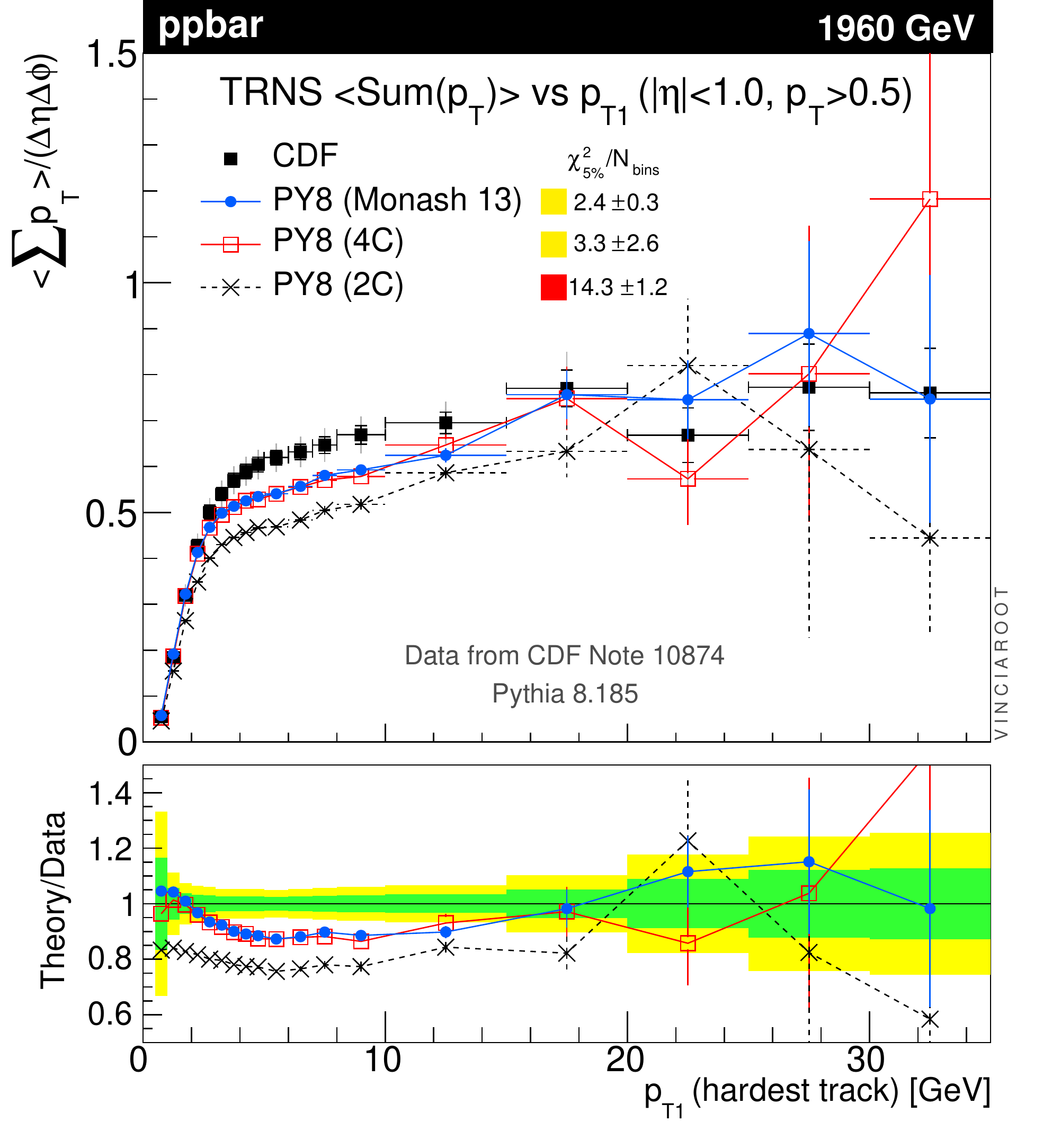}
\includegraphics*[scale=0.31]{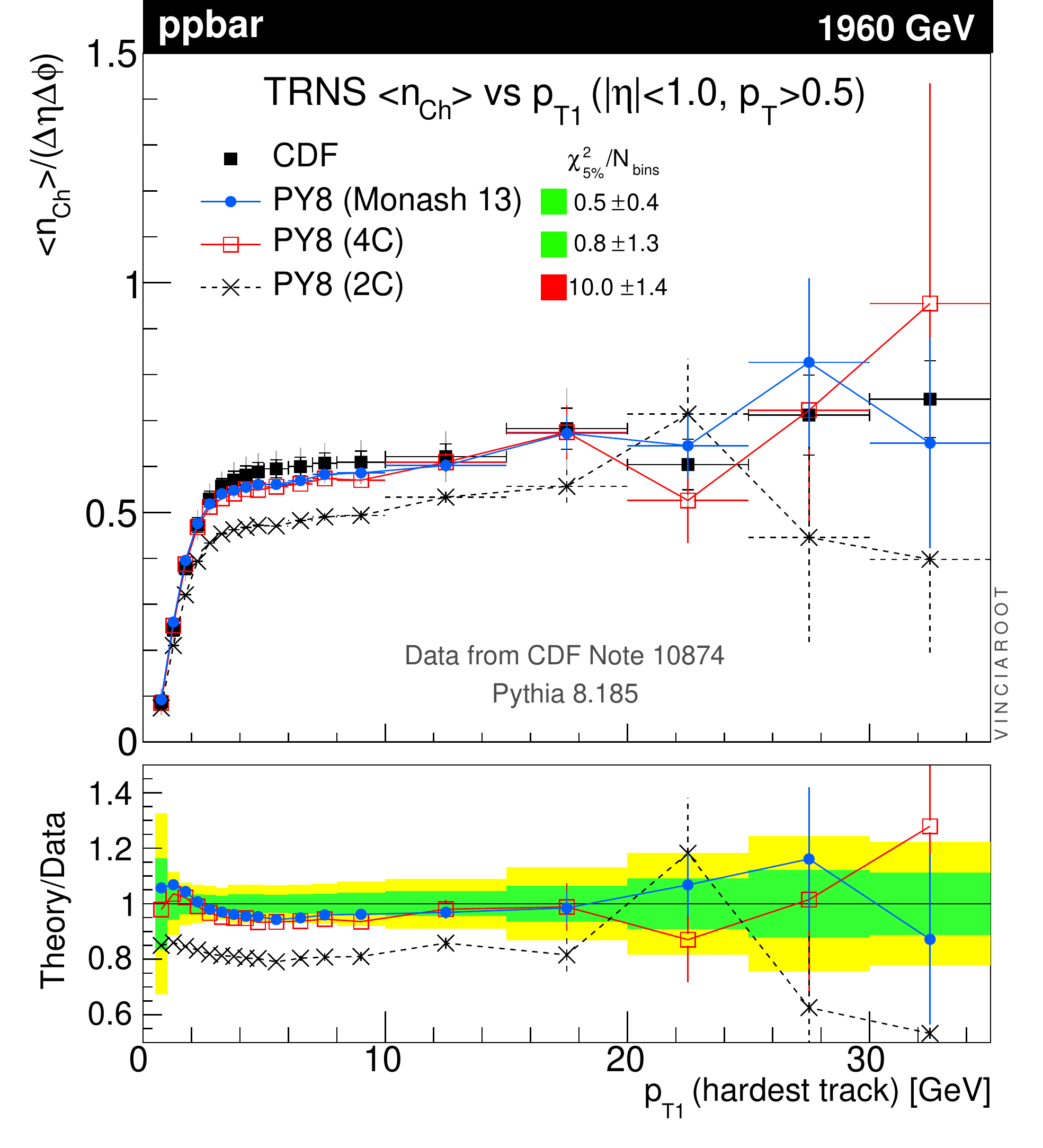}
\includegraphics*[scale=0.31]{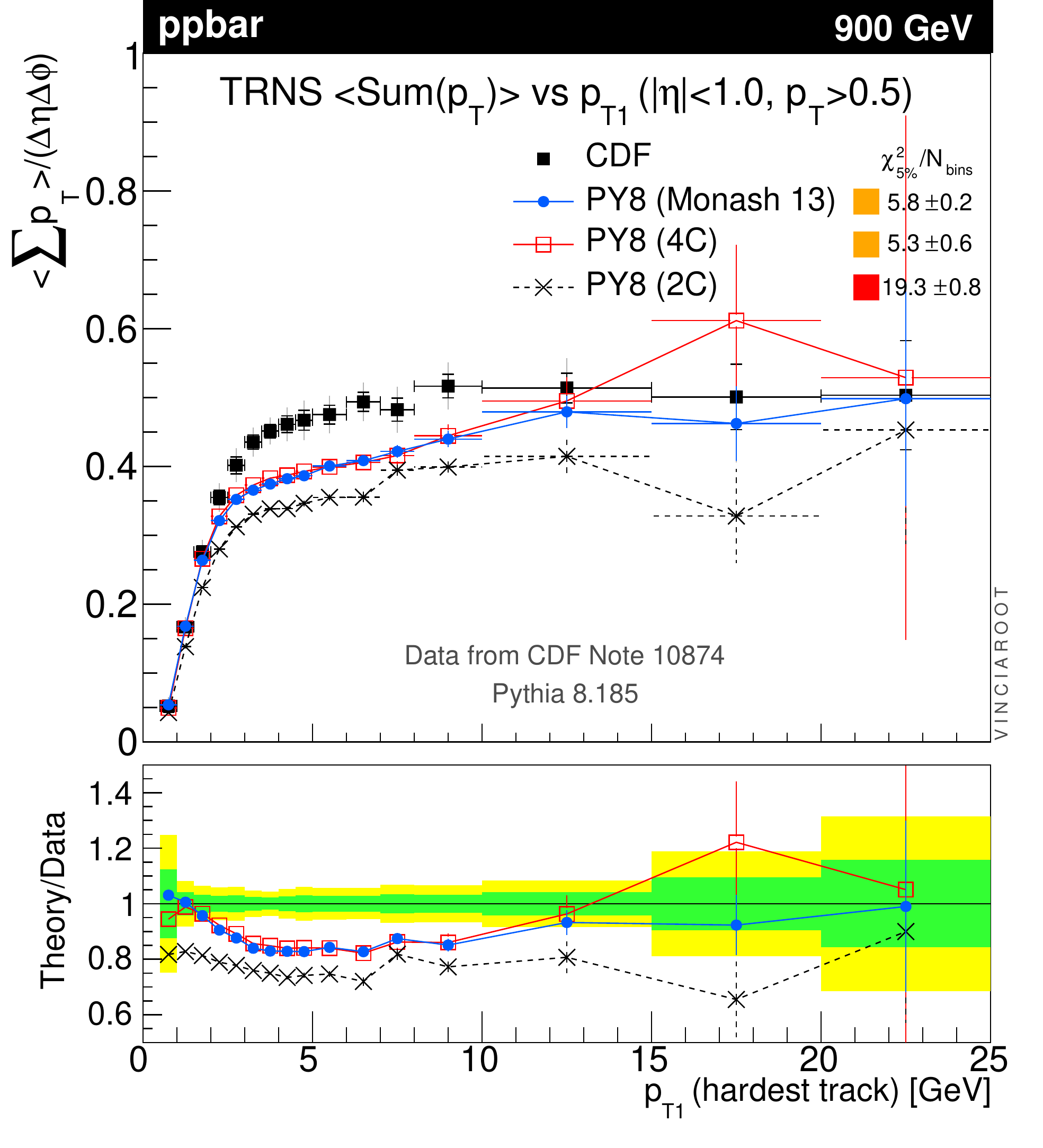}
\includegraphics*[scale=0.31]{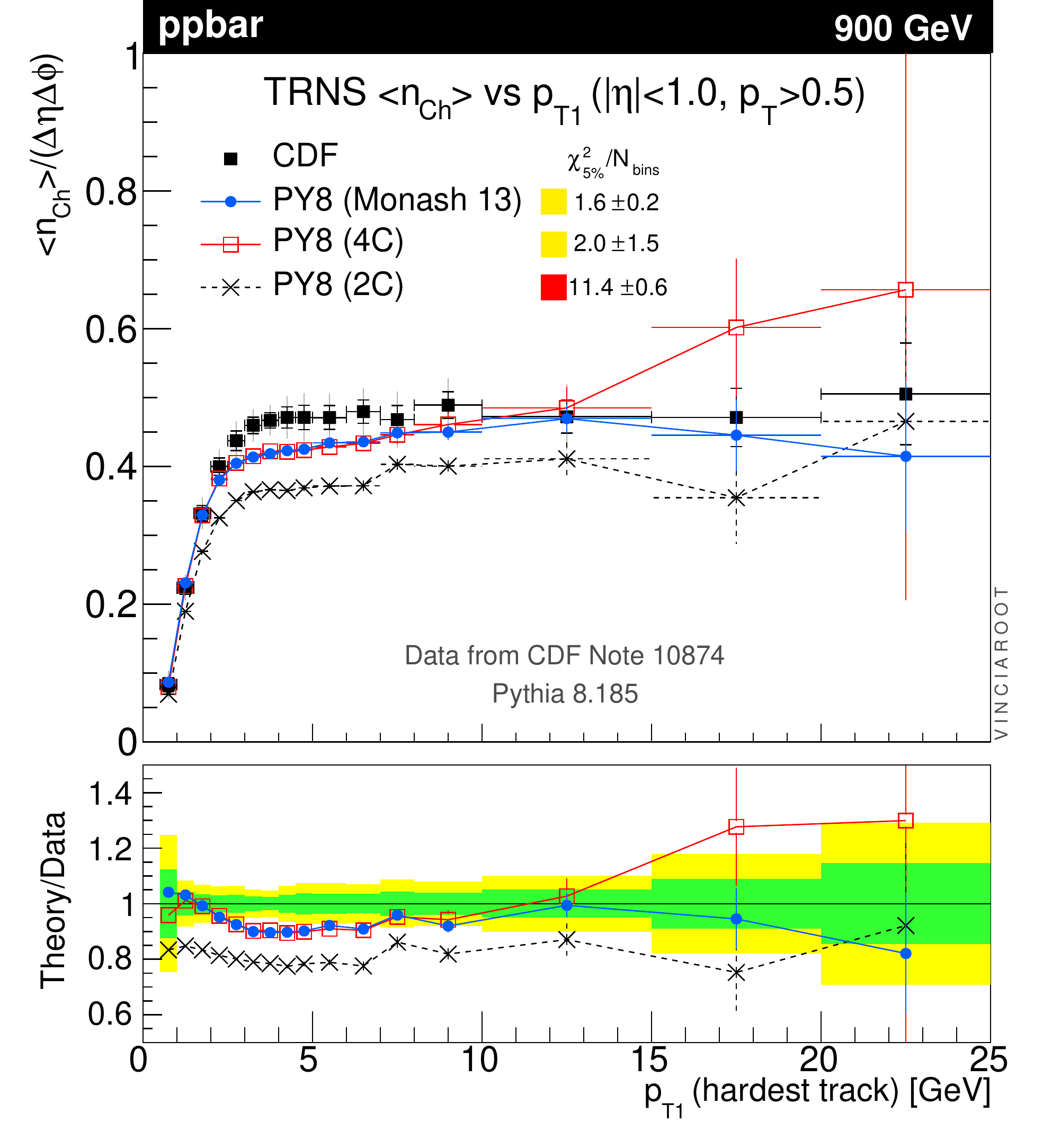}
\includegraphics*[scale=0.31]{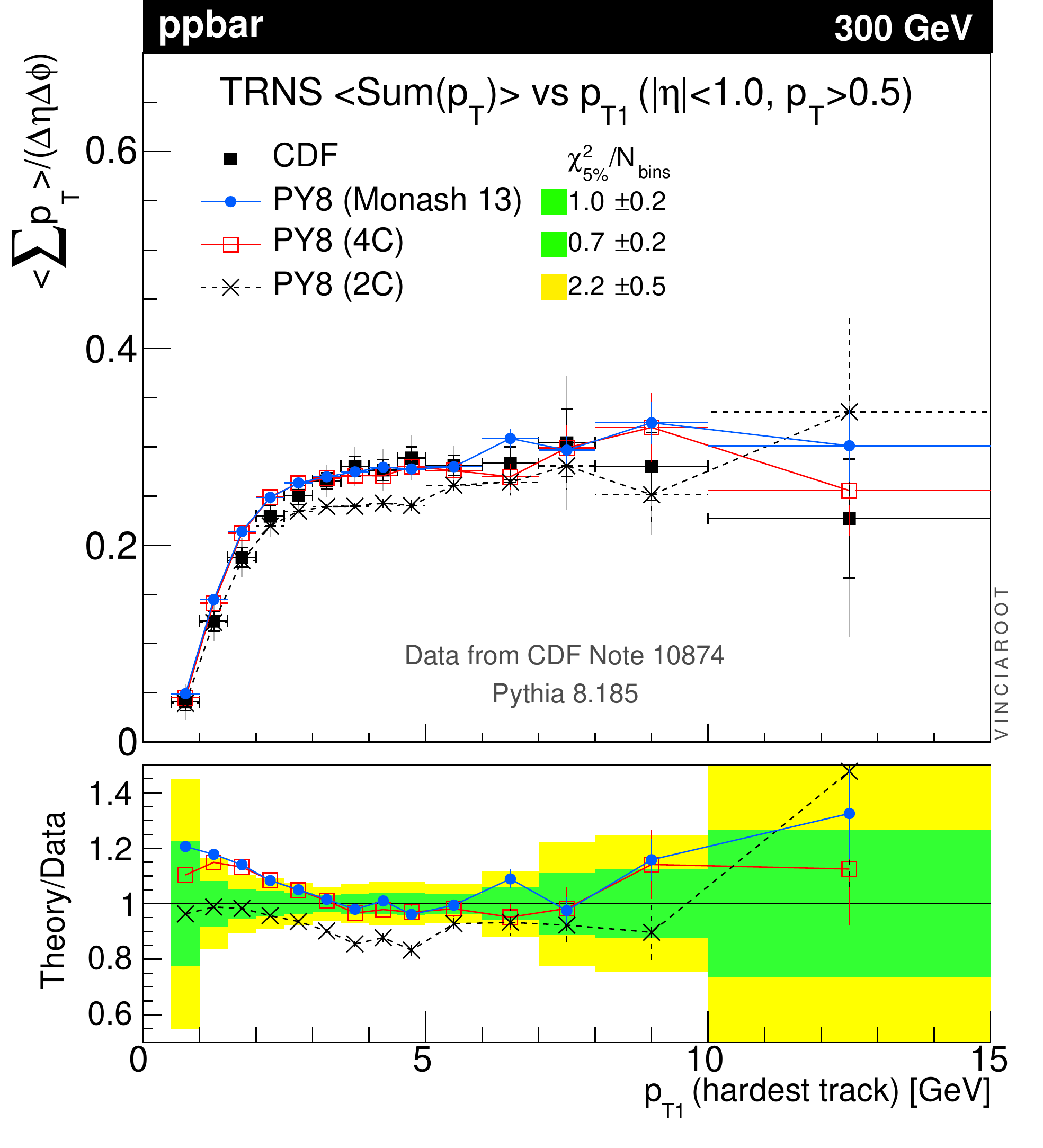}
\includegraphics*[scale=0.31]{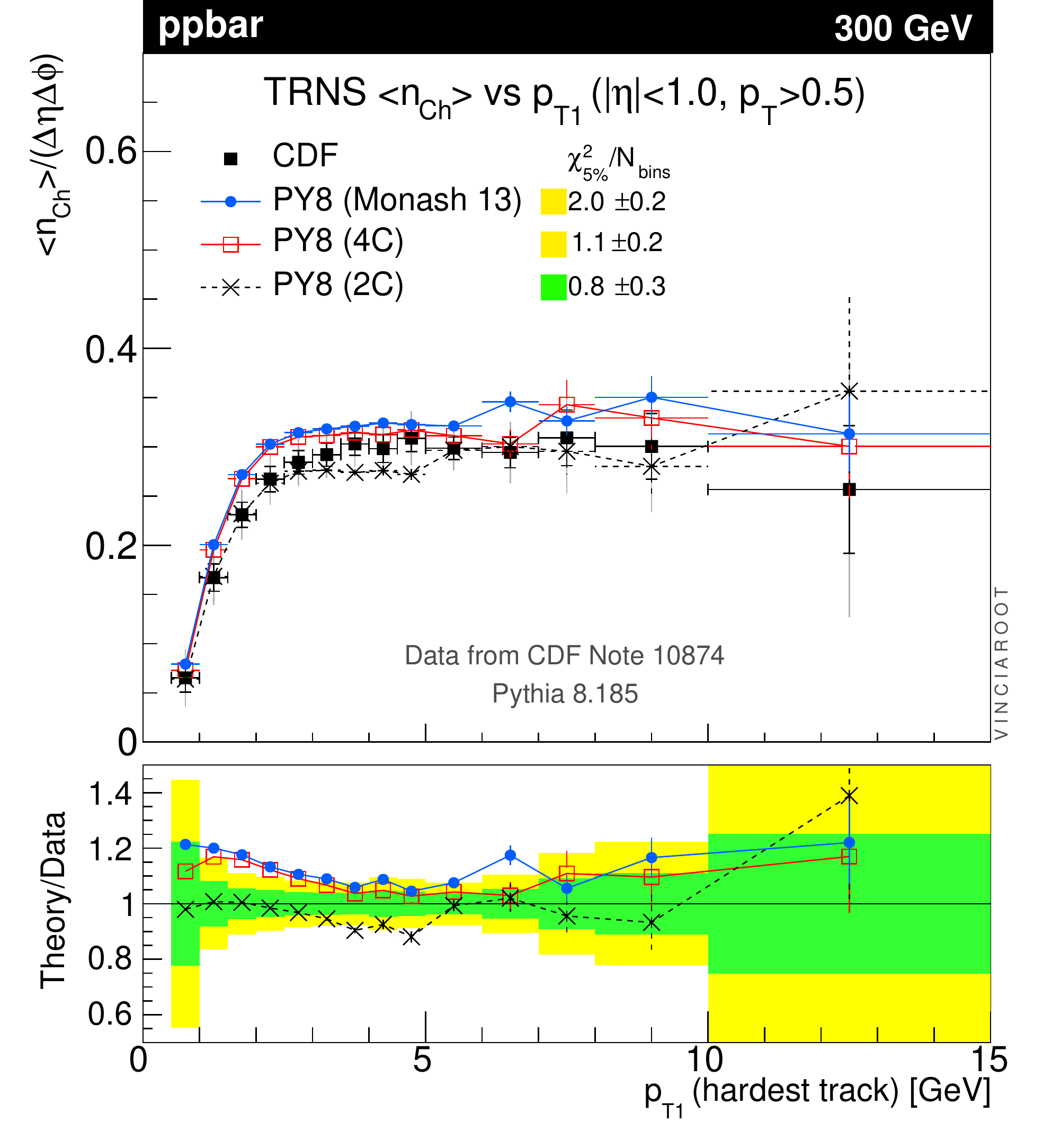}
\caption{The Tevatron energy scan. The underlying event ({\sl left:}
  average summed-$p_\perp$ density and {\sl right:} average track
  density, in the transverse region, as function of leading-track
  $p_\perp$) for 
  $p\bar{p}$ collisions at 300
  (bottom row), 900 GeV (middle row), and 1960 GeV (top row). 
\label{fig:scalingUE}}
\end{figure}
Finally, in \figRef{fig:scalingUE}, we compare to the underlying event
measured in the highly useful energy scan that was performed at the
Tevatron in the last days before its
shutdown~\cite{CDFnote10874,rickscan}, during which extremely 
high min-bias statistics were collected at 300 and
900 GeV CM energy over a period of a few days. As was already noted in
\secRef{sec:hadronColliders}, the UE at 7 TeV is slightly larger in
the Monash 2013 tune than in tune 4C. As can be seen from the plots
here, the two tunes give comparable results for all the Tevatron
energies. Interestingly, the UE plateau region at 900 and 1960 GeV 
is reached sooner in
these models than in the data, translating to a roughly 10\% - 20\%
too low UE level for leading-track $p_\perp$ values in the
neighbourhood of the transition from the rise to the plateau (roughly
for leading-track $p_\perp$ values $2<p_{T1}<10$ GeV). This indicates
that the energy scaling of the UE modeling and in particular 
the details of its transition between central and peripheral
collisions, is still not satisfactorily understood. 

\section{Conclusions and Exhortation \label{sec:proposals}}

We have presented a reanalysis of the constraints on fragmentation in
$ee$ collisions, and applied the results to update the final-state 
fragmentation parameters in \Py~8. We combine these parameters with
a tune to hadron-collider data, using a new NNPDF 2.3 LO QCD+QED PDF
set, which has been encoded so it is available as an internal PDF set
in \Py~8, independently of \cname{LHAPDF}~\cite{Bourilkov:2006cj}. 

In
this PDF set as well as in our tune, the value of the strong 
coupling for hard-scattering matrix elements is fixed to be  
$\alpha_s(M_Z)=0.13$, consistent with other LO determinations of it. 
For initial- and final-state radiation, our tune 
uses the effective value $\alpha_s(M_Z)=0.1365$. The difference is
consistent with an effective translation between the
$\overline{\mrm{MS}}$ and CMW schemes.   
We note that alternative (LO, NLO, and
NNLO) NNPDF 2.3 QCD+QED sets with $\alpha_s(M_Z)=0.119$ are also available in
the code, for people who want to check the impact of using a different
$\alpha_s(M_Z)$ value and/or higher-order PDF sets on hard-scattering
events. For the purpose of such studies, we point out that it is
possible, in \Py~8, to preserve most of the features of the shower-
and underlying-event tuning by changing only the PDF for the
hard-scattering matrix elements, leaving the PDF choice for the shower
evolution and MPI framework unaltered (see the \Py~8 HTML manual's PDF
section, under \ttt{PDF:useHard}). 

The updated
parameters are available as an option starting from \Py 8.185, by
setting\begin{center}
\texttt{Tune:ee = 7} and \texttt{Tune:pp = 14}~.
\end{center}

By no means do we claim that this should be regarded as the final word
in tuning the \Py~8 Monte Carlo model. First of all, the model 
continues to evolve. For instance, developments foreseen for the near
future include updates of colour reconnections, diffraction, and the
treatment of $g\to q\bar{q}$ splittings. Any of these should in
principle be accompanied by a reevaluation of the model constraints. 

Moreover, despite
the comprehensive view of collider data we have attempted to take in
this study, there still remains several issues that were not
addressed, 
including: initial-final interference and coherence
effects~\cite{Skands:2012mm,Winter:2013xka} (probably more a 
modelling issue than a tuning one); reliable estimates of theoretical
uncertainties~\cite{Buckley:2009bj,Skands:2010ak,Giele:2011cb,Schulz:2011qy,Richardson:2012bn,Hartgring:2013jma}; 
 diffraction\footnote{In particular, the constraints on fragmentation
  mainly come from SLD and LEP, where the non-perturbative parameters
  are clearly defined at the shower cutoff scale, $Q_\mrm{had}$, whereas
  diffraction is dominated by soft physics, for which the definition
  of the effective hadronization scale is less clear. The
  amount of MPI in hard diffractive events also requires
  tuning.}~\cite{Schuler:1993td,Navin:2010kk,Ciesielski:2012mc} and other
colour-singlet phenomena such as onium production; long-distance
(e.g., forward-backward, forward-central, and ``ridge''-type) 
correlations~\cite{Ansorge:1988fg,Khachatryan:2010gv,Wraight:2011ej,soegaard,Sicking:2012hga,Feofilov:2013kna,Abelev:2013sqa}; 
$B$-hadron decays~\cite{Lange:2001uf}; and tuning in the presence of
matrix-element matching, at LO and
NLO (see
\cite{Corke:2010zj,Cooper:2011gk,Richardson:2012bn,Hartgring:2013jma,Hoeche:2014lxa}
for recent phenomenological studies). 
Especially 
in the latter context of matrix-element matching, we expect that in
many cases \Py~8 will be used together with codes such as
ALPGEN~\cite{Mangano:2002ea}, MADGRAPH~\cite{Alwall:2011uj}, 
aMCatNLO~\cite{Hirschi:2011pa}, or POWHEG~\cite{Alioli:2010xd}, either
using the matching algorithms of those programs themselves, 
or via any of \Py's several internal (LHEF-based~\cite{Alwall:2006yp})
implementations of matching schemes (POWHEG~\cite{Frixione:2007vw},
CKKW-L~\cite{Catani:2001cc,Lonnblad:2001iq,Lavesson:2005xu}, 
MLM~\cite{Mangano:2006rw,Mrenna:2003if}, 
UMEPS~\cite{Lonnblad:2012ng}, NL3~\cite{Lavesson:2008ah},
UNLOPS~\cite{Lonnblad:2012ix}). The impact of such corrections on MC
tuning depends on the details of the matching scheme (especially its
treatment of unitarity), and there is in general a
non-negligible possibility of ``mis-tuning'' when combining a
stand-alone tune with ME corrections. A simple example illustrating
this is the effective value of $\alpha_s(M_Z)$, which for a
leading-order tune is typically of order $0.13$, while a consistent NLO 
correction scheme should be compatible with values closer to
$0.12$~\cite{Hartgring:2013jma}. There is also the question of the
running order of $\alpha_s$. The propagation of such changes from
the level of hard matrix elements through the shower and
hadronization tuning process are still not fully explored, and hence we
advise users to perform simple cross-checks, such as checking the
distributions presented in this paper, before and after applying
matrix-element corrections. Parameters that appear on both sides of
the matching, such as $\alpha_s$, should also be checked for
consistency~\cite{Cooper:2011gk}.

We noted several issues concerning the $ee$ data used to
constrain the fragmentation modelling, that it would be good to
resolve. In particular, we find some tensions between the
identified-particle rates extracted from 1) HEDPATA, 2) Sec.~46 of the
PDG, and 3) the $Z$ boson summary table in the PDG, as discussed in
more detail in \secRef{sec:hadronization}, and concerning which we
made some (subjective) decisions to arrive at a set of hopefully
self-consistent constraints for this work. We also note that the
overall precision of the fragmentation constraints could likely be
significantly improved by an FCC-ee type machine, such as Tera-Z, a
possibility we hope to see more fully explored in the context of
future $ee$ QCD phenomenology studies.  

We conclude that the new parameter set does improve
significantly on the previous default values in several respects,
including better agreement with data on:
\begin{enumerate}
\item the net strangeness fraction (has been increased by 10\%,
  reflected not only in improved kaon and hyperon yields, but also in
  the $D_s$ and $B_s$ fractions), 
\item the ultra-hard fragmentation tail (has been softened,
especially for leading baryons and for $D$ and $B$ hadrons), 
\item the $p_{TZ}$ spectrum (softened at low $p_{TZ}$), 
\item the minimum-bias charged multiplicity in the forward region (has
increased by 10\%), 
\item the underlying event at 7 TeV (is very slightly higher than before).
\end{enumerate}

Some questions that remain open include the following. 
We see a roughly 20\% excess of very soft kaons in both $ee$ and $pp$ environments,
cf.~\figsRef{fig:idSpectra} and \ref{fig:LHCK0S}, despite the
overall kaon yields being well described, and the overall baryon
yields at LHC appear to be underestimated by at least $30\%$ despite
good agreement at LEP. The momentum spectra of
heavier strange particles are also poorly reproduced, in particular at
LHC. It is interesting and exciting that some of the LHC spectra appear
to be better described by allowing collective flow in a fraction of events
(cf.~the EPOS model~\cite{Pierog:2013ria}), though we believe the jury is still out
on whether this accurately reflects the underlying physics. For
instance, it has been argued that colour reconnections can mimick flow
effects~\cite{Ortiz:2013yxa}, and they may also be able to 
modify the yield of baryons if the creation/destruction of string junctions is
allowed~\cite{Sjostrand:2002ip}. We look forward to future discussions
on these issues.

We round off with an exhortation for follow-ups on this study to 
provide:
\begin{itemize}
\item \textbf{Not only central tunes}: experiments
and other user-end colleagues need more than central
descriptions of data; there is an increasing need for serious uncertainty
estimates. In the context of tune variations, it is important
to keep in mind that the modelling uncertainties are often intrinsically 
non-universal. Therefore, the constraints obtained by considering data 
uncertainties only (e.g., in the spirit of \cname{Professor}'s
eigentunes~\cite{Buckley:2009bj}) can at most
constitute a lower bound on the theoretical uncertainty (similarly to
the case for PDFs). A serious uncertainty estimate includes some
systematic modelling variation, irrespectively of, and in addition to, what data
allows (e.g., in the spirit of the Perugia set of tunes for
\Py~6~\cite{Skands:2010ak}). We therefore hope the future will see
more elaborate combinations of data- and theory-driven approaches to
systematic tune uncertainties;
\item \textbf{Not only global tunes}: the power of
MC models lies in their ability to \emph{simultaneously} describe a
large variety of data, hence we do not mean to imply 
that one should give up on universality and tune 
to increasingly specific corners of phase space, disregarding (or de-emphasizing, with lower weights)
all others. However, as proposed in \cite{Schulz:2011qy},  
one can obtain useful explicit tests
of the universality of the underlying physics model by performing
independent tunes on separate ``physics windows'', say in the forward
vs.\ central regions, for different event-selection criteria, at
different collider energies, or even for different collider types. In
this connection, just making one global ``best-fit'' tune may obscure
tensions between the descriptions of different complementary data
sets. By performing independent tunes to each data set separately, and
checking the degree of universality of the resulting parameters, one
obtains a powerful cross check on the underlying physics model. 
If all sets produce the same or similar parameters, then universality
is OK, hence a global tune makes very good sense, and the remaining 
uncertainties can presumably be reliably estimated from data
alone. If, instead, some data sets result in significantly  
different tune parameters, one has a powerful indication that the
universality of the underlying modeling is breaking down, which can
lead to several productive actions: 1) it can be taken into account in
the context of uncertainty variations, 2) the nature of the data sets
for which non-universal tune parameters are obtained can implicitly
indicate the nature of the problem, leading to more robust conclusions
about the underlying model than merely whether a tune can/cannot fit
the data, and 3) the observations can be communicated to the model authors
in a more unambiguous way, hopefully resulting in a speedier cycle of
model improvements. 
\end{itemize}

We hope that the Monash 2013 tune parameters may serve as a useful
starting point for phenomenology studies and for future \Py~8 tuning efforts. 

\subsection*{Acknowledgments}
We thank S.~Mrenna, M.~Ritzmann, and T.~Sj\"ostrand for comments on
the manuscript and  
L.~de~Nooij for pointing us to the ALICE $K^*$ and $\phi$
measurements~\cite{Abelev:2012hy} and to the ATLAS $\phi$ measurements 
in~\cite{Aad:2014rca}. The work of J.~R.\ is supported by an STFC
Rutherford Fellowship ST/K005227/1. This work was supported in part by
the Research Executive Agency (REA) of the European Commission under
the Grant Agreements PITN-GA-2012-315877 (MCnet).

\appendix
\section{Monash 2013 Tune Parameters \label{app:tunes}}
In \tabsRef{tab:FSR} -- \ref{tab:mpi}, we list the FSR, fragmentation, 
parameters for the Monash tune of \Py. For
reference, we compare them to the current default parameters.

\begin{table}[tp]
\centering\small 
\begin{tabular}{llll}
\bf FSR Parameters & \bf Monash 13 & \bf (Default) & \bf Comment
\\\toprule 
TimeShower:alphaSvalue   &= 0.1365 & = 0.1383  &! Effective alphaS(mZ) value
\\TimeShower:alphaSorder   &= 1   & = 1      &! Running order
\\TimeShower:alphaSuseCMW  &= off & = off    &! Translation from
$\overline{\mbox{MS}}$ to CMW
\\TimeShower:pTmin         &= 0.50 & = 0.40  &! Cutoff for QCD radiation
\\TimeShower:pTminChgQ     &= 0.50 & = 0.40  &! Cutoff for QED radiation
\\TimeShower:phiPolAsym    &= on   & = on    &! Asymmetric azimuth distributions
\\\bottomrule
\end{tabular}
\caption{Final-state radiation (FSR) parameters. \label{tab:FSR}}
\end{table} 

\begin{table}[tp]
\centering\small 
\begin{tabular}{llll}
\bf HAD Parameters & \bf Monash 13 & \bf (Default) & \bf Comment
\\\toprule 
\multicolumn{4}{l}{\# String breaks: pT and z distributions}
\\StringPT:sigma          &= 0.335 &= 0.304  &! Soft pT in string breaks (in GeV)
\\StringPT:enhancedFraction &= 0.01  &= 0.01 &! Fraction of breakups with enhanced pT
\\StringPT:enhancedWidth    &= 2.0   &= 2.0  &!  Enhancement factor
\\StringZ:aLund            &= 0.68 &= 0.3  &! Lund FF a (hard fragmentation supp)
\\StringZ:bLund            &= 0.98 &= 0.8  &! Lund FF b (soft fragmentation supp)
\\StringZ:aExtraSquark     &= 0.0 &=  0.0 &! Extra a when picking up an s quark
\\StringZ:aExtraDiquark    &= 0.97 &= 0.50 &! Extra a when picking up a
diquark 
\\StringZ:rFactC &= 1.32 &=  1.00  &! Lund-Bowler c-quark parameter
\\StringZ:rFactB &= 0.855 &=  0.67 &! Lund-Bowler b-quark parameter
\\\# Flavour composition: mesons
\\StringFlav:ProbStoUD     &= 0.217  &= 0.19 &! Strangeness-to-UD ratio
\\StringFlav:mesonUDvector &= 0.5  &= 0.62 &! Light-flavour vector suppression
\\StringFlav:mesonSvector  &= 0.55 &= 0.725 &! Strange vector suppression
\\StringFlav:mesonCvector &= 0.88 &= 1.06   &! Charm vector suppression
\\StringFlav:mesonBvector &= 2.2  &= 3.0   &! Bottom vector suppression
\\StringFlav:etaSup        &= 0.60  &= 0.63  &! Suppression of eta mesons
\\StringFlav:etaPrimeSup   &= 0.12  &= 0.12  &! Suppression of eta' mesons
\\\# Flavour composition: baryons
\\StringFlav:probQQtoQ     &= 0.081  &= 0.09 &! Diquark rate (for baryon production)
\\StringFlav:probSQtoQQ    &= 0.915  &= 1.000 &! Strange-diquark suppression   
\\StringFlav:probQQ1toQQ0  &= 0.0275 &= 0.027 &! Vector diquark suppression
\\StringFlav:decupletSup   &= 1.0     &= 1.0 &! Spin-3/2 baryon suppression
\\StringFlav:suppressLeadingB &= off &= off &! Optional leading-baryon suppression
\\StringFlav:popcornSpair  &= 0.9 &= 0.5 &! 
\\StringFlav:popcornSmeson &= 0.5 &= 0.5 &! 
\\\bottomrule
\end{tabular}
\caption{String-breaking parameters. \label{tab:string-breaking}}
\end{table}

\begin{table}[tp]
\centering\small 
\begin{tabular}{llll}
\bf PDF and ME Parameters & \bf Monash 13 & \bf (Default) & \bf Comment
\\\toprule 
PDF:pSet                  &= 13     & = 8 &! PDF set for the proton
\\SigmaProcess:alphaSvalue  &= 0.130  & 0.135 & ! alphaS(MZ) for matrix elements
\\MultiPartonInteractions:alphaSvalue &= 0.130 & 0.135 & ! alphaS(MZ)
for MPI
\\\bottomrule
\end{tabular}
\caption{Parton-distribution (PDF) and Matrix-Element (ME) parameters. \label{tab:PDFME}}
\end{table} 

\begin{table}[tp]
\centering\small 
\begin{tabular}{llll}
\bf ISR Parameters & \bf Monash 13 & \bf (Default) & \bf Comment
\\\toprule 
SpaceShower:alphaSvalue   &= 0.1365 & = 0.137  &! Effective alphaS(mZ) value
\\SpaceShower:alphaSorder   &= 1   & = 1      &! Running order
\\SpaceShower:alphaSuseCMW  &= off & = off    &! Translation from
$\overline{\mbox{MS}}$ to CMW
\\SpaceShower:samePTasMPI   &= off  & = off   &! ISR cutoff type
\\SpaceShower:pT0Ref        &= 2.0  & = 2.0   &! ISR pT0 cutoff 
\\SpaceShower:ecmRef        &= 7000.0 & = 1800.0 &! ISR pT0 reference
ECM scale
\\SpaceShower:ecmPow        &= 0.0  & = 0.0 &! ISR pT0 scaling power 
\\SpaceShower:rapidityOrder &= on   & = on    &! Approx coherence via y-ordering
\\SpaceShower:phiPolAsym    &= on   & = on    &! Azimuth asymmetries from gluon pol
\\SpaceShower:phiIntAsym    &= on   & = on    &! Azimuth asymmetries from interference
\\TimeShower:dampenBeamRecoil &=         on &= on  & ! Recoil
dampening in final-initial dipoles
\\BeamRemnants:primordialKTsoft &= 0.9 & = 0.5 &! Primordial kT for
soft procs
\\BeamRemnants:primordialKThard &= 1.8 & = 2.0 &! Primordial kT for
hard procs
\\BeamRemnants:halfScaleForKT &= 1.5 &= 1.0 &! Primordial kT soft/hard
boundary
\\BeamRemnants:halfMassForKT &= 1.0 &= 1.0 &! Primordial kT soft/hard
mass boundary
\\\bottomrule
\end{tabular}
\caption{Initial-state radiation (ISR) and primordial-$k_T$
  parameters. \label{tab:ISR}} 
\end{table} 

\begin{table}[tp]
\centering\small 
\begin{tabular}{llll}
\bf MPI Parameters & \bf Monash 13 & \bf (Default) & \bf Comment
\\\toprule 
MultipartonInteractions:pT0Ref &= 2.28  & = 2.085 & ! MPI pT0 IR regularization scale 
\\MultipartonInteractions:ecmRef &= 7000.0 & = 1800.0 & ! MPI pT0
reference ECM scale
\\MultipartonInteractions:ecmPow &= 0.215  & = 0.19 & ! MPI pT0 scaling power
\\MultipartonInteractions:bProfile &= 3 & = 3 & ! Transverse matter
overlap profile 
\\MultipartonInteractions:expPow &= 1.85 &= 2.0 &! Shape parameter
\\BeamRemnants:reconnectRange &= 1.8 & = 1.5 &! Colour Reconnections
\\SigmaTotal:zeroAXB &=                  on &= on  & ! Carried over
from 4C
\\SigmaDiffractive:dampen &=             on &= on  & ! Carried over
from 4C
\\SigmaDiffractive:maxXB &=              65.0  &= 65.0 &! Carried over
from 4C
\\SigmaDiffractive:maxAX &=              65.0  &= 65.0 &! Carried over
from 4C
\\SigmaDiffractive:maxXX &=              65.0  &= 65.0 &! Carried over
from 4C
\\Diffraction:largeMassSuppress &= 4.0 &= 2.0 &! High-mass diffraction
suppression power
\\\bottomrule
\end{tabular}
\caption{Multi-Parton-Interaction (MPI), Colour-Reconnection (CR), and
  Diffractive parameters. \label{tab:mpi}}
\end{table} 

\clearpage
\section{Additional Plots \label{app:plots}}
\subsection{LEP Event-Shape Distributions \label{app:LEP}}

\begin{figure}[h!p]
\vskip-3mm\centering
\includegraphics*[scale=0.31]{vincia03-Monash-T.pdf}
\includegraphics*[scale=0.31]{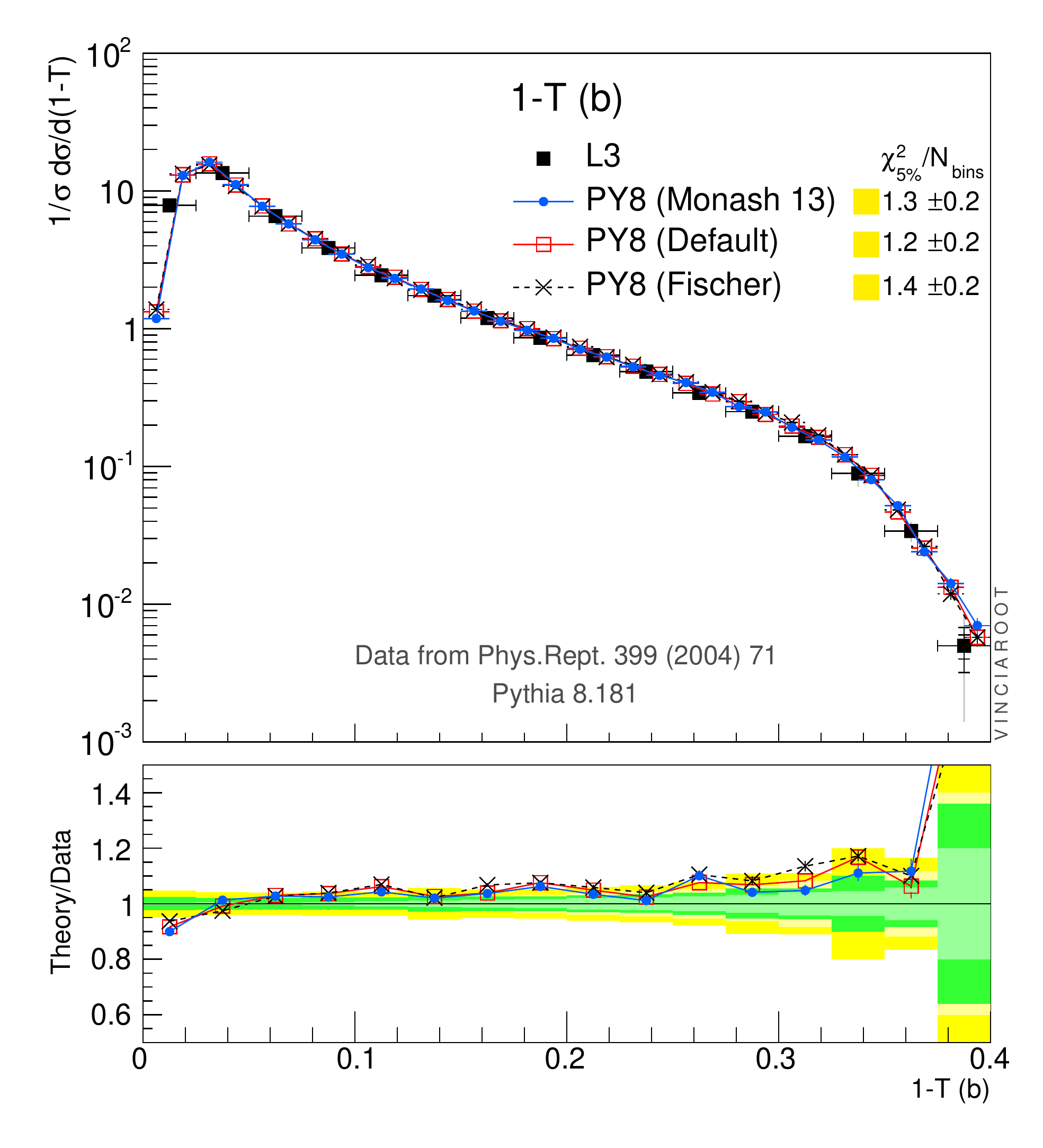}

\includegraphics*[scale=0.31]{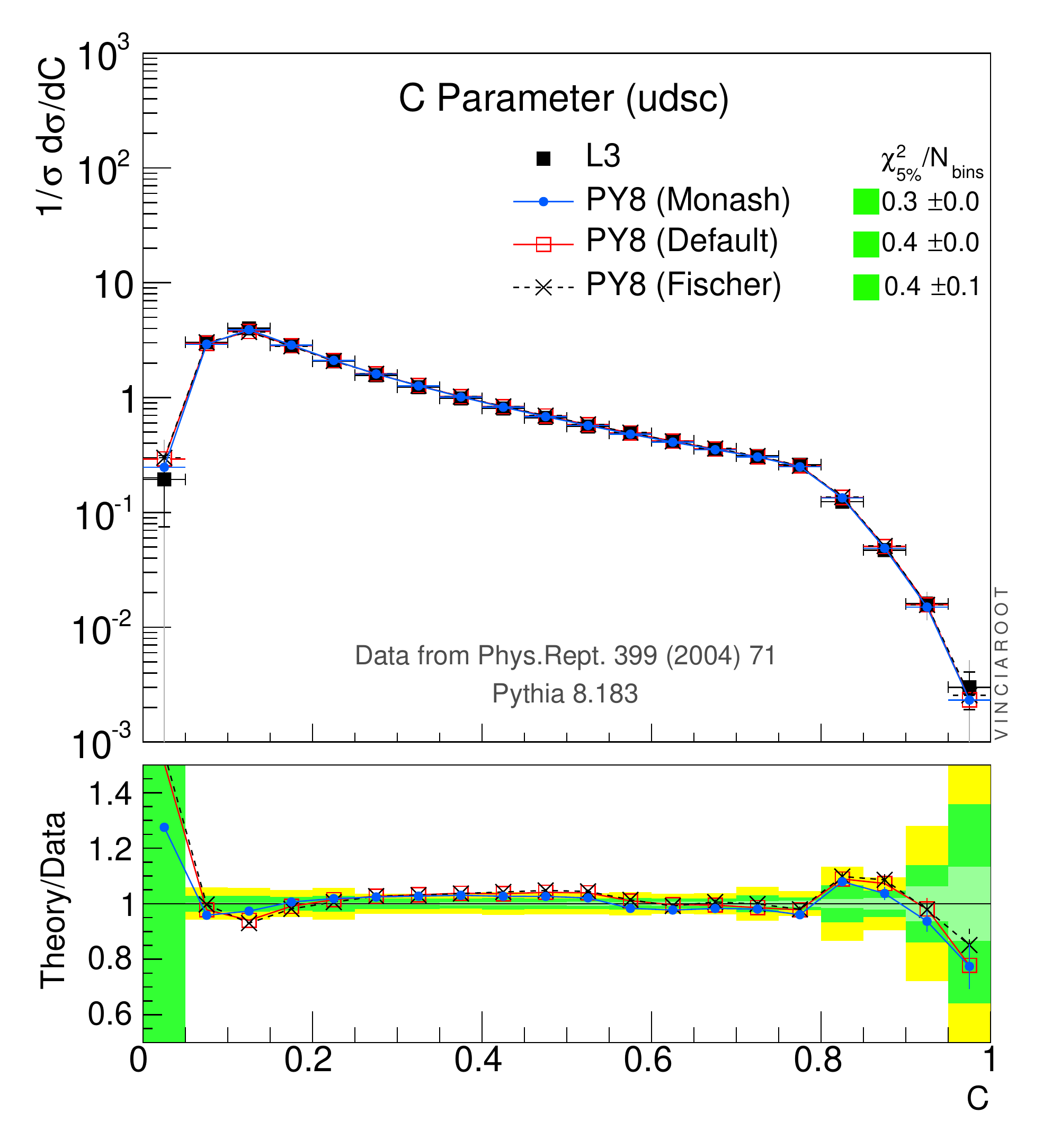}
\includegraphics*[scale=0.31]{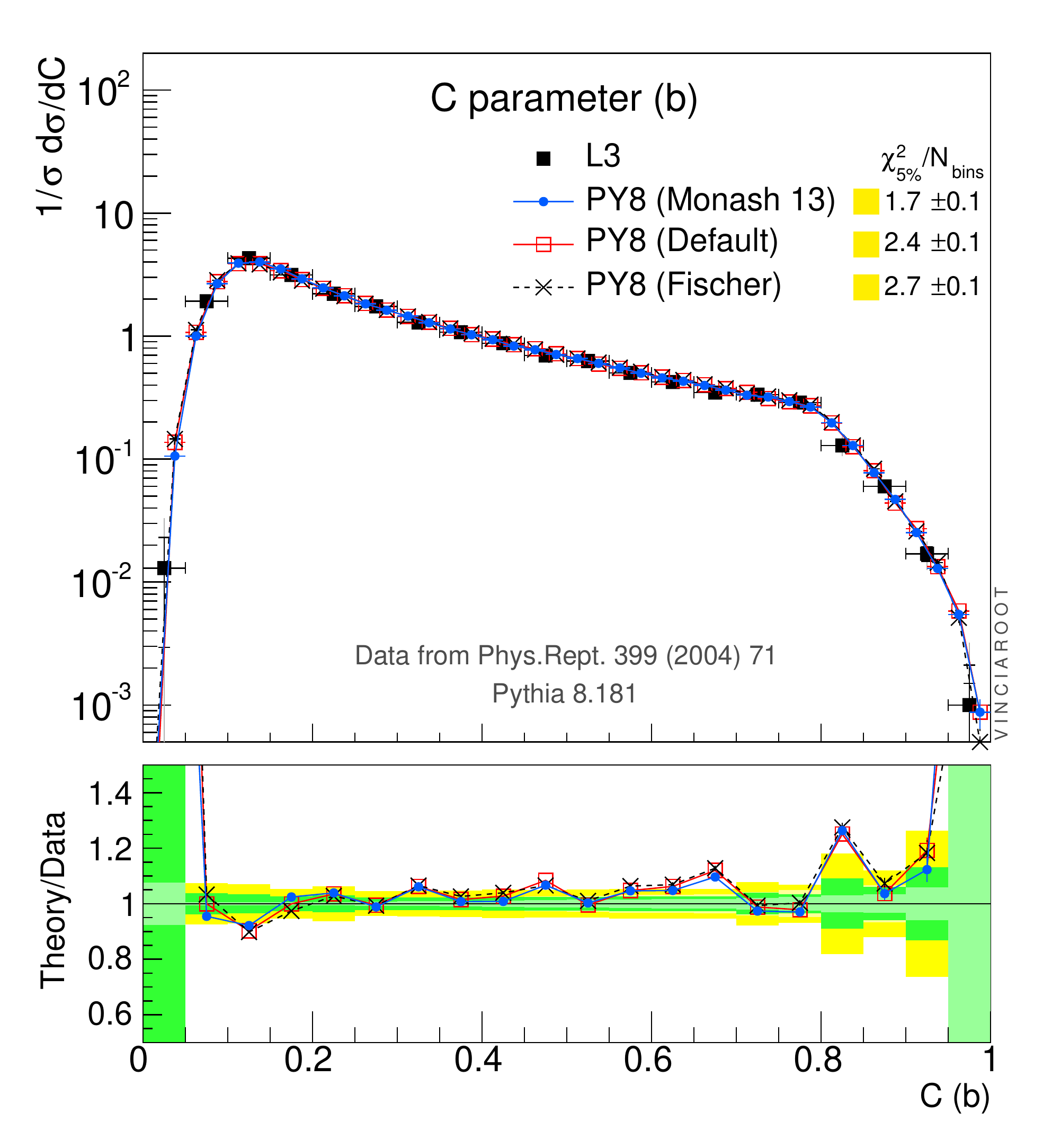}

\includegraphics*[scale=0.31]{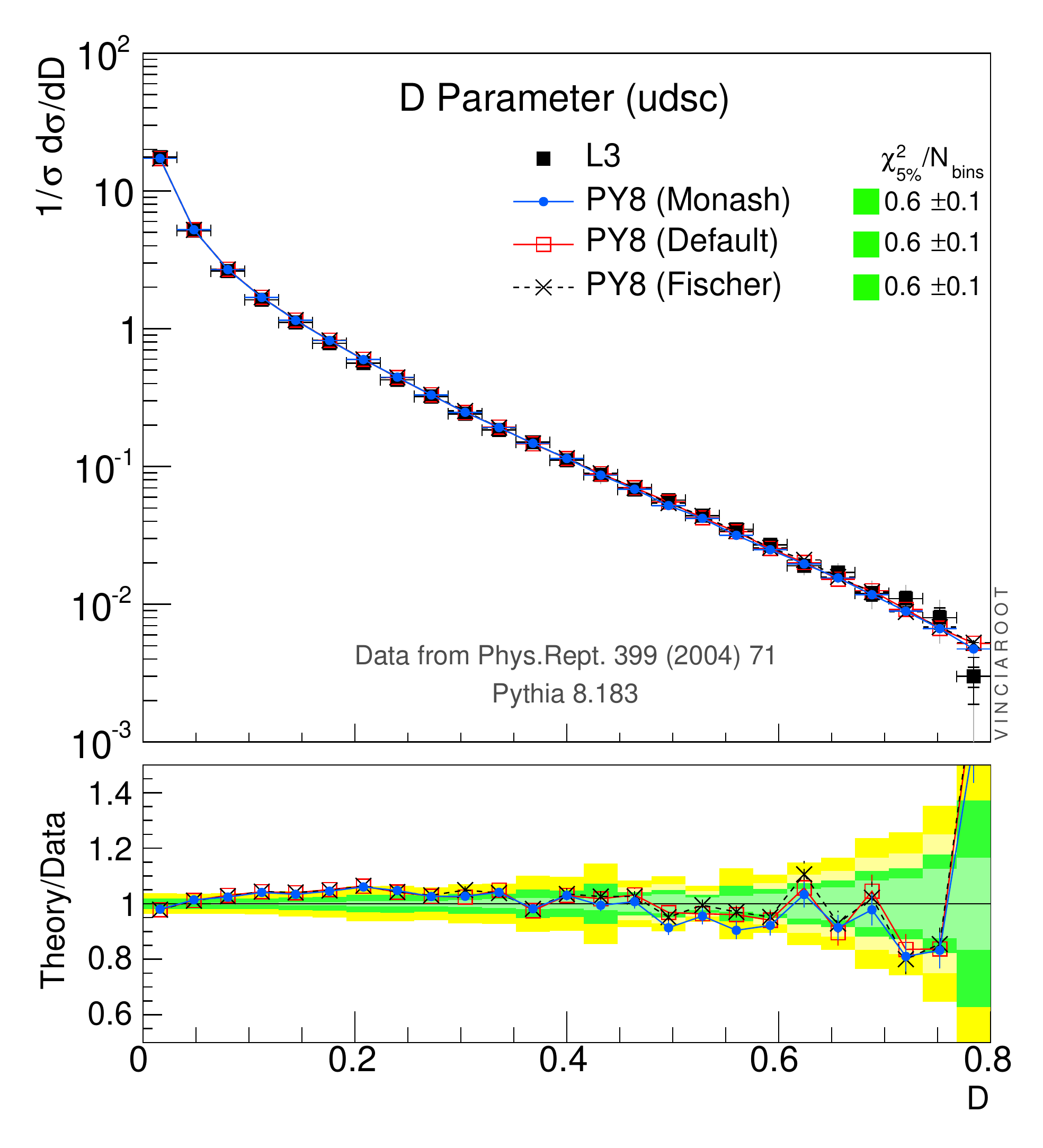}
\includegraphics*[scale=0.31]{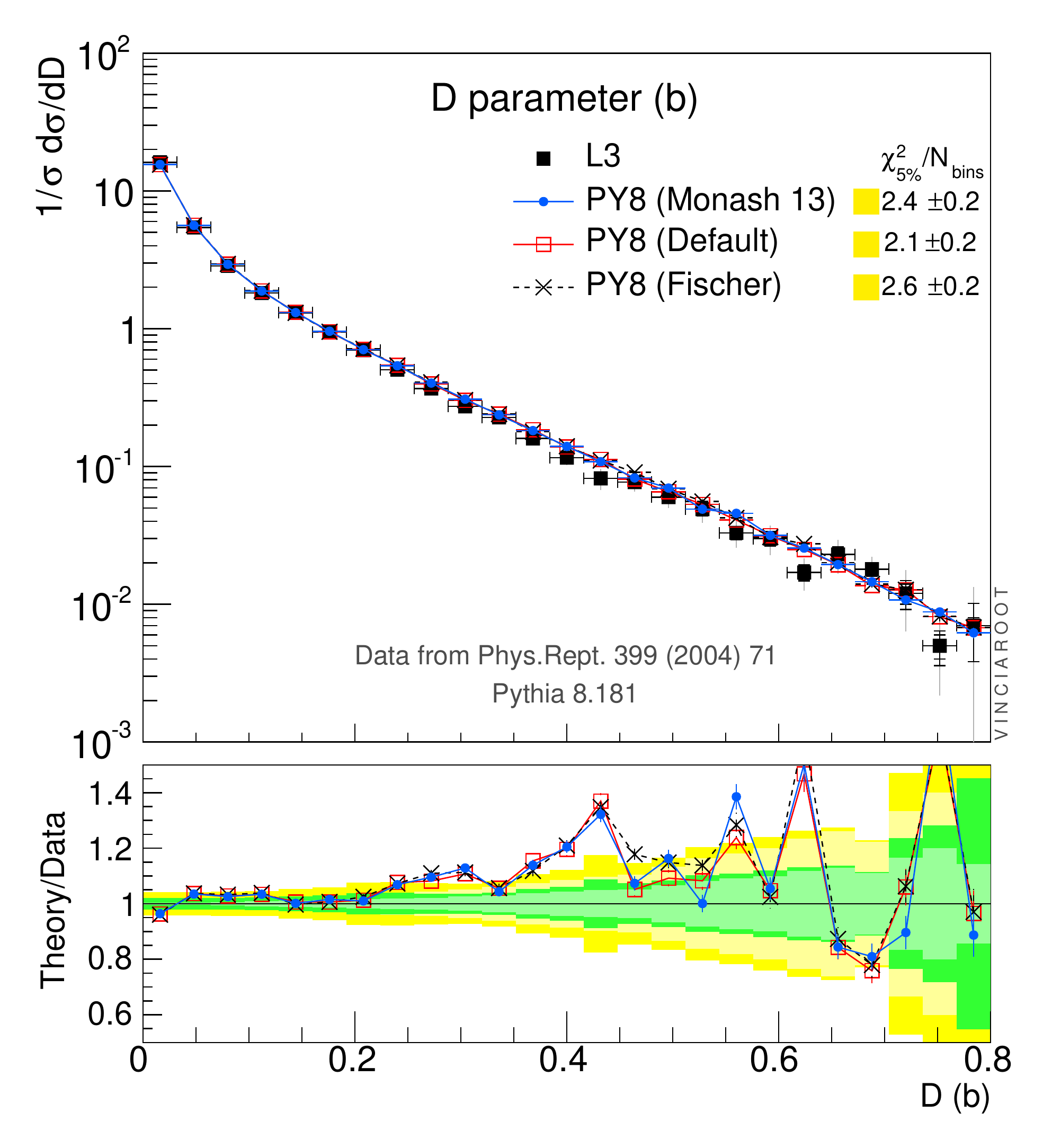}
\caption{Hadronic $Z$ decays at $\sqrt{s}=91.2\GeV$. The $T$, $C$, and $D$
  event-shape parameters, as measured by 
  L3~\cite{Achard:2004sv}, 
for light-flavour (left) and $b$-tagged (right) events, respectively.
\label{fig:LEPshapes1}
}
\end{figure}

\begin{figure}[h!p]
\centering\vskip-3mm
\includegraphics*[scale=0.31]{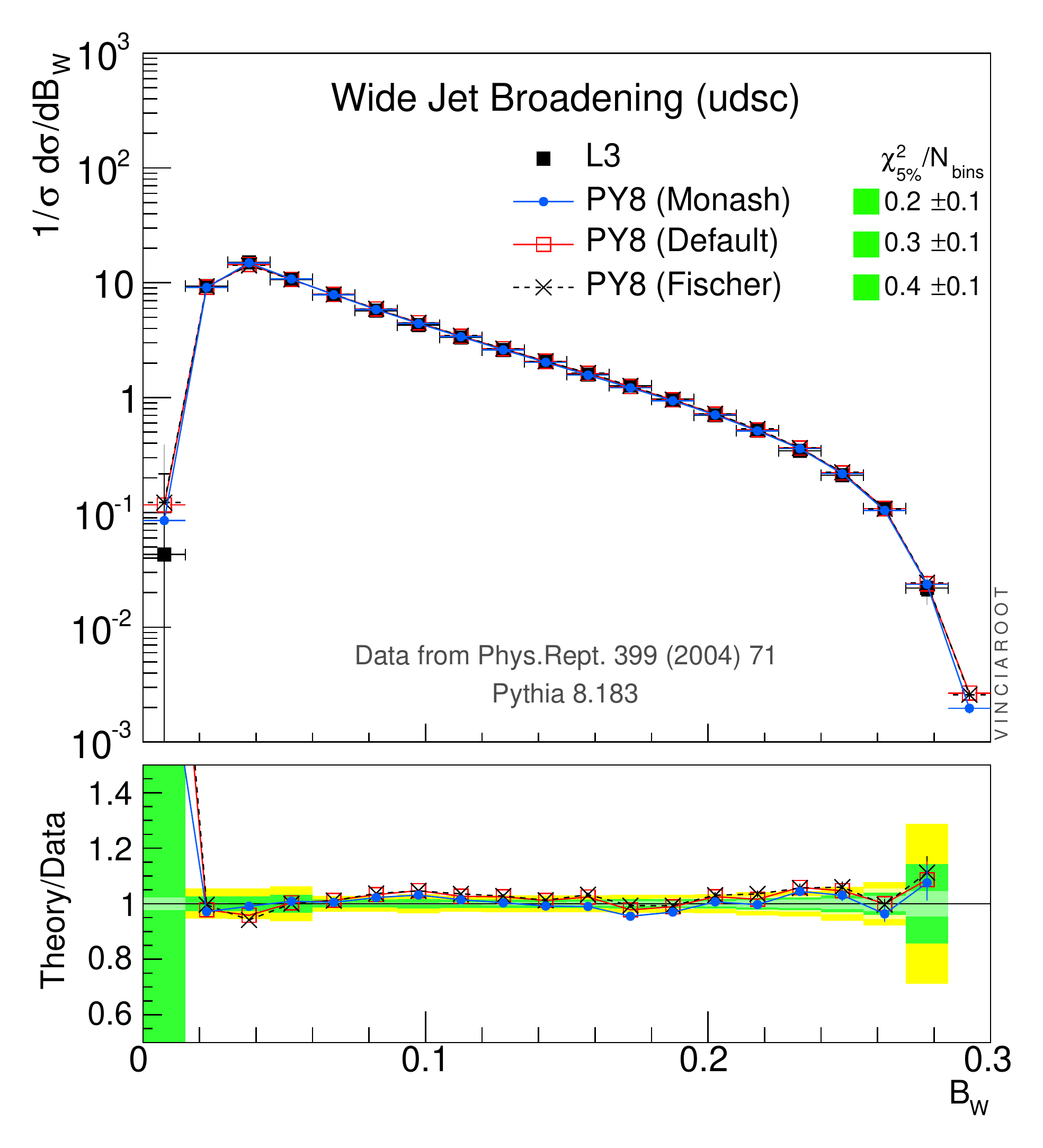}
\includegraphics*[scale=0.31]{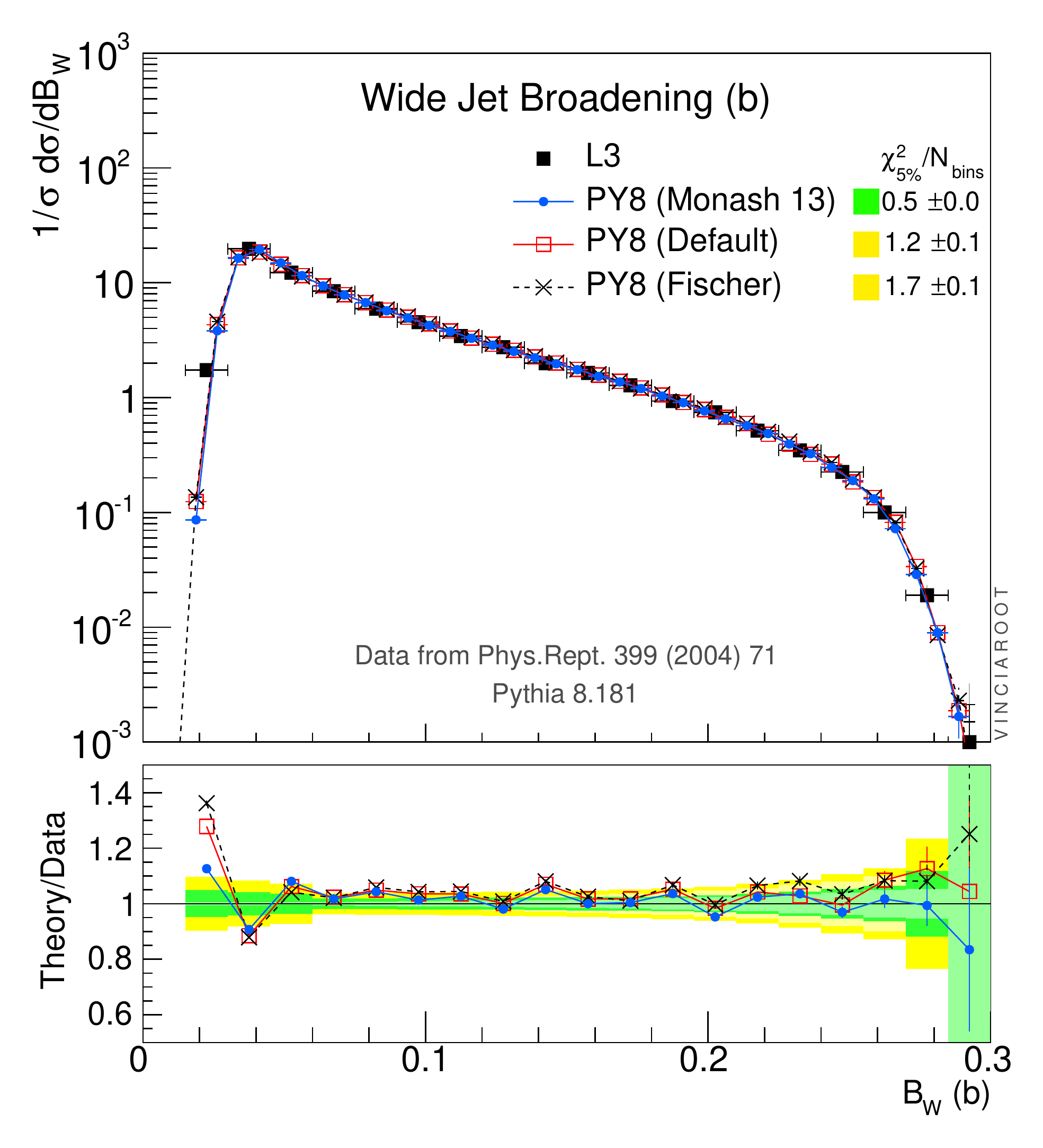}

\includegraphics*[scale=0.31]{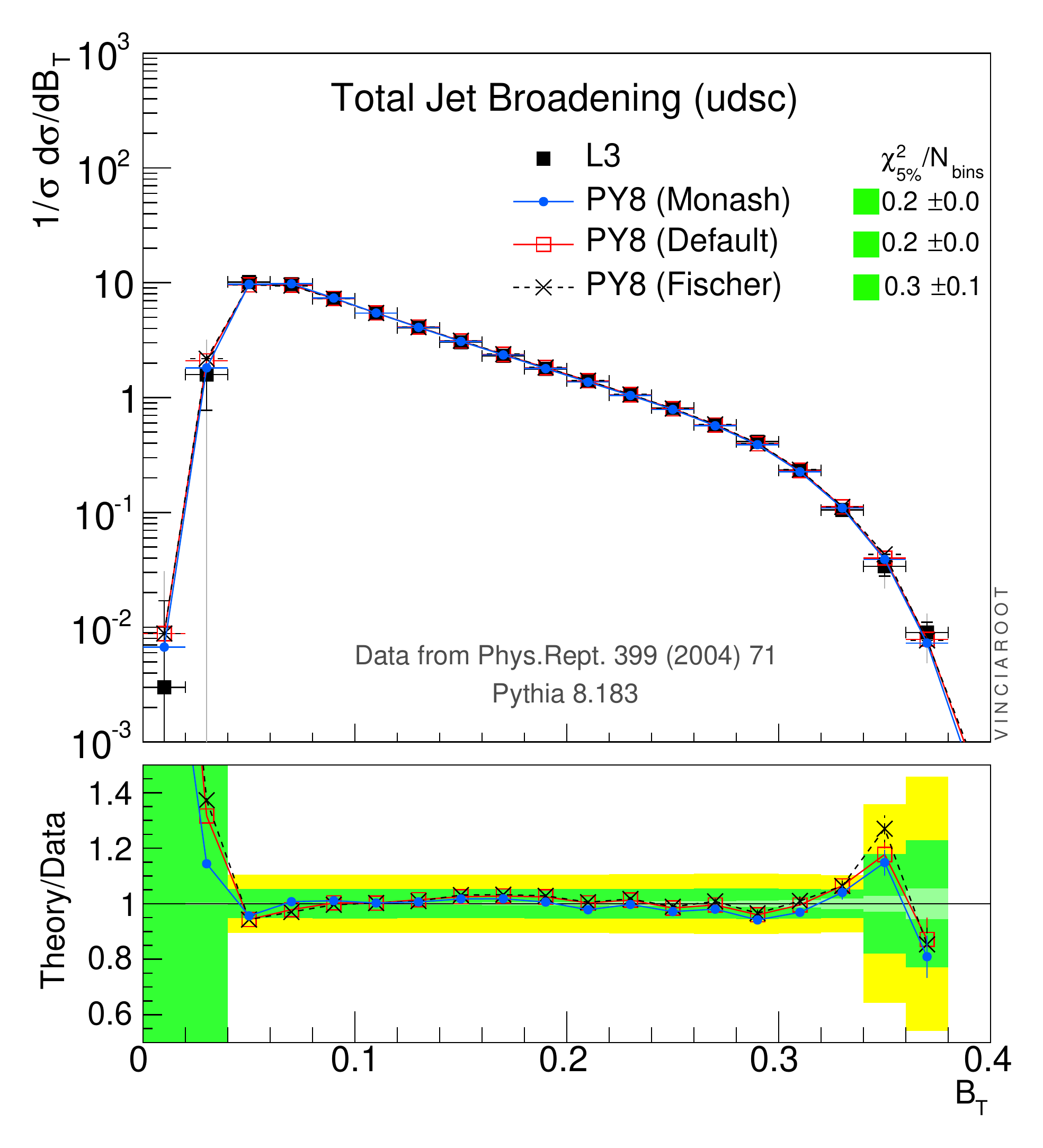}
\includegraphics*[scale=0.31]{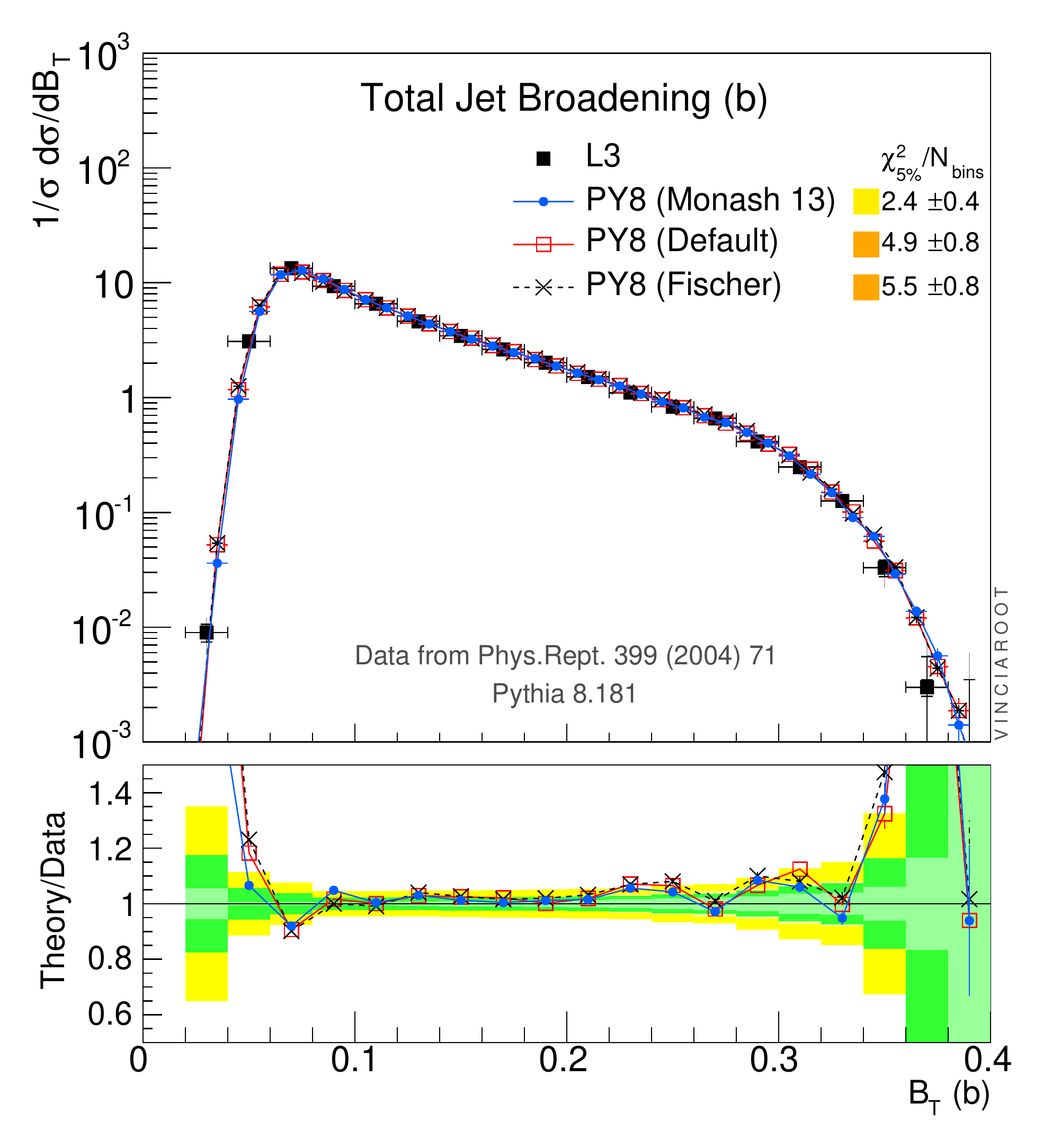}
\caption{Hadronic $Z$ decays at $\sqrt{s}=91.2\GeV$. The $B_W$ and $B_T$
  event-shape parameters, as measured by 
  L3~\cite{Achard:2004sv}, 
for light-flavour (left) and $b$-tagged (right) events, respectively.
\label{fig:LEPshapes2}
}
\end{figure}

To keep the main body of the paper as uncluttered as possible, we 
collect various plots of event-shape distributions in
\figsRef{fig:LEPshapes1} and \ref{fig:LEPshapes2}, separated into
light-flavour and $b$-tagged events on the left and right,
respectively. 

The experimental results come from the L3
experiment~\cite{Achard:2004sv}. However, since the data points are only
available with 3-digit precision, some of the least populated bins
contain artifacts like uncertainties being reported as exactly zero,
etc. Thus, we have been forced to make the following modifications to
the data set. 

The statistical uncertainty was reported as
zero for the last two bins of light-flavour Thrust as well as for the
last bin of the $C$, $D$, and $B_T$ parameters. Uncertainties
$<10^{-3}$ were derived 
using an approximate statistical scaling based on the contents and
uncertainties of the other bins. Likewise, the systematical
uncertainty for the last bin of Thrust was given as zero, which we
have replaced by the upper limit, $5\times 10^{-4}$. The last bin of
$B_T$ quoted a measured $y$ value of zero; removed in this study. 

For the heavy-flavour tagged event shapes, more significant rounding
issues were present. Thus, several of the first and last bins of each
distribution either quoted zero (statistical and/or systematic)
uncertainties, or ones with only a single digit of precision (such as
0.001, for which the rounding error could be up to $\sim$ 50\%). We
have interpreted all such values conservatively, inserting by hand 
a fourth digit on the uncertainties as large as could be consistent with
rounding. 

\subsection{Additional Particle Spectra \label{app:id}}

\begin{figure}[t!p]
\centering
\includegraphics*[scale=0.36]{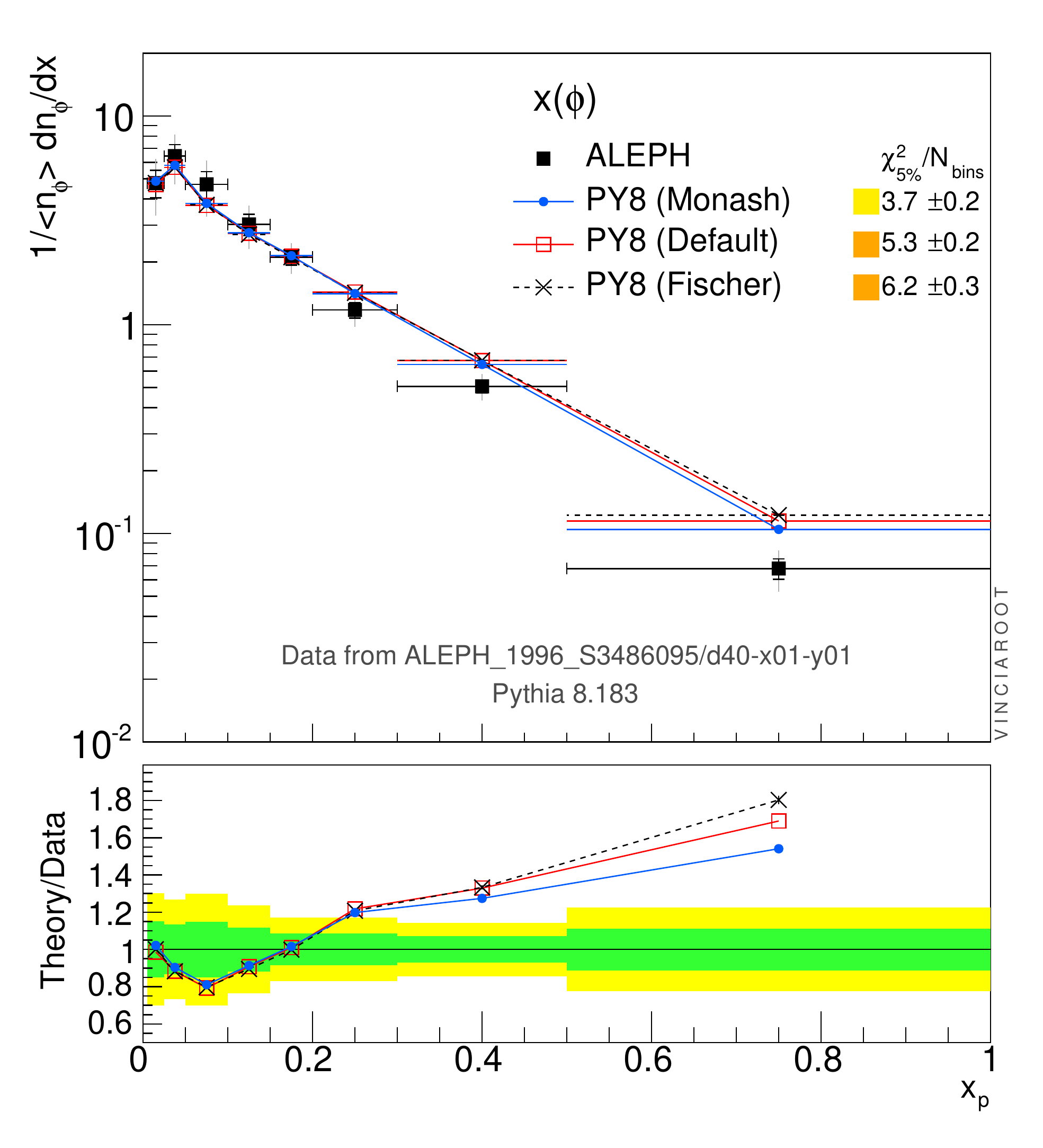}
\caption{Hadronic $Z$ decays at
  $\sqrt{s}= 91.2\GeV$. $\phi$ meson $x$ spectrum.
\label{fig:LEPphi}} 
\end{figure}

\begin{figure}[t!p]
\centering
\includegraphics*[scale=0.36]{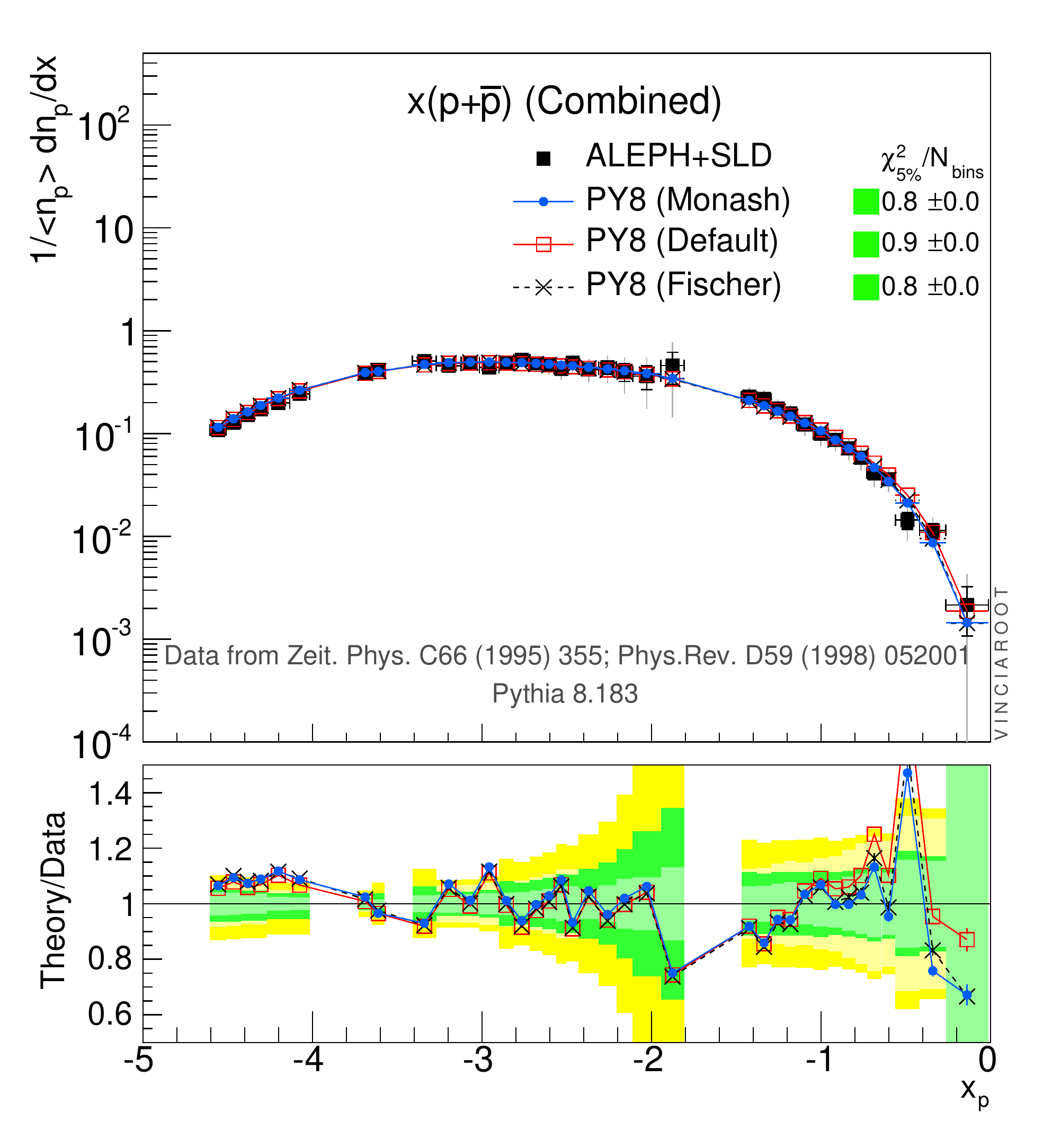}
\includegraphics*[scale=0.36]{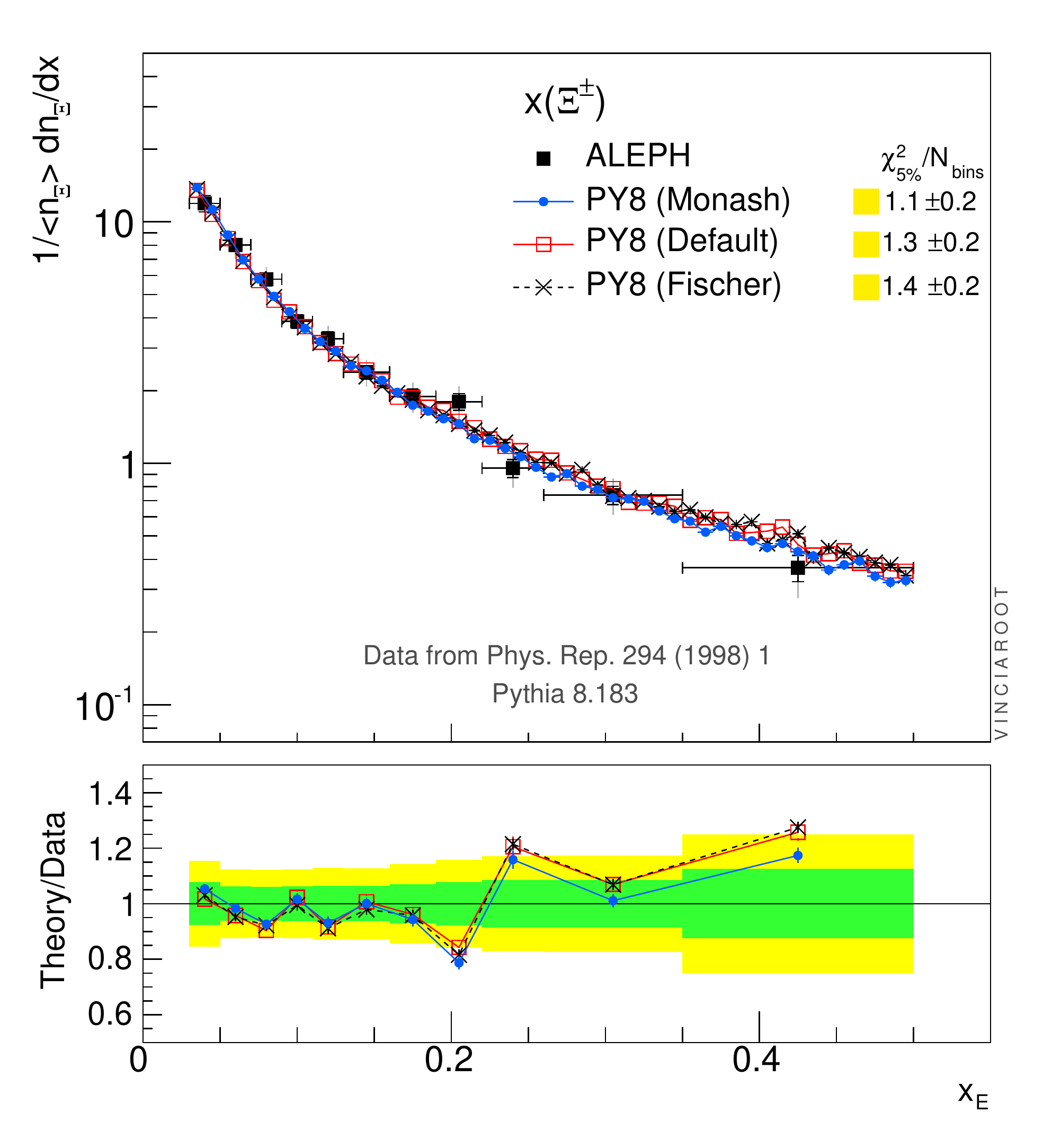}
\caption{Hadronic $Z$ decays at $\sqrt{s} = 91.2\GeV$.
$p^\pm$ and $\Xi^\pm$ $x$ spectra.
\label{fig:LEPpXi}} 
\end{figure}

\begin{figure}[t!p]
\centering
\includegraphics*[scale=0.36]{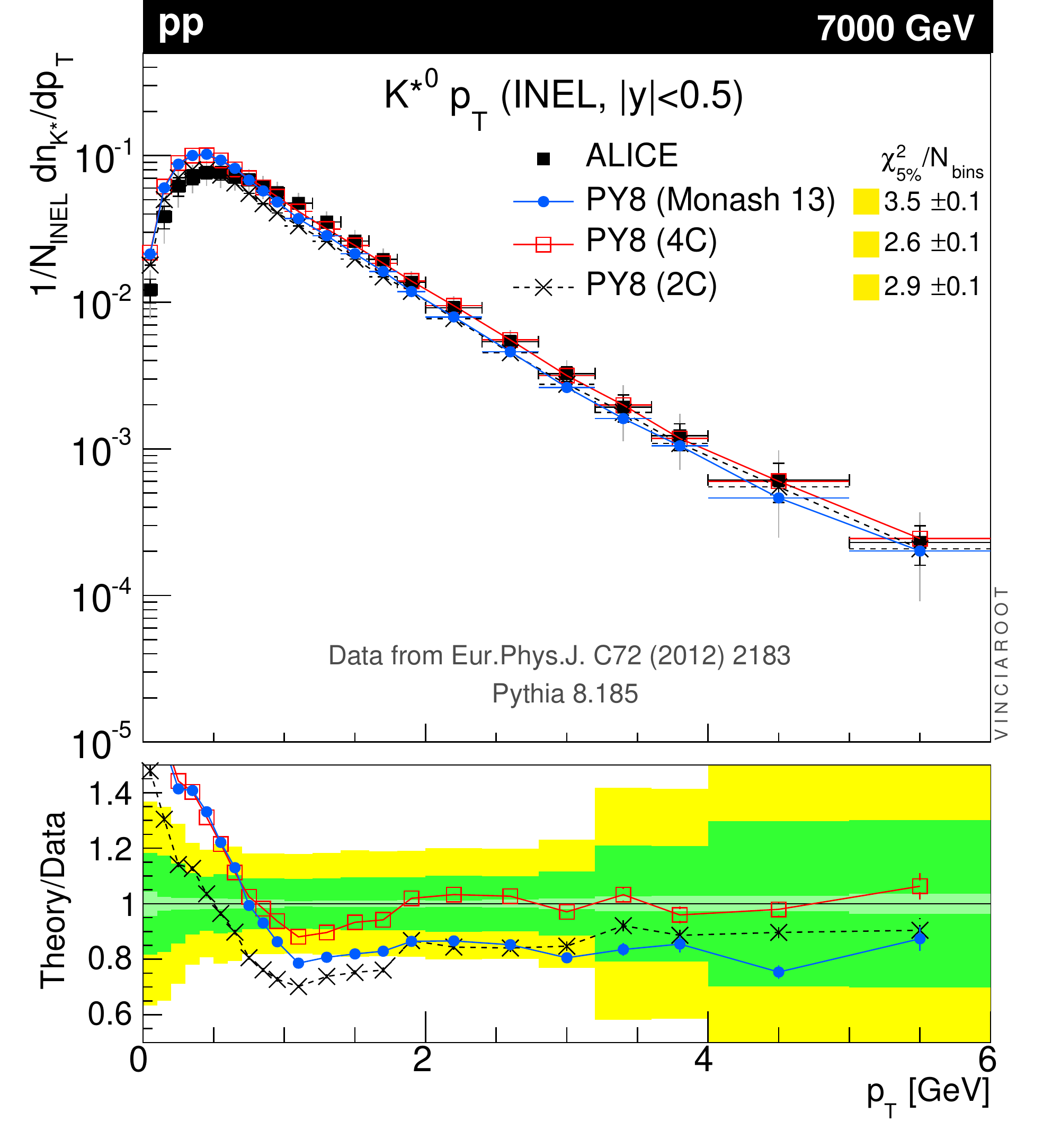}
\includegraphics*[scale=0.36]{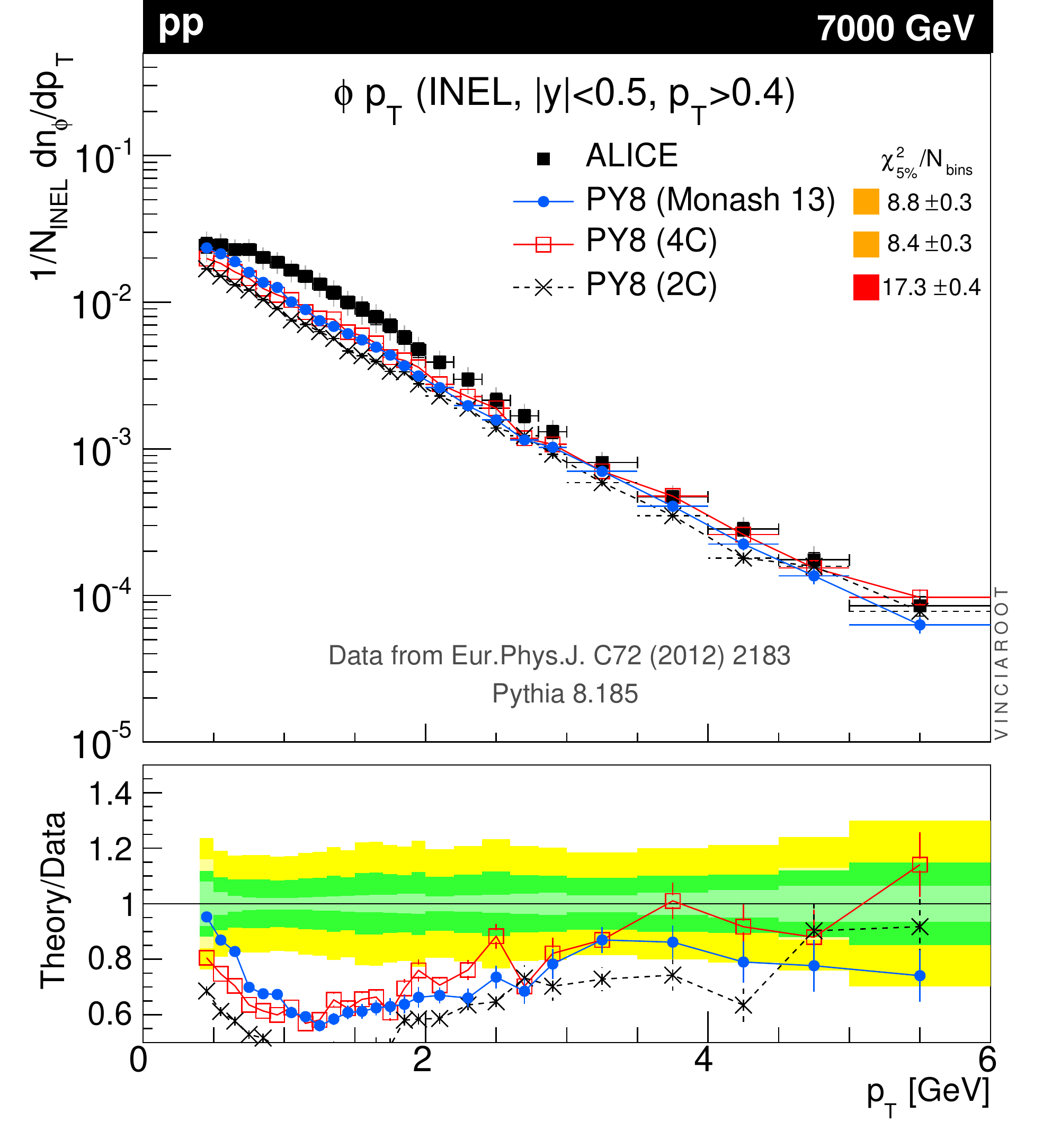}\\
\includegraphics*[scale=0.36]{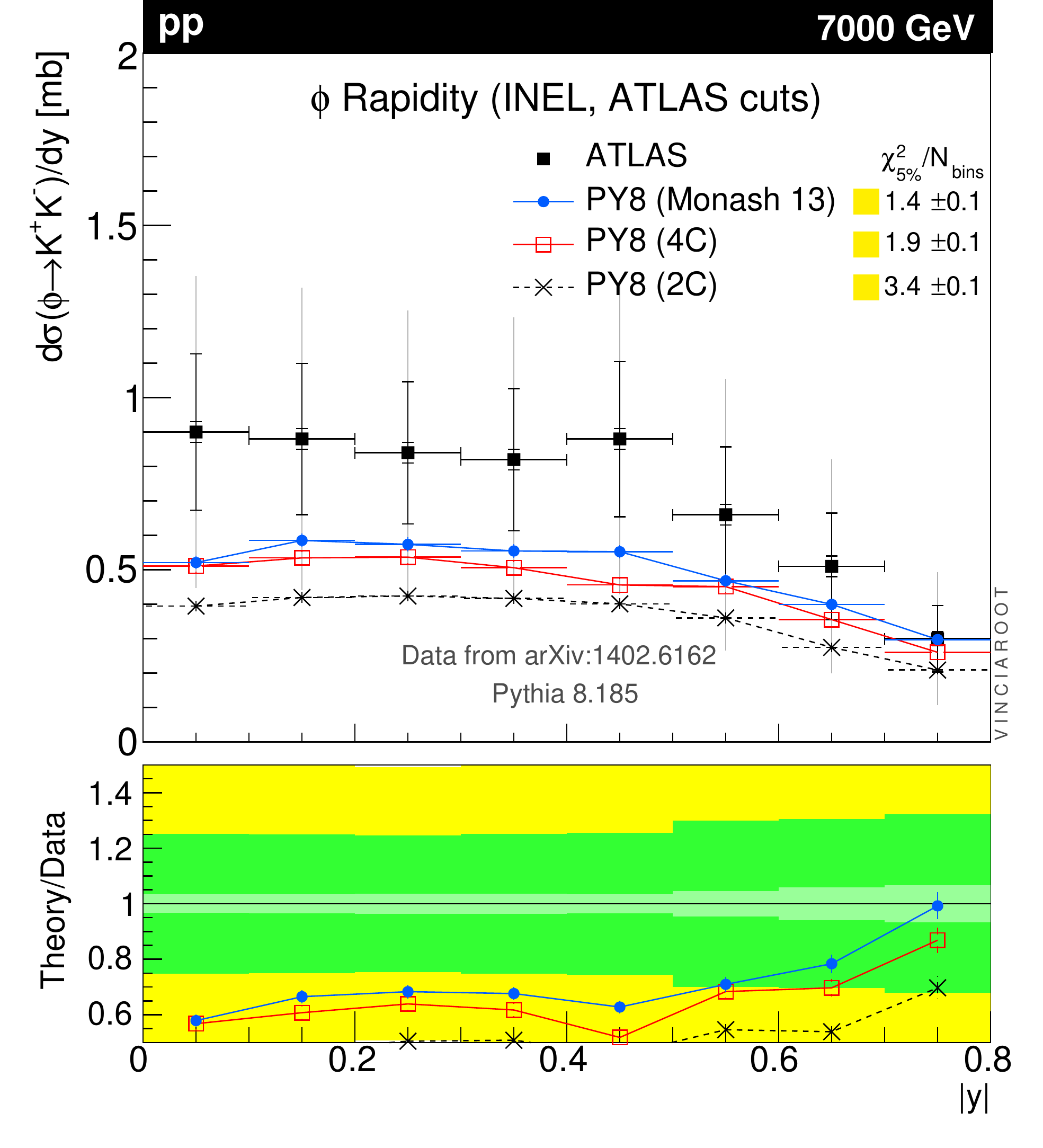}
\includegraphics*[scale=0.36]{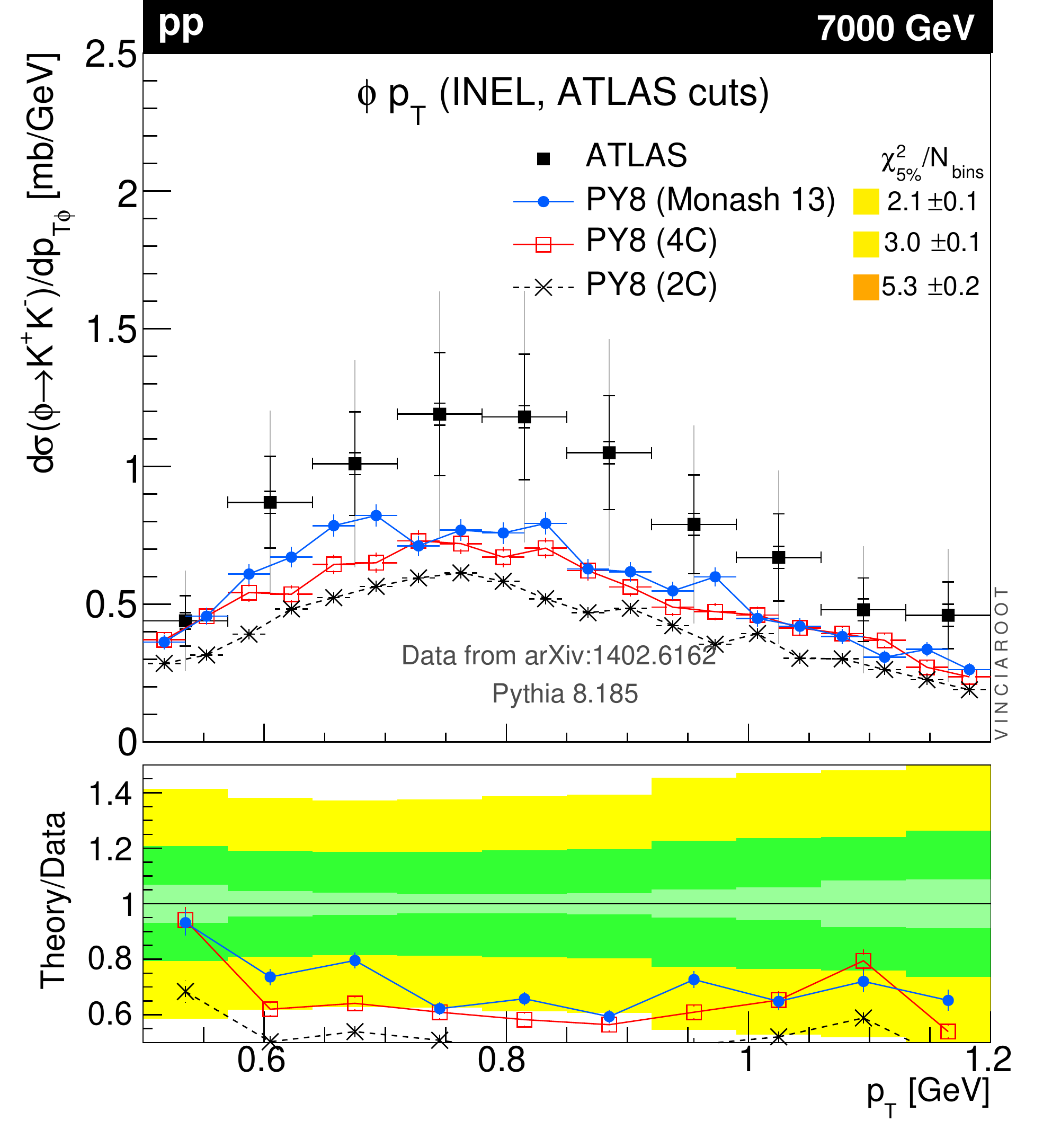}
\caption{$pp$ collisions at 7 TeV. {\sl Top row:} $K^*$ and $\phi$
  $p_\perp$ spectra, compared with ALICE data~\cite{Abelev:2012hy}.
{\sl Bottom row:} $\phi$ rapidity and $p_\perp$ spectrum, compared
with ATLAS data~\cite{Aad:2014rca}. The ATLAS cuts are $\phi \to K^+K^-$, 
$p_{\perp\phi} \in [0.5,1.2]\,\GeV$, $|y(\phi)|<0.8$, $p_{\perp K} >
0.23\,\mrm{GeV}$, $|p_K| < 0.8\,\mrm{GeV}$. 
\label{fig:LHCKstPhi}} 
\end{figure}

\begin{figure}[t!p]
\centering
\includegraphics*[scale=0.36]{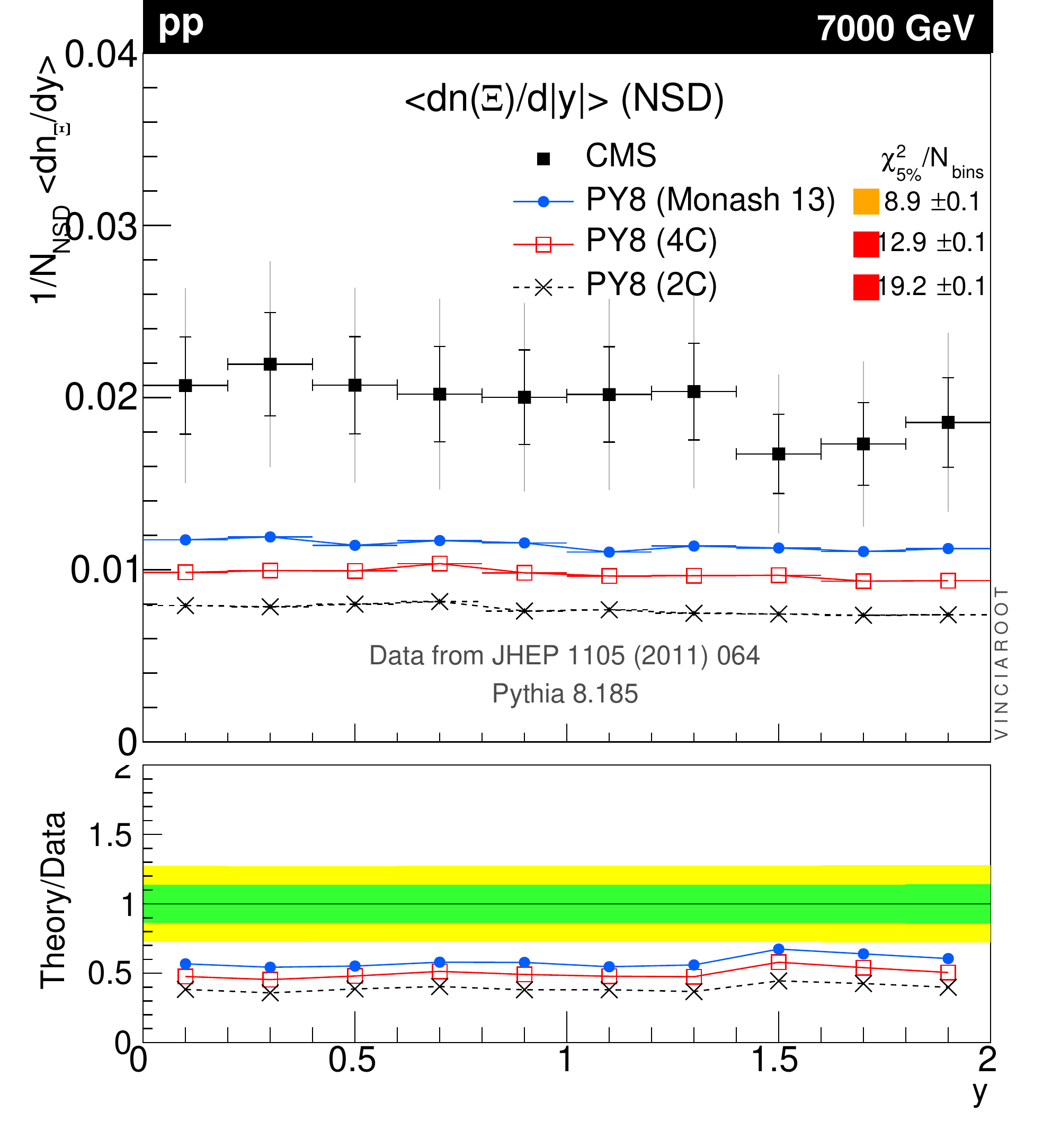}
\caption{$pp$ collisions at 7 TeV. 
$\Xi^-$ rapidity spectrum , compared with CMS data~\cite{Khachatryan:2011tm}.
\label{fig:LHCXi}} 
\end{figure}

In addition to the $K$ and $\Lambda$ spectra shown in the main
body of the paper (\secsRef{sec:idLEP} and \ref{sec:id}), 
we here include for reference the $x$ spectra of $\phi$ mesons, 
protons and $\Xi$ baryons at LEP in \figsRef{fig:LEPphi} and
\ref{fig:LEPpXi}, the $p_T$ spectrum of $K^*$ mesons and the rapidity and
$p_T$ spectra of $\phi$ mesons at LHC in
\figRef{fig:LHCKstPhi} (with absolute normalizations, 
to the number of inelastic events), and the rapidity spectrum of $\Xi$
baryons at LHC, in \figRef{fig:LHCXi}.

The transverse-momentum spectra of $K^*$ and $\phi$ mesons in
\figRef{fig:LHCKstPhi} exhibit the same 
qualitative behaviour as that of the $K_S$ mesons
(\figRef{fig:LHCK0S}), namely an excess
at very soft momenta below $\sim$ 500 MeV and a
depletion at slightly higher momenta between 1 and 2 GeV. As discussed in
\secRef{sec:id}, we did not find a way to
remove these undesirable features in the Monash 2013 tune, suggesting
that this is an issue that further theoretical modeling will be needed
to resolve.

The rapidity spectrum of $\Xi$ baryons, \figRef{fig:LHCXi}, shows
that, although the Monash tune does produce more $\Xi$
baryons overall (as expected also from the relative increase of $\Xi$
production at LEP, cf.~\figRef{fig:idParticles}), 
there is still a significant deficit of $\Xi$ baryons at the LHC, 
almost a factor 2 compared with the data. This is qualitatively
similar to the situation for $\Lambda$ baryons (\figRef{fig:LHCLam})
discussed in \secRef{sec:id}. Since new physics mechanisms may be
required to ``explain'' the missing baryons, we conclude that further
measurements and better precision on both the $\Lambda$ and $\Xi$
sectors (in addition to any other baryons that may be accessible)
would be highly interesting. More explicit recommendations can be
found in \secsRef{sec:id} and \ref{sec:proposals}.

\clearpage
\subsection{Energy Scaling of $\sigma_{2\to 2}(p_{T\mrm{min}})$ vs
  $\sigma_\mrm{inel}$ from 200 GeV to 100 TeV\label{app:sigma}}

\begin{figure}[htp]
\centering
\includegraphics*[scale=0.36]{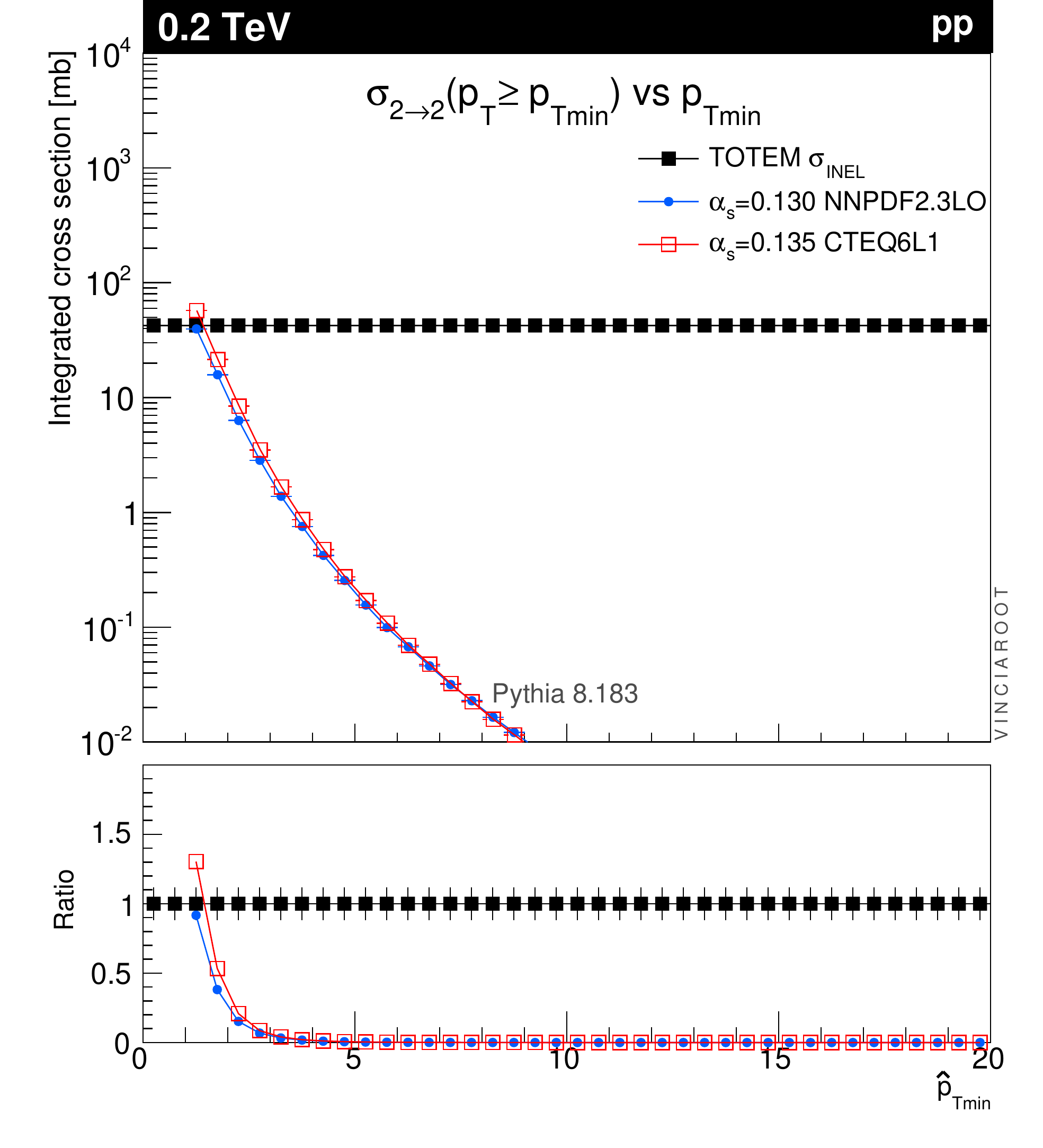}
\includegraphics*[scale=0.36]{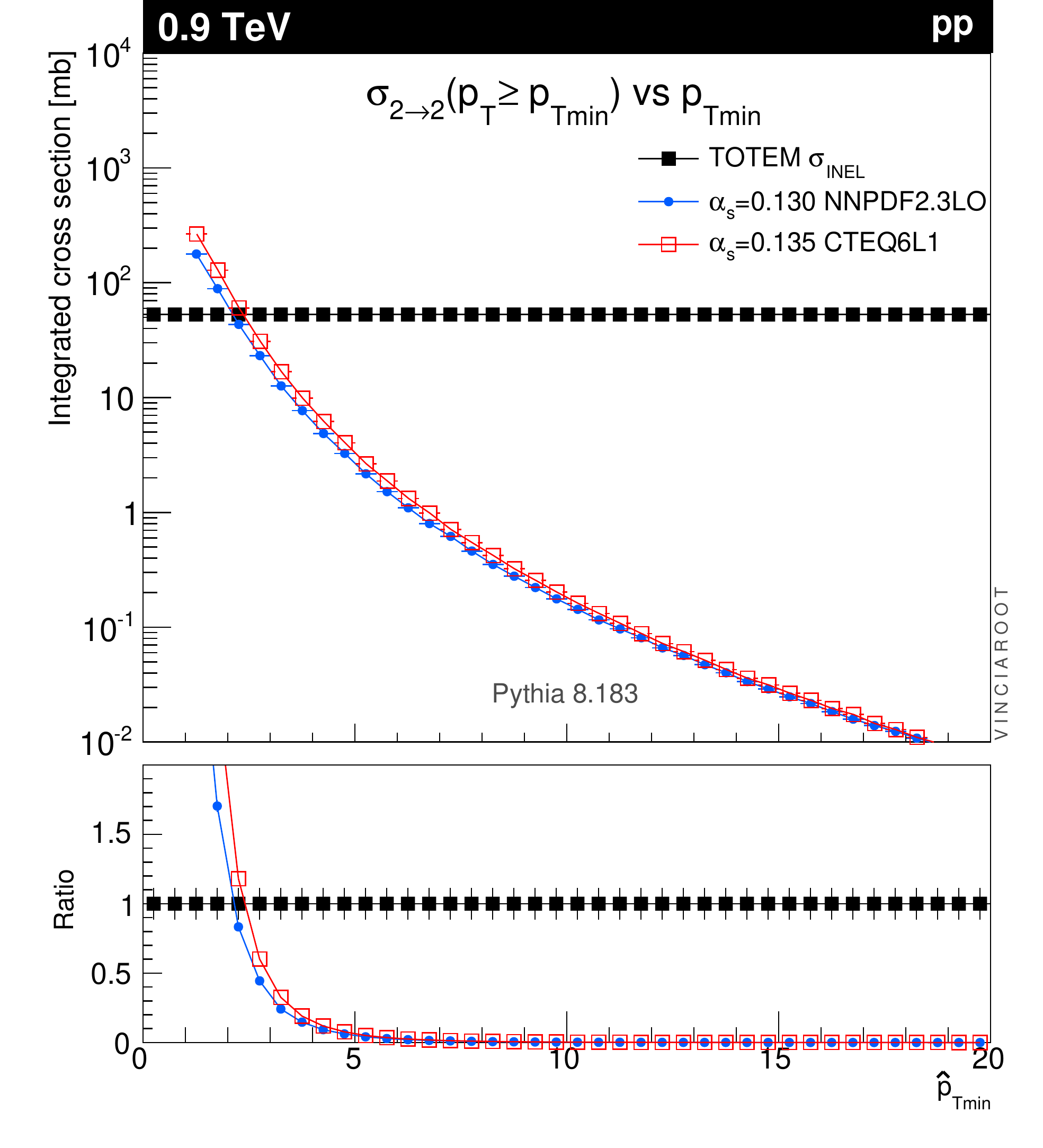}
\includegraphics*[scale=0.36]{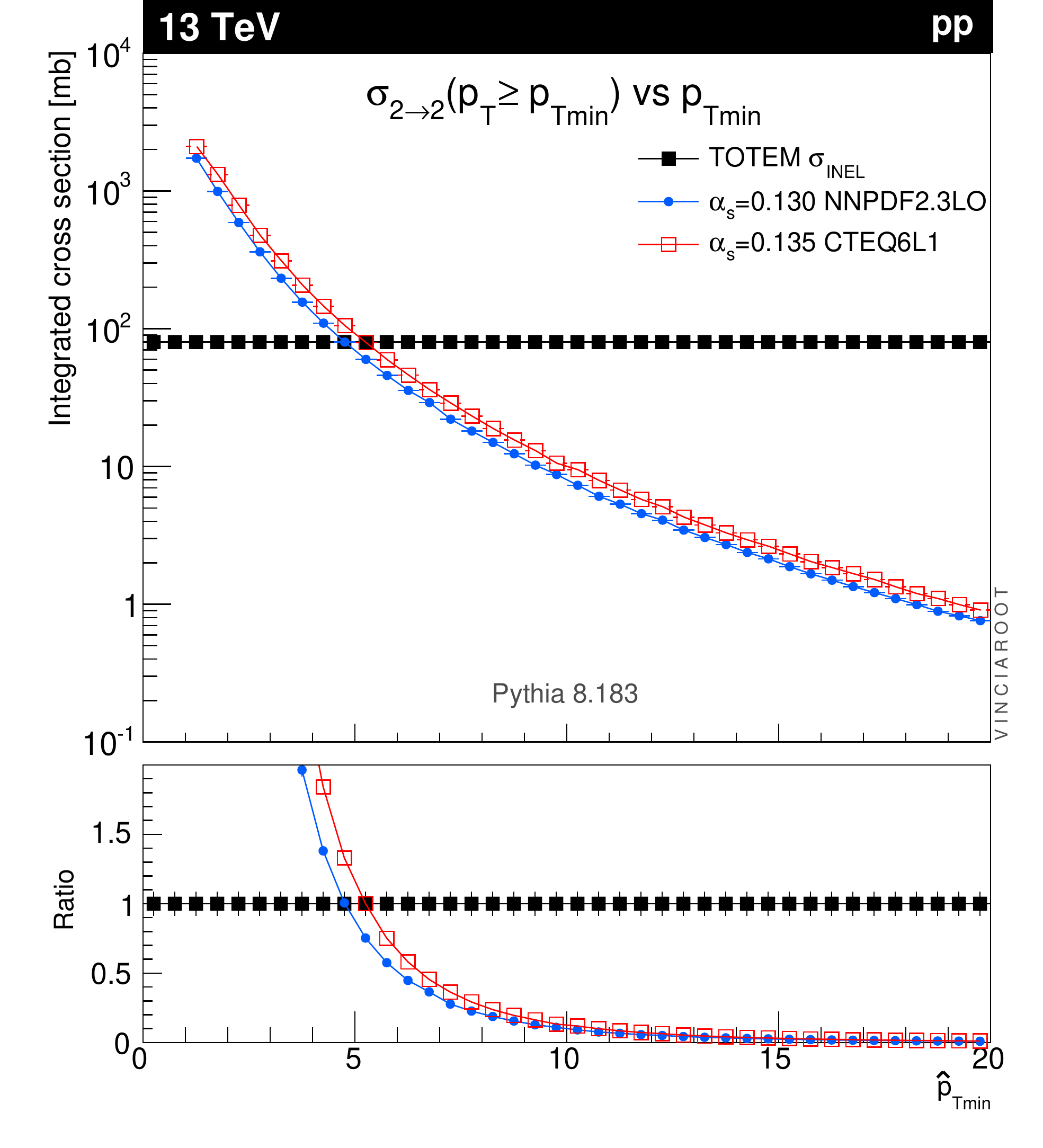}
\includegraphics*[scale=0.36]{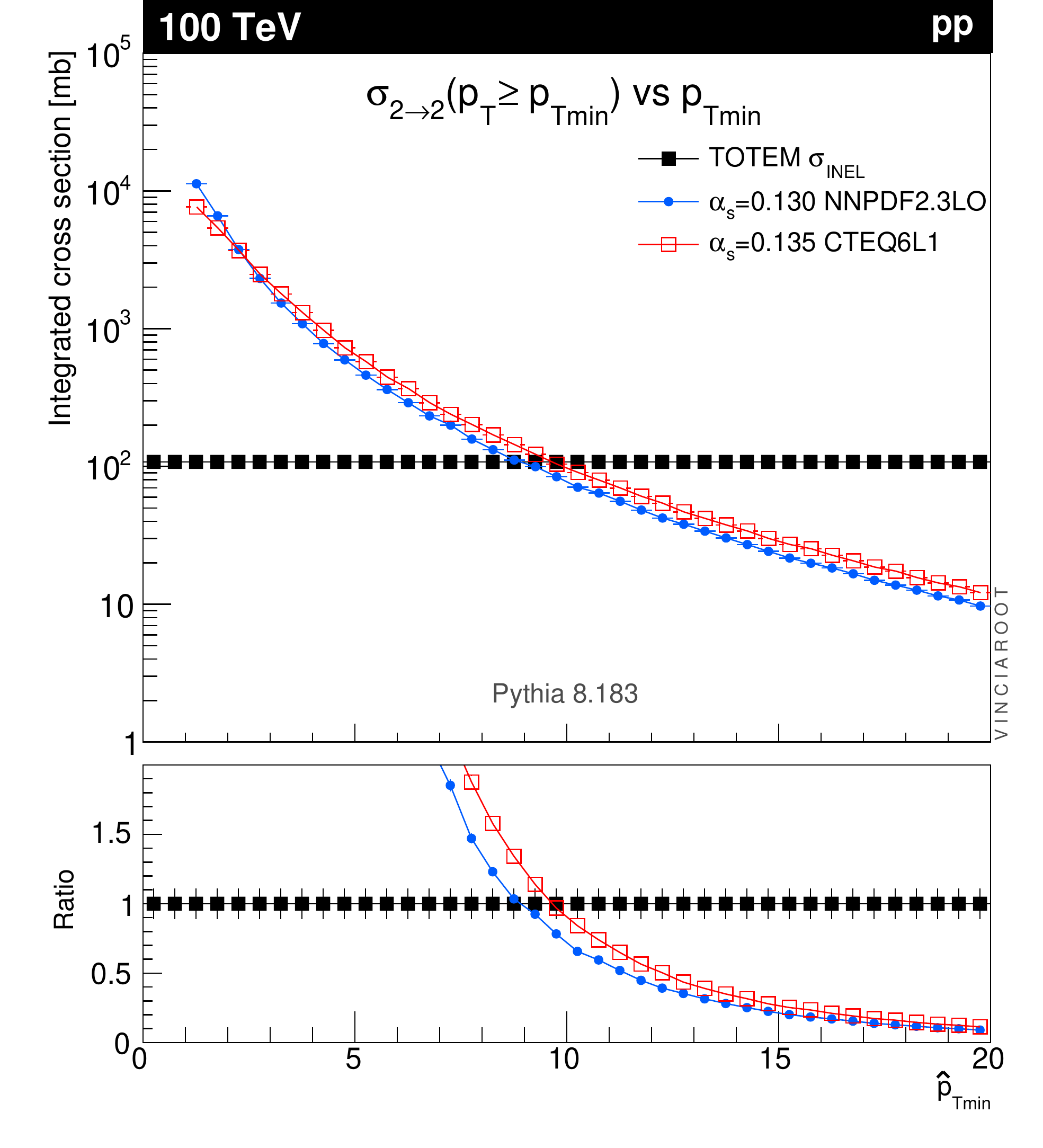}
\caption{$pp$ collisions at 4 different
  CM energies. Integrated QCD $2\to 2$ cross section above $p_{T\mrm{min}}$,
  as a function of $p_{T\mrm{min}}$. {\sl Top Left:}
  200 GeV; {\sl Top Right:} 900 GeV; {\sl 
    Bottom Left:} 13 TeV; {\sl Bottom Right:} 100 TeV.
\label{fig:sigma}} 
\end{figure}

In \figRef{fig:sigma}, we show the LO QCD $2\to 2$ cross
section, integrated above $p_{T\mrm{min}}$, 
  as a function of $p_{T\mrm{min}}$, in $pp$ collisions at 4 different
  CM energies, complementing and expanding on the 
8-TeV CM energy shown in the main body of the
  paper. We compare two
  different $\alpha_s$ and PDF choices, corresponding to those made in
  tunes Monash 13 (blue filled dots) and 4C (red open squares),
  respectively. As a reference for the total inelastic cross section
  at each energy,
  we base ourselves on the best-fit curve in the TOTEM cross-section
  measurement paper~\cite{Antchev:2013paa}, which in turn represents 
a fit produced by the COMPETE
  collaboration~\cite{Cudell:2002xe}. Uncertainties are rough
  conservative estimates based on 
  the plot in the TOTEM paper, but they are in any case 
  too small to significantly affect conclusions about the scale at
  which the partonic cross section saturates the hadronic one.

We observe that the $p_{T\mrm{min}}$ value for  
which the LO QCD $2\to 2$ partonic cross section formally becomes equal to the
total inelastic cross section (strongly suggesting that every event has
at least one such mini-jet pair) rises from values around 1 -- 2 GeV at energies
$\sqrt{s}<1\,\mrm{TeV}$, to 5 GeV at $\sqrt{s} = 13\,\mrm{TeV}$, and
finally 10 GeV at $\sqrt{s} = 100\,\mrm{TeV}$. 

\cleardoublepage\small
\bibliographystyle{utphys}
\bibliography{monash-tune}

\end{document}